\documentclass[rmp,aps,preprint,nofootinbib,endfloats]{revtex4}

\usepackage{graphics}
\usepackage{epsfig}
\begin{document}
\title{The Casimir force between real materials: experiment
and theory}

\author{G. L. Klimchitskaya}
\email{Galina.Klimchitskaya@itp.uni-leipzig.de}
\affiliation{Center of Theoretical Studies and Institute for Theoretical
Physics, Leipzig University, Postfach 100920, D-04009, Germany}
\affiliation{North-West Technical
University, Millionnaya St. 5, St.Petersburg 191065, Russia }
\author{U. Mohideen}
\email{Umar.Mohideen@ucr.edu}
\affiliation{Department of Physics and Astronomy, University of California,
Riverside, California 92521, USA }
\author{V. M. Mostepanenko}
\email{Vladimir.Mostepanenko@itp.uni-leipzig.de}
\affiliation{Center of Theoretical Studies and Institute for Theoretical
Physics, Leipzig University, Postfach 100920, D-04009, Germany}
\affiliation{Noncommercial Partnership ``Scientific Instruments'',
Tverskaya St.{\ }11, Moscow 103905, Russia}

\begin{abstract}
The physical origin of the Casimir force is connected with the existence of
zero-point and thermal fluctuations.  The Casimir effect is very general and
finds applications in various fields of physics. This review is limited to
the rapid progress at the intersection of experiment and theory that has
been achieved in the last few years.  It includes a critical assessment of
the proposed approaches to the resolution of the puzzles arising in the
applications of
the Lifshitz theory of the van der Waals and Casimir forces
to real materials.  All the primary experiments on the measurement of the
Casimir force between macroscopic bodies and the Casimir-Polder force
between an atom and a wall that have been performed in the last decade are
reviewed, including the theory needed for their interpretation. The
methodology for the comparison between experiment and theory in the
force-distance measurements is presented.  The experimental and theoretical
results described here provide a deeper understanding of the phenomenon of
dispersion forces in real materials and offer guidance for the application
of Lifshitz theory for the interpretation of the measurement results.
\end{abstract}

%\date{May 2001}
\maketitle
\tableofcontents

\section{ INTRODUCTION}
\label{sec:intro}

\subsection{ Fluctuations and the physical origin of the van der Waals and
Casimir force}
\label{sec:fluc}

Long-range forces that are different
from gravity but act between
electrically neutral atoms, or an atom and a macrobody or between two
macrobodies have been discussed for centuries. However, only after the
development of quantum mechanics and quantum field theory has the
physical picture of these forces become clear and the
first quantitative results have been obtained. The origin of
both the van der Waals and Casimir forces is connected with the existence
of quantum fluctuations. For a nonpolar atom the mean value of the
operator of the dipole moment in the ground state is equal to zero.
However, due to quantum fluctuations, the mean value of the square
of the dipole moment is not equal to zero. This leads to the
existence of what is referred to as {\it dispersion forces}
(Mahanty, Ninham, 1976; Parsegian, 2005)
which is the generic name for both the van der
Waals and Casimir forces.

For atomic separations between a few angstroms to a few nanometers
(which are much less than the characteristic absorption wavelength)
the retardation effects are negligible. In this separation region
the dispersion force is usually called the {\it van der Waals force}.
This is a nonrelativistic quantum phenomenon and its theory was
pioneered by London (1930). In the separation region of the
van der Waals interaction, a virtual photon emitted by one atom
can be absorbed by another atom during the life time of this photon
as determined by the
Heisenberg uncertainty relation. At relatively large atomic
separations, of order or larger than the characteristic
absorption wavelength, relativistic retardation effects play an
important role. At such separations the dispersion forces are
usually called {\it Casimir-Polder} (for atom-atom and atom-wall
interactions) or  {\it Casimir} forces (for interaction between
two macroscopic bodies). These are both relativistic and
quantum-mechanical phenomena
first described by Casimir and Polder (1948) and by Casimir (1948),
respectively. For atomic separations
$|\mbox{\boldmath$r$}_2-\mbox{\boldmath$r$}_1|$
of order the characteristic absorption wavelength, the virtual
photon emitted by one atom cannot reach the second one during its life
time. However, the operators of the quantized electric field at
the points $\mbox{\boldmath$r$}_1$ and $\mbox{\boldmath$r$}_2$
in the vacuum state are correlated such that
\begin{equation}
\langle E_k(t,\mbox{\boldmath$r$}_{1,2})\rangle=0\quad
\mbox{but}\quad
\langle E_k(t,\mbox{\boldmath$r$}_{1})\,
E_j(t,\mbox{\boldmath$r$}_{2})\rangle\neq 0.
\label{eqI1}
\end{equation}
\noindent
As a result, the atoms situated at points
$\mbox{\boldmath$r$}_1$ and $\mbox{\boldmath$r$}_2$ are characterized
by fluctuating dipole moments and these fluctuations are correspondingly
correlated leading to the Casimir-Polder and Casimir forces.

The theoretical approach to the atom-wall and the wall-wall interaction
developed by Casimir and Polder used  ideal metal
walls at zero temperature. The finite Casimir energy per unit area
of two infinitely large
parallel ideal metal walls separated by a distance $a$
(and the respective pressure)
was found as
a difference between the energies of zero-point (vacuum)
oscillations of the electromagnetic field in the presence and
in the absence of walls as
\begin{equation}
E_0(a)=-\frac{\pi^2}{720}\,\frac{\hbar c}{a^3}, \quad
P_0(a)=-\frac{\pi^2}{240}\,\frac{\hbar c}{a^4}.
\label{Cas}
\end{equation}
\noindent
 Ideal metals are characterized by perfect reflectivity at all
frequencies which means that
the  absorption wavelength is zero. Thus, the results (\ref{Cas})
are universal
and valid at any separation distance. They do not
transform to the nonrelativistic London forces at
short separations. Due to the difference in these early theoretical
approaches to the description of the dispersion forces the van der Waals
force and the Casimir-Polder (Casimir) force
were thought of as two different kinds of forces
rather than two limiting cases of a single physical phenomena,
as they are presently understood.

A unified theory of both the van der Waals and Casimir forces
between plane parallel material plates in thermal equilibrium
separated by a vacuum gap
was developed by Lifshitz (1956). Lifshitz's theory describes
dispersion forces between dissipative media
as a physical phenomenon caused by the
fluctuating electromagnetic field which is always present in both
the interior and the exterior of any medium.
According to the fluctuation-dissipation theorem, there is a
connection between the spectrum of fluctuations of the
physical quantity in an
equilibrium dissipative medium and the generalized susceptibilities
of this medium which describe its reaction to an
external influence.
Using the fluctuation-dissipation
theorem, Lifshitz derived the general formulas for the free energy and
force of the dispersion interaction. In the limit of dilute bodies these
formulas describe the dispersion forces acting between atoms and
molecules. In the framework of the Lifshitz theory material properties
are represented by the frequency-dependent dielectric permittivities
and atomic properties by the dynamic atomic polarizabilities.
In the limiting cases of short and large separation distances,
in comparison with the characteristic absorption wavelength, the
Lifshitz theory reproduces the respective results obtained by
London and by Casimir and Polder. It also describes the transition
region between the nonrelativistic and relativistic areas.
In fact the fluctuating electromagnetic field in the Lifshitz theory
is some classical analogue of vacuum (zero-point) oscillations
in the field-theoretical approach developed by Casimir.
Van Kampen et al. (1968), Ninham et al. (1970), Gerlach (1971) and
Schram (1973) narrowed the distinction between the Casimir and Lifshitz
approaches. They have obtained the Lifshitz formulas for the free
energy and force  between two
nondissipative material plates as the difference between the free
energies of zero-point and thermal oscillations in the presence and
in the absence of plates. The eigenfrequencies of these oscillations
were found from the standard continuity boundary conditions for
the electric field and magnetic induction on the surfaces of the dielectric
plates. Later Barash and Ginzburg (1975) generalized this approach for the
case of plates made of dissipative materials
in thermal equilibrium with a heat reservoir.
This generalization is presented also in the books by Milonni (1994) and by
Mostepanenko and Trunov (1997). The applicability of the Lifshitz
formula to dissipative materials was also demonstrated using the
scattering approach (Genet et al., 2003).
The assumption of a thermal equilibrium is basic for the Lifshitz theory.
We will repeatedly discuss the role of this assumption below.

Thus, the theoretical foundations of the Casimir interaction are
based on two approaches.
In one case the theory of the Casimir effect is based
on the theory of an equilibrium electromagnetic fluctuations in the media.
In the second case the Casimir effect is a vacuum quantum
effect resulting from the influence of external conditions and described
by quantum field theory. In this case, boundary conditions are
imposed (in place of material boundaries) which restrict the quantization
volume and affect the spectrum of zero-point and thermal
oscillations.
In fact the two different approaches can be reconciled.
There are derivations of the Lifshitz formulas and more general results
in different media where the dispersion force is viewed as a vacuum
quantum effect (Toma\v{s}, 2002; Raabe, Welsch, 2006).
The common roots in the theory of electromagnetic
oscillations relate the Casimir effect to other fluctuation
phenomena, such as the radiative heat transfer through a
vacuum gap (Volokitin, Persson, 2007). However, its origin in
quantum field theory relates the
Casimir effect to other quantum vacuum effects as the Lamb shift
and the anomalous magnetic moment of an electron, where virtual
particles play an important role (Jaffe, 2005).

During the last few years, far reaching generalizations of the
Lifshitz formulas were obtained which express the Casimir energy and
force between two separated bodies of arbitrary shape in terms of
matrices of infinite dimensions. This is often referred to as the
representation of the Casimir energy in terms of the functional
determinants (Emig et al., 2006; Kenneth, Klich, 2006, 2008) or
in terms of the scattering matrices (Bulgac et al., 2006;
Lambrecht et al., 2006). Several applications of new representations
related to the subject of this review are discussed in the respective
sections.

\subsection{Material and geometric properties in the theory of
dispersion forces}
\label{sec:real}

Both Casimir and Lifshitz considered dispersion forces in the configuration
of two closely spaced plane parallel plates. However, real bodies may
possess quite different geometrical properties. The case of an isolated
body is also of much interest for the physics of dispersion forces.
One of the unique features of the Casimir force that has attracted
widespread attention is its exotic geometry dependence. For example,
it was first realized that the Casimir force can be repulsive for an
ideal metal spherical shell (Boyer, 1968). Another striking example
of the geometric dependence was found in ideal metal rectangular boxes,
where the repulsive or attractive nature of the force depends on
the ratio of the size of the sides (Lukosz, 1971). Both these results
led to an explosion in theoretical studies of the shape dependence
of the Casimir force with ideal boundaries extensively reviewed in the
book literature (Milonni, 1994; Mostepanenko, Trunov, 1997;
Milton, 2001). This subject is, however, outside the scope of our
review which deals only with effects pertaining to real material bodies.

In addition to the remarkable shape dependence mentioned above,
there are more staid but no less important ones of interest to
experimental measurements of the Casimir force between real bodies.
Chief among these are the role of the roughness of the surface on
Casimir force, the role of curvature in the most often used
sphere-plate configuration, finite size effects of the bodies used in
the measurement and the lateral Casimir force from corrugated
surfaces. The last in particular exhibits {\it diffraction-like}
effects which can be experimentally observed. All these aspects are
covered in the review.

Another feature of real material bodies differentiating them from
ideal metal plates at zero temperature, as considered by Casimir,
are connected with the account of realistic material properties
and thermal effects.
As was mentioned in Sec.~I.A, in the Lifshitz theory real material
properties are taken into account through the frequency-dependent
dielectric permittivity $\varepsilon(\omega)$.
In its turn $\varepsilon(\omega)$ is expressed in terms of the
complex index of refraction as
$\varepsilon(\omega)=n^2(\omega)$.
Both quantities are measurable over wide frequency
regions. The magnitude of $\varepsilon(\omega)$ depends on the
nature and structure of the material including its electrical and
optical properties.
In the Lifshitz theory of dispersion forces the free energy
and other thermodynamic quantities are expressed as functionals
of $\varepsilon({\rm i}\xi_l)$, where the Matsubara frequencies
$\xi_l=2\pi k_BTl/\hbar$ are defined for all $l=0,\,1,\,2,\,\ldots$
($k_B$ is the Boltzmann constant, $T$ is the temperature).
In the alternative but mathematically equivalent formulation
the free energy is represented as a functional of
$\varepsilon(\omega)$ defined along the real frequency axis
(see Sec.\ II.A). The Lifshitz theory does not make a distinction
between the physical processes contributing to the value of
the dielectric permittivity. However, the applicability conditions of this
theory, in particular thermal equilibrium, must be followed.

As is clear from the above, we call materials {\it real} (real metals,
real dielectrics etc.) to underline that this review deals with
physical surfaces rather than with idealized ones like ideal
metals or dielectrics with constant dielectric permittivity
having perfect geometric shape.

In the last 10 years significant progress has been made in
the measurement of the Casimir and Casimir-Polder forces.
This has added substantial information to the application of the Lifshitz
theory to real materials. Historically the Lifshitz theory
was proposed for ideal dielectrics, i.e., for materials having zero
conductivity at any temperature. At nonzero temperature
dielectrics are characterized by a small but physically real
dc conductivity. From the start, the first classic papers
(Lifshitz, 1956; Dzyaloshinskii et al., 1961)
neglected this conductivity
 in the Lifshitz theory,
and at zero frequency dielectric materials were characterized by
a finite static dielectric permittivity
$\varepsilon_0=\varepsilon(0)$.
If, however, the dc conductivity of dielectrics at
$T\neq 0$ is included, $\varepsilon(\omega)$
would go to infinity in the limit of zero frequency (see Sec.\ II.D).
Geyer et al. (2005b) have shown that the inclusion of
even a negligibly small dc
conductivity for the configuration of two parallel dielectric
plates leads to an enormously large thermal correction to the Casimir
free energy and pressure.
If a physical effect is negligibly small, including it in
a general theory must not significantly influence
the results.
Thus, this thermal correction is surprising and it suggests that
perhaps something is being done incorrectly.
One might think that
the dc conductivity of dielectric materials {\it is not}
a minor effect that can be neglected in the theory of dispersion forces and
that the discovered large thermal correction is physically real but
was overlooked by the founders of the Lifshitz theory.
However, as was shown by Geyer et al. (2005b), the
inclusion of the dc conductivity in the model of the dielectric
response violates the Nernst heat theorem and, thus, makes the
Lifshitz theory inconsistent with thermodynamics. Moreover, a large
thermal correction resulting from the simple inclusion of the dc conductivity
was excluded experimentally in the experiments by Obrecht et al. (2007)
and Chen et al. (2007a, 2007b) (see Secs.\ I.C, V.B and VI.A).
Below we illustrate by many examples that the dc conductivity of
dielectric materials is unrelated to dispersion forces and its inclusion
contradicts the applicability conditions of the Lifshitz theory.

For the Casimir force between real metals the situation is closely
analogous to that of the dielectrics. Bostr\"{o}m and Sernelius (2000)
applied the Lifshitz theory to describe the free energy and force
in the configuration of two parallel metal plates at a temperature
$T$. The dielectric permittivity of the metal was obtained by using the
tabulated optical data for the complex index of refraction
extrapolated to zero frequency by means of the Drude model
(see Sec.\ III.A). This takes into account both the drift current
of  the conduction electrons including relaxation inherent to it
and displacement currents connected with plasma oscillations of free
electrons and interband transitions of core electrons. As a result, an
enormously large thermal correction to the Casimir free energy and
force was obtained from the Lifshitz theory at short separations
below $1\,\mu$m in qualitative disagreement with the case of ideal
metal plates. The Casimir force between plates made of ideal
metals at nonzero temperature
was first treated independently by Brown and Maclay (1969) using
thermal quantum field theory. Schwinger et al. (1978) demonstrated
that the Lifshitz theory is
in agreement with the case of ideal metals
if the limit $\varepsilon({\rm i}\xi)\to\infty$ is
taken first before the limit $\xi\to 0$.
Bezerra et al. (2002b, 2004) have shown that the inclusion of relaxation
processes of conduction electrons in the Lifshitz theory violates the
Nernst heat theorem for metals with perfect crystal lattices. This is also
in violation of the classical limit at large separations
(Feinberg et al., 2001; Scardicchio, Jaffe, 2006).
The large
thermal correction predicted by Bostr\"{o}m and Sernelius (2000) was
found to be inconsistent with the measurement data of the experiments
by Lamoreaux (1997) and by Decca et al. (2003a, 2005b, 2007a, 2007b)
(see Secs.\ I.C, IV.B and IV.C.1).
At the same time, the use of the free electron plasma model for the
characterization of a metal (Dzyaloshinskii et al., 1961;
Hargreaves, 1965; Schwinger et al., 1978; Mostepanenko, Trunov, 1985)
leads to small thermal corrections at short separations in
qualitative agreement with the case of ideal metal plates
(Genet et al., 2000; Bordag et al., 2000b). The results
obtained by neglecting the role of the conduction electrons
connected with the drift current are in agreement with thermodynamics
and with the classical limit (Bezerra et al., 2004).
As is argued below, the inclusion of a drift current is beyond the
applicability conditions of the Lifshitz theory.

\subsection{Modern Casimir force experiments}
\label{sec:modern}

There were many measurements of the van der Waals and Casimir force
made before 1990, of which only the experiment by van Blockland
and Overbeek (1978) with metallic surfaces can be considered as an
unambiguous demonstration
[a brief historical survay has been made by
Bordag et al. (2001)]. The modern stage in the Casimir experiments
is characterized by the use of a new generation of precise laboratory
techniques for the measurement of small forces and short separation
distances permitting one to determine the experimental precision and
compare data with theory. The experiment by Lamoreaux (1997)
using a torsion balance to measure the Casimir force
between a gold coated
spherical lens and a plate was the first in this more recent series.
The precision from 5 to 10\% that
might be achieved at separations of about $1\,\mu$m
(see Sec.\ IV.C.1 for more details) leads to the
conclusion that the Drude model is an inadequate
description in the theory
of the thermal Casimir force.

A series of measurements of the Casimir force with increased
sensitivity between a metallized sphere and a plate was performed
by using the atomic force microscope (Mohideen, Roy, 1998;
Roy et al., 1999; Klimchitskaya et al., 1999; Harris et al., 2000;
Chen et al., 2004a).
These experiments introduced the idea of using metallized polystyrene
spheres which have very low mass, as one of the interacting surfaces.
Even the first experiment in this series
(Mohideen, Roy, 1998) demonstrated the role of the skin depth
and the surface
roughness corrections to the Casimir force at separations
from 120\,nm to 300\,nm. Note that the corrections to the skin depth
are often referred to as the {\it finite conductivity
corrections}. It was shown that the data are in disagreement with the
ideal metal Casimir force whereas the inclusion of the finite
conductivity and roughness corrections leads to excellent
agreement between the data and theory (thermal corrections being
negligibly small in this experiment). The experimental error in
the measurement of the Casimir force
between 1 and 2\% depending on the confidence level was
justified at the shortest separations (Chen et al., 2004a).

The gradient of the Casimir pressure between two parallel metallic plates
was measured dynamically by Bressi et al. (2002). In this
experiment small oscillations were induced on one of the plates
at the resonant frequency and the frequency shift due
to the Casimir force was measured. This frequency
shift is proportional to the derivative of the Casimir force with
respect to the separation distance between the plates (see in Sec.\ IV.C.3).

Chen et al. (2002a, 2002b) have measured the lateral Casimir
force between a metallized sphere and a plate covered with
longitudinal co-axial sinusoidal corrugations of equal periods (Sec.~VII.B).
This force which was theoretically predicted by Golestanian and Kardar (1997,
1998) is a harmonic function of the phase shift between the
corrugations on the sphere and the plate. The demonstrated phenomenon
of the lateral Casimir force is promising for nanotechnology
where, together with the normal Casimir force, it allows one to actuate
any translations of a device element entirely due to the
presence of the zero-point
vacuum oscillations.

The most precise measurements of the Casimir force between
metallic surfaces were performed in a series of experiments by
Decca et al. (2003a, 2003b, 2004, 2005b, 2007a, 2007b)
using a micromechanical
torsional oscillator. This new technique was first used in Casimir
experiments by Chan et al. (2001a, 2001b) as a demonstration of the
actuation of micro- and nanomechanical devices by the Casimir force
(see Sec.\ IV.C.2). Although Decca et al. exploited the sphere-plate
configuration, the application of the dynamic measurement mode and
the proximity force approximation allowed them to determine the
Casimir pressure between two parallel plates. In the last of
this series of experiments (Decca et al., 2007a, 2007b) a 0.2\%
total experimental error was reported at a separation of 160\,nm.
The systematic error was shown to dominate
 the random one in the total experimental error.
This has been achieved here for the first time
in the Casimir force measurements. The measurements of
the Casimir pressure using the micromechanical torsional
oscillator exclude the prediction of large thermal
effects at a 99.9\% confidence level for a wide separation
region below $1\,\mu$m (see Sec.\ IV.B).

The next series of recent Casimir experiments
was devoted to the investigation of the Casimir interaction
between a metallized sphere and a semiconductor plate
(Chen et al., 2005, 2006a, 2006b, 2007a, 2007b).
In the early stages of this research it was demonstrated that the
change of the charge carrier concentration in
a semiconductor plate changes
the value of the Casimir force. In the experiments by Chen et al.
(2007a, 2007b) the difference Casimir force between the sphere and
the plate was measured in the presence and in the absence of incident
laser light on the plate. This allowed one to perform a
fundamental test on the role of semiconductor conductivity
properties in the Casimir force. The measurement data were found
to be in an excellent agreement with the Lifshitz theory if the
dc conductivity of the semiconductor plate in the dark phase
(no incident light)
is disregarded. If, however, the dc conductivity of the
plate in the dark phase is included, the theory was found to be
inconsistent with data at a confidence level of 95\%
(see Sec.\ V.B).

Another important experiment was performed by Obrecht
et al. (2007). This is the first measurement of the thermal
Casimir-Polder force made both in thermal equilibrium and in the
nonequilibrium case. The Casimir-Polder force between Rb atoms and
a dielectric substrate changes the frequency of dipole oscillations
excited in a Bose-Einstein condensate separated from a wall by
a distance of a few micrometers. The fractional difference of this
frequency calculated on the basis of the Lifshitz theory by
disregarding the dc conductivity of the wall material was
found to be in excellent agreement with data (Obrecht et al., 2007).
Klimchitskaya and Mostepanenko (2008b) have shown, however,
that with inclusion of small but the physically real dc conductivity
for the dielectric substrate the theory is inconsistent with data
(see Sec.\ VI.A).

The geometry effects beyond the proximity force approximation were
demonstrated in the experiment by Chan et al. (2008) where the gradient
of the Casimir force acting between an Au-coated sphere and Si plate
with deep rectangular trenches was measured (Sec.~V.D).

\subsection{Applications of the Casimir effect from fundamental
physics to nanotechnology}
\label{sec:appl}

Both the van der Waals and Casimir forces find many
applications spanning the range from the
purely scientific to the mostly technological. The reason lies in
the fluctuation nature of dispersion forces and the universal
role played by fluctuations in different physical phenomena.
Here we only briefly list some of the main areas where the Casimir
effect is important. For a detailed discussion of the applications
of the Casimir effect see the books by Milonni (1994),
Krech (1994), Mostepanenko and Trunov (1997), Milton (2001),
Parsegian (2005) and
the review papers by Plunien et al. (1986), Mostepanenko and
Trunov (1988), Kardar and Golestanian (1999), Bordag et al. (2001),
Milton (2004), Lamoreaux (2005),  Klimchitskaya and
Mostepanenko (2006) and by Milonni (2007).

A number of applications of the Casimir effect belong to
the field of condensed
matter physics (Krech, 1994) and nanotechnology (Buks, Roukes, 2001, 2002;
Chumak et al., 2004).
Areas of impact are
multilayered structures, wetting phenomena, colloids, critical
systems, adhesion of microelements
in nanoelectromechanical systems and absorption of
different gases by nanostructures. In elementary particle
physics the role of the Casimir effect is very important.
It is included in the calculation of hadron masses and provides an
effective mechanism for the compactification of extra spatial
dimensions in multi-dimensional physics (Mostepanenko, Trunov, 1997).
 In gravitation and cosmology the Casimir effect results
in a nonzero vacuum energy in spaces with non-Euclidean topology,
can drive the inflation process and leads to interesting effects in
brane models of the Universe (Saharian, 2006).
The Casimir effect has been actively used for obtaining stronger
constraints on the hypothetical long-range interactions predicted in
many theoretical schemes beyond the standard model (Bordag et al., 1998,
1999, 2000a; Long et al., 1999; Mostepanenko, Novello, 2001;
Fischbach  et al., 2001; Decca et al., 2003b, 2005a, 2005b, 2007a, 2007b;
Mostepanenko et al. 2008)).

Additional physical phenomena where the Casimir effect
is important are the quantum reflection of
atoms on different surfaces and Bose-Einstein condensation.
Casimir-Polder forces depend on real atomic and material properties
and influence the scattering amplitudes and condensation conditions
(Babb et al., 2004; Antezza et al., 2004).  Even in biophysics, a
proper account of the Casimir force is required for the understanding
of interaction of biological membranes through a liquid layer.

The diverse applications of the Casimir effect in both fundamental
and applied physics demand reliable theoretical methods allowing the
calculation of the Casimir force between real material bodies used in
experiments. Below we demonstrate that the Lifshitz theory
can be adapted to accomplish these ends.

\subsection{Structure of the review}
\label{sec:structure}

This review is devoted to the experimental and theoretical results
obtained primarily in the last 10 years and related to the Casimir
force between real materials. It is not aimed to cover not less important
but alternative results and methods related to the Casimir effect in
ideal configurations and more complicated geometries. The latter will
be touched only if they provide important guidelines for the
experimentally oriented theoretical approaches applicable to real
materials.

In Sec.~II we briefly present main results of the Lifshitz theory
for the configurations of two planar plates and an atom near a plate.
Some approximate methods applicable to nonplanar boundary surfaces
are also considered. Section III discusses the problem of how to compare
theory and experiment in the Casimir effect. Here, we consider the
modelling of the optical properties of real materials and corrections to
the Casimir force due to surface roughness and the finite size of the
interacting bodies. Special attention is paid to the quantitative
measure of agreement between experiment and theory in the
force-distance measurements.  Section IV presents the
main Casimir experiments with metallic test bodies including those
using an atomic force microscope and a micromechanical oscillator.
The obtained results are compared with different theoretical
computations using the Lifshitz theory.
The prospects for measuring the thermal Casimir force
between metals are also discussed. In Sec.\ V the measurements of
the Casimir force between a metallic sphere and a semiconductor plate
are presented. The discussion starts with the crucial experiment on
the optically modulated Casimir force.
The experiments with heavily doped semiconductors having
different charge carrier densities are also discussed.
Then we consider the recent experiment involving a semiconductor plate
with deep rectangular trenches.
The proposed
experiments on the change in Casimir forces for a dielectric-metal phase
transition, between a sphere and a plate with patterned geometry,
and with the pulsating Casimir force are considered. Section VI is
devoted to the measurements of the Casimir-Polder force. Here,
the experiment on the first measurement of the thermal Casimir-Polder
force at large separations is presented. Section VII deals with the
lateral Casimir force and Casimir torques. After a brief introduction
to the subject, the first experiment on the measurement of
the lateral Casimir force between corrugated surfaces is considered.
This is followed by the consideration of the Casimir torque
between corrugated surfaces at small angles and between anisotropic
surfaces. Section VIII contains our conclusions.

\section{ LIFSHITZ' THEORY OF THE THERMAL VAN DER WAALS AND CASIMIR
FORCES}
\label{sec:lifshitz}

\subsection{ Casimir interaction between two planar plates}
\label{sec:casimir}

Let us consider two thick dissimilar plane parallel plates
(semispaces) in thermal equilibrium at equal temperature $T$,
separated by an empty gap of width $a$. The Lifshitz formula
(Lifshitz, 1956; Dzyaloshinskii et al., 1961) represents the
van der Waals and Casimir free energy and force per unit area
(i.e., the pressure) in terms of the reflection coefficients
$r_{\rm TM}^{(n)}(\omega,k_{\bot})$ and
$r_{\rm TE}^{(n)}(\omega,k_{\bot})$ for the two independent
polarizations of the electromagnetic field. Here, $\omega$ is
the frequency and $k_{\bot}$ is the magnitude of the projection
of the wave vector onto the plane of the plates ($z$-axis is
perpendicular to the plates). Index $n=1,\,2$ labels the plates.
The transverse magnetic polarization (TM) means that the
magnetic field is perpendicular to the plane formed by
$\mbox{\boldmath$k$}_{\bot}$ and the $z$-axis, while for
the transverse electric polarization (TE) the electric
field is perpendicular to this plane. There are many
different derivations of the Lifshitz formula in the
literature based on different approaches: in the framework
of quantum statistical physics, thermal quantum field
theory in the Matsubara formulation, scattering theory etc.
[see, e.g., Schram (1973), Milonni (1994), Zhou, Spruch
(1995), Bordag et al. (2001), Genet et al. (2003b)].
In all the derivations the condition of thermal
equilibrium is used.
The final result is represented in one
of two equivalent forms: as a summation over the Matsubara
frequencies along the imaginary frequency axis or as an
integral over real frequencies. The first representation is
used more often as it is more convenient for computations.
Here, the Casimir free energy per unit area is given by
\begin{equation}
{\cal F}(a,T)=\frac{k_BT}{2\pi}\sum_{l=0}^{\infty}
{\vphantom{\sum}}^{\prime}
\Phi_E(\xi_l),
\label{eq1}
\end{equation}
\noindent
where the prime on the summation sign means that the term for
$l=0$ has to be multiplied by 1/2 and
\begin{equation}
\Phi_E(x)\equiv\!
\int_0^{\infty}\!\!\!\!k_{\bot}dk_{\bot}
\sum_{\alpha}\ln\bigl[1-r_{\alpha}^{(1)}({\rm i}x,k_{\bot})
r_{\alpha}^{(2)}({\rm i}x,k_{\bot}){\rm e}^{-2aq}\bigr].
\label{eq1a}
\end{equation}
\noindent
Here, $\alpha$ denotes TM or TE,
$\xi_l$ are the Matsubara frequencies, and
\begin{equation}
q\equiv q({\rm i}x,k_{\bot})=
\sqrt{k_{\bot}^2+\frac{x^2}{c^2}}.
\label{eq2}
\end{equation}
\noindent
Note that the function $\Phi_E$ in Eq.~(\ref{eq1}) depends on the real
argument $x=\xi_l$.
In the Lifshitz theory, the material media are described by the
dielectric permittivity depending only on the frequency (Lifshitz,
1956; Dzyaloshinskii et el., 1961; Lifshitz, Pitaevskii, 1980).
The description of the dielectric properties by
$\varepsilon(\omega)$ fully accounts for temporal dispersion
but neglects the possible contributions of spatial dispersion
to the van der Waals and
Casimir force. In the case of homogeneous
nonmagnetic media, the reflection coefficients are
\begin{eqnarray}
&&
r_{\rm TM}^{(n)}({\rm i}x,k_{\bot})=
\frac{\varepsilon^{(n)}({\rm i}x)q-
k^{(n)}}{\varepsilon^{(n)}({\rm i}x)q+k^{(n)}},
\nonumber \\
&&
r_{\rm TE}^{(n)}({\rm i}x,k_{\bot})=
\frac{q-k^{(n)}}{q+k^{(n)}},
\label{eq3}
\end{eqnarray}
\noindent
where
\begin{equation}
k^{(n)}\equiv k^{(n)}({\rm i}x,k_{\bot})=
\sqrt{k_{\bot}^2+\varepsilon^{(n)}({\rm i}x)\frac{x^2}{c^2}}.
\label{eq4}
\end{equation}
\noindent
We underline that the reflection coefficients (\ref{eq3}) are the
standard Fresnel reflection coefficients calculated, however, along the
imaginary axis ${\rm i}x$ where they are real.
They are derived from the standard continuity conditions for the
tangential and normal components of the electric field, magnetic
induction and electric displacement on the boundary planes,
$E_{1t}=E_{2t}$, $B_{1n}=B_{2n}$, $D_{1n}=D_{2n}$, and
$B_{1t}=B_{2t}$ (where the subscripts $t$ and $n$ denote the
tangential and normal components, respectively).
Note that these conditions can be identically
reformulated in Casimir spirit such that one deals with fields only
external to the plates (Emig, B\"{u}scher, 2004).

The reflection properties of the electromagnetic waves on metal
surfaces are often described in terms of the Leontovich surface
impedance (Landau et al., 1984). The corresponding expressions
for the reflection coefficients along the imaginary frequency
axis can be found, for instance, in the papers by Bezerra et al.
(2002c) and Geyer et al. (2003).

For plates of finite thickness or plates made of
several layers, the reflection coefficients have a more complicated
form depending on layer thicknesses [see, e.g., Zhou, Spruch
(1995), Klimchitskaya et al. (2000), Toma\v{s} (2002)].
In the literature, one can find the reflection
coefficients for planar multilayer magnetodielectrics with
$\mu\neq 1$ (Buhmann et al., 2005; Toma\v{s}, 2005).
The reflection coefficients have a more
complicated form if the plate material is
anisotropic.

The Casimir pressure between two plates  in
thermal equilibrium at a temperature $T$ is determined from Eq.\ (\ref{eq1})
\begin{equation}
{P}(a,T)=-\frac{\partial{\cal F}(a,T)}{\partial a}
=-\frac{k_BT}{\pi}\sum_{l=0}^{\infty}
{\vphantom{\sum}}^{\prime}
\Phi_P(\xi_l),
\label{eq5}
\end{equation}
\noindent
where
\begin{equation}
\Phi_P(x)\equiv\!
\int_0^{\infty}\!\!\!\!k_{\bot}dk_{\bot}q
\sum_{\alpha}\left[
\frac{{\rm e}^{2aq}}{r_{\alpha}^{(1)}({\rm i}x,k_{\bot})
r_{\alpha}^{(2)}({\rm i}x,k_{\bot})}-1\right]^{-1}\!\!\!.
\label{eq5a}
\end{equation}

As was mentioned above, Eqs.~(\ref{eq1}) and (\ref{eq5}) are
very useful for practical computations. For some purposes
(for example, to investigate the comparative role of propagating
and evanescent waves), the
following form of the Casimir free
energy  expressed as integrals along the real
frequency axis is convenient (Lifshitz, 1956; Bezerra et
al., 2007):
\begin{equation}
{\cal F}(a,T)=\frac{\hbar}{4\pi^2}
\int_0^{\infty}d\omega\coth\frac{\hbar\omega}{2k_BT}\,
{\rm Im}\Phi_E(-{\rm i}\omega).
\label{eq6}
\end{equation}
\noindent
Here, the function $\Phi_E$ is defined by Eq.~(\ref{eq1a}) with
$x=-{\rm i}\omega$ where $\omega$ is the real frequency.
The propagating waves correspond to $k_{\bot}<\omega/c$, i.e.,
to a pure imaginary $q$. The evanescent waves correspond
to $k_{\bot}\geq\omega/c$, i.e., to a real $q$.
These problems, however, are beyond the scope of
this review. They are considered, for instance, by Henkel et al.
(2004), Intravaia and Lambrecht (2005), Bimonte (2006a, 2006b)
and by Intravaia et al. (2007).

Using the Abel-Plana formula (Mostepanenko, Trunov, 1997) or
the Poisson summation formula (Mehra, 1967; Schwinger et al., 1978),
one can present the free energy per unit area and the pressure
in the form:
\begin{eqnarray}
&&
{\cal F}(a,T)=E(a)+\Delta{\cal F}(a,T),
\nonumber \\
&&
P(a,T)=P(a)+\Delta P(a,T).
\label{eq7}
\end{eqnarray}
\noindent
Here, $E(a)$ is given by
\begin{equation}
{E}(a)=\frac{\hbar}{4\pi^2}\int_{0}^{\infty}d\xi\,\Phi_E(\xi).
\label{eq8}
\end{equation}
\noindent
Similarly, the quantity $P(a)$ in the second equality
in Eq.~(\ref{eq7}) is equal to
\begin{equation}
{P}(a)=-\frac{\hbar}{2\pi^2}\int_{0}^{\infty}d\xi\,\Phi_P(\xi).
\label{eq9}
\end{equation}
\noindent
The second terms on the right-hand sides of equalities
(\ref{eq7}) are given by
\begin{eqnarray}
&&
\Delta{\cal F}(a,T)=\frac{{\rm i}k_BT}{2\pi}\int_{0}^{\infty}
dt\frac{\Phi_E({\rm i}\xi_1t)-\Phi_E(-{\rm i}\xi_1t)}{{\rm e}^{2\pi t}-1},
\nonumber \\
&&\label{eq9a} \\
&&
\Delta{P}(a,T)=-\frac{{\rm i}k_BT}{\pi}\int_{0}^{\infty}
dt\frac{\Phi_P({\rm i}\xi_1t)-\Phi_P(-{\rm i}\xi_1t)}{{\rm e}^{2\pi t}-1}.
\nonumber
\end{eqnarray}

Note that the quantities $E(a)$ and $P(a)$ in Eqs.~(\ref{eq8})
and (\ref{eq9}) are often referred to in the literature as the Casimir
energy per unit area and pressure at zero temperature, and
$\Delta{\cal F}(a,T)$ and $\Delta P(a,T)$ in Eq.~(\ref{eq9a})
are referred to as the thermal corrections to them.
This terminology is, however, correct only for plate materials
with temperature-independent properties. In this case the Casimir
free energy and pressure depend on the temperature only through the
Matsubara frequencies and the thermal corrections defined as
\begin{eqnarray}
&&
\Delta_T{\cal F}(a,T)\equiv {\cal F}(a,T)-{\cal F}(a,0),
\nonumber \\
&&
\Delta_TP(a,T)\equiv P(a,T)-P(a,0)
\label{eq10}
\end{eqnarray}
\noindent
are the same as
$\Delta{\cal F}(a,T)$ and $\Delta P(a,T)$ in Eq.~(\ref{eq7}).
In this case $E(a)={\cal F}(a,0)$ and $P(a)=P(a,0)$.
If, however, the properties of the medium (for instance, the
dielectric permittivity) depend on the temperature,
then Eqs.~(\ref{eq8})--(\ref{eq9a}) also contain a parametric
dependence on the temperature. Thus, the quantities
$E(a)$ and $P(a)$ are mixed quantities without a definite
physical meaning (Bezerra et al., 2002b).
They do not match ${\cal F}(a,T)$ and $P(a,0)$, respectively.
As a consequence, the thermal
corrections $\Delta_T{\cal F}(a,T)$ and $\Delta_T P(a,T)$
do not coincide with
$\Delta{\cal F}(a,T)$ and $\Delta P(a,T)$ in Eq.~(\ref{eq9a}).
Thus, the so-called {\it zero-temperature Lifshitz formulas}
(\ref{eq8}) and (\ref{eq9}) can be used for the approximate
calculation of the free energy and pressure only at rather small
separation distances, where the computational results using
Eq.~(\ref{eq1}) and  Eq.~(\ref{eq8}) practically coincide [respectively,
(\ref{eq5}) and  (\ref{eq9})].
Usually this is the case for the calculation of the nonrelativistic
van der Waals forces (Parsegian, 2005).

\subsection{ Approximations for nonplanar boundary surfaces}
\label{sec:appr}

The Lifshitz theory was formulated for the case of two parallel
plates. Experimentally it is hard to maintain two
plates parallel at short separations.
Because of this, most experiments (see
Secs.\ IV,\,V,\,VII) have been performed using the configuration
of a sphere above a plate. The configuration of a cylinder
above a plate also presents some advantages in comparison with
the case of two parallel plates (see Sec.\ IV.D).
Unfortunatelly, for many years it was not possible to obtain exact
expressions for the Casimir force in these configurations.
Thus, the approximate method of Derjaguin
(1934), later called
the proximity force approximation (PFA) (Blocki
et al., 1977), was used to compare experiment with theory.
According to this method, the Casimir energy in the gap between two
smooth curved surfaces at short separation can be calculated approximately
as a sum of energies between pairs of small parallel
plates corresponding to the curved geometry of the gap.
Specifically, under the condition
$a\ll R$, where $R$ is the sphere or cylinder radius, the
Casimir forces between the sphere or cylinder (per unit length)
and a plate are given by
\begin{equation}
F_s(a,T)=2\pi R{\cal F}(a,T),
\quad
F_c(a,T)=\frac{15\pi}{16}\sqrt{\frac{2R}{a}}{\cal F}(a,T),
\label{eq11}
\end{equation}
\noindent
respectively,
where ${\cal F}(a,T)$ is the Casimir free energy per unit area in the
configuration of two parallel plates, as defined in Eq.\ (\ref{eq1}).

Within the PFA it is not possible to control the error of the
approximation in Eq.\ (\ref{eq11}). From dimensional
considerations it is evident (Bordag et al., 2001) that the
relative error in Eq.\ (\ref{eq11}) should be of the order
of $a/R$, but the numerical coefficient of this ratio is
unknown. In fact, a rigorous determination of the error, introduced
by the application of the PFA, requires a comparison of
the approximate results in Eq.\ (\ref{eq11})
with the exact analytic results or with some precise numerical
computations in the respective configurations.

As was discussed in Sec.~I.B, during the last few years
the finite representation for the Casimir energy for two separated
bodies $A$ and $B$ in terms of the functional determinants was obtained.
In this representation the Casimir energy can be written in the form
(Kenneth, Klich, 2006, 2008)
\begin{eqnarray}
E(a)&=&\frac{1}{2\pi}\int_{0}^{\infty}d\xi\,
{\rm Tr}\ln\bigl(1-{\cal T}^A{\cal G}_{\xi,AB}^{(0)}
{\cal T}^B{\cal G}_{\xi,BA}^{(0)}\bigr)
\label{eqn16a} \\
&=&
\frac{1}{2\pi}\int_{0}^{\infty}d\xi\,
\ln{\rm det}\bigl(1-{\cal T}^A{\cal G}_{\xi,AB}^{(0)}
{\cal T}^B{\cal G}_{\xi,BA}^{(0)}\bigr).
\nonumber
\end{eqnarray}
\noindent
Here, ${\cal G}_{\xi,AB}^{(0)}$ is the operator for the free space
Green function with the matrix elements
$\langle\mbox{\boldmath$r$}|{\cal G}_{\xi,AB}^{(0)}|
\mbox{\boldmath$r$}^{\prime}\rangle$ where
$\mbox{\boldmath$r$}$ belongs to the body $A$ and
$\mbox{\boldmath$r$}^{\prime}$ to $B$.
${\cal T}^A\,({\cal T}^B)$ is the operator of the $T$-matrix for a body
$A$ and $B$, respectively. The latter is widely used in light
scattering theory, where it is the basic object for expressing the
properties of the scatterers (Bohren, Huffmann, 2004).
Using such representations, Emig et al. (2006) obtained
the analytic results for the electromagnetic Casimir energy
for an ideal metal cylinder above an ideal metal plane.
Eventually, the result is expressed through the determinant of an
infinite matrix with elements given in terms of the Bessel
functions. The analytic
asymptotic behavior of the exact Casimir energy at short
separations was found by Bordag (2006). It results in the
following expression for the Casimir force at $a\ll R$:
\begin{equation}
F_c(a,0)=-\frac{\pi^3}{384\sqrt{2}}\sqrt{\frac{R}{a}}
\,\frac{\hbar c}{a^3}\left[1-\frac{1}{5}\left(\frac{20}{\pi^2}-
\frac{7}{12}\right)\frac{a}{R}\right].
\label{eq12}
\end{equation}
\noindent
The PFA result in this case matches with the first term on the
right-hand side of (\ref{eq12}).
It can be obtained by replacing in the second equality in
Eq.\ (\ref{eq11}) the free energy ${\cal F}(a,0)$ with
the zero-temperature Casimir
energy in the configuration of two ideal metal plates $E_0(a)$ from
Eq.\ (\ref{Cas}).

Equation (\ref{eq12}) is very important. It demonstrates
that the relative error of the electromagnetic Casimir force
between a cylinder and a plate calculated using the PFA
is equal to $0.2886\,a/R$. Thus, for typical
experimental parameters of
$R=100\,\mu$m and $a=100\,$nm this error is approximately
equal to only 0.03\%.

For a sphere above a plate made of ideal metals at $T=0$ PFA
(\ref{eq11}) leads to the result
\begin{equation}
F_s(a,0)=-\frac{\pi^3R\hbar c}{360a^3}.
\label{eq13}
\end{equation}
\noindent
For this configuration
the exact analytic solution in the
electromagnetic case has not yet been obtained. The
scalar Casimir energy for a sphere above a plate was found by
Bordag (2006) and by Bulgac et al. (2006). The scalar Casimir energies
for both a sphere and a cylinder above a plate have also been
computed numerically using the wordline algorithms
(Gies, Klingm\"{u}ller, 2006a, 2006b), but it was noted that
the Casimir energies for the Dirichlet scalar field should
not be taken as an estimate for those in the electromagnetic
case.
Another numerical approach applicable in the electromagnetic case was
developed by Rodrigues, A. et al. (2007a). This approach has not yet been
applied at short separation distances of experimental interest.
For an ideal metal sphere above an ideal metal plane a
correction of order $a/R$ beyond the PFA was computed numerically
by Emig (2008) for $a/R\geq 0.075$ and by Maia Neto et al. (2008)
for $a/R\geq 0.15$. In both cases the extrapolation of the
obtained results to smaller $a/R$ leads to a coefficient
near $a/R$ approximately equal to 1.4.

In addition, the validity of the PFA for a sphere above a plate
has been estimated experimentally (Krause et al., 2007) and
the error introduced from the use of this
approximation was shown to be less than $a/R$
(see Sec.~IV.B for details). This is in disagreement with
the extrapolations made by Emig (2008) and Maia Neto et al. (2008).
To solve this contradiction, it is desirable to find
the analytical form of the first correction beyond the PFA
for a sphere above a plane,
like in Eq.~(\ref{eq12}) for the cylinder-plane configuration.

\subsection{ Casimir-Polder atom-plate interaction
and Bose-Einstein condensation}
\label{sec:CasPol}

The interaction of atoms with a wall has long been investigated
in different physical, chemical, and biological processes
including absorption and scattering from various surfaces
[see, e.g., the book by Mahanty and Ninham (1976)].
The general expression for the free energy of the atom-wall
interaction can be obtained from Eq.\ (\ref{eq1}) by
considering one of the plates as a rarified dielectric
(Lifshitz, 1956; Lifshitz, Pitaevskii, 1980; Milonni, 1994).
In doing so one expands the dielectric permittivity of a
rarified dielectric in powers of the number of atoms per unit
volume $N$, keeping only the first-order contribution
\begin{equation}
\varepsilon({\rm i}\xi)=1+4\pi\alpha({\rm i}\xi)N+O(N^2).
\label{eq14}
\end{equation}
\noindent
Here, $\alpha(\omega)$ is the dynamic polarizability of an atom.
We consider only atoms in the ground (or metastable) state.
The discussion of the interaction of excited atoms with a wall
[see, e.g., Buhmann, Welsch (2007)
and references therein] exceeds the scope
of this review.

As a result, the free energy of atom-wall interaction
 with the wall temperature $T$ in thermal equilibrium can
be presented in the form (Caride et al., 2005; Mostepanenko
et al., 2006a)
\begin{equation}
{\cal F}(a,T)=-{k_BT}\sum_{l=0}^{\infty}
{\vphantom{\sum}}^{\prime}
\alpha({\rm i}\xi_l)\int_{0}^{\infty}
\frac{k_{\bot}\,dk_{\bot}}{q_l}
\Phi_A(\xi_l,k_{\bot}),
\label{eq15}
\end{equation}
\noindent
where
\begin{equation}
\Phi_A(\xi_l,k_{\bot})\equiv
{\rm e}^{-2aq_l}\left[
\left(2q_l^2-\frac{\xi_l^2}{c^2}\right)r_{\rm TM}({\rm i}\xi_l,k_{\bot})-
\frac{\xi_l^2}{c^2}r_{\rm TE}({\rm i}\xi_l,y)\right].
\label{eq15a}
\end{equation}
\noindent
The reflection coefficient $r_{\rm TM}(0,k_{\bot})$ in Eq.\ (\ref{eq15a})
takes different values for different wall materials.
For dielectrics with finite static dielectric
permittivity $\varepsilon_0$ we get
$r_{\rm TM}(0,k_{\bot})=(\varepsilon_0-1)/(\varepsilon_0+1)$, and
for metals $r_{\rm TM}(0,k_{\bot})=1$.

At zero temperature Eq.\ (\ref{eq15}) results in
\begin{equation}
E(a)=-\frac{\hbar }{2\pi }\int_{0}^{\infty}d\xi\,
\alpha({\rm i}\xi)
\int_{0}^{\infty}\frac{k_{\bot}\,dk_{\bot}}{q}\,\Phi_A(\xi,k_{\bot}).
\label{eq17a}
\end{equation}
\noindent
In the nonrelativistic limit Eq.\ (\ref{eq17a}) leads to the well known
result (Lifshitz, Pitaevskii, 1980)
\begin{equation}
E(a)\approx -\frac{\hbar}{4\pi a^3}\int_{0}^{\infty}
\alpha({\rm i}\xi)\frac{\varepsilon({\rm i}\xi)-
1}{\varepsilon({\rm i}\xi)+1}\,d\xi\equiv -\frac{C_3}{a^3}.
\label{eq18}
\end{equation}
\noindent
If we consider an atom characterized by the frequency-independent
(static) polarizability $\alpha({\rm i}\xi_l)=\alpha(0)$
near an ideal metal wall at
$T=0$, Eq.\ (\ref{eq17a}) results in
\begin{equation}
E(a)=-\frac{3\hbar c}{8\pi a^4}\,\alpha(0)\equiv -\frac{C_4}{a^4}.
\label{eq19}
\end{equation}
\noindent
This is the famous result obtained by Casimir and Polder
(1948).

Calculations performed by using the exact expression (\ref{eq15})
show that the approximation (\ref{eq18}) is valid only at very short
separation distances $a<3\,$nm. To calculate $C_3$ in this case
one should use
the highly accurate $N$-oscillator model (Shih, Parsegian, 1975)
\begin{equation}
\alpha({\rm i}\xi)=\frac{e^2}{m}\sum_{n=1}^{N}
\frac{f_{0n}}{\omega_{0n}^2+\xi^2}
\label{eq20}
\end{equation}
\noindent
for the dynamic polarizability of an atom
(Caride et al., 2005).
In Eq.\ (\ref{eq20}) $m$ and $e$ are the electron mass and
charge, $f_{0n}$ and $\omega_{0n}$ are the oscillator strength
and transition frequency from the $n$th excited state to the ground-state
transition, respectively.
With the increase in the atom-wall distance the retardation effects
become important. Starting from distances of several tens of
nanometers, however, a more simplified single-oscillator
model for the atomic polarizability is applicable.
The description of an atom with the help of a
frequency-independent static polarizability works well only
at rather large separation distances $a>2\,\mu$m
(Babb et al., 2004). At room temperature, however, the
temperature corrections in Eq.\ (\ref{eq15}) come into play
at $a\geq 3\,\mu$m. Thus, the approximation (\ref{eq19})
at room temperature works well only within a rather narrow
separation region from 2 to $3\,\mu$m.

{}From Eq.~(\ref{eq15}), the Casimir-Polder force acting on an
atom situated near a wall a distance $a$ apart can be represented in
the form (Babb et al., 2004)
\begin{equation}
{F}(a,T)=-2{k_BT}\sum_{l=0}^{\infty}
{\vphantom{\sum}}^{\prime}
\alpha({\rm i}\xi_l)\int_{0}^{\infty}
k_{\bot}\,dk_{\bot}\,
\Phi_A(\xi_l,k_{\bot})
\label{eq21}
\end{equation}

Calculations of the energy of atom-wall interactions and of the
Casimir-Polder force play important role for the interpretation
of experiments on quantum reflection, i.e., above-barrier
reflection of slow atoms, with incident kinetic energy
exceeding the barrier height [see, e.g., C\^{o}t\'{e} et
al. (1998), Druzhinina, DeKieviet (2004), Pasquini et al.
(2004), Oberst et al. (2005a, 2005b), and references therein].
Results on Casimir-Polder
forces from such experiments are discussed in Sec. VI.B.

The Casimir-Polder interaction leads also to a change
of center-of-mass oscillation frequency $\omega_0$ of
a Bose-Einstein condensate (Antezza et al., 2004)
\begin{eqnarray}
\omega_0^2-\omega_z^2&=&-\frac{\omega_0}{\pi Am_a}\int_{0}^{2\pi/\omega_0}
\!\!\!\!\!d\tau\cos(\omega_0\tau)
\label{eq22} \\
&&\times
\int_{-R_z}^{R_z}\!\!\!d{z}
n_{z}({z})F[a+{z}+A\cos(\omega_0\tau),T].
\nonumber
\end{eqnarray}
\noindent
Here, the Casimir-Polder force in thermal equilibrium is given by
Eq.\ (\ref{eq21}), $m_a$ is the mass of atoms of the Bose-Einstein
condensate under consideration. The averaging procedure includes
the averaging of time over the period of oscillations
in the $z$-direction with an
amplitude $A$  ($z$ is perpendicular
to the plate) and the density of gas with a distribution
function
\begin{equation}
n_{z}({z})=\frac{15}{16R_z}\left(1-\frac{{z}^2}{R_z^2}
\right)^2,
\label{eq24}
\end{equation}
\noindent
where $R_z$ is the Thomas-Fermi radius in the $z$ direction.

The measurement of the change in the center-of-mass oscillation
frequency (Harber et al., 2005; Obrecht et al., 2007)
is a very sensitive test of the Casimir-Polder
force. Results of the recent experiment (Obrecht et al., 2007),
where the thermal effect was measured for the first time
in Casimir physics,
are discussed below in Sec.\ VI.A.

\subsection{ Puzzles in the application of the Lifshitz
theory to real materials}
\label{sec:pec}

\subsubsection{Real metals}

In the framework of the Lifshitz theory of the van der Waals and
Casimir force, discussed above, the calculational results depend
strongly on the model of dielectric permittivity used to
describe real material. Different problems arise for
metals and dielectrics. The source of the discrepancy is in the different
contributions from the zero frequency term
[i.e., from the term with $l=0$ in Eq.\ (\ref{eq1})].
For metals, problems result from the contribution of the
transverse electric mode. For ideal metal plates the Casimir
free energy was found independently of the Lifshitz theory within
the frames of thermal quantum field theory with electrodynamic
boundary conditions $E_t=B_n=0$ on the surface of plates
(Mehra, 1967; Brown, Maclay, 1969). As was mentioned in the
Introduction, the Lifshitz theory agrees with the
case of an ideal metal
if one takes the limit of infinite
dielectric permittivity, $\varepsilon({\rm i}\xi_l)=\infty$,
 before putting the frequency equal to
zero in the temperature sum (\ref{eq1}) (Schwinger et al., 1978).
Then from Eqs.\ (\ref{eq3}) and (\ref{eq4})  one obtains
\begin{equation}
r_{\rm TM}({\rm i}\xi_l,k_{\bot})=1, \quad
r_{\rm TE}({\rm i}\xi_l,k_{\bot})
=-1
\label{eq25}
\end{equation}
\noindent
for any $l$ including $l=0$.
After the substitution in Eq.\ (\ref{eq1}) this leads to the same
result for the Casimir free energy, as was obtained independently
using thermal quantum field theory.
The sequence of limiting transitions used to obtain Eq.~(\ref{eq25})
is called the {\it Schwinger prescription}.

The ideal metal, however, can be obtained as the limiting case of
real metals when the conductivity goes to infinity.
It is well known (Landau et al., 1984)
that for real metals $\varepsilon({\rm i}\xi)\sim 1/\xi$ when $\xi\to 0$
(the Drude model is one particular case).
Then from Eq.\ (\ref{eq3}) we get for real metals
\begin{equation}
r_{\rm TM}(0,k_{\bot})=1, \quad
r_{\rm TE}(0,k_{\bot})=0.
\label{eq26}
\end{equation}
\noindent
In the limit of infinite conductivity one  again obtains Eq.~(\ref{eq25})
for the reflection coefficients with $l\geq 1$, but for the
reflection coefficients with $l=0$ Eq.~(\ref{eq26}) remains valid.
This would lead to a different result for ideal metals than is obtained from
the Schwinger prescription. As shown below, this contradiction arises
from the use of the dielectric permittivity $\sim 1/\xi$ which is
outside the applicability conditions of the Lifshitz theory.

{}Conceptually the Lifshitz theory
provides a way for obtaining all the necessary results for any
real material.
 One may describe the free electrons in metals by the
plasma model [as it was suggested in the first papers devoted
to the calculation of the finite conductivity corrections to the
Casimir result (Lifshitz, 1956; Hargreaves, 1965;
Schwinger et al., 1978)]
\begin{equation}
\varepsilon(\omega)=1-\frac{\omega_p^2}{\omega^2}.
\label{eq27}
\end{equation}
\noindent
Here $\omega_p$ is the plasma frequency.
The dielectric permittivity (\ref{eq27}) is applicable in the frequency
region of the infrared optics (Lifshitz, Pitaevskii, 1981).
At room temperature all nonzero Matsubara frequencies belong to this region.
Because of this, the plasma model leads to rather accurate results
for the Casimir force at separation distances above the plasma wavelength.
Below the generalized plasma-like dielectric permittivity is also
considered which takes into account dissipation due to interband
transitions of core electrons and can be applied at shorter separations
(see Sec.~III.A.1).
The results obtained for real metals using the Lifshitz theory
with the permittivity (\ref{eq27})
are qualitatively close to those
obtained for ideal metals using thermal quantum field theory
(Bordag et al., 2000b; Genet et al.,
2000). Thus, in the low-temperature limit $T\ll T_{\rm eff}$
($k_BT_{\rm eff}=\hbar c/2a$) the asymptotic expression
for the thermal correction to the
energy for two similar plates, as defined in Eq.\ (\ref{eq10}), is given by
(Bordag et al., 2000b; Geyer et al., 2001)
\begin{eqnarray}
\Delta_T{\cal F}^{(p)}(a,T)&=&-\frac{\pi^2\hbar c}{720a^3}
\left\{1+\frac{45\zeta(3)}{\pi^3}
\left(\frac{T}{T_{\rm eff}}\right)^3-
\left(\frac{T}{T_{\rm eff}}\right)^4\right.
\nonumber \\
&&
+\left.\frac{2c}{\omega_pa}\left[
\frac{45\zeta(3)}{\pi^3}
\left(\frac{T}{T_{\rm eff}}\right)^3-\frac{1}{2}
\left(\frac{T}{T_{\rm eff}}\right)^4\right]+\ldots\right\}.
\label{eq28}
\end{eqnarray}
\noindent
Here, $\zeta(z)$ is the Riemann zeta function. Note that in this
case the quantity
$\Delta{\cal F}(a,T)$, as defined in Eq.\ (\ref{eq9a}),
coincides with the thermal correction (\ref{eq10}). For an ideal
metal $\omega_p\to\infty$ and Eq.\ (\ref{eq28}) reproduces the
low-temperature asymptotic behavior
obtained with thermal quantum field theory (Brown, Maclay, 1969).
In the high-temperature limit $T\gg T_{\rm eff}$ (at room
temperature $T=300\,$K it is achieved at $a\geq 6\,\mu$m)
the asymptotic expression for the  Casimir
free energy is (Bordag et al., 2000b)
\begin{equation}
{\cal F}(a,T)=-\frac{k_BT\zeta(3)}{8\pi a^2}\left(1-
\frac{2c}{\omega_pa}+\,\cdots\right).
\label{eq28a}
\end{equation}
\noindent
Here, $c/\omega_p=\delta_0$ is the skin depth,
and $\delta_0/a$ is a small parameter.
The finite conductivity correction in Eq.\ (\ref{eq28a}) is
negligibly small. Thus, the high-temperature limit for
real metals described by the plasma model dielectric
permittivity (\ref{eq27}) coincides with that obtained
for ideal metals (Brown, Maclay, 1969; Feinberg et al.,
2001).
The agreement between the Lifshitz theory combined with the plasma
model and the case of ideal metals, as described by thermal quantum
field theory, has a simple explanation. This is because for the
plasma model with $\omega_p\to\infty$, it is not Eq.\ (\ref{eq26}) but
Eq.\ (\ref{eq25}) which is satisfied for the reflection coefficients at
zero frequency.

A different situation comes about when  the Drude dielectric function
\begin{equation}
\varepsilon_D(\omega)=1-\frac{\omega_p^2}{\omega\bigl[
\omega+{\rm i}\gamma(T)\bigr]},
\label{eq29}
\end{equation}
\noindent
where $\gamma(T)$ is the relaxation parameter,
is substituted into the Lifshitz formula
(Bostr\"{o}m, Sernelius, 2000; H{\o}ye et al., 2003, 2006;
Brevik et al., 2005).
In this case the zero-frequency values of the reflection
coefficients are given by Eq.\ (\ref{eq26}), and the
high-temperature limit is equal to one half
of the corresponding value for ideal metals,
regardless of how large a conductivity for the real metal is used
(Klimchitskaya, Mostepanenko, 2001).

The low-temperature thermal correction, calculated with the
dielectric permittivity (\ref{eq29}), can be presented
in the form (Bezerra et al., 2004)
\begin{eqnarray}
&&
\Delta_T{\cal F}^{(D)}(a,T)=\Delta_T{\cal F}^{(p)}(a,T)+
{\cal F}^{(\gamma)}(a,T)
\nonumber \\
&&~~~
-\frac{k_BT}{16\pi a^2}\int_{0}^{\infty}ydy
\ln\left[1-\left(\frac{cy-\sqrt{4a^2\omega_p^2+c^2y^2}}{cy+
\sqrt{4a^2\omega_p^2+c^2y^2}}\right)^2{\rm e}^{-y}\right].
\label{eq30}
\end{eqnarray}

With the condition that
\begin{equation}
\gamma(T)<\xi_1
\label{eq31}
\end{equation}
\noindent
the low-temperature behavior of the second term on the right-hand
side of Eq.\ (\ref{eq30}) is
${\cal F}^{(\gamma)}(a,T)\sim\gamma(T)\ln(T/T_{\rm eff})$
(Bezerra et al., 2004).
For metals with perfect crystal
lattices $\gamma(T)\to 0$ as $T^2$ (Lifshitz et al., 1973),
 and the inequality
(\ref{eq31}) is satisfied down to $T=0$. For metals with
impurities there exists some residual value
$\gamma_{\rm res}\neq 0$ (Kittel, 1996).
As a result, at a $T$ of about $10^{-4}\,$K typical residual
relaxation becomes equal to $\xi_1$ and for smaller $T$
Eq.\ (\ref{eq31}) is violated.

For room temperature, at separation distances below $1\,\mu$m
the third term on the right-hand side of
Eq.\ (\ref{eq30}) is dominant. It describes a rather
large, linear in temperature thermal correction to the
Casimir free energy and pressure, as predicted by the Drude model.
This thermal correction is positive, i.e., it decreases the magnitude
of the Casimir free energy and pressure. As we will see in Sec.\ IV.B,
such a thermal correction is excluded by experiments at a high
confidence level.

{}From a theoretical basis the thermal correction (\ref{eq30})
is not appropriate
because it
is inconsistent with thermodynamics for metals with perfect
crystal lattices. To verify the consistency with thermodynamics,
one should consider the Casimir entropy
(Bezerra et al., 2002a, 2002b, 2004)
\begin{equation}
S(a,T)=-\frac{\partial{\cal F}(a,T)}{\partial T}.
\label{eq33}
\end{equation}
\noindent
In the case of metals described by the plasma model dielectric
permittivity (\ref{eq27}) $S^{(p)}(a,T)\sim T^2\to 0$ when $T\to 0$
(Bezerra et al., 2002b). In the case of Drude metals with
perfect crystal lattices, the Casimir entropy has a nonzero
negative value at $T=0$ following from (\ref{eq30})
(Bezerra et al., 2002a, 2002b, 2004, 2006; Mostepanenko et al., 2006b)
\begin{eqnarray}
&&
S^{(D)}(a,0)=\frac{k_B}{16\pi a^2}\int_{0}^{\infty}ydy
\ln\left[1-\left(\frac{cy-\sqrt{4a^2\omega_p^2+c^2y^2}}{cy+
\sqrt{4a^2\omega_p^2+c^2y^2}}\right)^2{\rm e}^{-y}\right]
\nonumber \\
&&~~~~~~
=-\frac{k_B\zeta(3)}{16\pi a^2}
\left(
1-\frac{4c}{\omega_pa}
+12\frac{c^2}{\omega_p^2a^2}-\,\cdots\right)<0.
\label{eq34}
\end{eqnarray}
Note that the use of the Drude model in the Lifshitz formula is usually
justified by the fact that it provides smooth transition between the
regions of the normal skin effect and that of infrared optics. This ignores
the region of the anomalous skin effect where the Lifshitz formula
in terms of $\varepsilon(\omega)$ is not applicable. With the decrease
of $T$ the application region of the normal skin effect becomes more
narrow and the application region of the anomalous skin effect widens.
However, at any $T>0$, there exists a frequency region including zero
frequency, where the normal skin effect is applicable. Thus, it is
worthwhile to use the Drude model at low $T$ when discussing the
thermodynamic consistency of the Lifshitz theory.

It was shown (Bostr\"{o}m, Sernelius, 2004;
H{\o}ye et al., 2007; Brevik et al., 2008), that
for Drude metals with impurities the Casimir
entropy jumps abruptly to zero  at $T<10^{-4}\,$K
starting from a negative value (\ref{eq34}).
The respective analytic expression for the low-temperature behavior
of the entropy has been obtained
(H{\o}ye at al., 2007; Brevik et al., 2008) under
the condition $\zeta_l\ll\gamma_{\rm res}$ which is
opposite to the inequality (\ref{eq31}).
The application region and the coefficients of this analytic
expression were determined incorrectly because an
overestimated value of $\gamma_{\rm res}=\gamma(T=300K)$
was used. The correct values of the coefficients were found by
Klimchitskaya and Mostepanenko (2008a) who also demonstrated
that the application region of the obtained expressions is
at temperatures below $T=10^{-5}\,$K.

According to H{\o}ye et al. (2007) and Brevik et al. (2008),
the formal satisfaction of the Nernst theorem for the Drude
model, as applied to metals with impurities, resolves
the problem of thermodynamic consistency of that model
combined with the Lifshitz theory. Klimchitskaya and
Mostepanenko (2008a), however, have remarked that  a
perfect crystal lattice is a truly equilibrium system with
a nondegenerate dynamical state of the lowest energy.
At zero temperature any part of the system must be in the
ground quantum state (Landau, Lifshitz, 1980; Rumer, Ryvkin, 1980).
As a result, the Casimir entropy computed for a perfect crystal
lattice, must be equal to zero, a direct consequence
of quantum statistical physics. Thus, they pointed out that
the Drude model combined with
the Lifshitz formula violates the Nernst heat theorem for a perfect
crystal lattice and is in contradiction with quantum statistical physics.

H{\o}ye et al. (2008) have argued that the Drude
model combined with the Lifshitz formula is thermodynamically
consistent in the case of a perfect crystal lattice also.
To justify this statement they introduced the definition of a perfect
lattice as the limiting case of a crystal lattice with nonzero residual
relaxation $\gamma_{\rm res}$ when $\gamma_{\rm res}\to 0$.
However, the common definition of a perfect lattice used in condensed matter
physics (Kittel, 1996) defines it as a lattice with perfect crystal
structure without impurities, i.e., with $\gamma_{\rm res}=0$ from
the outset. This common definition allows one to develop the quantum
theory of electron-phonon interactions which is based on the symmetry
properties of the lattice. Specifically, the relaxation parameter turns
out to be temperature-dependent and vanishes with vanishing
temperature.
The definition of a perfect lattice introduced by H{\o}ye
et al. (2008)
violates lattice symmetry properties. For example,
it would not be possible to reproduce the standard results of the theory
of elementary excitations on the basis of this definition.
Recently, Intravaia and Henkel (2008) independently
verified that the Drude model used in the Lifshitz formula
violates the Nernst heat theorem  for metals with perfect
crystal lattices as commonly understood
[$\gamma(T)\to 0$ at $T\to 0$], in accordance with original
statement by Bezerra et al. (2002a, 2002b, 2004).

Note that the substitution of the Drude dielectric function (\ref{eq29})
into the Lifshitz formula (\ref{eq1})
for metal plates of finite size is somewhat contradictory
(Parsegian, 2005; Geyer et al., 2007; Mostepanenko,
Geyer, 2008).
As pointed out by Parsegian (2005),
the Drude model is derived from the Maxwell equations applied to an
infinite metallic medium (semispace) with no external sources,
zero induced charge density and with nonzero drift current.
In such a medium there are no walls limiting the flow of charges.
For real metal plates of finite size, however, the applicability
conditions of the Drude model are violated.
If one admits that the electric field
of the zero-point oscillations with vanishingly
small frequencies (almost zero) creates a
short-lived current of conduction electrons,
this would lead to the
formation of almost constant charge densities $\pm\Sigma$ on
the side walls of the Casimir plates. As a result, both the
electric field and the current inside the plate would vanish
due to screening, in a very short
time interval of order of $10^{-18}\,$s (Geyer et al., 2007).
Because of this behavior, for electromagnetic fields
at very low frequency, the finite metal plates
should be described not by the Drude dielectric function
(\ref{eq29}), but  by the plasma model
which allows for displacement currents only (see Sec.\ III.A).
The role of finite size effects of the conductors was also
illustrated in the case of two wires  of finite length described
by the Drude model and interacting through the inductive
coupling between Johnson currents (Bimonte, 2007).
It was shown that in the thermal interaction between the wires the
Nernst theorem is followed or violated depending on whether
the capacitive effects associated with the end points of the wires
are taken or not taken into account.

Another approach to the resolution of the puzzles arising
 in the application of the Lifshitz theory to real materials
includes the effect of  screening on the reflection coefficients
(Dalvit, Lamoreaux, 2008).
This approach takes into account both the drift and diffusion currents
of free charge carriers through use of the transport Boltzmann equation.
It uses the standard Lifshitz formulas (\ref{eq1}) and
(\ref{eq15}) for the free energy of wall-wall and atom-wall
interaction with the TE reflection coefficient, $r_{\rm TE}$,
as defined in Eq.~(\ref{eq3}) within the Drude model approach,
but with the modified TM reflection
coefficient
\begin{equation}
r_{\rm TM}^{\rm mod}({\rm i}x,k_{\bot})=
\frac{\varepsilon({\rm i}x)q-k-\frac{k_{\bot}^2}{\eta({\rm i}x)}\,
\frac{\varepsilon({\rm i}x)-
\varepsilon_c({\rm i}x)}{\varepsilon_c({\rm i}x)}}{\varepsilon({\rm i}x)q
+k+\frac{k_{\bot}^2}{\eta({\rm i}x)}\,
\frac{\varepsilon({\rm i}x)-
\varepsilon_c({\rm i}x)}{\varepsilon_c({\rm i}x)}}.
\label{eq35}
\end{equation}
\noindent
Here, the dielectric permittivity is given by
\begin{equation}
\varepsilon({\rm i}x)=\varepsilon_c({\rm i}x)+
\frac{\omega_p^2}{x(x+\gamma)},
\label{eqn40a}
\end{equation}
\noindent
where $\varepsilon_c({\rm i}x)$ is the permittivity of the bound
core electrons
\begin{equation}
\varepsilon_c({\rm i}x)=1+\sum_{j}
\frac{g_j}{\omega_j^2+x^2+\gamma_jx}
\label{eqn40b}
\end{equation}
\noindent
with nonzero oscillator frequencies $\omega_j\neq 0$, the oscillator
strengths $g_j$ and the damping parameters $\gamma_j$.
The quantity $\eta({\rm i}x)$ is defined as
\begin{equation}
\eta({\rm i}x)=\left[k_{\bot}^2+\kappa^2
\frac{\varepsilon_c(0)}{\varepsilon_c({\rm i}x)}\,
\frac{\varepsilon({\rm i}x)}{\varepsilon({\rm i}x)-
\varepsilon_c({\rm i}x)}\right]^{1/2},
\label{eqn40c}
\end{equation}
\noindent
where $1/\kappa$ is the screening length.
If the charge carriers of density $n$ are described by the classical
Maxwell-Boltzmann statistics, as was supposed by Dalvit  and Lamoreaux (2008),
one gets the Debye-H\"{u}ckel screening length
$R_D=1/\kappa=(\varepsilon_0 k_BT/4\pi e^2n)^{1/2}$. Assuming the
Fermi-Dirac statistics, one arrives at the Thomas-Fermi screening length
$R_{TF}=1/\kappa=(\varepsilon_0 E_F/6\pi e^2n)^{1/2}$
(Landau, Lifshitz, 1980; Chazalviel, 1999).
Here, $e$ is the electron charge,
$\varepsilon_0=\varepsilon_c(0)$ is the dielectric
constant due to the core electrons, and $E_F=\hbar\omega_p$ is the
Fermi energy.
It was shown, however, that the
modification of the TM reflection coefficient in accordance with
Eq.~(\ref{eq35}) leads to the
violation of the Nernst heat theorem for those dielectric materials
whose charge carrier density does not vanish when $T\to 0$ and
conductivity vanishes due to the vanishing mobility
(Klimchitskaya, 2008; Klimchitskaya et al., 2008b).
For metals with perfect
crystal lattices this approach violates the Nernst theorem
as discussed above (Mostepanenko et al., 2009).
Note that Dalvit and Lamoreaux (2008) apply their
approach to intrinsic semiconductor media. Bearing in mind,
however, that the classical Drude dielectric permittivity is
exploited, it is justified to apply it for metals and doped
semiconductors as well by substituting the respective screening length.
As we will see in Secs.\ IV.B and
V.B, the computational results using the modified
TM reflection coefficient are
excluded by the measurement data of two precise experiments (Decca et
al., 2007a, 2007b; Chen et al., 2007a,b).

As we have seen above, the description of real metals at low frequencies
by a seemingly obvious dielectric permittivity of the Drude model
leads to serious difficulties. The attempts of solving this problem by
taking into account the screening effects due to finite size of the plates
or diffusion currents of free charge carriers also failed.
The latter approach, if successful, would be in fact a recognition that
the standard reflection coefficients, as used in the
Lifshitz theory, expressed in terms of the dielectric
permittivity are incorrect and one of them must be modified using an
additional microscopic quantity, the density of free charge carriers.

The origin of the problem, however, can be understood if one takes into
consideration the fact
that the inclusion  of the drift current, as described by the
Drude model, violates the basic applicability condition of the Lifshitz
theory. This is the demand that the dielectric media and the fluctuating
electromagnetic field should be in thermal equilibrium with the heat
reservoir. The point is that the existence of the drift electric
current leads to a violation of time reversal symmetry and the
introduction of the interaction  between the system and the heat
reservoir violating the state of thermal equilibrium (Bryksin, Petrov, 2008).
This is seen from the fact that the Drude model is derived from Maxwell
equations with the drift current of conduction electrons on the right-hand
side (see above). In its turn, the drift current leads to the emergence
of Joule losses, i.e., to a heating of the crystal lattice
(Geyer et al., 2003).  In order to preserve the temperature constant, the
unidirectional flux of heat from the Casimir plates to the heat reservoir
must inevitably arise. Such an interaction between the system and the
heat reservoir cannot exist in the state of thermal equilibrium as it is
in contradiction with its definition (Rumer, Ryvkin, 1980;
Kondepudi, Prigogine, 1998).

Thus, just not any dielectric permittivity can be considered in combination
with the Lifshitz formula. Specifically, the dielectric permittivity
of the Drude model (or any other inversely proportional to the first
power of the frequency at low frequencies) does not belong to the application
region of the Lifshitz theory. In a similar way, the approach based on
the transport Boltzmann equation,
in addition to the drift current, introduces
the diffusion current determined by a nonequilibrium distribution of free
charge carriers. This exacerbates the violation of thermal
equilibrium. Thus, the modified reflection coefficient (\ref{eq35})
obtained from this approach cannot be substituted into the Lifshitz
formula. On the other hand, the dielectric permittivity of the plasma model
(\ref{eq27}) and its generalized version taking into account the interband
transitions [see Eq.~(\ref{eq45a}) in Sec.~III.A.1] do not admit the
drift current, but only the displacement one. Thus, they are compatible with
the demand of thermal equilibrium and can be substituted into the Lifshitz
formulas.

\subsubsection{Real dielectrics}

As mentioned in the Introduction,
puzzling results were obtained in the literature
on the application of the
Lifshitz theory to real dielectrics.
It is well known that all dielectric materials
(insulators and intrinsic semiconductors
characterized by a band gap of different width, Mott-Hubbard
dielectrics and doped semiconductors with doping
concentration below the critical value) have a zero value of the dc
conductivity at zero temperature
[see, e.g., Mott (1990), Shklovskii,
Efros (1984)]. At nonzero temperature all these materials
have a nonzero value of dc conductivity. Sometimes this
conductivity is rather large, as, for instance, for doped
semiconductors with a doping concentration only slightly
below the critical value. In the commonly used
application of the Lifshitz theory the
dc conductivity of dielectrics is neglected.
Their conductivity is assumed to be equal to
zero at any temperature. In this case the dielectric
permittivity is represented in the form [see, e.g.,
Parsegian (2005)]
\begin{equation}
\varepsilon(\omega)=\varepsilon_c(\omega)=1+\sum_{j}
\frac{g_j}{\omega_j^2-\omega^2-{\rm i}\gamma_j\omega},
\label{eq36}
\end{equation}
\noindent
where $\omega_j\neq 0$ are the oscillator frequencies,
$g_j$ are the oscillator strengths and $\gamma_j$ are
the damping parameters [note that when calculating
$\varepsilon({\rm i}\xi)$ along the imaginary frequency
axis the damping parameters are often neglected and
some approximate formulas are used (Hough, White, 1980;
Bergstr\"{o}m, 1997; Parsegian, 2005)].
{}From (\ref{eq36}), the dielectric permittivity
at zero frequency is given by
\begin{equation}
\varepsilon_0\equiv\varepsilon(0)=1+\sum_{j}
\frac{g_j}{\omega_j^2}<\infty
\label{eq37}
\end{equation}
\noindent
and the reflection coefficients at zero frequency are equal to
\begin{equation}
r_{\rm TM}(0,k_{\bot})=
\frac{\varepsilon_0-1}{\varepsilon_0+1}\equiv r_0, \quad
r_{\rm TE}(0,k_{\bot})=0.
\label{eq37a}
\end{equation}
\noindent
However, if one takes into account the existence of free charge
carriers at nonzero temperature, the dielectric permittivity
can be represented in the form
\begin{equation}
\tilde\varepsilon(\omega,T)=\varepsilon(\omega)+{\rm i}
\frac{4\pi\sigma_0(T)}{\omega},
\label{eq38}
\end{equation}
\noindent
where $\sigma_0(T)$ is the dc conductivity at $T\neq 0$ and
$\varepsilon(\omega)$ is given by Eq.\ (\ref{eq36})
(Palik, 1985; Zurita-S\'{a}nches et al., 2004).
Then instead of Eq.~(\ref{eq37a}), the reflection coefficients
at $\xi=0$ are given by Eq.~(\ref{eq26}).

The analysis of the low-temperature asymptotic behavior of the
Casimir free energy (\ref{eq1})
calculated with the dielectric permittivity (\ref{eq36})
for two similar plates
shows that (Geyer et al., 2005b, 2006; Klimchitskaya, Geyer, 2008)
\begin{equation}
{\cal F}(a,T)=E(a)-\frac{\hbar c}{8a^3}\left[
\frac{b(a){\rm Li}_2(r_0^2)}{3(\varepsilon_0^2-1)}
\left(\frac{T}{T_{\rm eff}}\right)^2
+\frac{\zeta(3)r_0^2(\varepsilon_0+1)}{4\pi}
\left(\frac{T}{T_{\rm eff}}\right)^3\right],
\label{eq39}
\end{equation}
\noindent
where ${\rm Li}_n(z)$ is the polylogarithm function, and the
following notation is used:
\begin{equation}
b(a)=\sum_{j}\frac{g_j\gamma_jc}{2a\omega_j^4}.
\label{eq40}
\end{equation}
\noindent
{}From Eq.~(\ref{eq39}) the Casimir entropy is given by
the expression
\begin{equation}
{S}(a,T)=\frac{k_B}{4a^2}\,\frac{T}{T_{\rm eff}}\left[
\frac{2b(a){\rm Li}_2(r_0^2)}{3(\varepsilon_0^2-1)}
+\frac{3\zeta(3)r_0^2(\varepsilon_0+1)}{4\pi}\,
\frac{T}{T_{\rm eff}}\right].
\label{eq41}
\end{equation}
\noindent
As is seen from Eq.~(\ref{eq41}), $S(a,T)$ goes to zero
when $T\to 0$, i.e., the Nernst heat theorem is satisfied
when the dielectric permittivity is given by Eq.~(\ref{eq36})
with a finite static value (\ref{eq37}).

A completely different result is obtained if one takes into account the
conductivity of dielectrics at zero
frequency that arises at $T>0$. Note that in all cases this
conductivity vanishes exponentially,
$\sigma_0(T)\sim\exp(-C/T)$, when $T\to 0$ (Mott, 1990;
Shklovskii, Efros, 1984).
The reflection coefficients at zero frequency calculated with
the dielectric permittivity (\ref{eq38}) do not satisfy
Eq.\ (\ref{eq37a}) but Eq.\ (\ref{eq26}). This means that the
TM reflection coefficient has a discontinuity when the
dc conductivity of the dielectric material is taken into account.
Substituting the dielectric
permittivity (\ref{eq38}) into the Lifshitz formula (\ref{eq1}),
we arrive at (Geyer et al., 2005b)
\begin{equation}
\tilde{\cal F}(a,T)={\cal F}(a,T)-\frac{k_BT}{16\pi a^2}
\bigl[\zeta(3)-{\rm Li}_3(r_0^2)\bigr]+R(T),
\label{eq42}
\end{equation}
\noindent
where the low-temperature behavior of ${\cal F}(a,T)$ is given by
Eq.\ (\ref{eq39}) and $R(T)$ decreases exponentially when $T$
vanishes.

{}From Eq.~(\ref{eq42}), the Casimir entropy at $T=0$ is given
by
\begin{equation}
\tilde{S}(a,0)=\frac{k_B}{16\pi a^2}
\bigl[\zeta(3)-{\rm Li}_3(r_0^2)\bigr]>0
\label{eq43}
\end{equation}
\noindent
in violation of the Nernst theorem.
Thus, the inclusion of a nonzero conductivity arising at $T>0$
into the model of the dielectric response leads
to a contradiction between the Lifshitz theory and
thermodynamics for all materials which have a
zero conductivity at zero temperature, i.e., for all dielectrics.
The same happens if we include the dc
conductivity of the dielectric at $T>0$ when describing
the Casimir force between a metal and a dielectric plate
(Geyer et al., 2005a; 2006, 2008a).
In this case the Casimir entropy at $T=0$ is given
by Eq.\ (\ref{eq43}) where ${\rm Li}_3(r_0^2)$ should
be replaced with ${\rm Li}_3(r_0)$.
The above puzzling results are in line with those obtained for real
metals described by the Drude model. They are explained by the fact that
the dielectric permittivity (\ref{eq38}) is outside the application
region of the Lifshitz theory. Thus,
 the nonzero conductivity arising at $T>0$
for all dielectric materials
must be disregarded in theoretical computations using
the Lifshitz formula (\ref{eq1}). This rule has been
already confirmed experimentally (see Secs.\ V.B and
VI.A ).

Similar to the case of two dielectric plates, for an atom near
a dielectric wall  the Lifshitz
theory is thermodynamically consistent if the dc conductivity is neglected.
However, the inclusion of the
dc conductivity of the wall material in accordance with Eq.\ (\ref{eq38})
leads to the violation of the Nernst heat theorem (Klimchitskaya et al.,
2008b). An attempt to solve this problem beyond the scope
of the standard Lifshitz theory was undertaken by Pitaevskii (2008).
This attempt makes use of the effect of Debye-H\"{u}ckel screening
already mentioned above in connection with metals. The effect of
screening is taken into account only for the static field. It leads to the
replacement of the transverse magnetic reflection coefficient at zero
frequency, as given by Eq.\ (\ref{eq37a}), with
\begin{equation}
r_{\rm TM}^{\rm mod}(0,k_{\bot})=
\frac{\varepsilon_0\sqrt{k_{\bot}^2+\kappa^2}-
k_{\bot}}{\varepsilon_0\sqrt{k_{\bot}^2+\kappa^2}+k_{\bot}},
\label{eq43a}
\end{equation}
\noindent
where $\kappa=1/R_D$ is defined in Sec.~II.D.1.
 When the total density of charge
carriers $n=0$, Eq.\ (\ref{eq43a}) leads to the same result as
Eq.\ (\ref{eq37a}). For $n\to\infty$, at fixed $T\neq 0$,
$r_{\rm TM}^{\rm mod}(0,k_{\bot})=1$, as is in the standard Lifshitz
theory with the inclusion of the dc conductivity.
This solves the peculiarity of the small $\sigma$ limit, i.e., the
abrupt jump from Eq.~(\ref{eq37a}) to Eq.~(\ref{eq26}) arising from an
arbitrarily small conductivity. According to Pitaevskii (2008),
``the above-mentioned peculiarity of the small $\sigma$ limit
exists only for the $l=0$ term''. This permits one to assume
that all the coefficients $r_{\rm TM,TE}({\rm i}\xi_l,k_{\bot})$
with $l\geq 1$ under some conditions remain unchanged. This assumption
was confirmed by Dalvit and Lamoreaux (2008) who obtained both
the TM and TE reflection coefficients with account of the screening
effects at arbitrary Matsubara frequencies. The TE reflection coefficient
was found to be coinciding with the standard one
using the Drude model approach
and the modified TM
reflection coefficient is given by Eq.~(\ref{eq35}).
It is easily seen that for materials with sufficiently small charge
carrier density (for instance for Si with
$n\leq 10^{17}\,\mbox{cm}^{-3}$) the modified reflection coefficient,
Eq.~(\ref{eq35}), at room temperature $T=300\,$K
takes approximately the same values as the standard
one in Eq.~(\ref{eq3}) at all nonzero Matsubara frequencies.
If one takes into account the fact
that the reflection coefficient (\ref{eq43a})
obtained by Pitaevskii (2008) is just the reflection coefficient
(\ref{eq35}) at $x=0$, this concludes the proof of the above
assumption.

The reflection coefficient (\ref{eq43a}) can be obtained
by the formal replacement of the dielectric permittivities depending
only on frequency with the permittivities depending both on frequency
and the wave vector in the Fresnel reflection coefficient
of an uniaxial crystal.
 To confirm that this is the case, one should introduce
dissimilar permittivities $\varepsilon_x=\varepsilon_y$ and
$\varepsilon_z$ and replace the coefficient $r_{\rm TM}$ in Eq.\ (\ref{eq3})
with
\begin{equation}
r_{\rm TM}^{\rm mod}({\rm i}x,k_{\bot})=
\frac{\sqrt{\varepsilon_x({\rm i}x)\varepsilon_z({\rm i}x)}q-
k_{z}}{\sqrt{\varepsilon_x({\rm i}x)\varepsilon_z({\rm i}x)}q+k_{z}}.
\label{eq43b}
\end{equation}
\noindent
Here, $k_z$ is defined by Eq.\ (\ref{eq4}) with the replacement of
$\varepsilon({\rm i}x)$ for $\varepsilon_z({\rm i}x)$. With
$\varepsilon_x$ and $\varepsilon_z$ depending only on the frequency,
Eq.\ (\ref{eq43b}) is commonly used for the description of uniaxial
crystals (Greenaway et al., 1969; Blagov et al., 2005).
Then Eq.\ (\ref{eq43a}) follows from Eq.\ (\ref{eq43b}) if one puts
\begin{equation}
\varepsilon_x(0)=\varepsilon_0, \quad
\varepsilon_z(0)=\varepsilon_z(0,k_{\bot})=\varepsilon_0
\left(1+\frac{\kappa^2}{k_{\bot}^2}\right),
\label{eq43c}
\end{equation}
\noindent
where now $\varepsilon_z(0)$ depends on the wave vector.
Thus, to obtain (\ref{eq43a}) one substitutes into the reflection
coefficient (\ref{eq43b}) derived from the standard continuity boundary
conditions the dielectric permittivity depending on a wave vector
(see Sec.~III.A.3 for reasons why this is inappropriate).

It was verified (Klimchitskaya et al., 2008b) that for pure insulators
and intrinsic semiconductors [i.e., for materials for which $n(T)$
exponentially decays to zero with vanishing temperature] the Lifshitz
theory with the modified reflection coefficient (\ref{eq43a}) satisfies the
Nernst heat theorem. However, if $n$ does not go to zero when $T$ goes
to zero (this is true, for instance, for dielectric materials, such as
semiconductors doped below critical doping concentration or solids with
ionic conductivity), the Lifshitz theory with the reflection
coefficient (\ref{eq43a}) leads to a positive
Casimir-Polder entropy at $T=0$, i.e.,
violates the Nernst heat theorem (Klimchitskaya et al., 2008b).
Note that here the entropy of the fluctuating electromagnetic field
is nonzero at $T=0$ and depends on separation, whereas the entropy
of the plates is separation-independent.
Because of this, the entropy of the plates at $T=0$ cannot compensate
the nonzero entropy of the fluctuating field so as to make the total
entropy independent of the parameters of the
system in accordance with the Nernst theorem.
In fact, the conductivity is given by $\sigma_0(T)=n|e|\mu$, where $\mu$
is the mobility of charge carriers (Ashcroft, Mermin, 1976).
Although $\sigma_0(T)$ goes to zero exponentially fast for all
dielectrics when $T$ goes to zero, for most of them this happens not due
to the vanishing $n$ but due to the vanishing mobility. One such example is
a fused silica (SiO${}_2$) considered in Sec.\ VI.A in connection with
experiments on the Casimir-Polder force. Its conductivity is ionic in
nature and is determined by the concentration of impurities (alkali ions)
which are always present as trace constituents.
Two more examples are doped semiconductors with dopant concentration
below critical and dielectric like semimetals.

As we will see in Sec.\ V.B, the modification of the transverse magnetic
reflection coefficient at zero frequency in accordance with Eq.\ (\ref{eq43a})
is in contradiction with the data of the experiment on
the optically modulated
Casimir forces (Chen et al., 2007a, 2007b).
This is in line with the fact that, being the particular case of
Eq.~(\ref{eq35}), the modified reflection coefficient (\ref{eq43a})
violates the applicability conditions of the Lifshitz theory.

\section{  HOW TO COMPARE THEORY AND EXPERIMENT}
\label{sec:how}

\subsection{ Modelling of the optical properties of real materials}
\label{sec:mod}

\subsubsection{Kramers-Kronig relations}

To compare theory and experiment one first needs reliable
theoretical results. These results can be obtained by using the
Lifshitz formula (\ref{eq1}) with an appropriate dielectric
permittivity for the plate material along the imaginary frequency
axis. In general,
it can be obtained from the imaginary part of complex
dielectric permittivity
$\varepsilon(\omega)=\varepsilon^{\prime}(\omega)+{\rm i}
\varepsilon^{\prime\prime}(\omega)$ and the Kramers-Kronig
relations. The imaginary part of the permittivity is equal to
$\varepsilon^{\prime\prime}(\omega)=2n^{\prime}(\omega)
n^{\prime\prime}(\omega)$ where $n^{\prime}(\omega)$
and $n^{\prime\prime}(\omega)$
are the real and imaginary parts of the
complex index of refraction.
The tabulated optical data for $n^{\prime}(\omega)$
and $n^{\prime\prime}(\omega)$ for the different materials can be
found, e.g., in the handbook
by Palik (1985). It should be noted
that the form of Kramers-Kronig relations is different depending
on the analytic properties of the dielectric
permittivity under consideration.
If $\varepsilon(\omega)$ is a regular function at
$\omega=0$ [the case of dielectrics with dielectric permittivity
given by Eq.\ (\ref{eq36})], the Kramers-Kronig relations take
the simplest form (Jackson, 1999)
\begin{eqnarray}
&&
\varepsilon^{\prime}(\omega)=1+\frac{1}{\pi}P\int_{-\infty}^{\infty}
\frac{\varepsilon^{\prime\prime}(\xi)}{\xi-\omega}\,d\xi,
\nonumber \\
&&
\varepsilon^{\prime\prime}(\omega)=-\frac{1}{\pi}P\int_{-\infty}^{\infty}
\frac{\varepsilon^{\prime}(\xi)}{\xi-\omega}\,d\xi,
\label{eq44}
\end{eqnarray}
where the integrals are understood as a principal value. The third
dispersion relation, expressing the dielectric permittivity along
the imaginary frequency axis, in this case is
\begin{equation}
\varepsilon({\rm i}\xi)=1+\frac{2}{\pi}\int_{0}^{\infty}
\frac{\omega\varepsilon^{\prime\prime}(\omega)}{\xi^2+\omega^2}\,d\omega.
\label{eq45}
\end{equation}
\noindent
The optical data for dielectrics are available in a wide
frequency range $[\omega_{\min},\omega_{\max}]$
(Palik, 1985).
Near $\omega_{\min}$,
$\varepsilon^{\prime\prime}(\omega)$
is almost zero. Bearing in mind that $\omega_{\min}$ is usually below
all the relevant resonances of the respective dielectric medium,
it is possible
to extrapolate  $\varepsilon^{\prime\prime}(\omega)$ as zero
in the region $0\leq\omega\leq\omega_{\rm min}$.
Regarding $\omega_{\max}$, it is usually sufficiently high that no
extrapolation to region $\omega>\omega_{\max}$ is required.
 There are also many approximate analytic
expressions for the dielectric permittivity (\ref{eq36})
[see, e.g., Hough, White (1980), Bergstr\"{o}m (1997),
Parsegian (2005)] suggested for the calculation of the Hamaker
constant at short separation distances in the region of the
nonretarded van der Waals force. These expressions can also be
used for calculations of the retarded Casimir forces using
the Lifshitz formula (\ref{eq1}) (Parsegian, 2005).

The description of the dielectric properties of
metals is much more complicated. In calculations of the
zero-temperature Casimir energy and
pressure [Eqs.\ (\ref{eq8}) and (\ref{eq9})] the plasma model
dielectric permittivity (\ref{eq27}) is often used following
Lifshitz (1956). This model is, however,
applicable only at  separation distances larger
than the plasma wavelength, as it does not take into account
the absorption bands due to the core electrons. To address
this problem, there are two approaches suggested in the
literature.
In the first approach the tabulated optical data
for the complex index of refraction of
the metal under consideration are used
to obtain the imaginary part of the dielectric permittivity.
The latter is extrapolated to the
region of low frequencies with the help of the Drude dielectric
function (Lamoreaux, 1999; Lambrecht, Reynaud, 2000;
Klimchitskaya et al., 2000). In the second approach
the core electrons are described by a set of oscillators
with nonzero oscillator frequencies [similar to Eq.\ (\ref{eq36})
(Jackson, 1999)]
but the free conduction electrons are described using the plasma
model (Klimchitskaya et al., 2007a; Decca et al., 2007b;
Geyer et al. 2007; Mostepanenko, Geyer, 2008).
In this case we obtain the so-called
{\it generalized plasma-like dielectric permittivity}
\begin{equation}
\varepsilon_{gp}(\omega)=\varepsilon_{c}(\omega)
-\frac{\omega_p^2}{\omega^2}=
1-\frac{\omega_p^2}{\omega^2}+
\sum_{j}\frac{g_j}{\omega_j^2-\omega^2-{\rm i}\gamma_j\omega}.
\label{eq45a}
\end{equation}
\noindent
The values of the oscillator parameters of core electrons are found
from the tabulated optical data for the complex index of refraction
(Palik, 1985). Keeping in mind that the Casimir effect is a broad
band phenomenon, it is of utmost importance to use as exact data
as possible determined over a wide frequency range in the
interband absorption region $\omega>2.5\,$eV.
Presently such data (and also data for lower frequencies)
can be obtained using an ellipsometer. Thus, Svetovoy et al. (2008)
have found precise optical properties of several Au films in the
frequency region from 0.73\,eV to 8.86\,eV using a vacuum
ultraviolet ellipsometer (and from 0.038\,eV to 0.65\,eV
using the infrared variable angle spectroscopic ellipsometer).

In both approaches the dielectric permittivity along the imaginary
frequency axis can be obtained by means of the Kramers-Kronig
relations.
Regardless of which approach is used,
the dielectric permittivity of a metal is not
regular at $\omega=0$ but has a pole. This changes the form of the
Kramers-Kronig relations (\ref{eq44}).
If $\varepsilon(\omega)$ has a simple pole at $\omega=0$,
$\varepsilon(\omega)\approx 4\pi{\rm i}\sigma_0/\omega$,
the first two Kramers-Kronig relations are
(Landau et al., 1984)
\begin{eqnarray}
&&
\varepsilon^{\prime}(\omega)=1+\frac{1}{\pi}P\int_{-\infty}^{\infty}
\frac{\varepsilon^{\prime\prime}(\xi)}{\xi-\omega}\,d\xi,
\label{eq46} \\
&&
\varepsilon^{\prime\prime}(\omega)=-\frac{1}{\pi}P\int_{-\infty}^{\infty}
\frac{\varepsilon^{\prime}(\xi)}{\xi-\omega}\,d\xi
+\frac{4\pi\sigma_0}{\omega}.
\nonumber
\end{eqnarray}
\noindent
The third dispersion relation in this case is given by Eq.\ (\ref{eq45}),
i.e., remains the same as
for a dielectric permittivity regular at $\omega=0$.

Let us now consider the generalized plasma-like
dielectric permittivity  (\ref{eq45a}) having a
second-order pole at zero frequency [i.e., an asymptotic
behavior $\varepsilon(\omega)\approx -\omega_p^2/\omega^2$
when $\omega\to 0$].
The standard derivation (Jackson, 1999; Landau et al., 1984),
when applied to this case, leads to a one more form
of the Kramers-Kronig relations (Klimchitskaya et al., 2007a)
\begin{eqnarray}
&&
\varepsilon^{\prime}(\omega)=1+\frac{1}{\pi}P\int_{-\infty}^{\infty}
\frac{\varepsilon^{\prime\prime}(\xi)}{\xi-\omega}\,d\xi
-\frac{\omega_p^2}{\omega^2},
\nonumber \\
&&
\varepsilon^{\prime\prime}(\omega)=-\frac{1}{\pi}P\int_{-\infty}^{\infty}
\frac{\varepsilon^{\prime}(\xi)+\frac{\omega_p^2}{\xi^2}}{\xi-\omega}\,d\xi.
\label{eq47}
\end{eqnarray}
\noindent
In this case the third Kramers-Kronig relation (\ref{eq45}) is also
replaced with (Klimchitskaya et al., 2007a)
\begin{equation}
\varepsilon({\rm i}\xi)=1+\frac{2}{\pi}\int_{0}^{\infty}
\frac{\omega\varepsilon^{\prime\prime}(\omega)}{\xi^2+\omega^2}\,d\omega
+\frac{\omega_p^2}{\xi^2},
\label{eq48}
\end{equation}
\noindent
i.e., it acquires an additional term.

\subsubsection{Calculation for Au}

Let us now discuss the application of the above two approaches to
calculate the dielectric permittivity along the imaginary
frequency axis
for Au, a metal, widely used in the recent Casimir force measurements
(see Secs.\ IV,\,\,V and VII).
As discussed above, both approaches
use the tabulated optical data for the complex index
of refraction (Palik, 1985).
The calculation results are presented in Fig.\ 1.
The dashed line in this figure is obtained using the
Kramers-Kronig relation (\ref{eq45}) with
$\varepsilon^{\prime\prime}(\omega)$ given by the complete
tabulated optical data extrapolated to low frequencies
by the imaginary part of the Drude dielectric function with
$\omega_p=9.0\,$eV and $\gamma=0.035\,$eV (Lambrecht,
Reynaud, 2000; Klimchitskaya et al., 2000).
The solid line in Fig.\ 1 is obtained using the
Kramers-Kronig relation (\ref{eq48}) adapted for the generalized
plasma-like
model  with
$\varepsilon^{\prime\prime}(\omega)$ describing  the
contribution of core electrons.
In this case $\varepsilon^{\prime\prime}(\omega)$
is given by the tabulated data with the
contribution of free conduction electrons subtracted (Decca et
al., 2007b) using $\omega_p=9.0\,$eV. Note that for
frequencies $\xi<10^{16}\,$rad/s the simple analytic
six-oscillator model can be also used
for the description of core
electrons (Decca et al., 2007b; Mostepanenko, Geyer, 2008).
 As is seen in
Fig.\ 1, the computational results for $\varepsilon({\rm i}\xi)$
using the  two approaches coincide at $\xi\geq 10^{15}\,$rad/s,
but have different behavior at lower frequencies.

In Fig.~2 we present the computational results
for the correction factors to the Casimir energy per unit area (a)
and pressure (b) obtained by using the dielectric permittivities
presented in Fig.\ 1 by the solid and dashed lines.
These correction factors are defined as
\begin{equation}
\kappa_E(a,T)=\frac{{\cal F}(a,T)}{E_0(a)}, \quad
\kappa_P(a,T)=\frac{{P}(a,T)}{P_0(a)},
\label{eq48a}
\end{equation}
\noindent
where $E_0(a)$ and $P_0(a)$ are the Casimir energy and pressure
between ideal metal planes defined in Eq.\ (\ref{Cas}).
The solid lines represent the computational results using the
generalized plasma-like model [bold lines are obtained
using Eqs.\ (\ref{eq1}) and (\ref{eq6}) at $T=300\,$K whereas
fine lines using the zero-temperature
Eqs.\ (\ref{eq8}) and (\ref{eq9}); in the case of the Casimir
pressure bold and fine solid lines practially coincide at
all $a\leq 1\,\mu$m]. The dashed lines correspond to
computations using the tabulated optical data extrapolated by the
Drude model [long-dashed lines are obtained
using Eqs.\ (\ref{eq1}) and (\ref{eq6}) at $T=300\,$K whereas
short-dashed lines using the zero-temperature
Lifshitz formulas (\ref{eq8}) and (\ref{eq9}) with the dielectric
permittivity at room temperature]. As is seen in Fig.\ 2a,b,
there are large deviations between the zero-temperature
short-dashed and $T=300\,$K long-dashed
lines at all separation distances above 100\,nm.
This means that the zero-temperature formulas (\ref{eq8}) and (\ref{eq9})
with substituted room-temperature values of the dielectric
permittivity accounting for the relaxation processes of conduction
electrons can only be approximately used at
short separation distances, i.e., in the region of the
van der Waals forces and in the beginning of the transition region
to the Casimir forces (Parsegian, 2005).
At larger separation distances Eqs.\ (\ref{eq8}) and (\ref{eq9})
do not reproduce the values of the Casimir free energy.
By using
this model of the dielectric permittivity, all calculations
at separations above 100\,nm at room temperature should
be done with the thermal Lifshitz formulas (\ref{eq1})
 and (\ref{eq6})
for the Casimir free energy per unit
area and pressure. We emphasize,
however, that the theoretical results obtained in this way are
 excluded by experiment at a high confidence level (see
 Section.\ IV.B).

 The preceding sheds light on the problem of sample-to-sample dependence
 of the optical data. Actually, this dependence is mostly
 determined by the relaxation processes of free conduction
 electrons at infrared and optical frequencies [e.g., due to different
 characteristic sizes of grains in thin films (Sotelo et
 al., 2003)]. Uncertainty of the Casimir pressure due to
 the variation of optical properties was investigated by
 Pirozhenko et al. (2006) and Svetovoy et al. (2008)
 using the zero-temperature Lifshitz formula
 (\ref{eq9}). For this purpose both the real and imaginary parts of
the dielectric permittivity of Au, as given by the optical data in
the infrared region collected by different authors, were
approximated by the real and imaginary parts of the Drude permittivity
with an additional polarization term.
It is easily seen that all sets of the obtained Drude parameters
are excluded by the experiment on the measurement of the Casimir
pressure [see the general method of the verification of hypotheses,
as applied to this case, in Sec.\ III.C.2 and Fig.~11(b) in Sec.\ IV.B).
 In contrast, the generalized plasma-like dielectric
 permittivity is based on the characteristics of materials
 determined by the structure of a crystal cell ($\omega_p$
is one such factor).
These characteristics are not
 sensitive to sample-to-sample variations of the optical
properties of the deposited
 polycrystal films. Large variations in the values of $\omega_p$
 obtained by Pirozhenko et al. (2006) for different Au samples can
 be explained by the fact that the authors used
 the Drude model with frequency-independent relaxation
 parameter, whereas at high frequencies the relaxation parameter
 of conduction electrons depends on the frequency (Lifshitz et
 al., 1973). As discussed above, the
 drift current of conduction electrons is
 not compatible with the Lifshitz theory of dispersion forces
 (Parsegian, 2005; Geyer et al., 2007; Mostepanenko, Geyer, 2008).
Because of this,
 sample-to-sample variations of scattering processes connected with this
 current should be disregarded in the theory of the Casimir force.

\subsubsection{Problem of spatial dispersion}

 Another problematic issue is connected with the role of spatial dispersion.
 In particular, it is well known that in the frequency
 region of the anomalous skin effect, metals cannot be described by the
 dielectric permittivity depending only on the frequency (Lifshitz et
 al., 1973; Lifshitz, Pitaevskii, 1981). The Lifshitz formulas
 (\ref{eq1}), (\ref{eq3}), (\ref{eq6}), (\ref{eq8}) and (\ref{eq9}), however,
 are derived only for materials which can be described by
 $\varepsilon(\omega)$. The application range of the Lifshitz
 theory can be widened using the Leontovich surface impedance
 boundary condition (Landau et al., 1984).
 Kats (1977) pioneered the application of
this approach to describe
 metals in the region of the anomalous skin effect at zero temperature.
The Leontovich approach makes it possible to take into account real
material properties without considering the interior of the plates.
In this approach the standard continuity boundary conditions applicable
to interfaces between spatially local media are not followed.
 The Leontovich impedance approach now
appears interesting
 for the description of the thermal Casimir force.
It was suggested (Bezerra et al., 2002c;  Geyer et al., 2003)
to extrapolate the impedance for
 the frequency region around the characteristic frequency
 $\omega_c=c/(2a)$ to all frequencies.
Some other authors, however,
for separation distances related to infrared optics
have used the form of impedance for the normal (Torgerson, Lamoreaux, 2004)
or the anomalous (Svetovoy, Lokhanin, 2003) skin effect. In these
latter cases
 the theoretical results turn out
 to be in violation of the Nernst heat theorem and
in contradiction with experiment (Geyer et al., 2003, 2004;
 Bezerra et al., 2007).

 There is  another approach to
 account for spatial dispersion in the literature
by the substitution of
the dielectric permittivity depending on both the frequency and
wave vector, $\varepsilon(\omega,\mbox{\boldmath$k$})$,
 into the usual Lifshitz formulas
(\ref{eq1}), (\ref{eq5}) or (\ref{eq8}), (\ref{eq9})
with some modified reflection coefficients.
This approach is, however, unjustified.
The review by Barash and Ginzburg (1975) contains a few references
to incorrect results obtained by different authors using such
substitutions. In the last few years, more papers
have been published
(Esquivel et al., 2003; Esquivel, Svetovoy, 2004;
Contreras-Reyes, Moch\'{a}n, 2005; Sernelius, 2005; Svetovoy,
Esquivel, 2005) employing the same approach in an attempt
to solve the problem resulting from using the Drude
dielectric permittivity in the Lifshitz formula.
It was demonstrated (Klimchitskaya, Mostepanenko, 2007) that
these attempts are not warranted because the
modified reflection coefficients
are derived using phenomenological (additional) boundary conditions
and dielectric permittivities
$\varepsilon(\omega,\mbox{\boldmath$k$})$ which are ill-defined
quantities in the presence of a gap between the plates as it violates
the assumption of
translational invariance in the $z$-direction. In the
presence of spatial dispersion, the boundary conditions
following from the Maxwell equations differ from the standard
continuity boundary conditions
due to the the inclusion of the induced charge and current densities
(Agranovich, Ginzburg, 1984;  Ginzburg, 1985).
It was shown also that in the presence of the boundary surface the
dielectric permittivities depending on
$\omega,\>\mbox{\boldmath$k$}$
can be approximately used for the description of the remainder of
the medium except for a layer adjacent to the boundary surface
(Agranovich, Ginzburg, 1984). However, it is unlikely that this
phenomenological approach would be applicable for the calculation
of the Casimir force between metallic surfaces where the boundary
effects are of prime importance.
The generalization of the Lifshitz formula in terms of the
functional determinants and scattering matrices [see Eq.~(\ref{eqn16a})]
opens opportunities to include the effects of spatial dispersion.
For the configuration of two plane parallel plates the matrix elements
of the operator ${\cal G}_{\xi,AB}^{(0)}$ reduce to a factor
$\exp(-qa)$, and the same holds for ${\cal G}_{\xi,BA}^{(0)}$
(Lambrecht et al., 2006). To describe the plates including the
spatial dispersion, one should find the respective
$T$-matrices of the operators ${\cal T}^{A}$ and ${\cal T}^{B}$.
In principle, this could be done by solving Maxwell's equations for
a given Casimir configuration made of a nonlocally responding
material without using the dielectric permittivity depending on
$\omega$ and $\mbox{\boldmath$k$}$. This problem awaits  its resolution.

\subsection{ Corrections to the Casimir force due to the imperfect
geometry of interacting bodies}
\label{sec:cor}

\subsubsection{Surface roughness}

The problem of roughness corrections to dispersion forces has
long attracted the attention of researchers [see, e.g., Maradudin,
Mazur (1980), Mazur, Maradudin (1981), Derjaguin et al. (1988),
Rabinovich, Churaev (1989), where roughness corrections to
the nonretarded van der Waals force were investigated]. These corrections
can be calculated using perturbation theory and Green's functions
(Balian, Duplantier, 1977, 1978), functional
integration (Golestanian, Kardar, 1997, 1998; Emig et al., 2001,
2003) or by the phenomenological methods of pairwise summation
or geometrical averaging  (Bordag et al., 1994, 1995a, 1995b;
Klimchitskaya et al., 1999).
In doing so some small parameters
should be introduced which characterize the deviation from
basic geometry.

Let us consider two plates whose surfaces possess small
deviations from the plane geometry. The surfaces of these plates
can be described by the equations
\begin{equation}
z_1^{(s)}=A_1f_1(x_1,y_1),\quad
z_2^{(s)}=a+A_2f_2(x_2,y_2),
\label{eq49}
\end{equation}
\noindent
where $a$ is the mean value of the separation distance between the plates.
The values of the amplitudes are chosen in such a way that
$\max|f_i(x_i,y_i)|=1$. The averaging of equations (\ref{eq49})
over the total area of the plates with appropriately chosen origin
on the $z$-axis leads to
\begin{eqnarray}
&&
\langle z_1^{(s)}\rangle=A_1\langle f_1(x_1,y_1)\rangle
=0,
\label{eq50} \\
&&
\langle z_2^{(s)}\rangle=a+A_2\langle f_2(x_2,y_2)\rangle
=a.
\nonumber
\end{eqnarray}
\noindent
As mentioned above, we consider small deviations from the
basic geometry, i.e., $A_i\ll a$.

The topography of the surfaces (\ref{eq49}) can be obtained
experimentally as an AFM image of a rather large area
(see, e.g., Fig.\ 8 in Sec.\ IV.A). AFM images of the surfaces
of test bodies used in the experiments described in Secs.\ IV.A,\,B, V, VII
show that the roughness is represented by stochastically
distributed distortions of different height with a typical
lateral correlation length of about 200 to 400\,nm.
If the correlation effects can be neglected (below we discuss why
they are small),
 the surface topography is approximately characterized by
a discrete set of pairs $(v_i^{(1,2)},h_i^{(1,2)})$ where
$v_i^{(1,2)}$ is the fraction of the surface area
with heights $h_i^{(1,2)}$ on the surfaces 1 or 2,
respectively. These data allow one to determine the zero
roughness levels $H_0^{(1,2)}$ relative to which the mean values
of the functions $f_{1,2}$ in Eq.\ (\ref{eq49}) are equal to
zero:
\begin{equation}
\sum_{i=1}^{K^{(k)}}\bigl(H_0^{(k)}-h_i^{(k)}\bigr)v_i^{(k)}=0,
\quad
k=1\, ,2.
\label{eq51}
\end{equation}
\noindent
Note that the  separation distances between the test bodies in the
Casimir force measurements are measured just between these
zero roughness levels.

The Casimir pressure between the two ideal metal plates
with rough surfaces (\ref{eq49}) at zero temperature
can be approximately represented in the
form
\begin{equation}
P(a)=-\frac{\pi^2\hbar c}{240 a^4}\,\kappa_P^{(r)},
\label{eq52}
\end{equation}
\noindent
where $\kappa_P^{(r)}$ is the correction factor due to surface roughness.
In the framework of the pairwise summation method,
which is based on the additive summation of the Casimir-Polder
interatomic potentials with the subsequent normalization of the obtained
interaction constant,
this correction
factor calculated up to the fourth perturbation order in relative
amplitudes $A_i/a$ is (Bordag et al., 1995a)
\begin{eqnarray}
&&
\kappa_P^{(r)}=1+\frac{10}{a^2}\left(
\langle f_1^2\rangle A_1^2-2\langle f_1f_2\rangle A_1A_2
+\langle f_2^2\rangle A_2^2\right)
\label{eq53} \\
&&~~
+\frac{20}{a^3}
\left(
\langle f_1^3\rangle A_1^3-3\langle f_1^2f_2\rangle A_1^2A_2
+3\langle f_1f_2^2\rangle A_1A_2^2
-\langle f_2^3\rangle A_2^3\right)
\nonumber \\
&&~~
+\frac{35}{a^4}
\left(
\langle f_1^4\rangle A_1^4-4\langle f_1^3f_2\rangle A_1^3A_2
+6\langle f_1^2f_2^2\rangle A_1^2A_2^2
-4\langle f_1f_2^3\rangle A_1A_2^3
+\langle f_2^4\rangle A_2^4\right).
\nonumber
\end{eqnarray}
\noindent
The mixed terms in Eq.\ (\ref{eq53}) represent interference
behavior.

For example, in the case of two plane plates with a small angle $\alpha$
between them, it follows from Eq.\ (\ref{eq53})
\begin{equation}
\kappa_P^{(r)}=1+\frac{10}{3}\left(\frac{\alpha L}{a}\right)^2+
7\left(\frac{\alpha L}{a}\right)^4,
\label{eq54}
\end{equation}
\noindent
where $2L$ is the characteristic length of the plate.
As is easily seen from Eq.\ (\ref{eq54}), for plates
with $L\approx 1\,$mm at a separation $a\approx 500\,$nm
the correction due to nonparallelity of the plates is less
than 1\% if $\alpha$ does not exceed $2.7\times 10^{-5}\,$rad.
This is the reason why the experiments with two plane plates demand
a high degree of parallelity.
Note also that the small average separations needed for measurements
cannot be achieved if the plates are tilted, as they might  make
contact at the edges.
For the above example the contact of the plate edges occurs at a tilt angle
equal to $1.72'=5\times 10^{-4}\,$rad.
The correction factor $\kappa_E^{(r)}$ to the Casimir energy
between the two ideal metal plates
\begin{equation}
E(a)=-\frac{\pi^2\hbar c}{720 a^3}\,\kappa_E^{(r)}
\label{eq58}
\end{equation}
\noindent
is connected to (\ref{eq53}) by the equation
\begin{equation}
\kappa_P^{(r)}=\kappa_E^{(r)}-\frac{1}{3}\,
\frac{d\kappa_E^{(r)}}{da}.
\label{eq54a}
\end{equation}

In practical computations it is often convenient not to consider
the complete roughness topography of surfaces (\ref{eq49}),
but to
describe the roughness as a stochastic process by its rms
variance
\begin{equation}
\delta_{{\rm st},k}^2=\sum_{i=1}^{K^{(k)}}
\bigl(H_0^{(k)}-h_i^{(k)}\bigr)^2v_i^{(k)}.
\label{eq55}
\end{equation}
\noindent
If roughness can be characterized by the rms variance,
the correction factor to the Casimir pressure between the two
parallel plates takes a simpler form (Bordag et al., 1995b)
\begin{equation}
\kappa_P^{(r)}=1+\frac{10}{a^2}\bigl(
\delta_{{\rm st},1}^2+\delta_{{\rm st},2}^2\bigr)
+\frac{105}{a^4}\bigl(
\delta_{{\rm st},1}^2+\delta_{{\rm st},2}^2\bigr)^2.
\label{eq56}
\end{equation}
\noindent
For specially prepared surfaces in modern Casimir experiments, the
roughness corrections calculated using
Eqs.\ (\ref{eq53}) and (\ref{eq56}) usually lead to the same result.
The correction factor to the Casimir energy
due to stochastic roughness takes the form (Bordag et al., 1995b)
\begin{equation}
\kappa_E^{(r)}=1+\frac{6}{a^2}\bigl(
\delta_{{\rm st},1}^2+\delta_{{\rm st},2}^2\bigr)
+\frac{45}{a^4}\bigl(
\delta_{{\rm st},1}^2+\delta_{{\rm st},2}^2\bigr)^2.
\label{eq59}
\end{equation}

For the configuration of a sphere (spherical lens) above a plate,
which is preferred from an experimental point of view, the
investigation of roughness corrections to the Casimir force
in the framework of the pairwise summation method was performed by
Klimchitskaya and Pavlov (1996). It was demonstrated that if
the characteristic lateral scale of surface roughness on a plate,
$\Lambda_p$, and on a sphere, $\Lambda_s$, are rather small,
$\Lambda_p,\,\Lambda_s\ll\sqrt{aR}$,
the correction factor to the Casimir force between a sphere
and a plate made of ideal metals is the same as for the Casimir
energy between the two ideal  metal plates, i.e., is equal
to $\kappa_E^{(r)}$.

If the above inequalities are not fulfilled (e.g., there
are significant deviations of the lens surface from a spherical shape),
an additional term of the order $\sim 1/a$  results in the
correction factor. That is why it is necessary
to check carefully that the curvarture radius of the sphere
is constant. The role of roughness in atom-wall interaction was
investigated by Bezerra et al. (2000b).

The dependence of the Casimir force between metallic
test bodies on the penetration depth of the electromagnetic
oscillations into the metal (i.e., on the skin depth)
was first demonstrated by Mohideen and Roy (1998).
This raised the question of how to take into account both the
conductivity and roughness corrections. The simplest approach
is a multiplicative one: to calculate the Casimir free energy
(\ref{eq1}) or the Casimir pressure (\ref{eq6}) using
real material properties, including the role of the skin depth,
and multiply the obtained results by the correction
factor for ideal metal plates, $\kappa_E^{(r)}$ or
$\kappa_P^{(r)}$, respectively [see, e.g., Mohideen, Roy (1998),
Chan et al. (2001a)]. This approach corresponds to the addition of
the respective errors if they are small. Thus, it is not
applicable when at  least one of the  corrections is large.
At short separation
distances $a$, where the conductivity corrections are not small
and strongly depend on $a$ [see Figs.\ 2(a) and 2(b)] more
sophisticated methods are needed.

The nonmultiplicative approach of the
geometrical averaging was elaborated  by Klimchitskaya et al.
(1999). According to this method, the Casimir pressure
between the two rough plates made of real materials can be calculated
as
\begin{equation}
P^{(r)}(a,T)=\sum_{i=1}^{K^{(1)}}\sum_{j=1}^{K^{(2)}}
v_i^{(1)}v_j^{(2)}P(a+H_0^{(1)}+H_0^{(2)}-h_i^{(1)}-h_j^{(2)},T).
\label{eq60}
\end{equation}
\noindent
Here, $P(z,T)$ is calculated by using Eq.~(\ref{eq5}) including
real material properties. Note that Eq.\ (\ref{eq60}) is not
reduced to a simple multiplication of the correction factors due
to finite conductivity and surface roughness but takes into
account their combined (nonmultiplicative) effect as well.

In a similar way, the Casimir free energy for two rough plates
can be calculated as
\begin{equation}
{\cal F}^{(r)}(a,T)=\sum_{i=1}^{K^{(1)}}\sum_{j=1}^{K^{(2)}}
v_i^{(1)}v_j^{(2)}{\cal
F}(a+H_0^{(1)}+H_0^{(2)}-h_i^{(1)}-h_j^{(2)},T),
\label{eq61}
\end{equation}
\noindent
where ${\cal F}(z,T)$ is presented in Eq.~(\ref{eq1}).
The method of geometrical averaging is widely used when
comparing theoretical and experimental results in experiments
on the measurements of the Casimir force
[see, e.g., Klimchitskaya et al. (1999), Chen et al. (2004),
Decca et al. (2003a, 2003b, 2005b, 2007a, 2007b), Lisanti et al. (2005),
Munday, Capasso (2007)].

Equations (\ref{eq60}) and (\ref{eq61}) are based, however, on the
proximity force
approximation  and do not take
into account the diffraction-like correlation effects. In
first order perturbation theory, these effects have been
investigated for ideal metal plates covered with periodic roughness
(Emig et al., 2001, 2003) and for
metals described by the plasma model covered
with stochastic roughness  (Genet
et al., 2003a; Maia Neto et al., 2005). As was demonstrated
in these papers, the correlation effects noticeably contribute
to the roughness correction only at relatively large separations
of order or much larger than the roughness period $\Lambda$ (or the
correlation length) [see also van
Zwol et al. (2007)]. The special analysis (Chen et al., 2004;
Decca et al., 2005) based on the results by Emig et al.
(2001, 2003) shows that at separation distances
$a$ a few times less than $\Lambda$ the role of correlation effects
is negligibly
small (about 0.03 to 0.04\% change in the magnitude of the roughness
correction  for roughness topography of test
bodies used in the respective experiments). With an increase of
separation distance the correlation effects become more
important, but at such large separations the complete roughness correction
including the correlation effects
is negligibly small.
Thus, an accurate account of surface roughness
corrections would need more sophisticated methods beyond the pairwise
summation approach or geometrical averaging for test bodies with roughness
having a  small correlation length.
In this case there would be large correlation effects even at relatively
short separations where the role of roughness is significant.
The situation may become even harder if the roughness amplitudes are
not small and one has to go beyond the lowest order
perturbation theory. With respect to the roughness corrections,
to date, however, there are no measurements of the Casimir force
which fall into this category.
The first experimental evidence for the effects beyond the proximity
force approximation was obtained by Chan et al. (2008) by
measuring the Casimir interaction between an Au sphere and a Si plate
with deep rectangular trenches. This experiment is discussed in
Sec.~V.D.

There is one more effect connected with surface roughness
which should be accounted for in Casimir force measurements.
It is conceivable that the spatial variations of the surface
potentials due to grains of the polycrystalline metal film
(the so-called {\it patch potentials}) simulate the Casimir
force. Speake and Trenkel (2003) have obtained general
expressions for the electrostatic free energy and pressure in the
configuration of two parallel plates, which result from random
variations of patch potentials. These expressions can be used
to analyze the role of patch effects in any performed experiment.
In particular, it was shown (Chen et al., 2004; Decca et al.,
2005) that in the most precise experiments measuring the
Casimir force between metallic test bodies (see Secs.\ IV.A
and IV.B) the patch effects are negligibly small even at
the shortest separation distances.

\subsubsection{Finite size and thickness}

The last point to be considered in this section is the influence
of the finite size and a thickness
of interacting bodies on the calculational
results. In fact, expressions (\ref{eq1}) and (\ref{eq6}) for
the Casimir free energy per unit area and pressure are
obtained for plates of infinite area. Using the proximity
force approximation, we obtain from (\ref{eq1}) the Casimir force
between a sphere of radius $R\gg a$ above an infinitely
large plate. In experimental configurations, however, the plate
always has a finite size, and, instead of a sphere, a spherical
lens of some thickness $H<R$ may be used.
It should be noted that recently, rigorous methods
based on quantum field theory first principles have
been developed which allow one to calculate the scalar and
electromagnetic Casimir force between arbitrarily shaped compact
objects [see, e.g., Gies, Klingm\"{u}ller (2006a, 2006b, 2006c), Emig
et al. (2007, 2008)]. However, the application of these methods
for calculation of
small corrections due to the finite sizes
of the test bodies in experimental
configurations, where the separation distances
are rather small, still remains a problem (especially in the
electromagnetic case). For this reason,
finite size corrections still need to be estimated using the simple method
of pairwise summation. For a spherical lens of radius $R$ and
thickness $H$ placed at a distance $a$ above a disk of radius $L$
the finite size correction factor to the Casimir
force is given by (Bezerra et al., 1997)
\begin{equation}
\kappa_{\rm fs}=1-\frac{a^3}{R^3}(1-T)^{-3}.
\label{eq62}
\end{equation}
\noindent
Here, $T\equiv\max[R/\sqrt{R^2+L^2},(R-H)/R]$.
Using Eq.\ (\ref{eq62}) one can estimate that in all experiments
performed to date (see Secs.\ IV,\ V) the correction due to the
finite size of the test bodies is negligibly small. Thus, it
is reasonable to compare experimental results with
the theoretical computations using Eqs.\ (\ref{eq1}) or
(\ref{eq6}).

The finite thickness of plates can be taken into account by the
replacement of reflection coefficients (\ref{eq3}) with those
obtained for multilayered structures (Zhou, Spruch, 1995;
Klimchitskaya et al., 2000). For metal plates of larger thickness
than the skin depth in the frequency region of infrared
optics ($\delta_0=c/\omega_p\approx 22\,$nm for Au)
the calculational results are the same
as those for semispaces. Thus, for typical metal films of
about 100\,nm thickness deposited on the surfaces of test bodies
in different experiments, the Lifshitz formulas for semispaces can
be reliably used (Klimchitskaya et al., 2000).
As an example, if in one case we have two Au semispaces at a separation of
200\,nm and in the other an Au semispace at the same separation from
a 100\,nm thick Au film made on an Si semispace substrate,
the relative difference in the Casimir energy is less than 0.01\%.
Recent experiments on the measurement of the Casimir force
between a metallic sphere and the semiconductor (Si) plate (Chen et al.,
2005, 2006a, 2006b, 2007a, 2007b) have created an interest in
the role of the thickness of Si plate (Duraffourg, Andreucci, 2006;
Lambrecht et al., 2007; Pirozhenko, Lambrecht, 2008a).
In all these papers the Lifshitz formula at zero temperature is used
in the computations and in the latter two, large separations up to 1\,mm were
considered. The dependences of the Casimir force on the
plate thickness $d$ and the doping concentration are investigated.
The obtained results are, however, not applicable at large
separation distances at the laboratory temperature $T=300\,$K.
For example, we plot in Fig.\ 3 the ratio of the Casimir free energy
for the configuration of a Si plate of thickness $d$ at a
separation $a$ from an Au semispace to the free energy in the
configuration of Si and Au semispaces, as a function of Si plate
thickness $d$. Solid lines 1 and 2 are computed at $T=300\,$K
at $a=1$ and $5\,\mu$m, respectively. Dashed lines 1 and 2
are computed at
$T=0$ at the same respective separations. As is seen in the figure,
at $a=5\,\mu$m the deviation between the solid and dashed lines
reaches 20\%. Thus, the role of finite thickness of the
plate at separations above $1\,\mu$m should be investigated using
the Lifshitz formula at nonzero temperature.

\subsection{Quantitative comparison between experiment
and theory in force-distance measurements}
\label{sec:quant}

\subsubsection{Experimental errors and precision}

When comparing experimental data from Casimir force experiments
with theoretical predictions, the important question on how to
quantitatively characterize the measure of agreement between
them has to be addressed. Previously in the literature the concept of the
root-mean-square deviation between theory and experiment has been
used to quantify the precision of measurements and the
agreement with theory [see, e.g., the review of early
experiments made by Bordag et al. (2001), and also some later
experiments (Bressi et al., 2002; Decca et al., 2003a; Chen
et al., 2004a)]. It is known, however, that this method is not
appropriate for strongly nonlinear quantities, such as the
Casimir force, which changes rapidly with separation distance
(Rabinovich, 2000). It was shown (Ederth, 2000) that the
calculation of the root-mean-square deviation between
experiment and theory leads in this case to different results
when applied in different measurements ranges, while no
better method was suggested. Later a rigorous approach to
the comparison of experiment and theory in
Casimir force measurements was presented (Decca et al.,
2005b; Klimchitskaya et al., 2006b;
Chen et al., 2006b), using statistical methods.

The first step in the application of
this approach is to characterize how precise
the experimental data are without any relation to the theory.
To do this, the absolute and relative total experimental
errors should be calculated as a combination of systematic
and random errors. Let us use in the error analysis
the notation $\Pi^{\rm expt}(a)$ for the measured quantity which is
either the Casimir pressure $P^{\rm expt}(a)$ in the
configuration of two parallel
plates or the Casimir force $F^{\rm expt}(a)$ between
a sphere and a plate, as a function of separation distance
$a$ between them.

In each experiment, there are several sources of the
absolute systematic errors $\Delta_i^{\! s}\Pi^{\rm expt}(a)$
and respective relative systematic errors
$\delta_i^{s}\Pi^{\rm expt}(a)=
\Delta_i^{\! s}\Pi^{\rm expt}(a)/|\Pi^{\rm expt}(a)|$,
where $1\leq i\leq J$ (see Secs.\ IV--V for the
description of specific experiments).
It is necessary to stress that both in metrology and in all
natural sciences (physics, chemistry, biology etc.) the term
{\it systematic error} is used in two different meanings (Rabinovich, 2000).
According to the first meaning, a systematic error is some bias
in the measurement which always makes the measured value
higher or lower than the true value. Such systematic errors in the
measurement results are usually removed using some known process,
i.e., through a calibration. They can be also taken into account as
corrections (see the description of the calibration procedure and an example
of correction in the measurement of the Casimir force discussed in
Sec.~IV.A). The systematic errors in this understanding are often
called {\it systematic deviations}. Below it is assumed that the
experimental data under consideration are already free of such deviations.

Another meaning, which is used below in the review, defines
the systematic errors
as the errors of a calibrated measurement device. The errors of some
theoretical formula used to convert the directly measured quantity into
the indirectly measured one (see Sec.~IV.B) are also considered as
systematic. In accordance with common understanding, the error of a calibrated
device is the smallest fractional division on the scale of this device.
In the limits of this range the
systematic errors are considered as random quantities
characterized by a uniform distribution. Because of
this, the total relative systematic error is
(Rabinovich, 2000)
\begin{equation}
\delta^{s}\Pi^{\rm expt}(a)=\min\left\{\sum_{i=1}^{J}
\delta_i^{s}\Pi^{\rm expt}(a),k_{\beta}^{(J)}\sqrt{\sum_{i=1}^{J}
\bigl[\delta_i^{s}\Pi^{\rm expt}(a)\bigr]^2}\right\},
\label{eq63}
\end{equation}
\noindent
where $\beta$ is the chosen confidence level, $k_{\beta}^{(J)}$
is the tabulated coefficient depending on $\beta$
and on the number of sources of
systematic errors $J$ in the experiment under consideration
(Rabinovich, 2000).
In precise experiments, errors should be determined at a
confidence level $\beta\geq 95$\%. However, in many Casimir force
measurements  only a 67\%
confidence level is used (see Sec.\ IV).
Equation (\ref{eq63}) can be written equivalently in terms
of the absolute errors.

In precise experiments (see Secs.\ IV, V) it is common
to perform several
sets of measurements, say $n$,  within the same
separation region ($a_{\min},a_{\max}$). This is done in order to
decrease the random error and to narrow the confidence
interval. Each set consists of pairs $[a_i,\Pi^{\rm expt}(a_i)]$,
where $1\leq i\leq i_{\max}$ ($i_{\max}$ is different
for different experiments). All measurement data can
be represented as pairs $[a_{ij},\Pi^{\rm expt}(a_{ij})]$, where
$1\leq j\leq n$. If separations with fixed $i$ are approximately the
same in all sets of measurement (i.e., $a_{ij}\approx a_i$),
the mean and the variance of the mean at each point $a_i$
are obtained in the standard way (Rabinovich, 2000)
\begin{eqnarray}
&&
\bar\Pi_i^{\rm expt}=\frac{1}{n}\sum_{j=1}^{n}\Pi^{\rm expt}(a_{ij}),
\label{eq64} \\
&&
s^2_{\bar\Pi_i}=\frac{1}{n(n-1)}\sum_{j=1}^{n}[\Pi^{\rm expt}(a_{ij})-
\bar\Pi_i^{\rm expt}]^2.
\nonumber
\end{eqnarray}
If separations with fixed $i$ are different in different sets of
measurements (i.e., with different $j$), the calculation of
variance demands a more sophisticated procedure (Decca et al.,
2005b; Klimchitskaya et al., 2006b).

Direct calculations show that the mean values $\bar\Pi_i^{\rm expt}$ are
uniform, i.e., change smoothly with the change of $i$.
The variances of the mean, $s_{\bar\Pi_i}$, are, however, not
uniform. To smooth them, a special procedure is used in
statistics (Brownlee, 1965; Cochran, 1954).
The application of this procedure allows one to replace
$s_{\bar{\Pi}_i}$ with the smooth function of separation
$s_{\bar{\Pi}}(a)$ (Klimchitskaya et al., 2006b).
Then the random absolute error  at a separation distance $a$ can
be calculated
at a chosen confidence level $\beta$
\begin{equation}
\Delta^{\! r}\Pi^{\rm expt}(a)=s_{\bar\Pi}(a)t_{(1+\beta)/2}(n-1).
\label{eq65}
\end{equation}
\noindent
Here, the value of $t_p(f)$ can be found in the tables for
Student's $t$-distribution [see, e.g., Brandt (1976)].

To find the total experimental error of
$\Pi^{\rm expt}(a)$, one should combine the random and systematic
errors.
In so doing it is assumed that the random error is characterized by the
Student distribution which approaches to the normal distribution
with the increasing number of the measurement sets. The systematic error
is assumed to be described by a uniform distribution.
There are different methods in statistics
to obtain this combination
(Rabinovich, 2000). A widely used
 method is based on the
value of the quantity
$r(a)=\Delta^{\! s}\Pi^{\rm expt}(a)/s_{\bar\Pi}(a)$.
According to this method, at all $a$, where $r(a)<0.8$,
the contribution from the systematic error is
negligible and
$\Delta^{\! t}\Pi^{\rm expt}(a)=\Delta^{\! r}\Pi^{\rm expt}(a)$
at a 95\% confidence level
[in so doing $\Delta^{\! r}\Pi^{\rm expt}(a)$ is also
calculated at 95\% confidence using Eq.\ (\ref{eq65})].
If the inequality
$r(a)>8$ is valid, the random error is negligible
and $\Delta^{\! t}\Pi^{\rm expt}(a)=\Delta^{\! s}\Pi^{\rm expt}(a)$ with
$\Delta^{\! s}\Pi^{\rm expt}(a)$ calculated at 95\% confidence.
In the separation region, where $0.8\leq r(a)\leq 8$, a
combination of errors is performed using the rule
\begin{equation}
\Delta^{\! t}\Pi^{\rm expt}(a)=
q_{\beta}(r)\bigl[\Delta^{\! r}\Pi^{\rm expt}(a)+
\Delta^{\! s}\Pi^{\rm expt}(a)\bigr],
\label{eq66}
\end{equation}
\noindent
where the coefficient $q_{\beta}(a)$ at a $\beta=0.95$ confidence
level varies between 0.71 and 0.81 depending on $r(a)$.
Being conservative, one can use $q_{0.95}(r)=0.8$.

It is important that the total experimental error
$\Delta^{\! t}\Pi^{\rm expt}(a)$
and the respective total relative error
$\delta^{t}\Pi^{\rm expt}(a)=
\Delta^{\! t}\Pi^{\rm expt}(a)/|\Pi^{\rm expt}(a)|$
characterize the {\it precision} of an experiment in its own without
comparison with any theory. Note also that in metrology an
experiment is called {\it precise} if the random error is
much smaller than the systematic error (Rabinovich, 2000).
In Casimir force measurements performed up to date, only one
experiment (Decca et al., 2007a, 2007b) satisfies this rigorous
criterion.

\subsubsection{Comparison of experiment with theory}

The theoretical values $\Pi^{\rm theor}(a)$ (the pressure between the
plates or the force between the sphere and the plate) computed using
the Lifshitz theory also contain some errors.
If the Casimir force between a sphere and a plate is calculated,
one of the theoretical
errors is caused by the use of the proximity force approximation.
To be conservative, one can set the respective relative
error to $a/R$, i.e., to its maximum possible value (see Sec.\ II.B).
Another error is due to the
uncertainty in the optical data used. It can be
conservatively estimated to be equal to 0.5\% over the entire
measurement range (Chen et al., 2004a). The total theoretical
error can be computed using Eq.\ (\ref{eq63}) adapted
for theoretical errors. Note that when
using different sets of optical data one can arrive at
different computational results
for $\Pi^{\rm theor}(a)$. This has resulted in
statements in the literature [see, e.g., Pirozhenko
et al. (2006), Capasso et al. (2007)] that if different sets
of optical data obtained for different samples lead to
deviations in the Casimir force equal to, say, around 5\%, then the comparison
of theory with experiment on a better level is impossible.
These statements are, however, incorrect from the point of view
of measurement theory (Decca et al., 2007b).
The point to note is that here we are dealing with methods for the
{\it verification of hypotheses}.
The procedures for such verifications are well developed
in statistics. Different hypotheses correspond to
using different theoretical approaches (for instance, based on
the Drude or plasma model description of conduction electrons
in metals or the inclusion or the neglect of the small dc conductivity
of dielectrics at nonzero temperature) with different sets of
optical data available in the literature. The comparison of
different hypotheses with experiment can be performed
by plotting the respective theoretical bands
as functions of separation (the width of the band is determined by the
total theoretical error)
and the experimental data  with their absolute errors
on one graph
(see, e.g., Fig.\ 11(a) in Sec.\ IV.B). If the theoretical band
does not overlap with the experimental data, including
their errors, over a wide separation
range, it is experimentally excluded  and the
corresponding hypothesis must be rejected.
It is important to note  that the data and theory are not compared at
just one point but over a wide separation region where the
distance dependence is nonlinear.
The set of optical
data which contradicts the experimental results, when combined
with any theoretical approach, should be considered as irrelevant
to the actual properties of the film used in the measurement.

Another method to compare theory with experiment is the
consideration of the confidence interval for the random
quantity $\Pi^{\rm theor}(a_i)-\bar{\Pi}^{\rm expt}(a_i)$
(Decca et al., 2005b; Chen et al., 2006b).
In doing so,
one should take into account that this comparison is
done at the separation distances $a_i$ measured
with some error $\Delta a$. Thus, the
theoretical value $\Pi^{\rm theor}(a_i)$ at this point can be
known only with the relative error $\alpha\Delta a/a_i$
(Iannuzzi et al., 2004a). Here,
$\alpha=3$ for the force between a sphere and a plate
and $\alpha=4$ for the pressure between the two plates.
This error, although not connected with the
theoretical approach used, should be included as part
of the theoretical errors in Eq.\ (\ref{eq63}).
Note that at short separations it results in a major contribution to
the total theoretical error $\Delta^{\! t}\Pi^{\rm theor}$.
Then, the absolute error of
the difference between theory and experiment,
$\Xi_{\Pi}(a)$, at a 95\% confidence can be calculated
by using Eq.\ (\ref{eq63}) with $J=2$ and
$k_{0.95}^{(2)}=1.1$:
\begin{eqnarray}
&&
\Xi_{\Pi}(a)=\min\left\{
\vphantom{\sqrt{\bigl[\Delta^{\! t}\Pi^{\rm expt}(a)\bigr]^2}}
\Delta^{\! t}\Pi^{\rm expt}(a)+\Delta^{\! t}\Pi^{\rm theor}(a),\right.
\label{eq67} \\
&&~~~~~
1.1\left.
\sqrt{\bigl[\Delta^{\! t}\Pi^{\rm expt}(a)\bigr]^2+
\bigl[\Delta^{\! t}\Pi^{\rm theor}(a)\bigr]^2}\right\}.
\nonumber
\end{eqnarray}
\noindent
The confidence interval for the quantity
$\Pi^{\rm theor}(a)-\bar{\Pi}^{\rm expt}(a)$
at a 95\% confidence level is given by
$\bigl[-\Xi_{\Pi}(a),\Xi_{\Pi}(a)\bigr]$.
When one compares a theoretical approach with
experimental data, the differences between the theoretical and
mean experimental values of $\Pi$ may or may not
belong to this interval. A theoretical approach for
which not less than 95\% of the differences
$\Pi^{\rm theor}(a)-\bar{\Pi}^{\rm expt}(a)$ belong to
the interval $\bigl[-\Xi_{\Pi}(a),\Xi_{\Pi}(a)\bigr]$
within any separation subinterval $[a_1,a_2]$ of the
entire measurement range is consistent with the experiment.
In this case the measure of agreement between experiment
and theory is given by $\Xi_{\Pi}(a)/|\bar{\Pi}^{\rm expt}(a)|$.
On the contrary, if for some theoretical approach
a subinterval $[a_1,a_2]$ exists, where almost all
differences
$\Pi^{\rm theor}(a)-\bar{\Pi}^{\rm expt}(a)$ are outside of
the confidence interval $\bigl[-\Xi_{\Pi}(a),\Xi_{\Pi}(a)\bigr]$,
this theoretical approach is excluded by experiment at
separations from $a_1$ to $a_2$ at a 95\% confidence level.
If the theoretical approach (hypothesis) is excluded by experiment at
a 95\% confidence level, the probability that it is true is at most 5\%.
It may happen that several theoretical approaches $i=1,\,2,\ldots$
are consistent with experiment, i.e., not less than 95\%
of the differences $\Pi_i^{\rm theor}(a)-\bar{\Pi}^{\rm expt}(a)$
($i=1,\,2,\ldots$) belong to the confidence interval
$\bigl[-\Xi_{\Pi}(a),\Xi_{\Pi}(a)\bigr]$ (such situations are considered
in Secs.~IV and V). The statistical criteria used do not allow
to indicate a probability of the event that one of these approaches or all
of them are false. The rejection of some of the experimentally
consistent approaches can be done on a theoretical basis only.
For example, if the measurement is performed at room temperature and
both theoretical computations done at $T=0\,$K and at $T=300\,$K are
consistent with the data, the computation at $T=300\,$K can be considered
as preferable. In fact to reliably discriminate between the two
experimentally consistent theoretical approaches,  more exact
measurements are desirable. {}From the above it becomes clear that
the described statistical criteria are somewhat asymmetric. They provide
solid grounds for rejection of the experimentally inconsistent
approaches at a high confidence level but do not permit to claim
the experimental confirmation of some consistent approach at the same
high confidence.
These general criteria are illustrated in
Secs.\ IV and V using the examples of different Casimir
force measurements (see Figs.\ 12, 13, \ref{fgVp6}).
It would be interesting to apply more sophisticated statistical methods,
Bayesian, for instance
(Carlin, Louis, 2000; Berger, 1993; Lehmann, Romano, 2005)
for the data analysis of force-distance relations resulting from the Casimir
force measurements. Until now, however, such methods have not been used
in the literature devoted to the experimental study of the Casimir effect.

It is notable that conclusions concerning consistency or rejection
of a hypothesis do not depend on the method of comparison used
(see discussion of different situations in Secs.\ IV--VI).
The half-width of the confidence interval $\Xi_{\Pi}(a)$
has very little dependence on the theoretical approach
used (i.e., on the hypothesis to be verified). It usually
results from the experimental errors in the
measurement of the separation distance and the Casimir force
or the pressure.

%%%%%%%%%%%%%%%%%%%%%%%%%%%%%%%%%%%%%%%%%%%%%%%%%%%%%%%%%%%%%
%%%%%%%%%%%%%%%%%%%%%%%%%%%%%%%%%%%%%%%%%%%%%%%%%%%%%%%%%%%%
%%%%%%%%%%%%%%%%%%%%%%%%%%%%%%%%%%%%%%%%%%%%%%%%%%%%%%%%%%%
\section{ CASIMIR EXPERIMENTS WITH METALLIC TEST BODIES}
\label{sec:met}

\subsection{ Measurements of the Casimir force
using an atomic force microscope}
\label{sec:afm}

Recent interest in the experimental investigation of the
Casimir effect has its beginning in the experiment by Lamoreaux (1997).
This experiment, however, contains several uncertainties that do not
permit one to consider it as precise and definitive (see the discussion
in Sec.~IV.C.1).
The first precise and definitive direct measurements of the Casimir force
between a metal coated sphere and a plate were performed by
Mohideen and Roy (1998), Roy et al. (1999) and Harris et al. (2000)
in three successive experiments using an atomic force microscope (AFM)
operated in vacuum. In these experiments the fundamental requirements
for the measurement of the Casimir force proposed by Sparnaay (1958, 1989),
such as (i) clean surfaces for the test bodies, (ii) precise and reproducible
measurement of the separation distance between them and (iii) low electrostatic
potential differences, were met for the first time [brief historical survey of
the Casimir force measurements performed before 1990 is given by
Bordag et al. (2001)]. The authors introduced the now standard use of metal
coated
polystyrene spheres, which have the unique advantages of being extremely light
weight, perfectly smooth (made from liquid phase) and very low eccentricity
(less
than parts per thousand) as one of the surfaces in Casimir force
measurements.  These measurements also pioneered the independent
determination of the roughness induced average separation distances
at contact, the cantilever calibration and of the potential
differences between the two surfaces using the electrostatic force. This
allowed a careful comparison to the theory and a precise study of the effect
of the metal conductivity (also in terms of the skin depth) and the
surface roughness which previously could not be done.

A schematic diagram of the experimental setup for the measurement of
the Casimir force using the AFM is shown in Fig.\ 4. The Casimir force
between a sphere of radius $R$ and a plate causes the cantilever
to flex. This flexing is detected by the deflection of a laser beam
leading to a difference signal $S_{\rm def}$ between photodiodes A and B
which was calibrated by means of the electrostatic force
between the sphere and the plate.
For this purpose various voltages $V$ were applied to the plate while
the sphere remained grounded. The measurements of the electrostatic
forces were performed at separations $a$ larger than a few micrometers
where the contribution from the Casimir force is negligibly small.
The exact expression for the electrostatic force in sphere-plate
configuration is given by (Smythe, 1950)
\begin{equation}
F_{\rm el}(a)=2\pi\epsilon_0(V-V_0)^2\sum_{n=1}^{\infty}
\frac{\coth\alpha-n\coth{n\alpha}}{\sinh{n\alpha}}\equiv
X(a)(V-V_0)^2,
\label{eq4p1}
\end{equation}
\noindent
where $\cosh\alpha=1+a/R$, $V_0$ is the residual voltage on the sphere,
and $\epsilon_0$ is the permittivity of the vacuum. {}From a
comparison of the theoretical expression (\ref{eq4p1}) with the
experimental data, the separation on contact $a_0$ and the residual potential
difference between the test bodies were determined. Finally the absolute
separations were found from
\begin{equation}
a=a_0+a_{\rm piezo}+S_{\rm def}m,
\label{dist}
\end{equation}
\noindent
where the movement $a_{\rm piezo}$ of the piezoelectric
tube on which the plate is mounted was calibrated by an optical fiber
interferometer (Chen, Mohideen, 2001). As the piezo was moved with
continuous periodic triangular voltages, the nonlinearities can be precisely
accounted for through interferometric calibration.  The last term
in (\ref{dist}) is due to the decrease in the separation distance from the
flexing
of the cantilever in response to any force. Note that $S_{\rm def}<0$ for
an attractive force. The deflection coefficient  $m$ was found from
electrostatic calibration.

To measure the Casimir force between the sphere and the plate, they are both
grounded together with the AFM. The plate is then moved towards the
sphere using the piezo and the corresponding
photodiode difference signal is measured. It is converted into the
respective deflection of the cantilever tip and using Hooke's law into
the values of the Casimir force $S_{\rm def}km$ where $k$ is the spring
constant
(see below for the results of electrostatic
calibration and the values of main parameters).
In the first two measurements
using the AFM by Mohideen, Roy (1998) and Roy et al. (1999), the
sphere and the plate were coated with about 300nm of Al. To prevent
rapid oxidation of Al it was sputter coated with thin Au/Pd layers.
In the first experiment (Mohideen, Roy, 1998; Klimchitskaya et al., 1999)
it was clearly demonstrated that the Casimir force between real metal
surfaces deviates significantly from the Casimir prediction made for
ideal metal surfaces. In Fig.\ 5 the data for measured Casimir force,
as a function of separation distance $a$, are shown as open squares.
The dashed line indicates the Casimir force (\ref{eq13}) between an ideal
metal plate and an ideal metal sphere of $R=100\pm 2\,\mu$m radius used in the
experiment. The solid line shows the theoretical Casimir force calculated with
account of the corrections due to the skin depth and the surface roughness.
This force was found using the fourth order perturbation theory in the
relative skin depth $\delta_0/a$ defined in Sec.\ II.D.1. The perturbation
theory (Klimchitskaya et al., 1999; Bezerra et al., 2000a), when applied
in the appropriate region of the plasma model, leads to approximately the
same results as the Lifshitz formula at zero temperature (first
measurements using the AFM were not of sufficient precision to
measure the thermal effect at $T=300\,$K).
The Casimir force including the effect of both the skin depth and roughness
corrections was calculated using Eq.\ (\ref{eq61}) and the PFA (\ref{eq11}).
As is seen in Fig.\ 5, the solid line is in a very good agreement
with data, thus, demonstrating the role of skin depth and roughness
corrections.

We now consider in more detail the most precise, third, measurement
of the Casimir force between metal surfaces by means of an AFM
(Harris et al., 2000). The use of a thin Au/Pd coating on the top
of the Al in the first two experiments prevented a complete theoretical
treatment of the properties of the real metal surfaces. Because of this,
in the third experiment, Au layers of $86.6\pm 2\,$nm thickness were
coated on both the plate and the spherical surfaces. Such a coating
is sufficient to reproduce the properties of an infinitely thick metal
(see Sec.\ III.B.2). The sphere diameter (including the metal coating)
was measured using the scanning electron microscope to be
$191.3\pm 0.5\,\mu$m (independent calibration of the scanning electron
microscope was done with  interferometrically calibrated AFM piezo
standards).
The cantilever was calibrated and the residual potential between
the surfaces was measured using
the electrostatic force at separations
larger than $3\,\mu$m with voltages
$V$ from 3\,V to --3\,V applied to the plate.
As a result the following values were obtained: $m=8.9\pm 0.3\,$nm per unit
deflection signal, $km=0.386\pm 0.003\,$nN per unit
deflection signal, $a_0=32.7\pm 0.8\,$nm, and the residual potential
on the grounded sphere $V_0=3\pm 3\,$mV. At separations of about 60\,nm
this residual potential leads to forces which are only 0.075\% of
the Casimir force. At $a=100\,$nm the contribution of the residual
electric force increases up to 0.17\% and at $a=160\,$nm up to 0.36\%
of the Casimir force.

The magnitudes of the residual potential and separation on contact
were shown to be independent on distances where the measurements of
the electric force were performed.
The electric force corresponding to the residual potential
difference has been subtracted from the measured total force
to obtain the pure Casimir force.
This is in fact a correction made to remove the systematic deviation due
to the residual potential difference, as discussed in Sec.~III.C.1.
{}From the above numbers it becomes clear
that the electrostatic residuals are negligible in comparison with the
Casimir force. The electric force due to the patch potentials contributes
only 0.23\% and 0.008\% of the Casimir force at separations
$a=62\,$nm and $a=100\,$nm, respectively (Chen et al., 2004).
Recently it was claimed that the residual potential $V_0$ from the
electrostatic calibration in sphere-plate configuration is separation
dependent (Kim et al., 2008). The authors used an Au-coated sphere
of 30.9\,mm radius at separations of a few tens nanometers
above an Au coated plate. On the basis of these
measurements a reanalysis of the independence of $V_0$ on separation
in the measurements of the Casimir force by means of an atomic force
microscope and a micromachined oscillator (see Sec.~IV.B) was invited.
Decca et al. (2008) demonstrated, however, that for a centimeter-size
spherical lens at such short separations from the plate, the
electrostatic force law used by Kim et al. (2008) is not applicable
due to the inevitable deviations from a perfect spherical shape of
the mechanically polished and ground surface and the presence of
dust or other impurities.
Because of this, the
observed anomalies in the electrostatic calibration are
not directly relevant
to the experimental results considered in the present and next sections.

The averaged Casimir force measured from 27 scans is reproduced in Fig.\ 6
as dots within the measurement range from 62 to 300\,nm. In the same
figure the solid line represents the Casimir force between ideal
metal interacting surfaces. This once again clearly demonstrates
the difference between ideal metal surfaces and real materials.

The experimental errors in the experiment by Harris et al. (2000)
were re-analyzed by Chen et al. (2004a). The maximum value of the
systematic error at a 95\% confidence level was shown to be
$\Delta^{\! s}F^{\rm expt}=2.7\,$pN and of the
random error $\Delta^{\! r}F^{\rm expt}=5.8\,$pN over the entire
measurement range. Using the combination rule in Eq.\ (\ref{eq66}),
the total experimental error of the force measurements is given by
$\Delta^{\! t}F^{\rm expt}=6.8\,$pN.
It does not depend on separation.
The respective relative error
varies from 1.5 to 2\% when the separation increases from 63 to 72\,nm.
It increases to 4.8 and 30\% when separation increases to 100 and 200\,nm,
respectively.
 The error in the
measurements of absolute separations was found to be $\Delta a=1\,$nm.
All errors are determined at a 95\% confidence level. The experimental
data over the separation region from 63 to 100\,nm are shown in Fig.\ 7
as crosses where the sizes of the errors in the
separation and force measurements are
given in true scales. To make the figure readable, only every second data
point is presented.
We underline that the measurement of the Casimir force described above is
independent in the sense that no fit to any theory of dispersion forces
has been used. The above procedures used a fit only to the well understood
electric force in order to find the values of some parameters like
the separation on contact $a_0$, the residual potential difference
$V_0$ and deflection coefficient $m$.

Theoretical calculation of the Casimir force by Harris et al. (2000)
was performed using the Lifshitz formula at zero temperature (\ref{eq8})
and the PFA (\ref{eq11}). The dielectric permittivity of Au along the
imaginary frequency axis was obtained using the first approach
described in Sec.\ III.A.1, i.e., with the optical data extrapolated to
low frequencies by means of the Drude model (the dashed line in Fig.\ 1).
The character of roughness was investigated from the analysis of the
AFM images of the surfaces of Au films. Typical AFM image of surface
topography is presented in Fig.\ 8. As is seen in this figure, the
roughness is mostly represented by the stochastically distributed
distortions with the typical heights of about 2--4\,nm, and rare pointlike
peaks with the typical heights up to 16\,nm. Zero roughness levels on
both the plate and the sphere were found from Eq.\ (\ref{eq51}) to be
$H_0=2.734\,$nm (Chen et al., 2004a).
The rms variance defined in Eq.\ (\ref{eq55}) is equal to
$\delta_{\rm st}=1.18\,$nm. The role of roughness in this experiment
is very small. Even at the shortest separation $a=62\,$nm roughness
contributes only 0.24\% of the Casimir force. The diffraction-type
and correlation effects discussed in Sec.\ III.B.1 were shown to
contribute less than one tenth of this value (Chen et al., 2004a).

The comparison of the data with the Lifshitz theory at $T=0$ taking the
surface roughness into account demonstrated very good agreement
over the entire measurement range (Harris et al., 2000; Chen et al., 2004a).
However, as was discussed in Sec.\ II.D.1, the use of the Drude model
at low frequencies presents serious difficulties. In addition, according
to Sec.\ III.A.2, the use of the zero-temperature Lifshitz formula for
the interpretation of the experiment performed at $T=300\,$K is
open to discussion. In Sec.\ III.A.1 a second approach
to the calculation of Au dielectric permittivity along the imaginary
frequency axis, proposed in the literature, was described.
It is based on the plasma-like dielectric permittivity with the
inclusion of the relaxation due to the core electrons alone (the solid line
in Fig.\ 1). This approach can be used in combination with the Lifshitz
formula at $T\neq 0$ with no thermodynamic inconsistencies.
The calculations of the Casimir force in the experimental configuration
of Harris et al. (2000) at $T=300\,$K using the generalized plasma model
(i.e., with the dielectric permittivity given by the solid line in Fig.\ 1)
were performed by Klimchitskaya et al. (2007a). The obtained Casimir
force, as a function of separation, is shown by the solid band in Fig.\ 7.
The width of this band indicates the total theoretical error
calculated at a 95\% confidence level as described in Sec.\ III.C.2
taking into account the error of the PFA and uncertainties in the
optical data of Au related to the interband transitions.
As is seen in Fig.\ 7, the theoretical approach using the generalized
plasma-like permittivity is also in a very good agreement with the data.
Within the separation region from 63 to 72\,nm, where the relative total
experimental error determined at a 95\% confidence level varies from
1.5 to 2\%, all experimental crosses overlap with the theoretical
band. {}From this one can conclude that the measure of agreement between
experiment and theory at separation distances from 63 to 72\,nm also
varies from 1.5 to 2\% of the measured force.

Thus both theoretical approaches using the complete
optical data for Au extrapolated to low frequencies by the Drude model
and the Lifshitz formula at $T=0$, or the generalized plasma-like
dielectric permittivity and the Lifshitz formula at $T=300\,$K are
consistent with the experimental data by Harris et al. (2000).
This is because at separations below 100\,nm, where the precision
of this experiment is relatively high, the theoretical approaches used
lead to almost coincident results. At such short separations approximately
the same theoretical results are obtained also by using the complete
optical data extrapolated to zero frequency by the Drude model and the
Lifshitz formula at $T=300\,$K. However, at larger separation distances
the calculational results obtained using the above two approaches at nonzero
temperature become significantly different. Because of this, high precision
experiments at separations of a few hundred nanometers, considered in the
next section, are very important for the understanding of
underlying physics.

\subsection{ Precise determination of the Casimir pressure using
a micromachined oscillator}
\label{sec:mmo}

Microelectromechanical systems are well adapted for the investigation
of small forces acting between closely spaced surfaces.
One such system, a {\it micromachined oscillator}, was first used
by Chan et al. (2001a, 2001b) to demonstrate the influence of the
Casimir force on the static and dynamic properties of micromechanical
systems (see Sec.\ IV.C.2).

Precise determination of the Casimir pressure between two parallel
metallic plates by means of
a micromachined oscillator was performed in three successive
experiments by Decca et al.
These experiments do not use the configuration of two parallel plates.
The first experiment was made with a
Au-coated sphere and Cu-coated plate (Decca et al., 2003a, 2003b, 2004).
The second experiment with several improvements, used both
an Au-coated sphere and a plate (Decca et al., 2005b;
Klimchitskaya et al., 2005).
Further improvements implemented in the third experiment
with an Au-coated sphere and Au-coated plate
(Decca et al.,
2007a, 2007b) made it the most precise and reliable measurement with
metallic test bodies ever performed in the Casimir force measurements
to date. Here, we briefly discuss the
measurement scheme and the main physical results following from the
third experiment by Decca et al. (2007a, 2007b).

In metrological terms this is an {\it indirect measurement}
(Rabinovich, 2000) of the Casimir pressure between two Au-coated
parallel plates.
Note that the results of {\it direct measurements} are found just from
the experiment. The results of {\it indirect measurements} are obtained
with the help of calculations using the known equations
which relate the quantity under consideration with some quantities
measured directly.
The Casimir force per unit area (pressure)
was determined in these experiments dynamically by means
of a micromechanical torsional oscillator consisting of a
$500\times 500\,\mu\mbox{m}^2$ heavily doped polysilicon plate
suspended along one central planar axis by serpentine springs, and a sphere
of $R=151.3\,\mu$m radius above it attached to an optical fiber (see
Fig.\ 9). During the measurements, the separation
distance between the sphere and the plate was varied harmonically,
$\tilde{a}(t)=a+A_z\cos(\omega_rt)$,  where $\omega_r$ is the resonant
angular frequency of the oscillator under the influence of the
Casimir force $F_s(a,T)$ from the sphere, and $A_z/a\ll 1$.
The frequency $\omega_r$ is related to the natural
angular frequency of the oscillator
$\omega_0=2\pi\times(713.25\pm 0.02)\,$Hz by (Chan et al., 2001a, 2001b;
Decca et al., 2003a, 2003b)
\begin{equation}
\omega_r^2=\omega_0^2\left[1-\frac{b^2}{I\omega_0^2}\,
\frac{\partial F_s(a,T)}{\partial a}\right],
\label{eq4p2}
\end{equation}
\noindent
where $b$ is the lever arm, $I$ is the moment of inertia,
$b^2/I=1.2432\pm 0.0005\,\mu\mbox{g}^{-1}$.

The frequency shift $\omega_r-\omega_0$ is directly measured and
from Eq.\ (\ref{eq4p2}) it leads to the
calculated values of the Casimir force gradient
$\partial F_s/\partial a$. Using the PFA (\ref{eq11}) this gradient can be
expressed through the effective Casimir pressure in the configuration of
two parallel plates
\begin{equation}
P(a,T)=-\frac{\partial{\cal F}(a,T)}{\partial a}=
-\frac{1}{2\pi R}\,\frac{\partial F_s(a,T)}{\partial a}.
\label{eq4p3}
\end{equation}
\noindent
Because of this, a direct measurement of the frequency shift caused by the
Casimir force between a sphere and a plate results in an indirect measurement
of the Casimir pressure in the configuration of two parallel plates.

The separation distance between the zero roughness levels of Au layers on the
plate and on the sphere (see Sec.\ III.B.1)
is given by
\begin{equation}
a=z_{meas}-(D_1+D_2)-b\theta.
\label{eq4p4}
\end{equation}
\noindent
Here, $z_{meas}$ is the separation between the end of the cleaved fiber and
the platform, $\theta$ and $D_{1,2}$ are defined in Fig.\ 9. The lever
arm $b$ was measured optically. The value of $\theta$ was determined
by measuring the difference in capacitance between the plate and the
right and left electrodes shown in Fig.\ 9. The value $D_1+D_2$ was measured
as a part of the system calibration. This was done by the application of
voltages
$V$ to the sphere while the plate was grounded. The electric force
between a sphere and a plate was measured in the static regime, with no
harmonic variations of the separation distance between them, at large
separations $a>3\,\mu$m. The residual potential difference $V_0$ was
found to be independent of separation [details of calibration procedures
are presented by Decca et al. (2005b, 2008)]. We emphasize that in the second
(Decca et al., 2005b) and third (Decca et al., 2007a, 2007b) experiments
of this series, contact between the sphere and the plate has
not been achieved and was not needed for the determination of absolute
separations.

A set of 120 curves of $F_{el}(a)$ was then used to fit the quantity
$D_1+D_2$. Finally absolute separations $a$ were measured with
an absolute error $\Delta a=0.6\,$nm determined at a 95\% confidence.
In contrast to previous measurements (Decca et al., 2003a, 2003b, 2004,
2005b), in this experiment $n=33$ sets of measurements within a
separation region from 162 to 746\,nm were performed at almost
the same intermediate separations $a_i$ ($1\leq i\leq 293$) in each set.
This was made possible due to about 7\% improvement in the
vibrational noise, and an improvement in the interferometric technique
used to yield the distance $z_{meas}$ (see Fig.\ 9). The use of a two-color
fiber interferometer yielded an error $\Delta z_{meas}=0.2\,$nm, and
for every repetition of the Casimir pressure measurement it was possible
to reposition the sample to within $\Delta z_{meas}$.

The mean values of the Casimir pressure (\ref{eq64})
were determined from the frequency shifts
by using Eqs.\ (\ref{eq4p2}), (\ref{eq4p3}) from data obtained
in 33 sets of measurements.
They are plotted in Fig.\ 10 as a function of separation. The random
experimental error was found from Eqs.\ (\ref{eq64}) and  (\ref{eq65}).
It varies from $\Delta^{\! r}P^{\rm expt}=0.46\,$mPa at $a=162\,$nm to
$\Delta^{\! r}P^{\rm expt}=0.11\,$mPa at $a=300\,$nm. This value remains
constant for separations up to $a=746\,$nm.
The systematic error in the indirect measurement of the pressure is
determined by errors in the measurements of the resonant frequency,
radius of the sphere, and also by the error introduced by using the PFA.
The latter is now related to experiment rather than to theory, because
in this dynamic measurement PFA is a part of the experimental procedure
of the determination of the Casimir pressure between the two parallel plates.
According to the results presented in Sec.\ II.B, the error due to the use
of PFA is less than $a/R$. Because of this, Decca et al. (2007a, 2007b)
estimated this error conservatively as $a/R$. By combining all the
above $J=3$ systematic errors at a 95\% confidence using the
statistical rule (\ref{eq63}), the resulting systematic error was obtained.
It is equal to $\Delta^{\! s}P^{\rm expt}=2.12\,$mPa at $a=162\,$nm,
decreases to 0.44\,mPa at $a=300\,$nm, and then to 0.31\,mPa at
$a=746\,$nm. Finally, the total experimental error
$\Delta^{\! t}P^{\rm expt}(a)$ at a 95\% confidence level was obtained
using the statistical rule formulated in Sec.\ III.C.1.
As a result,
$\Delta^{\! t}P^{\rm expt}(a)=\Delta^{\! s}P^{\rm expt}(a)$ within
the entire measurement range, i.e., the total experimental error
is determined by only the systematic one.
This means that the experiment by
Decca et al. (2007a, 2007b) satisfies one of the main requirements
imposed on {\it precise} experiments in metrology (Rabinovich, 2000).
For now no other experiment in Casimir physics satisfies this
requirement. The total relative experimental error
varies from 0.19\% at $a=162\,$nm, to 0.9\% at $a=400\,$nm, and
to 9.0\% at the largest separation $a=746\,$nm.

To conclusively compare the experimental data with theory,
the topography of metallic coatings both on the plate (1) and on the
sphere (2) was investigated using an AFM. {}From the AFM images, the
fraction of each surface area $v_i^{(1,2)}$ with the height $h_i^{(1,2)}$
was determined. It was found that for the plate
$h_i^{(1)}$ varies from 0 to 18.35\,nm and for the sphere
 $h_i^{(2)}$ varies from 0 to 10.94\,nm. Using Eq.\ (\ref{eq51})
the zero roughness levels on the plate and on the sphere are equal to
$H_0^{(1)}=9.66\,$nm and $H_0^{(2)}=5.01\,$nm. The contribution
of correlation and diffraction-like effects in the roughness
correction was shown to be negligibly small (Decca et al., 2005b).
The overall contribution of roughness was shown to be only 0.5\%
of the Casimir pressure at $a=162\,$nm and it decreases with the
increase of separation. To conclusively compare experimental data
with theory, the resistivity of the Au layers as a function of
temperature was also measured in the region from $T_1=3\,$K to 400\,K.
This has led to a slightly modified values of the Drude parameters,
$\omega_p=8.9\pm 0.1\,$eV, $\gamma=0.0357\,$eV to be used in
theoretical calculations (compare with $\omega_p=9.0,$eV,
$\gamma=0.035\,$eV in Sec.\ III.A.2).

The theoretical Casimir pressure as a function of separation between
the plates was calculated using Eqs.\ (\ref{eq60}) and (\ref{eq5})
with different approaches to the determination of the dielectric
permittivity along the imaginary frequency axis (see Sec.\ III.A.2)
and using the Leontovich surface impedance (Sec.\ II.A). The comparison
of experiment with theory was performed in two different ways
described in Sec.\ III.C.2 (Decca et al., 2007a, 2007b).
In Fig.\ 11(a,b) the mean experimental pressures are plotted as crosses
within the separation region from 500 to 600\,nm. For other separation
regions situation is quite similar (Decca et al., 2007a). The arms of
the crosses show the total experimental errors of separation and
Casimir pressures in true scales determined at a 95\% confidence level.
The light-gray band in Fig.\ 11(a) shows the theoretical results
computed using the generalized plasma-like dielectric permittivity,
i.e., with $\varepsilon_{Au}({\rm i}\xi)$, as given by the solid line
in Fig.\ 1. The width of the theoretical band in the vertical direction
indicates the total theoretical error (Sec.\ III.C.2).
It is equal to 0.5\%  of  the calculated pressure and arises from
the variations of the optical data of core electrons. Other factors,
such as patch potentials, were shown to be negligible (Decca et al., 2005b).
Note that when the experimental data with the errors, shown as crosses,
are compared
with the theoretical band computed for the entire measurement range,
the errors $\Delta a$ in the measurement of separations are irrelevant
to theory. The dark-gray band in Fig.\ 11(a) shows the theoretical
results computed using the complete tabulated data extrapolated to low
frequencies by means of the Drude model, i.e., with
$\varepsilon_{Au}({\rm i}\xi)$ shown as the dashed line in Fig.\ 1.
As is seen in Fig.\ 11(a), all data crosses overlap with the light-gray
band, but are separated by a gap from the dark-gray band. This means
that the theory using the generalized plasma-like permittivity is
consistent with experiment within the limits of the experimental
error. At the same time, a theory using the complete optical data
extrapolated by the Drude model is excluded by the data at a 95\%
confidence level.

As mentioned in Sec.~III.C.2, there are statements in the literature
that the choice of different sets of Drude parameters may lead to an
up to 5\% variation in the theoretical
values of the Casimir pressure (Pirozhenko et al., 2006).
According to Sec.\ III.C.2, the hypothesis that the thin films used in
the experiment possess such unusual Drude parameters are not supported by the
data. Nevertheless, the disagreement between the Drude model approach
and the experimental data only increases if some of the alternative
Drude parameters, as considered by Pirozhenko et al. (2006), are used.
As an illustration, in Fig.\ 11(b) the computation results obtained
using the Drude model approach with $\omega_p$ varying from 6.85\,eV
(Pirozhenko et al., 2006) to 9.0\,eV are plotted as the dark-gray band.
Note that the values of the relaxation parameter only slightly
influence the Casimir pressure. As is seen in Fig.\ 11(b), the use of any
alternative value of $\omega_p$ makes the disagreement between
the Drude model approach and the data even more acute.

Another way to compare theory with experiment considered in Sec.\ III.C.2
is illustrated in Fig.\ 12, where the differences
$P^{\rm theor}(a_i)-P^{\rm expt}(a_i)$ are shown as dots and solid lines
indicate the 95\% confidence interval $[-\Xi_P(a),\Xi_P(a)]$ for
the quantity $P^{\rm theor}(a_i)-P^{\rm expt}(a_i)$ (Decca et al., 2007b).
The values of $P^{\rm theor}$ in Fig.\ 12(a) are computed using the
generalized plasma-like model, whereas the values
of $\tilde{P}^{\rm theor}$ in Fig.\ 12(b) using the Leontovich surface
impedance (Decca et al., 2005b). As is seen in Fig.\ 12, both theoretical
approaches are consistent with the data. Note that the values of
theoretical pressures in Fig.\ 12 are computed at the experimental
separations $a_i$. As a result, these values are associated with the
additional errors $4\Delta a/a_i$ discussed in Sec.\ III.C.2.
They are the primary
theoretical errors at short separations in this approach to comparing theory
with experiment. The respective width of the confidence interval
$[-\Xi_P(a),\Xi_P(a)]$ is overestimated. The actual agreement of the
theoretical approach with data can be characterized by the
deviation of the differences $P^{\rm theor}(a_i)-P^{\rm expt}(a_i)$
from zero. At the shortest separations this deviation exceeds the total
experimental error but (for the theory using the generalized plasma-like
permittivity) at $a>300\,$nm both are approximately equal.
The measure of agreement between experiment and theory given by
$\Xi_P/|\bar{P}^{\rm expt}|$, as defined in Sec.~III.C.2, is equal
to 1.9\% and 1.8\% at separations $a=162$ and 400\,nm, respectively.

In Fig.~13 the same method of comparing theory with experiment is applied to
the theoretical approach
based on the Drude model. The solid and dashed lines indicate
the limits of the 95\% and 99.9\% confidence intervals, respectively
(Decca et al., 2007b). The differences
$P_D^{\rm theor}(a_i)-P^{\rm expt}(a_i)$ are shown as dots. As can be seen
in the figure, the theoretical approach using the extrapolation of
the optical data to low frequencies by means of the Drude model is
experimentally excluded at a 95\% confidence level for the entire
measurement range from 162 to 746\,nm. Also note that the
differences $P_D^{\rm theor}(a_i)-P^{\rm expt}(a_i)$ are outside
the 99.9\% confidence interval at separations from 210 to 620\,nm.
Finally Decca et al. (2007a, 2007b) concluded that their experiments cannot
be reconciled with the Drude model approach to the thermal Casimir force.
They also remarked that the experiments with the micromachined
torsional oscillator are not of sufficient precision to measure the small
thermal effect between the two metal bodies at separations of a few hundred
nanometers, as predicted by the generalized plasma-like model.
The proposed experiments capable of this goal are considered in Sec.\ IV.D.

The experimental data by Decca et al. (2007a, 2007b) were also compared
with the theoretical results obtained using the
modification of the transverse magnetic reflection coefficient in
accordance with Eq.~(\ref{eq35}) (Dalvit, Lamoreaux, 2008).
A charge carrier density
$n\approx 5.9\times 10^{22}\,\mbox{cm}^{-3}$ at $T=300\,$K and
the Thomas-Fermi screening length were used
in the computations for Au.
In Fig.\ \ref{fgLam1} the differences between the computed theoretical
Casimir pressures and the experimental data by
Decca et al. (2007a, 2007b) are shown as dots. As is seen in the figure,
the data exclude the theoretical approach using the modification
(\ref{eq35}) at a 95\% confidence level within the entire measurement
region from 160 to 750\,nm and at a 99.9\% confidence level within
the measurement region from 160 to 640\,nm.
Note that for metals the same theoretical results, as in the approach
by Dalvit and Lamoreaux (2008), are obtained in the approaches by
Pitaevskii (2008) and Svetovoy (2008). Thus, the last two approaches
are also inconsistent with the experimental data by Decca et al.
(2007a, 2007b).

Another experiment performed by means of the micromechanical
torsional oscillator was the test of corrections to the PFA
(Krause et al., 2007). Taking these corrections into account,
the Casimir force between a sphere of radius $R$ and a plate can be
presented in the form (Scardicchio, Jaffe, 2006)
\begin{equation}
F_s(a,R)=2\pi RE(a)\left[1+\beta\,\frac{a}{R}+
O\left(\frac{a^2}{R^2}\right)\right].
\label{eqn84a}
\end{equation}
\noindent
Here $E(a)$ is the Casimir energy per unit area of two parallel plates and
$\beta$ is a dimensionless parameter characterising the lowest order
deviation from the PFA. The constraints on the parameter $\beta$ in
Eq.~(\ref{eqn84a}) can be obtained from the static measurements of the
Casimir force. In these measurements the separation distance between
the sphere and the plate is not varied harmonically and the force
$F_s(a,R)$ is a directly measured quantity.

Dynamic measurements of the Casimir pressure described above in this section
are more precise than the static ones. Substituting the Casimir force
(\ref{eqn84a}) into the right-hand side of Eq.~(\ref{eq4p3}),
one obtains the following expression for the effective Casimir pressure
(Krause et al., 2007)
\begin{equation}
P^{\rm eff}(a,R)=P(a)\left[1+\tilde\beta(a)\,\frac{a}{R}+
O\left(\frac{a^2}{R^2}\right)\right],
\label{eqn84b}
\end{equation}
\noindent
where $P(a)$ is the Casimir pressure between two parallel plates and
the dimensionless quantity $\tilde\beta(a)$ is given by
\begin{equation}
\tilde\beta(a)=\beta\left[1-\frac{E(a)}{aP(a)}\right].
\label{eqn84c}
\end{equation}
\noindent
Note that for ideal metal bodies $\tilde\beta(a)=2\beta/3={\rm const}$.

To obtain constraints on $\beta$ and $\tilde\beta$ a series of both
static and dynamic measurements has been performed with Au coated
plate and spheres with radii $R=10.5$. 31.4, 52.3, 102.8 and $148.2\,\mu$m.
The static measurement of the Casimir force between the sphere and the
plate was performed at separations from 160\,nm to 750\,nm in 10\,nm
steps. The dynamic determination of the effective Casimir pressure
between two parallel plates was done at separations from 164 to 986\,nm
with 2\,nm steps. The influence of the effects of the nonzero skin depth and
surface roughness on the dominant first-order correction to the PFA
was estimated to be of order 10\% and 1\%, respectively.
The comparison between data and theory at separations $a<300\,$nm
leads to the result $|\tilde\beta(a)|<0.4$ at a 95\% confidence level.
In the same separation region, $|\beta|<0.6$ was obtained
(Krause et al., 2007). These constraints are compatible with the exact
results for a cylinder-plate configuration, but are not in
agreement with  the
extrapolations made for a sphere above a plate (see the
discussion in Sec.~II.B).

\subsection{Other experiments on the Casimir force between metal bodies}
\label{sec:other}

\subsubsection{Torsion pendulum experiment}

Chronologically, the first experiment, in the more recent series
of the Casimir force measurements was performed by Lamoreaux (1997).
While this experiment rekindled interest
in the investigation of the Casimir force and stimulated further
development of the field, the obtained results contain several
uncertainties. The Casimir force between an Au coated spherical lens
and a flat plate was measured using a torsion pendulum.
A lens with a radius $R=11.3\pm 0.1\,$cm [later corrected to
$12.5\pm 0.3\,$cm (Lamoreaux, 1998)] was mounted on a piezo stack and
a plate on an arm of the torsion balance in vacuum. The other arm
of the torsion balance formed the center electrode of a dual parallel plate
capacitor. The positions of this arm and consequently the angle of
the torsion pendulum were controlled by the application of voltages to the
plates
of the dual capacitor. The Casimir force between the plate and the lens
would result in a torque, leading to a change in the angle of the torsion
balance. This change results in changes of the capacitances. Then
compensating voltages were applied to these capacitances to counteract
the change in the angle of the torsion balance. These compensating
voltages were a measure of the Casimir force.

The calibration of the measurement system was done electrostatically.
When the lens and plate surfaces were grounded, a ``shockingly large''
(Lamoreaux, 1997) residual potential difference $V=430\,$mV was
measured. The respective large electrostatic force was
compensated with application of voltage to the lens. However, there
appears to have been a large residual electrostatic force
even after this compensation, which
was determined only by fitting to the total force
including the Casimir force above $1\,\mu$m.
``Typically, the Casimir force had magnitude of at least 20\% of
the electrical force at the point of closest approach'' (Lamoreaux, 1997).
The uncertainty in the measurement of absolute separations $\Delta a$
``was normally less than $0.1\,\mu$m'' (Lamoreaux, 1997).

Data were compared with  theory for the ideal metal lens
and plate. The obtained conclusion that there is an agreement at
the level of 5\% in the 0.6 to $6\,\mu$m range (Lamoreaux, 1997)
was found to be incompatible with the magnitude of the thermal correction
to the
Casimir force (Bordag et al., 1998).
Thus, at separations of 4, 5 and $6\,\mu$m
the thermal correction is 86\%, 129\% and 174\% of the zero-temperature
force, respectively. The data, however, were found to be ``not of
sufficient accuracy to demonstrate the finite temperature correction''
(Lamoreaux, 1997). {}From this it follows that the agreement of the data
with the theory at the level of
5\%--10\% may exist only
at separations of about $1\,\mu$m. Here, the thermal correction is
relatively small, whereas the larger corrections due to the
skin depth and surface roughness have the opposite sign and partially
compensate each other. Keeping in mind that theoretical forces calculated
at experimental separations are burdened with an additional error of
about $3\Delta a/a\approx 30$\% at $a=1\,\mu$m [see Sec.\ III.C.2 and
the original paper by Iannuzzi et al. (2004a)], the errors in
the differences $F^{\rm expt}(a_i)-F^{\rm theor}(a_i)$, as shown in Fig.\ 4,
bottom (Lamoreaux, 1997), are significantly underestimated.

The results of this experiment at about $1\,\mu$m separation were used
(Torgerson, Lamoreaux, 2004; Lamoreaux, 2005) to exclude the theoretical
approach to the thermal Casimir force which uses the Drude model at
low frequencies. The latter predicts --19\% thermal correction at
$a=1\,\mu$m (to be compared with a 1.2\% thermal correction in the case of
ideal metals) which was not experimentally observed. The
conclusion that the Drude model approach is inconsistent with data
is in agreement with the same conclusion obtained at a high confidence
level in the much more precise experiments
of Decca et al. (2003b, 2004, 2005b, 2007a, 2007b) (see Sec.\ IV.B).

\subsubsection{Micromechanical devices actuated by the Casimir force}

Chan et al. (2001a) used a micromachined torsional device actuated
by the Casimir force. A more advanced
version of this device shown in Fig.\ 9
was later used in the most precise measurements of the Casimir force
(see Sec.\ IV.B). When the sphere in Fig.\ 9 was moved closer to the
plate in a vacuum with a pressure of less than 1\,mTorr, the Casimir force,
acting on the plate, tilted it about its central axis towards the sphere.
Thus, vacuum oscillations of the electromagnetic field led to the
mechanical motion of the plate demonstrating the first micromechanical
device driven by the Casimir force. Using a micromachined oscillator,
Chan et al., (2001a) measured the Casimir force between an Au coated
polystyrene sphere and a plate.
The comparison of the measurement data with
the ideal metal Casimir result again demonstrated
the influence of skin depth and surface roughness on the
Casimir force between real material
bodies [previously demonstrated by Mohideen and Roy (1998) and
Harris et al. (2000), see Figs.\ 5 and 6].

A similar device was used to investigate the influence of the Casimir
force on the oscillatory behavior of micromachines (Chan et al., 2001b).
The simple model of the Casimir oscillator is shown in Fig.\ 15.
It consists of a movable metallic plate which
is subject to both the restoring
force of a spring, and the Casimir force from the interaction with
a fixed metallic sphere. The force from the spring is linear
in the movement of the plate $\Delta a$, whereas the Casimir force is
nonlinear in $\Delta a$.
The potential energy of this microdevice possesses a local
and a global minima separated by the potential barrier.
The Casimir force changes the resonant frequency of oscillations around
the local minimum, and makes
the oscillations anharmonic (Chan et al., 2001b).
These properties may be useful in future micromechanical systems driven
by the Casimir force.

\subsubsection{The experiment using the parallel plate configuration}

The only experiment in the recent series of Casimir force measurements
which uses the
original configuration of two plane plates was performed by Bressi et al.
(2002). A Si cantilever and a thick plate rigidly connected to a
frame (source) both covered with a Cr layer with adjustable separation distance
between them were used as two plates. The coarse separation distance
was adjusted with a dc motor and fine tuning was achieved
using a linear piezo-electric transducer
attached to the frame. Calibration was performed using the
electrostatic force. As a result, the error in absolute separation was
found to be $\Delta a=35\,$nm. Small oscillations induced on the source
by the application of a sinusoidal voltage to the piezo induce
oscillations of the cantilever through the Casimir force.
Thus, this experiment is dynamical like those considered in Sec.IV.B.
The motion of
the resonator placed in the vacuum was detected by means of a fiber optic
interferometer. After subtracting the electrostatic forces, the
residual frequency shift is given by
\begin{equation}
\Delta\nu_{\rm expt}^2(a)=\nu^2-\nu_0^2=-\alpha
\frac{\partial P(a)}{\partial a}.
\label{eq4p5}
\end{equation}
\noindent
Here, $\nu_0=138.275\,$Hz is the natural frequency in the absence of the
Casimir pressure $P$ and
$\alpha=S/(4\pi^2m_{\rm eff})\approx 0.0479\,\mbox{m}^2/\mbox{kg}$ ($S$
being the area of a capacitor formed by the plates,
$m_{\rm eff}$ is the effective mass).

Since the measurement is a dynamic one, the directly
measured quantity is the frequency shift (\ref{eq4p5})
arising from the effect of the Casimir force. This frequency shift
is related to the gradient of the Casimir pressure using
Eq.\ (\ref{eq4p5}). Thus, though this experiment,
uses the configuration of two parallel plates, it is an indirect
measurement of the gradient of the
the Casimir pressure between the plates, similar to  the
experiments using the sphere-plate configuration in the dynamic regime
for the same purpose (see Sec.\ IV.B).
Bressi et al. (2002) did not aim to restore the Casimir pressure
from the pressure gradient. Instead, they fitted the experimental data
$\Delta\nu_{\rm expt}^2(a)$ to $\Delta\nu_{\rm theor}^2(a)$
computed from the theoretical dependence of the Casimir pressure
between two ideal metal plates, $-K_C/a^4$, with a free parameter
$K_C$. The best fit resulted in
$K_C=(1.22\pm 0.18)\times 10^{-27}\,\mbox{N\,m}^2$. This was compared
with the exact Casimir coefficient for ideal metal plates in Eq.\ (\ref{Cas})
$K_C=\pi^2\hbar c/240=1.3\times 10^{-27}\,\mbox{N\,m}^2$.
The conclusion was drawn that the related force coefficient is determined
at the 15\% precision level (Bressi et al., 2002).

{}From the point of view of a general method for the comparison of
experiment with theory, which was developed after this experiment was
performed, it would be reasonable not to use any fit but
to compare $\Delta\nu_{\rm expt}^2(a)$
with the exact expression for $\Delta\nu_{\rm theor}^2(a)$ with no
adjustable parameters like $K_C$. In Fig.\ 16(a) the results of such
comparison are presented where the experimental data are shown as crosses
and the solid line shows $\Delta\nu_{\rm theor}^2$ computed
from the exact expression for the Casimir pressure between two ideal metal
plates, as given in Eq.\ (\ref{Cas}). As can be seen in the figure, at
separations below $1\,\mu$m the experimental crosses only touch the
solid line. This can be explained by the role of the nonzero skin depth.
If instead of $P(a)=P_0(a)$, one uses the Casimir pressure with
the first and
second order corrections due to the nonzero skin depth (Bezerra et al., 2000a)
\begin{equation}
P(a)=-\frac{\pi^2\hbar c}{240 a^4}\left(1-\frac{16}{3}\,\frac{c}{\omega_pa}+
24\frac{c^2}{\omega_p^2a^2}\right),
\label{eq4p6}
\end{equation}
\noindent
the agreement with data improves. As an illustration,
$\Delta\nu_{\rm theor}^2$ is recalculated using the theoretical Casimir
pressure (\ref{eq4p6}) with the $\omega_p\approx 13\,$eV for Cr.
The results are shown as the solid line in Fig.\ 16(b). It is seen that
the Casimir pressure taking the skin depth into account is in much better
agreement with the data than the ideal metal Casimir pressure.
Suggestions on how to improve the sensitivity of this experiment,
as proposed in the literature, are discussed in Sec.\ IV.D.

\subsubsection{The Casimir force between thin metallic films}

Lisanti et al. (2005) reported the observation of the skin-depth effect
on the Casimir force between metallic surfaces. The Casimir
force between a thick plate and a $100\,\mu$m radius polystyrene sphere
coated with metallic films of different thicknesses was measured.
The sphere was positioned above a micromachined torsional balance.
The Casimir attraction between the sphere and the top plate of the balance
induced a rotation angle which was measured as a function of the separation
between the surfaces. Without the indication of errors and confidence
levels, it was reported that the Casimir attraction between the metallic
plate and the sphere with coatings thinner than the skin-depth is
smaller than that of the same plate and a sphere with thick metal coating.
Physically, this is in fact the same effect as is demonstrated in
Figs.\ 5 and 6 in Sec.\ IV.A where the Casimir forces acting between ideal
and between real metals were compared with corrections based on the
plasma frequency (finite conductivity corrections). Ideal metals are better
reflectors and
the magnitude of the Casimir force between them is larger than between real
metals. In a similar manner thick real metal films are better reflectors
than thin real metal films.

Lisanti et al. (2005) compared their experimental results with the
Lifshitz theory at zero temperature adapted for the description of
layered structures (Zhou, Spruch, 1995; Klimchitskaya et al., 2000;
Toma\u{s}, 2002). It was shown that the experimental forces obtained
for films of thickness smaller than the skin depth have smaller
magnitudes than those computed for such films using the Lifshitz theory.
This result is not surprising because, as
it was argued in the literature, for films of
small thickness the effects of spatial dispersion should be taken into
account which are not included in the Lifshitz theory (Klimchitskaya
et al., 2000).

\subsubsection{Dynamic measurement using an atomic force microscope}

Jourdan et al. (2009) have recently performed a measurement of the Casimir
force
between an Au-coated sphere of $R=20\,\mu$m radius and a plate using
an AFM. No direct sphere-plate contact was used to determine
absolute separation distances between the surfaces. The AFM cantilever
with the glued sphere were considered as a harmonic oscillator with
the natural resonant frequency $\omega_0=2\pi\times 50182\,$rad/s
which is modified by the effect of the Casimir force.
In these two aspects (direct contact between the two surfaces was avoided and
the dymanic measurement mode was used) the experimental approach
resembles the one used by Decca et al. (2005b, 2007a, 2007b) in the
measurements of the Casimir pressure by means of a micromechanical
torsional oscillator. The uncertainty in the absolute separations,
$\Delta a=2\,$nm, was, however, more than 3 times larger than in
experiments with a micromachined oscillator (see Sec.\ IV.B).
The experimental data for the force gradient were compared with the
Lifshitz theory at zero temperature. The dielectric permittivity along
the imaginary frequency axis was found using the complete optical
data (Palik, 1985) extrapolated to zero frequency by means of the
Drude model with $\omega_p=9.0\,$eV and $\gamma=0.035\,$eV (Lambrecht,
Reynaud, 2000). The correction due to surface roughness was not taken
into account. The discrepancy between the force gradient measurements
and the results of the theoretical computations
described above were found to be within 3\%
of the theoretical force at separations between 100 to 200\,nm.
The measured Casimir force gradient was compared also with that computed
using ideal metal surfaces and a deviation was reported.
On this basis it was concluded that the experimental data demonstrate
again the finite conductivity (skin-depth) effects on the Casimir force.

\subsubsection{Ambient Casimir force measurements}

In this recent period two other ambient (open to air) experiments
measuring the
Casimir force were reported. Both did not use a vacuum environment and
reported the presence of water layers on the interacting
surfaces.  The first of these
was the experiment by Ederth  (2000), where the force was
measured between two cylindrical template stripped gold surfaces with
0.4 nm roughness, in a distance range from 20 to 100\,nm.
The 200\,nm gold films were fixed to 10\,mm
radius silica cylinders using a ``soft glue''. In addition a hydrocarbon
layer of hexadecanethiol was applied to the interacting
surfaces. It was noted that this top hydrocarbon layer was
necessary to preserve the purity of the gold surface in the
ambient environment.  This hydrocarbon organic layer prevented
a direct measurement of the electrostatic forces or an
independent determination of the surface separation. Also
the presence of the hydrocarbon layer means that the
experiment cannot be strictly classified as that between
two gold surfaces. One of the
cylinders was mounted on a piezo and the other on bimorph
deflection sensor.  The charge produced by the bimorph in
response to the deflection induced by a force on the
interacting surfaces was measured by an electrometer amplifier. The
soft glue led to a deformation which was estimated
to be 18-20\,nm for glue thicknesses used. To compensate for this
deformation the two measured force-distance curves were shifted, one
by 9 and the other by 12 nm respectively to overlap the
calculations. The author reports that ``it is not possible
to establish with certainty'' the validity of the
displacement due to the
deformation and that it ``diminishes the strength of the
measurement as a test of the Casimir force and also precludes
a quantitative assessment of the agreement between theory  and
experiment.''

A second experiment done in an ambient environment using the AFM
for separation distances between 12 and 200\,nm
was reported recently (van Zwol et al., 2008).  Here a
Au-coated sphere of $R=20\,\mu$m radius was fixed to a gold
coated AFM cantilever. The Au coated plate was mounted on the
piezo. Both the sphere and plate were coated with 100\,nm of Au.
Here the optical properties of the Au coated plate were measured with an
ellipsometer in
the wavelength region from
137\,nm to $33\,\mu$m and fit to obtain the plasma frequency of
$7.9\pm 0.2\,$eV and a relaxation frequency of $0.048\pm 0.005\,$eV since
the finite conductivity corrections for the separation
range considered are large. The roughness or the water
layer was not taken into account in the fit.
The errors in the cantilever spring constant
and the deflection coefficient were reported to be 4\% and 3\%
respectively, which together were reported to lead to errors
of 4\%--10\%.
The calibration errors were reported to
lead to an overall error of about 5\%--35\%.
The electrostatically measured contact potential
$10\pm 10\,$mV was reported to lead to a 10\% error.
The authors report that they were not able to independently
determine the separation on contact of the two surfaces
due to the
stiff cantilevers employed. Based on the roughness they
estimate a 1\,nm error in the contact separation
``leading therefore to a $~$28\% relative
error at the smallest separations.'' However, a general 10\% agreement with
the theory was reported below 100\,nm separation.
Given the ambient nature of the experiment,
``typically a few nanometers'' water layer on both surfaces were
present but not treated in the theoretical comparison or systematic errors.
Repeating the experiments in a vacuum environment should
allow for a more definitive comparison.

\subsubsection{Related measurements}

An interesting preliminary test of the Lifshitz theory for three-layer
systems is the measurement of the attractive Casimir force between
an Au-coated sphere and a plate immersed in ethanol using an AFM
(Munday, Capasso, 2007). The obtained experimental data were compared
with the Lifshitz theory taking into account the frequency dependence
of the dielectric functions of Au and ethanol, and the correction
due to surface roughness. Consistency of the obtained data with
Lifshitz's theory was claimed although at separations below 50\,nm
a disagreement has been observed which increases with the decrease of
separation. However, as was commented in the literature (Geyer et al., 2008b),
the theoretical computations of the Casimir force between the smooth Au
surfaces separated by ethanol done according to the method
provided by Munday and Capasso (2007) [i.e., by the use of the Kramers-Kronig
relations and tabulated optical data (Palik, 1985)]
lead to a discrepancy up to 25\%
with respect to the reported theoretical results.
The latter can be reproduced if one uses at all imaginary frequencies the
Drude dielectric function (\ref{eq29}) for both
the sphere and plate materials.
A second drawback is that the effect
of the residual potential difference between the sphere and the plate was
calculated incorrectly and significantly underestimated by a factor of 590.
Finally, the possible interaction between the double layer formed
in liquids due to salt impurities,
which would decrease the electrostatic force, was not taken
into account without any justification. The resulting electrostatic
force is of the same magnitude as the Casimir force to be measured.
All this makes the interpretation of this experiment uncertain
(Geyer et al., 2008b). In the reply (Munday, Capasso, 2008) it was
recognized that the original paper (Munday, Capasso, 2007) did in fact use
the Drude model. It was also recognized that the equation originally used
to estimate the residual electrostatic force ``is not strictly correct'' and
that salt contaminants exist even in the purest solutions leading to
the screening of electrostatic interaction.
In a later work
(Munday et al. 2008) the effect of electrostatic forces and Debye
screening on the measurement of the Casimir force in fluids was further
investigated. The electrostatic force with account of Debye screening
was calculated as outlined by Geyer et al. (2008b). The influence of
the concentration of salt impurities on the
Debye screening was investigated.
No measurable change in the force with or without grounding of the
sphere was reported.
The Casimir force was measured 51 times
with the standard deviation which varies from 130 to 90\,pN when
separation increases from 30 to 80\,nm (the respective variances are
less by a factor of $\sqrt{51}$). As a result, the relative
random experimental
error of the Casimir force measurements in this experiment determined at
a 67\% confidence level is equal to approximately 7\% at $a=30\,$nm
and increases to 60\% at $a=80\,$nm. Bearing in mind that for
$a<30\,$nm the authors recognize deviations between the theory and the
measurement data, this experiment can be considered as only a qualitative
demonstration of the Casimir force in fluids.

Another experiment used an adaptive holographic interferometer to
measure periodical nonlinear deformations of a thin pellicle caused by
an oscillating Casimir force due to a spherical lens
(Petrov et al., 2006). Both test bodies were coated with a thin Al film and
placed in a vacuum chamber. The lens was mounted on a vibrating piezodriver.
As a result, the oscillations of the lens position led to a periodic
modulation of the Casimir force. The experimental data were found to be
in only qualitative agreement with the theory based on ideal metal
boundaries at separation distances of a few hundred nanometers.
Corrections due to the finite skin depth and surface roughness
were not provided.
Also the use of Al whose surface undergoes rapid oxidation
even in relatively high vacuum adds uncertainty to
the results of this experiment.
However, the new measurement technique used may be promising
for future measurements of the Casimir force.

\subsection{Future prospects to measure the thermal Casimir force}
\label{sec:therm}

As described in Secs.~IV.A-C, the measurements of the Casimir force
between macroscopic bodies performed to date were not of sufficient
precision to measure the magnitude of the thermal effect.
The experiments using the micromachined oscillator (Sec.\ IV.B) possess
the highest experimental precision at separations below $1\,\mu$m. They have
been used to exclude thermal effects, as predicted by models using the
Drude model at low frequencies. However, thermal effects predicted by
the generalized plasma-like model at short separations remain below the
experimental sensitivity. In this respect, large separation measurements
of the Casimir force would be of great interest. At separations of a few
micrometers
the thermal regime is reached where the Casimir free energy is
entirely of thermal origin. Calculations using the plasma model or the
generalized plasma-like model result in Eq.\ (\ref{eq28a}).
The Drude model approach to the thermal Casimir force leads to only
one half of the result for ideal metal plates [Eq.\ (\ref{eq28a}) with
$\omega_p\to\infty$]. Thus, large-separation measurements of the
Casimir force would bring direct information to bear on the magnitude
of the thermal effect between macroscopic bodies (such a measurement
in the atom-plate configuration has been performed already, see Sec.\ VI.A).

In Sec.~IV.C.1 it was shown that the experiment using the
torsion pendulum within the region from 0.6 to $6\,\mu$m is in fact
uncertain at separations above $2\,\mu$m where the thermal effects
begin to make a substantial contribution. An analysis (Lamoreaux, Buttler,
2005) shows that the torsion pendulum technique has the potential
to measure the Casimir force between a plate and a spherical lens
at $a=4\,\mu$m with a relative error of 10\%. Bearing in mind that at
$a=4\,\mu$m the thermal correction contributes as much as 86\% of the
zero-temperature Casimir force, such an experiment, if successfully
performed, holds great promise.

There is a proposal aimed at measuring the Casimir force in the
cylinder-plate configuration at separations around $3\,\mu$m
(Brown-Hayes et al., 2005). This geometry can be considered as a
compromise between the two parallel plate configuration (which is
connected with serious experimental difficulties associated with the
parallelity of the plates) and a sphere above
a plate. In addition, as discussed in Sec.\ II.B, the exact solution
for the ideal metal cylinder-plate configuration has been obtained recently.
This gives the possibility to determine the accuracy of the PFA and
to apply it to real materials with high reliability. Finally it was
concluded that using the dynamic measuring scheme it is possible to
measure the Casimir force between a plate and a cylinder at
separations of about $3\,\mu$m with a precision of a few percent
(Brown-Hayes et al., 2005).

Another proposal suggests to use a highly-sensitive torsion balance in
the separation range from 1 to $10\,\mu$m to measure the Casimir force
in the configuration of two parallel plates (Lambrecht et al., 2005).
The construction of the balance is similar to that used in the
E\"{o}tvos-type experiments aimed to test the equivalence principle.
It is planned to measure the thermal Casimir force with an accuracy
of a few percent and to discriminate between different
theoretical approaches
discussed in the literature.

One more experiment exploiting the two parallel plate configuration
is proposed at separations larger than a few micrometers (Antonini et al.,
2006). The experimental scheme is based on the use of a Michelson-type
interferometer and the dynamic technique with one oscillating plate.
Calibrations show that a force of $5\times 10^{-11}\,$N can be measured
in this setup with the relative error from about 10 to 20\%.
This would be sufficient to measure the thermal effect at a separation
of $5\,\mu$m (Antonini et al., 2006).

The thermal effect in the Casimir force can be measured at short separations
below $1\,\mu$m if the difference in the thermal forces $\Delta F$ at
different temperatures rather than the absolute value of the thermal
Casimir force is measured (Chen et al., 2003). For real metals, this
difference of the thermal Casimir forces (in contrast with the relative
thermal correction) does not increase but decreases with increasing
separation distance. This allows the observation of the thermal
effect on the Casimir force at small separations of about $0.5\,\mu$m where
the relative thermal correction, as predicted by the generalized
plasma-like model, is rather small. Preliminary estimation shows that
with a sphere of $R=2\,$mm radius attached to a cantilever of an AFM the
measurable changes of the force amplitude of order $10^{-13}\,$N are
achievable from a 50\,K change in the temperature. Such a temperature
difference
can be obtained by the illumination of the sphere and the
plate surfaces with laser pulses of $10^{-2}\,$s duration (Chen et al.,
2004b). The idea of exploiting difference force measurements to
probe the thermal Casimir effect in superconducting cavities was
proposed by Bimonte (2008).

{}From the above discussion it is clear, that experimental investigations
of thermal effects in the Casimir force appear feasible in the
near future.

\section{ CASIMIR FORCE BETWEEN A METALLIC SPHERE AND A
SEMICONDUCTOR PLATE}
\label{sec:semi}

\subsection{Motivation for use of semiconductors}
\label{sec:why}

A vital issue in many applications of the Casimir effect is how
to control the magnitude of the force by changing the parameters of
the system. A natural method for this control is to change the
material properties of the interacting bodies. Arnold et al. (1979) made
an attempt to modify the van der Waals and Casimir forces between
semiconductors with light. Attractive forces were measured between a glass
lens and a Si plate and also between a glass lens coated with
amorphous Si and a Si plate. The glass lens, however, is an insulator and
therefore the electric forces, such as due to work function potential
differences, could not be controlled. This might explain why Arnold
et al. (1979) found no force change occurred on illumination at separations
below 350\,nm, where it should have been most pronounced.
One more attempt to modify the Casimir force was made by Iannuzzi et al.
(2004b) when measuring the Casimir force acting between a plate and a sphere
coated with a hydrogen-switchable mirror that becomes transparent upon
hydrogenation. Despite expectations, no significant decrease of the
Casimir force owing to the increased transparency was observed.
The negative result is explained by the Lifshitz theory which requires
the change of the reflectivity properties within a wide range of
frequencies in order to markedly affect the magnitude of the Casimir force.
This requirement is not satisfied by hydrogenation.

The appropriate materials for the control, modification and fine tuning of
the Casimir force are semiconductors. The reflectivity properties of
semiconductor surfaces can be changed in a wide frequency range by
changing the carrier density through the variation of temperature,
using different kinds of doping, or, alternatively, via illumination
of the surface with laser light. At the same time, semiconductor
surfaces with reasonably high conductivity avoid accumulation of excess
charges and, thus, preserve the advantage of metals for Casimir
force measurements. In addition, as semiconductors are the basic fabrication
materials for nanotechnology, the use of semiconductor surfaces for the
control of the Casimir force will lead to many applications.

\subsection{ Optically modulated Casimir force}
\label{sec:opt}

The modification of the Casimir force through the radiation induced
change in the carrier density was first reliably demonstrated by
Chen et al. (2007a, 2007b). A high vacuum based AFM was employed to
measure the change in the Casimir force between an Au-coated sphere and
a Si membrane in the presence and in the absence of incident light.
An oil-free vacuum chamber with a pressure of around $2\times 10^{-7}\,$Torr
was used. A polystyrene sphere of diameter $2R=197.8\pm 0.3\,\mu$m
coated with an Au layer of $82\pm 2\,$nm thickness was mounted at the tip
of a $320\,\mu$m conductive cantilever (the general scheme of the experiment
is shown in Fig.\ \ref{fgVp1}). A specially fabricated Si membrane
[see Chen et al. (2007b) for preparation details] was mounted on top of
the piezo which is used to change the separation distance $a$ between the
sphere and the membrane from contact to $6\,\mu$m.
The excitation of the carriers in the Si membrane was done with 5\,ms
wide light pulses (50\% duty cycle). These pulses were obtained from
a cw Ar ion laser light at 514\,nm wavelength modulated at a frequency
of 100\,Hz using an acousto-optic modulator (AOM). The laser pulses were
focused on the bottom surface of the Si membrane. The Gaussian width of
the focused beam on the membrane was measured to be $w=0.23\pm 0.01\,$mm.
The resulting modification of the Casimir force in response to the
carrier excitation was measured with a lock-in amplifier
(see Fig.\ \ref{fgVp1}).
The same function generator signal used to generate the Ar laser
pulses was also used as a reference for the lock-in amplifier.

The illumination of the Si has to be done such that very little if any
light impinges on the sphere, as this would lead to a light induced
force from the photon pressure. As the Si membrane is illuminated from
the bottom, care should be taken that the fraction of light transmitted
through the membrane is negligibly small. The optical absorption depth for
Si at a wavelength of 514\,nm is equal to $1\,\mu$m. The thickness of
the membrane used was measured to be $4.0\pm 0.3\,\mu$m. The calculations
presented showed
that for the membrane thicknesses used, the force on the sphere due to
photon pressure
varies from 2.7 to 8.7\% of the difference of dispersion forces to be
measured when separation increases from 100 to 200\,nm.

The calibration of the setup, determination of the cantilever deflection
coefficient and the average separation on contact between the test
bodies are performed as in earlier experiments with metal test bodies
[see Sec.\ IV.A and original papers (Harris et al., 2000; Chen et al.,
2006b, 2007b; Chiu et al., 2008). The deflection coefficient was found
to be $m=137.2\pm 0.6\,$nm per unit deflection signal. The difference
in the value of $m$ from previous measurements (see Sec.\ IV.A) is due
to the use of the 514\,nm optical filter (see Fig.\ \ref{fgVp1}) which
reduces the cantilever deflection signal. This filter was used to prevent
the interference of the 514\,nm excitation light with the cantilever
deflection signal from the 640\,nm laser (see Fig.\ \ref{fgVp1}).
The separation distance on contact was determined to be $a_0=97\,$nm.
The uncertainty in the quantity $a_0+mS_{\rm def}$ was found to be 1\,nm.
This leads to practically the same error $\Delta a=1\,$nm in the absolute
separations (\ref{dist}), because the error in piezo calibration is
negligibly small.
An independent measurement of the lifetime of the carriers excited in
the Si membrane by pulses from the Ar laser was performed using a
noninvasive optical pump-probe technique (Sabbah, Riffe, 2002;
Nagai, Kuwata-Gonokami, 2002).

In the experiment under consideration it is not sufficient to determine
the residual potential difference $V_0$ between the sphere and the
membrane as described in Sec.\ IV.A. In the presence of a pulse train
the values of the residual potential difference can be different:
$V_0^l$ and $V_0$ during the bright and dark phases of a laser pulse
train, respectively. Both values are generally different
from the residual potential difference determined in the absence of
pulse train. In the measurement procedure, the voltages $V^l$ and $V$
are applied to the membrane during the bright and dark phases, respectively.
The difference in the total force (electric and Casimir) for the
states with and without carrier excitation was measured. Using
Eq.\ (\ref{eq4p1}), this difference can be written in the form:
\begin{equation}
\Delta F_{\rm tot}(a)=X(a)\left[(V^l-V_0^l)^2-(V-V_0)^2\right]+
\Delta F(a).
\label{eq5p2}
\end{equation}
\noindent
Here $\Delta F(a)$ is the difference of the Casimir forces for the
states with and without light. By keeping $V={\rm const}$ and
changing $V^l$, the parabolic dependence of $\Delta F_{\rm tot}$
on $V^l$ was measured. The value of $V^l$ where this function
reaches a maximum [recall that $X(a)<0$ in Eq.\ (\ref{eq4p1})] is
$V_0^l$. Then by keeping $V^l={\rm const}$ and changing $V$, the
parabolic dependence of $\Delta F_{\rm tot}$
on $V$ was measured, which allows one to find $V_0$.
Both procedures were repeated at different separations and the values
$V_0^l=-0.303\pm 0.002\,$V and $V_0=-0.225\pm 0.002\,$V were found to be
independent of separation in the range from 100 to 500\,nm.
These values were shown to change only within the resolution error. It
was stated that the independence of the residual potential difference
of separation is a basic and necessary condition for every Casimir
force measurement. The dependence of $V_0$ of separation indicates
the presence of electrostatic surface impurities, space charge effects
and/or electrostatic
inhomogeneities on the sphere or plate surface (Chiu et al., 2008).
The small
change in the residual potential difference between the sphere and
the membrane in the presence and in the absence of excitation light is
primarily due to the screening of surface states by few of the optically
excited electrons and holes. The value of this difference of around
78\,mV is equal to the change in band bending at the surface.
It is consistent
with the fact that almost flat bands are obtained at the surface with the
surface passivation technique used for the preparation of the Si membrane
(Chen et al., 2007b).

Then, other voltages ($V^l,V$) were applied to the Si membrane and the
difference in the total force $\Delta F_{\rm tot}$ was measured as
a function of separation within the interval from 100 to 500\,nm.
{}From these measurement results, the difference in the Casimir force,
$\Delta F^{\rm expt}(a)$, was determined from Eq.\ (\ref{eq5p2}).
This procedure was repeated with some number of pairs ($J$) of different
applied voltages $(V^l,V)$ and at each separation the mean value
$\langle \Delta F^{\rm expt}(a)\rangle$ was found.
The measurements were performed for different absorbed laser powers:
$P^{\rm eff}=9.3\,$mW ($J=31$), 8.5\,mW ($J=41$), and 4.7\,mW ($J=33$),
corresponding to the incident powers of 15.0, 13.7 and 7.6\,mW, respectively.
The experimental data for $P^{\rm eff}=9.3\,$mW and 4.7\,mW are shown
by dots in Figs.\ \ref{fgVp3}(a) and (b), respectively. As expected,
the magnitude of the Casimir force difference has the largest values at
the shortest separations and decreases with the decrease of separation.
It also decreases with the decrease of the absorbed laser power.

The analysis of the experimental errors performed in line with Sec.\ III.C.1
(Chen et al., 2007b) shows that the total experimental error in this
experiment is equal to the random one. The relative experimental error
at a 95\% confidence level varies from 10\% to 20\% at a separation
$a=100\,$nm and  from 25\% to 33\% at a separation
$a=180\,$nm for different absorbed laser powers. As an example, in
Fig.\ \ref{fgVp4} the experimental data are shown with their absolute
errors calculated at a 95\% confidence level for the absorbed power
4.7\,mW in the separation region from 150 to 200\,nm [each third dot from
Fig.\ \ref{fgVp3}(b) is shown].
Thus, the modulation of the dispersion force with light is demonstrated
by Chen et al. (2007a, 2007b) at a high reliability and confidence.

For the comparison of experimental results with theory, the difference of
dispersion forces between a sphere and a plate was calculated according
to the first equality in Eq.\ (\ref{eq11}) with the free energy given
by the Lifshitz formula (\ref{eq1}) (Chen et al., 2007a, 2007b).
The surface roughness was taken into account by using geometrical
averaging (\ref{eq61}). It was shown that the contribution from the
roughness correction to the calculated value of $\Delta F(a)$ is very
small. At the shortest separation $a=100\,$nm it is equal to only 1.2\%
and decreases to 0.5\% at $a=150\,$nm.
The calculations were done at the laboratory temperature $T=300\,$K.

Both cases of the dielectric permittivity of the high-resistivity
($10\,\Omega\,$cm) Si in
the absence of light were considered, i.e.,
 the inclusion or the neglect of the
dc conductivity (see Sec.\ II.D.2). The dielectric permittivity of the
dielectric Si (no light and the dc conductivity is disregarded) along
the imaginary frequency axis was found from the tabulated optical data
(Palik, 1985). It is shown in Fig.\ \ref{fgVp5} as the long-dashed line.
Here, $\varepsilon_0=11.66$. Note that the use of the analytic
approximation for $\varepsilon_{Si}$ (Inui, 2003) leads to about 10\%
error in the magnitudes of the Casimir force (Chen et al., 2006b) and,
thus, is not suitable for the comparison with precise measurements.
The dielectric permittivity of the high-resistivity Si with dc conductivity
included (the charge carrier density
$\tilde{n}\approx 5\times 10^{14}\,\mbox{cm}^{-3}$)
is shown in Fig.\ \ref{fgVp5} by the short-dashed line.

On irradiation of the Si membrane with light, an equilibrium value of
the carrier density is rapidly established during a period of time much
shorter than the duration of a laser pulse (note that the lifetimes
were independently measured). This allows one to assume
that there is an equilibrium concentration of pairs (electrons and holes)
when the light is incident. In this case the dielectric permittivity along
the imaginary frequency axis is commonly represented in the form
[see, e.g., Palik (1985), Vodel et al. (1992), Inui (2004, 2006)]
\begin{equation}
\varepsilon_{Si}^{(l)}({\rm i}\xi)=\varepsilon_{Si}({\rm i}\xi)+
\frac{\omega_{p(e)}^2}{\xi[\xi+\gamma_{(e)}]}+
\frac{\omega_{p(p)}^2}{\xi[\xi+\gamma_{(p)}]},
\label{eq5p3}
\end{equation}
\noindent
where $\omega_{p(e,p)}$ and $\gamma_{(e,p)}$ are the plasma frequencies
and the relaxation parameters of electrons and holes, respectively
[$\gamma_{(e)}\approx 1.8\times 10^{13}\,$rad/s,
$\gamma_{(p)}\approx 5.0\times 10^{12}\,$rad/s (Vogel et al., 1992),
the values of the plasma frequencies are given below].
To avoid violation of the thermal equilibrium
arising from the use of the Drude model (see Sec.\ II.D.1),
it was suggested (Mostepanenko, Geyer, 2008) to include the influence
of free charge carriers in metal-type semiconductors by means of the
generalized plasma-like model (\ref{eq45a}).
This rule should be also applied to all doped semiconductors with dopant
concentration above critical.
Then, instead of
Eq.\ (\ref{eq5p3}), the dielectric permittivity is given by
\begin{equation}
\varepsilon_{Si}^{(l)}({\rm i}\xi)=\varepsilon_{Si}({\rm i}\xi)+
\frac{\omega_{p(e)}^2}{\xi^2}+
\frac{\omega_{p(p)}^2}{\xi^2}.
\label{eq5p4}
\end{equation}

The plasma frequency can be calculated from
\begin{equation}
\omega_{p(e,p)}=\left(\frac{ne^2}{m_{e,p}^{\ast}\epsilon_0}\right)^{1/2},
\label{eq5p5}
\end{equation}
\noindent
where the effective masses are $m_p^{\ast}=0.2063m$,
$m_e^{\ast}=0.2588m$, and $n$ is the concentration of charge carriers
(Vogel et al., 1992). The value of $n$ for the different absorbed powers
can be calculated with the knowledge that at equilibrium the number
of charge-carrier pairs created per unit time per unit volume is equal to
the recombination rate of pairs per unit volume. This leads to
(Chen et al., 2007b)
\begin{equation}
n=\frac{4P_w^{\rm eff}\tau}{\hbar\omega d\pi w},
\label{eq5p6}
\end{equation}
\noindent
where $P_w^{\rm eff}=0.393P^{\rm eff}$ is the power in the central part of
the Gaussian beam focused on the membrane (see above), $\tau$ is the measured
life time of the charge carriers.
For the absorbed power $P^{\rm eff}=9.3\,$mW [Fig.\ \ref{fgVp3}(a)] it
follows from Eqs.\ (\ref{eq5p6}) and (\ref{eq5p5}) that
$n_a=(2.1\pm 0.4)\times 10^{19}\,\mbox{cm}^{-3}$,
$\omega_{p(e)}=(5.1\pm 0.5)\times 10^{14}\,$rad/s,
$\omega_{p(p)}=(5.7\pm 0.6)\times 10^{14}\,$rad/s  (Chen et al., 2007b).
Here and below all errors are found at a 95\% confidence level.
For $P^{\rm eff}=4.7\,$mW [Fig.\ \ref{fgVp3}(b)]
the same parameters are
$n_b=(1.4\pm 0.3)\times 10^{19}\,\mbox{cm}^{-3}$,
$\omega_{p(e)}=(4.1\pm 0.4)\times 10^{14}\,$rad/s,
$\omega_{p(p)}=(4.6\pm 0.4)\times 10^{14}\,$rad/s.
[Note that the original publication by Chen et al. (2007b) contains two
misprints corrected here.]
The corresponding dielectric permittivities are shown in Fig.\ \ref{fgVp5}
with the solid lines labeled $a$ and $b$. It should be
remarked that the dielectric
permittivities (\ref{eq5p3}) and (\ref{eq5p4}) almost overlap
at all nonzero Matsubara frequencies.

The dielectric permittivity of Au along the imaginary frequency axis is
presented in Fig.\ 1. For the dielectric permittivity of Si in the dark
phase, and
the bright phase when Eq.\ (\ref{eq5p3}) is used, the reflection
coefficient is $r_{\rm TE}^{Si}(0,k_{\bot})=0$. Thus, the
calculation results do not depend on whether the solid or the dashed line
in Fig.\ 1 for $\varepsilon_{Au}({\rm i}\xi)$ is used.
If in the presence of light $\varepsilon_{Si}^{(l)}({\rm i}\xi)$ is
given by Eq.\ (\ref{eq5p4}), then the Au dielectric permittivity in the
framework of the generalized plasma-like model should be used (the solid
line in Fig.\ 1).

The results of the numerical computations of the difference Casimir force,
$\Delta F^{\rm theor}(a)$, between rough surfaces with the dielectric
permittivity (\ref{eq5p4}) in the presence of light and by neglecting
dc conductivity in the absence of light are shown by the solid lines
in Fig.\ \ref{fgVp3} (for different absorbed powers). The results of
the analogous calculations with the dielectric
permittivity (\ref{eq5p3}) in the presence of light are shown
in Fig.\ \ref{fgVp3} with the dotted lines. As is seen from Fig.\ \ref{fgVp3},
both the solid and dotted lines are in good agreement with the
experimental data shown by dots. Thus, the experiment under consideration
does not allow one to discriminate between the dielectric permittivities
(\ref{eq5p3}) and (\ref{eq5p4}) used above for the description of charge
carriers in metal-type semiconductors. This is also seen in
Fig.\ \ref{fgVp6} where the dots present the values of the differences between
$\Delta F^{\rm theor}(a)$ and $\langle\Delta F^{\rm expt}(a)\rangle$ for
the absorbed power 4.7\,mW and the solid lines show the 95\% confidence
interval
calculated as described in Sec.\ III.C.2 (Chen et al., 2007b;
Mostepanenko, Geyer, 2008). Dots labeled 1 in Figs.\ \ref{fgVp6}(a) and
\ref{fgVp6}(b) are obtained with $\varepsilon_{Si}({\rm i}\xi)$ in the
dark phase but with the dielectric permittivities (\ref{eq5p3}) or
 (\ref{eq5p4}) in the bright phase, respectively. It can be observed that
the dots
labeled 1 are inside the confidence intervals in both Figs.\ \ref{fgVp6}(a)
and \ref{fgVp6}(b). To distinguish between the two models using the
dielectric permittivities (\ref{eq5p3}) and (\ref{eq5p4}) of metal-type
semiconductors more precise experiments are required (see Sec.\ V.D.2).

The calculation of the theoretical differences $\Delta F^{\rm theor}(a)$
using (in the absence of light) the dielectric permittivity with
the inclusion of dc conductivity leads to the results presented by the
dashed lines
in Figs.\ \ref{fgVp3} and \ref{fgVp4}. As is seen in the figures, this
theoretical prediction is excluded by the data. The same can also be observed
in Fig.\ \ref{fgVp6}(a) where the dots labeled 2 present the differences
between the theoretical results calculated with the small dc
conductivity of the Si membrane included in the absence of light and the
experimental
data. Almost all the dots labeled 2 are outside the 95\% confidence interval.
This means that a model which includes the dc conductivity of the
high-resistivity Si is excluded by the experimental data at a 95\%
confidence level within the separation region from 100 to 200\,nm.

The results of this experiment were applied (Klimchitskaya et al., 2008b)
to test the modification of the transverse magnetic reflection coefficient,
as given by Eq.~(\ref{eq35}) (Dalvit, Lamoreaux, 2008) and
Eq.\ (\ref{eq43a}) with the standard contributions of all
nonzero Matsubara frequencies (Pitaevskii, 2008).
The experimental data for the difference of the Casimir forces
in the presence and in the absence of laser light are shown in
Fig.\ \ref{fgVpLP} for the absorbed power (a) of 9.3\,mW and (b) of 4.7\,mW.
In contrast to Fig.\ \ref{fgVp4}, here the experimental errors are
presented at a 70\% confidence level. The computational results obtained
on the basis of the standard Lifshitz theory with the dc conductivity
neglected in the dark phase are shown by the solid lines. The dashed lines
are computed using Eq.\ (\ref{eq35}) for the modified TM reflection
coefficient at all $x=\xi_l$
with concentration of charge carriers $n=\tilde{n}$ and the
Debye-H\"{u}ckel length
in the dark phase, and $n=2n_a$ or $n=2n_b$ in the presence of light.
The computation of the dashed lines was repeated using Eq.\ (\ref{eq43a})
for the TM reflection coefficient at zero frequency.
At all nonzero Matsubara frequencies, the standard terms of the Lifshitz
formula were used.
In both cases practically coinciding computational resuts were obtained
shown as the dashed lines in Fig.~22(a,b).
As seen in
Fig.\ \ref{fgVpLP}(a,b), the experimental data are consistent with the
theoretical results computed on the basis of the standard Lifshitz theory
with the dc conductivity of dielectric Si neglected in the dark phase
(the solid lines). The theoretical results computed using Eq.\ (\ref{eq35})
at all $\xi_l$
or Eq.\ (\ref{eq43a}) with the standard contributions of all nonzero
Matsubara frequencies
are excluded by data at a 70\% confidence level (Klimchitskaya et al., 2008b).

According to Svetovoy (2008), the experimental data for the difference
Casimir force are equally consistent with the nonlocal approach using
the reflection coefficient (\ref{eq43a}) and the Lifshitz theory with
dc conductivity neglected in the dark phase. To prove this, the
experimental data of Fig.~22(a) at a 70\% confidence level were used,
but the dashed line was replaced with the theoretical band
whose width was
determined at a 95\% confidence level using the respective uncertainty
in charge carrier density $\Delta n=0.4\times 10^{19}\,\mbox{cm}^{-3}$.
Such a mismatched comparison of experiment with theory is irregular.
It can be easily seen that the theoretical bands related to the solid and
dashed lines in Fig.~22(a) do not overlap if one uses in computations
the uncertainty in charge carriers,
$\Delta n=0.3\times 10^{19}\,\mbox{cm}^{-3}$,
determined at the same
70\% confidence level as the experimental errors
(Mostepanenko et al., 2009).

The experimental results presented in this section support the conclusion
made in Sec.~II.D.2. In accordance with this conclusion, one should
disregard the role of free charge carriers in all dielectric materials
(in particular, in semiconductors with dopant concentration below
critical) when calculating dispersion forces in the framework of the
Lifshitz theory.

\subsection{Doped semiconductors with different charge carrier
densities}
\label{sec:doped}

\subsubsection{$p$-type silicon}

The most important materials used in nanotechnology are semiconductors with
conductivity properties ranging from the metallic to the dielectric.
As was mentioned in Sec.\ V.A, semiconductors with a relatively high
conductivity have an advantage that they avoid accumulation of residual
charges but, at the same time, possess a typical dielectric dependence of
the dielectric permittivity on frequency within a wide frequency range
(see, e.g., Fig.\ \ref{fgVp5}). This makes it possible to examine the
influence of doping concentration on the Casimir force. Chen et al.
(2005, 2006b) first measured the Casimir force between an Au coated
sphere of the diameter $2R=202.6\pm 0.3\,\mu$m (an Au layer of 105\,nm
thickness was used) and $5\times 10\,\mbox{mm}^2$ single crystal silicon
Si$\langle 100\rangle$ plate. The resistivity of this $p$-type B-doped
plate measured by using the four-probe technique was found to be
$\rho=0.0035\,\Omega\,$cm.

The Casimir force was measured with the help of an improved version of
the setup
previously used (Harris et al., 2000) for the two Au test bodies (see
Sec.\ IV.A). The main improvements in the experimental setup were the
use of much higher vacuum and the reduction of the uncertainty in the
determination of absolute separation $a$ [see Eq.\ (\ref{dist})].
As in Sec.\ V.B,
a much higher vacuum ($2\times 10^{-7}\,$Torr) is needed to maintain
the chemical purity of the Si surface which otherwise oxidizes rapidly
to SiO${}_2$. The high-vacuum system consists of oil-free mechanical
pumps, turbo pumps, and ion pumps. To maintain the lowest pressure during
data acquisition, only the ion pump is used. This helps to reduce the
influence of mechanical noise. The absolute error in the determination of
absolute separation was reduced to $\Delta a=0.8\,$nm in comparison
to $\Delta a=1\,$nm (see Sec.\ IV.A). This was achieved by using a piezo
capable of travelling a distance $6\,\mu$m from initial separation to
contact. More details on the improvements are provided by Chiu et al.
(2008).

In contrast to Au, the Si surface is very reactive. Because of this,
a special passivation procedure is needed
to prepare it for force measurements
(Chen et al., 2006b). To characterize the topography of both samples,
the Au coating on the sphere and the surface of the Si plate were investigated
using an AFM. Images resulting from the surface scan of the Au coating
show that the roughness is mostly represented by stochastically
distributed distortions of about 8--22\,nm height with
$\delta_{\rm st}^{(1)}=3.446\,$nm. The surface scan of the Si surface shows
much smoother distortions with typical heights from 0.4 to 0.6\,nm
and $\delta_{\rm st}^{(1)}=0.111\,$nm.

All calibrations and determination of the residual electrostatic force and
the separation on contact were done immediately before the Casimir force
measurements in the same high-vacuum apparatus (see the description of
this procedure in Sec.\ IV.A). The force calibration constant was
determined to be $1.440\pm 0.007\,$nN per unit cantilever deflection
signal, and the residual potential difference $V_0=-0.114\pm 0.002\,$V.

In Fig.~\ref{fgVp7}, the mean measured Casimir force is presented
(average of 65 measurements). The variance of this mean is found
to be approximately the same over the entire measurement range
$62.33\,\mbox{nm}\leq a\leq 349.97\,$nm and equal to 1.5\,pN. Thus, the
random experimental error at a 95\% confidence level is 3.0\,pN.
The systematic error in this experiment, determined at a 95\% confidence
level (in line with Sec.\ III.C.1) is equal to 1.17\,pN.
The total experimental error, calculated by using Eq.\ (\ref{eq66}) at
a 95\% confidence level, is $\Delta^{\! t}F^{\rm expt}\approx 3.33\,$pN.
Thus, the relative experimental error varies from 0.87\% at the shortest
separation to 3\% at $a=100\,$nm and to 64\% at $a=300\,$nm.

In Fig.~\ref{fgVp8}, the experimental data with their error bars
are plotted as crosses for a more narrow separation range from 75 to 90\,nm.
The solid line in Fig.\ \ref{fgVp8} shows the theoretical results calculated
using the Lifshitz theory with the dielectric permittivity (\ref{eq5p3})
[for the $p$-type plate used $\omega_{p(e)}=0$,
$\omega_{p(p)}=7\times 10^{14}\,$rad/s (Chen et al., 2006b)].
The dashed line in the same figure presents the calculation results for
dielectric Si (the permittivity is shown by the long-dashed line
in Fig.\ \ref{fgVp5}). As is seen from Fig.\ \ref{fgVp8}, the solid line is
consistent with the experiment whereas the dashed line is inconsistent.
Thus, for the Si plate with a rather high doping concentration
($n\approx 3\times 10^{19}\,\mbox{cm}^{-3}$) the presence of free charge
carriers markedly influences the Casimir force.

\subsubsection{$n$-type silicon}

Further investigation of the difference in the Casimir force for samples
with different charge-carrier densities was performed for $n$-type Si
doped with P (Chen et al., 2006a). The above same high vacuum based
AFM was used to measure the Casimir force between an Au coated sphere of
a diameter $2R=201.8\pm 0.6\,\mu$m and two $4\times 7\,\mbox{mm}^2$ size
Si plates with different charge carrier densities placed next to each
other. The thickness of the Au coating on the sphere was measured to be
$92\pm 2\,$nm.

The two Si samples chosen for this experiment were identically polished,
single crystals, of $500\,\mu$m thickness. The resistivity of the samples was
measured using the four-probe technique to be $\rho_a=0.43\,\Omega\,$cm.
Thus, the concentration of the charge carriers
$n_a\approx 1.2\times 10^{16}\,\mbox{cm}^{-3}$ was well below the
critical concentration corresponding to the dielectric-metal transition
in Si doped with P ($n_{\rm cr}\approx 1.3\times 10^{18}\,\mbox{cm}^{-3}$).
One of these samples was used as the first Si plate in the experiment.
The other one was subject to thermal diffusion doping to prepare the
second, lower resistivity plate, with
$\rho_b\approx 6.7\times 10^{-4}\,\Omega\,$cm
($n_{b}\approx 3.2\times 10^{20}\,\mbox{cm}^{-3}$).
Both plates of higher and lower resistivity were subjected to a special
passivation procedure to prepare their surfaces for the force measurements.

All calibration procedures were done as in the previous measurements
(see Sec.\ IV.A), separately for the samples of higher and lower
resistivity. For the sample of higher resistivity, the value of the
residual potential difference $V_0=-0.341\pm 0.002\,$V
was obtained.
The cantilever deflection constant was found to be $1.646\pm 0.004\,$nN
per unit
deflection signal, and the separation on contact $a_0=32.4\pm 1.0\,$nm.
For the sample of lower resistivity
$V_0=-0.337\pm 0.002\,$V, $a_0=32.3\pm 1.0\,$nm, and the
cantilever deflection constant $1.700\pm 0.004\,$nN per unit
deflection signal was found.
Note that the expression for the electric force (\ref{eq4p1}) used in
the calibration does not take into account possible influence of
screening effects (space-charge layer at the surface of Si).
According to Bingqian et al. (1999), for high-resistivity $n$-type Si with
the concentration of charge carriers of order $10^{16}\,\mbox{cm}^{-3}$,
the impact of this layer on the electrostatic force is negligible at
separations from 300--400\,nm to $2.5\,\mu$m where the calibration fit
was performed.
Obviously in the case of lower resistivity Si the influence of
space-charge layer is even more negligible.
These conclusions were experimentally confirmed through the demonstration
that both the residual potential difference $V_0$ and the separation on
contact $a_0$ are the same regardless of the separation distance where the
calibration fit is performed
(Chen et al., 2006a; Chiu et al., 2008).

The Casimir force as a function of the separation was measured for both
samples immediately after the calibration procedure was done for each
one. The mean values of the measured Casimir force,
$\bar{F}^{\rm expt}$, are presented in Fig.\ \ref{fgVp9}.
Dots labeled ``a'' show the results for the sample of higher resistivity
(average of 40 measurements). Dots labeled ``b'' correspond to the
sample of lower resistivity (average of 39 measurements). As is seen from
 Fig.\ \ref{fgVp9}, dots labeled ``a'' and ``b''
 are distinct from each other
demonstrating the effect of different charge-carrier densities in
the two Si plates used.

The error analysis performed, as described in Sec.\ III.C.1, shows
that the total experimental errors determined at a 95\% confidence level are
equal to the random ones in each measurement. Thus,
$\Delta^{\! t}F_a^{\rm expt}=8\,$pN, 6\,pN and 4\,pN at $a=61.19\,$nm,
70\,nm and $a\geq 80\,$nm, respectively.
The measurement for the sample of lower resistivity is slightly more noisy.
Here, $\Delta^{\! t}F_a^{\rm expt}=11\,$pN, 7\,pN and 5\,pN at $a=61.19\,$nm,
70\,nm and $a\geq 80\,$nm. {}From Fig.\ \ref{fgVp9} it is seen that the
deviation between the two sets of data is larger than the total experimental
error in the separation region from 61.12 to 120\,nm.

The force-distance relationship measured for the two Si samples was compared
with theory. For the description of the sample with lower resistivity
the dielectric permittivity (\ref{eq5p3}) was used [with
$\omega_{p(p)}=0$, $\omega_{p(e)}=2.0\times 10^{15}\,$rad/s].
The sample of higher resistivity was described by the dielectric
permittivity of a dielectric Si (the long-dashed line in  Fig.\ \ref{fgVp5}).

In  Fig.\ \ref{fgVp11} the difference of the measured Casimir forces for the
plates of the lower and higher resitivity,
$F_b^{\rm expt}(a)-F_a^{\rm expt}(a)$, versus separation are shown as dots.
In the same figure, the difference in the respective theoretically computed
Casimir forces is shown by the solid line. As is seen from
Fig.\ \ref{fgVp11}, the experimental and theoretical results as functions
of $a$ are in a very good agreement.

The experiments, described in this section, demonstrate the possibility
of modifying the Casimir force by changing the doping concentration of
semiconductor materials.

Note that recently, a theoretical investigation of the influence
of Si doping concentration on the Casimir force in the separation region
from 1\,nm to 5\,mm was performed using the zero-temperature Lifshitz
formula (Pirozhenko, Lambrecht, 2008a). The obtained results, however,
are reliable only at separations below $1\,\mu$m, because the use of
the zero-temperature Lifshitz formula with the room-temperature parameters at
separations above $1\,\mu$m is physically problematic (Bordag et al., 2001).

\subsection{Silicon plate with rectangular trenches}

Chan  et. al. (2008)
reported a measurement of the Casimir
force between a gold coated sphere of radius $R$
and a silicon surface that had been
structured with nanoscale rectangular corrugations (trenches).
Measurements were performed in the dynamic regime using a micromechanical
torsional oscillator (see Sec.~IV.B). This means that the immediately
measured quantity was the change of the resonant frequency of the
oscillator which is proportional to the derivative of the Casimir
force with respect to separation distance. In accordance with the
proximity force approximation given by Eq.~(\ref{eq4p3}), this
derivative is equal to
\begin{equation}
F_s^{\prime}(a)=-2\pi R P(a),
\label{21.x}
\end{equation}
\noindent
where $P(a)$ is the Casimir pressure between one rectangular corrugated
plate and one plane plate (see Fig.~27).
Note that the experiment was done at room temperature, but the obtained
results were compared with theoretical computations at $T=0$.

For an ideal metal case such a configuration was considered by
B\"{u}scher and Emig (2004) using exact methods presented in Sec.~II.B.
It was shown that in the limiting case $a\ll\Lambda$ (recall that
separation is measured between the zero corrugation level of the lower
plate and the upper plate) the Casimir pressure for the fractional
surface area of solid volume equal to 1/2 is given by
\begin{equation}
P(a)=-\frac{\pi^2\hbar c}{240}\,\frac{1}{2}\,\left[
\frac{1}{\left(a-\frac{H}{2}\right)^4}+
\frac{1}{\left(a+\frac{H}{2}\right)^4}\right].
\label{21.x1}
\end{equation}
\noindent
This is in fact the minimum value of the Casimir pressure between the
rectangular corrugated and plane plates.  Under the condition $a\ll\Lambda$
for ideal metals the same result is obtained by the application of the
proximity force approximation or pairwise summation methods
(see Sec.~III.B.1).

In the opposite limit $a\gg\Lambda$ the exact result is given by
(B\"{u}scher, Emig, 2004)
\begin{equation}
P(a)=-\frac{\pi^2\hbar c}{240}\,
\frac{1}{\left(a-\frac{H}{2}\right)^4},
\label{21.x2}
\end{equation}
\noindent
which is up to a factor of 2 larger in magnitude than the prediction of
the proximity force approximation and paiwise summation in Eq.~(\ref{21.x1}).
This is the maximum value of the Casimir pressure in the configuration
of the rectangular corrugated and plane plates made of ideal metals.

The trenches were fabricated in $p$-doped
silicon
(the density of charge carriers $2\times 10^{18}\,\mbox{cm}^{-3}$
was determined from the dc conductivity which was
equal to $0.028\,\Omega\,$cm).
Silicon oxide was used as the mask.  Deep UV lithography followed by
reactive ion etching was used to transfer the pattern.
Trenches of approximate depth $1\,\mu$m
were created. Three types of samples were made on the same wafer: sample $A$
with $\Lambda_A=1\,\mu$m period, sample $B$ with $\Lambda=400\,$nm period
and one with a flat surface.
  The fractional
areas of solid volume were found to be $p_A=0.478\pm 0.002$
and $p_B=0.510\pm 0.001$ for
the samples $A$ and $B$ using a scanning electron microscope.
The residual
hydrocarbons were removed by oxygen plasma etching and the oxide mask was
etched with HF.  Samples of size 0.7\,mm by 0.7\,mm were used for the force
measurement.  The corrugated silicon surface was prepared as was
done by  Chen {\it et al.} (2006a, 2006b)
using hydrogen passivation.  However, after this the silicon
chip was baked to 120$\,{}^{\circ}$C
to remove the residual water from the bottom of the
trenches, which might impair the passivation layer leading to
patch potentials.

The micromachined oscillator used in the measurements
consists of a $3.5\,\mu$m thick, $500\,\mu$m square
silicon plate.  Unlike in previous experiments with
micromachined oscillators (see Sec.~IV.B),
the spheres were attached to the plate. Two sputter gold
coated with thickness of about 400\,nm glass spheres of radius
$R=50\,\mu$m were attached on
top of each other to the torsional oscillator at a distance of
$b=210\,\mu$m
using conductive epoxy. Two spheres are used to provide a large distance
between the corrugated surface and the top of the torsional oscillator.
The resonant frequency ($\omega_0=2\pi\times 1783\,$Hz,
quality factor $Q= 32000$) was excited by
applying voltage on one of the bottom electrodes.  The oscillations were
detected with additional voltages with
an amplitude of 100\,mV and a frequency
of $2\pi\times 102\,$kHz which were applied to measure
the capacitance change between the top
plate and the bottom electrodes. A phase locked loop was used to detect the
change in the resonant frequency as a function of the distance between the
sphere and the corrugated plate.

The measurements were done at a vacuum of $10^{-6}\,$torr using a dry roughing
pump and a turbo pump.  The residual potential difference $V_0$
and the initial
separation between the surfaces  $a_0$ were determined
and the calibration was done using
electrostatic forces. No value of $a_0$ or
errors in its determination were provided.  A residual potential difference
$V_0\sim -0.43\,$V was found between a sphere and a flat Si plate
and noted to vary with 3\,mV within the separation
distance 100\,nm to $2\,\mu$m.
If the same was found for the corrugated Si surfaces
was not mentioned.  Voltages between
$V_0+245\,$mV to $V_0+300\,$mV were applied
and the calibration constant was found to be
$628\pm 5\,\mbox{m\,N}^{-1}\,\mbox{s}^{-1}$. Since no
analytic expression for the electrostatic force is available for the trench
geometry a 2$D$ numerical solution of the Poisson equation was used to
calculate the electrostatic energy between a flat plate and the trench
surface. This energy was then converted to a force between a sphere and the
trench surface using the PFA.

The Casimir force gradients
between the flat plate and samples $A$ and $B$ were measured
after the application of compensating voltages to the plates. The main
uncertainty in these measurements is reported as that
coming from thermomechanical noise with a value of about
$0.64\,\mbox{pN\,$\mu$m}^{-1}$ at $a=800\,$nm.
  The Casimir force gradient
between the flat plate and the
gold sphere $F_{\rm flat}^{\prime}$ was first measured.
Good agreement was found with the calculation results using the
Lifshitz theory. The tabulated data for gold and silicon (Palik 1985) were
used along with the modification corresponding to the carrier density of
silicon (Chen et al.,  2006a).  The roughness correction was taken into
account using a rms roughness of 4\,nm on the sphere and 0.6\,nm on the
silicon surface measured using an AFM (see Sec.~III.B.1).
Next the force gradients $F_{A,{\rm expt}}^{\prime}$
and $F_{B,{\rm expt}}^{\prime}$ were measured on the
corrugated surfaces using the same gold sphere.

For the configurations used by Chan et al. (2008) one can neglect
the contribution from the remote bottom parts of trenches. Then the proximity
force approximation leads to the following
force gradients for the samples
$A$ and $B$
\begin{eqnarray}
&&
F_{A,{\rm PFA}}^{\prime}(a)=-2\pi Rp_AP^{(r)}\left(a-\frac{H}{2}\right),
\nonumber \\
&&
F_{B,{\rm PFA}}^{\prime}(a)=-2\pi Rp_BP^{(r)}\left(a-\frac{H}{2}\right),
\label{21.x3}
\end{eqnarray}
\noindent
where $P^{(r)}(a)$ is the Casimir pressure between two noncorrugated plates
covered with a stochastic roughness calculated using the Lifshitz formula
as described in Sec.~III.B.1.

To compare the experimental data with theory, Chan et al. (2008)
considered the ratios
\begin{equation}
\rho_A=\frac{F_{A,{\rm expt}}^{\prime}}{F_{A,{\rm PFA}}^{\prime}}
\qquad
\rho_B=\frac{F_{B,{\rm expt}}^{\prime}}{F_{B,{\rm PFA}}^{\prime}}.
\label{21.x4}
\end{equation}
\noindent
It was shown that for sample $A$ there are deviations of $\rho_A$ from
unity up to 10\% over the measurement range from $a=650$ to 750\,nm
exceeding the experimental errors. For the sample $B$
 there are deviations of $\rho_B$ from
unity up to 20\% over the same measurement range.
This difference in the comparison of the experimental data for the
samples $A$ and $B$ with the PFA results is natural, as
$a/\Lambda_A=0.7$ and $a/\Lambda_B=1.75$ at a typical separation
distance considered $a=700\,$nm. Thus, for sample $B$ the applicability
condition of the PFA, $a/\Lambda\ll 1$, considered in Sec.~III.B.1
is violated to a larger extent
than for the sample $A$.

The measurements were repeated 3 times for each sample  and consistent
results  have been observed.  The data were also compared to
values from the exact calculations for ideal metal boundaries, which were
converted to the sphere-trenched plate case using the PFA.
  However, the measured deviations from PFA, as applied to the
rectangular corrugations, were found to be
50\% less than that expected for ideal metals.
This discrepancy was reported as quite natural due to the interplay
of nonzero skin depth and geometry effects.

Thus, the experiment by Chan et al. (2008)
reports the measurement of deviation
resulting from geometry for corrugated rectangular trenches of
relatively small periods.
The depth
of the corrugated trenches allowed good comparison to the
results obtained using the PFA taking
into account only the fractional area of the top surface. Deviations between
10-20\% from the PFA were reported.   At present no
theoretical computations
exist which would allow a definite comparison between experiment
and theory for spherical and
corrugated surfaces made of real metals at room temperature
[at $T=0$ such computations were performed by Lambrecht and
Marachevsky (2008)].

\subsection{Future prospects to measure the Casimir force
with semiconductor surfaces}

\subsubsection
{The dielectric-metal transition}
\label{sec:mott}

The exciting possibility for the modulation of the Casimir force due to
a change of charge carrier density is offered by semiconductor materials
that undergo the dielectric-metal transition with the increase of
temperature. {}From a fundamental point of view, the modulation of the
Casimir force due to the phase transitions of different kinds offers one
more precision test of the role of conductivity and optical properties
in the Lifshitz theory of the Casimir force.

An experiment was proposed (Castillo-Garza et al., 2007) to measure the
change of the Casimir force acting between an Au coated sphere and a
vanadium dioxide (VO${}_2$) film deposited on a sapphire substrate which
undergoes the dielectric-metal transition with the increase of temperature.
It has been known that VO${}_2$ crystals and thin films undergo an
abrupt transition from semiconducting monoclinic phase at room
temperature to a metallic tetragonal phase at 68${}^{\circ}$C
(Zylbersztejn, Mott, 1975; Soltani et al., 2004; Suh et al., 2004).
The phase transition causes the resistivity of the sample to decrease by
a factor of $10^4$ from 10 to $10^{-3}\,\Omega\,$cm. In addition,
the optical transmission for a wide region of wavelengths extending
from $1\,\mu$m to greater than $10\,\mu$m decreases by more than a factor
of 10 to 100.

The increase of temperature necessary for the phase transition can be
induced by laser light (Soltani et al., 2004; Suh et al., 2004).
Thus, a setup similar to the one employed in the demonstration of
optically modulated dispersion forces (see Sec.\ V.B) can be used.
In the initial stage of the experimental work, the procedures of film
fabrication and their heating were investigated (Castillo-Garza et al., 2007).
The preliminary theoretical results are also obtained (Castillo-Garza et al.,
2007; Pirozhenko, Lambrecht, 2008a) based on the Lifshitz theory and
optical data for VO${}_2$ films (Verleur et al., 1968).
The calculations of the difference Casimir force between an Au sphere and
VO${}_2$ film on sapphire substrate after and before the phase transition
show that the proposed experiment has much
promise for the understanding of the role of
free charge carriers in the Lifshitz
theory of dispersion forces.

Interesting results can also be obtained when investigating the change
of the Casimir free energy in the phase transition of a metal to the
superconducting state. The variation of the Casimir free energy
during this transition is very small (Mostepanenko, Trunov, 1997).
Nevertheless, the magnitude of this variation can be comparable to the
condensation energy of a semiconducting film and causes a measurable
increase in the value of the critical magnetic field (Bimonte et al.,
2005a, 2005b).

\subsubsection{Casimir forces between a sphere and a plate with
patterned geometry}
\label{sec:patt}

The difference force measurements are very sensitive to relatively small
variations of the Casimir force (see Sec.\ V.B). Recently an
experimental scheme was proposed (Castillo-Garza et al., 2007) which
promises a record sensitivity to a force difference at the level of 1\,fN.
The patterned Si plate with two sections of different doping
concentrations (see Fig.\ \ref{fgVp12}) is mounted on a piezo below
an Au coated sphere attached to the cantilever of an AFM. The piezo
oscillates in the horizontal direction causing the flexing of the
cantilever in response to the Casimir force above different regions of
the plate. Thus, the sphere is subject to the difference Casimir force
which can be measured using the static and dynamic techniques.
The patterned plate is composed of a single crystal Si specifically
fabricated to have adjacent sections of two different charge carrier
densities. A special procedure was developed for the preparation of the
Si sample with the two sections having different conductivities
(Castillo-Garza et al., 2007) where the $p$- and $n$-type dopants can be
used (B and P respectively). Sharp transition boundaries between the two
sections of the Si plate of width less than 200\,nm can be achieved.
Identically prepared but unpatterned samples can be used to measure
the properties which are needed for the theoretical computations (Hall
probes for measuring the charge carrier concentration, a four-probe
technique for measuring conductivity). The measurement of the difference
Casimir force is planned as follows. The Si plate is positioned such that
the boundary is below the vertical diameter of the sphere (see
Fig.\ \ref{fgVp12}). The distance between the sphere and Si plate $a$ is
kept fixed and the Si plate oscillates in the horizontal direction
using the piezo such that the sphere crosses the boundary in the
perpendicular direction during each oscillation [a similar approach was
exploited (Decca et al., 2005a) for constraining new forces from the
oscillations of the Au coated sphere above two dissimilar metals, Au
and Ge]. The Casimir force on the sphere changes as the sphere crosses
the boundary. This change corresponds to the differential force
\begin{equation}
\Delta F(a)=F_{\tilde{n}}(a)-F_n(a)
\label{eq5p8}
\end{equation}
\noindent
equal to the difference of the Casimir forces due to the different charge
carrier densities $\tilde{n}$ and $n$, respectively. This causes a
difference in the deflection of the cantilever. In order to reduce random
noise by averaging, the periodic horizontal movement of the plate
will be of an angular frequency $\Omega\sim 0.1\,$Hz. The amplitude of
 plate oscillations is limited by the piezo characteristics, but can be of
order $100\,\mu$m, much larger than the typical width of the transition
region equal to 200\,nm.

The proposed experiment holds promise for the investigation of the possible
variation of the Casimir force in the dielectric-metal transition in
semiconductors with the increase of doping concentration (Klimchitskaya,
Geyer, 2008). It has the potential to
distinguish between the two
models of the dielectric permittivity of semiconductors with the concentration
of charge carriers above the critical [see Eqs.\ (\ref{eq5p3}) and
(\ref{eq5p4}) in Sec.\ V.B].

\subsubsection{Pulsating Casimir force}

At present, a clear consensus has
been reached that the applications of the Casimir
force in the design, fabrication and actuation of micro- and
nanomechanical devices are very promising. When the characteristic sizes
of a device shrink below a micrometer, the Casimir force becomes
larger than typical electric forces. Considerable opportunities for
micromechanical design would be opened by pulsating Casimir plates
moving back and forth entirely due to the effect of the zero-point
energy, without the action of mechanical springs. This can be achieved
only through the use of both attractive and repulsive Casimir forces.
In connection with this, it should be noted that while the repulsive Casimir
froces for a single cube or a sphere are still debated, the Casimir
repulsion between two parallel plates is well understood.
Repulsion occurs when the plates with dielectric permittivities
$\varepsilon_1$ and $\varepsilon_2$ along the imaginary frequency axis
are immersed inside a medium with the dielectric permittivity
$\varepsilon_0$ such that
$\varepsilon_1<\varepsilon_0<\varepsilon_2$ or
$\varepsilon_2<\varepsilon_0<\varepsilon_1$
(Mahanty, Ninham, 1976). At short separations in the nonretarded van der
Waals regime this effect has been discussed for a long time and measurements
have been reported [see, e.g., review by Visser (1981) and one of the
later experiments by Meurk et al. (1997)]. At separations of about
30\,nm the same effect was measured by Munday et al. (2009).

Recently it was shown that the illumination of one (Si) plate in
a three-layer
systems Au-ethanol-Si, Si-ethanol-Si and
$\alpha$-Al${}_2$O${}_3$-ethanol-Si with laser pulses can change the
Casimir attraction to Casimir repulsion and vice versa (Klimchitskaya et
al., 2007b). The illumination can be performed in the same way as
described in Sec.\ V.B. The calculations show that in the system
 Au-ethanol-Si the force is repulsive at separations $a>160\,$nm.
The illumination of the Si plate, as in Sec.\ V.B, changes this repulsion to
attraction.  In the system Si-ethanol-Si the force between the Si plates is
attractive, however, with one Si plate illuminated the attraction is
replaced with repulsion at separations $a>175\,$nm.

In the systems mentioned above, the magnitudes of the repulsive forces are
several times less than the magnitude of the attractive forces at the same
separations. However, it is possible to design a system where the
light-induced Casimir repulsion is of the same order of magnitude as the
 attraction. A good example is given by the three-layer system
$\alpha$-Al${}_2$O${}_3$-ethanol-Si where the Si plate is illuminated with
laser pulses. The respective computational results for the Casimir
pressure versus separation are presented in Fig.\ \ref{fgVp13}.
The Casimir attraction (solid line 1) changes to repulsion at
separations $a>70\,$nm when the Si plate is illuminated (solid line 2).

Note that the observation of the pulsating Casimir force requires that
the plates be completely immersed in a liquid far away from any
air-liquid interfaces. This prevents the occurrence of capillary
forces. Surface preparation of the plates is necessary to bring about
an intimate contact between the plates and the liquid. The only
liquid-based force is the drag force due to the movement of the plates
in response to the change of the force. For pressure values of
around 10\,mPa and typical spring constants of
0.02\,N/m, the corresponding drag pressure from the plate movement would be
six orders of magnitude less in value.
Thus, in near future one may expect the experimental confirmation
for the possibility of Casimir repulsion.

\section{ EXPERIMENTS ON THE CASIMIR-POLDER FORCE}
\label{sec:expCP}

\subsection{Demonstration of the thermal Casimir-Polder force}

\subsubsection{ The force in thermal equilibrium}
\label{sec:eq}

As mentioned in Sec.~II.C, the Casimir-Polder interaction between an atom
and a wall leads to a change of the center-of-mass oscillation frequency
of the Bose-Einstein condensate in the direction perpendicular to the
wall (Antezza et al., 2004). This frequency shift can be measured
precisely, leading to an indirect measurement of the Casimir-Polder
force (Harber et al., 2005). Using this technique, the first measurement
of the thermal Casimir-Polder force at large separations between an atom
and a plate was performed by Obrecht et al. (2007). In that experiment
the dipole oscillations with the frequency $\omega_0$ were excited in
a ${}^{87}$Rb Bose-Einstein condensate separated by a distance of a few
micrometers from a fused-sulica substrate (wall). The Casimir-Polder
force (\ref{eq21}) between a Rb atom and a substrate changes the magnitude
of the oscillation frequency making it equal to some $\omega_z$
(the $z$-direction is perpendicular to the wall). The use of the
Bose-Einstein condensate is convenient because it provides a spatially
compact collection of a relatively large number of atoms
[$2.5\times 10^{5}$ atoms in the experiment by Obrecht et al. (2007)].
It is well-characterized by the Thomas-Fermi density profile (\ref{eq24})
with the radius $R_z=2.69\,\mu$m (corresponding to a magnetic trap
frequency in the radial direction $\omega_0=2\pi\times 229\,$Hz).
This profile is used for spatial averaging in Eq.\ (\ref{eq22}).

The detailed description of the experimental setup, calibration procedures
and measurements is presented in the papers by McGuirk et al. (2004) and
Harber et al. (2003, 2005). The directly measured quantity was the
relative frequency shift
\begin{equation}
\gamma_z=\frac{|\omega_0-\omega_z|}{\omega_0}=
\frac{|\omega_0^2-\omega_z^2|}{2\omega_0^2},
\label{eq6p1}
\end{equation}
\noindent
where $\omega_0^2-\omega_z^2$ is connected [see Eq.\ (\ref{eq22})] with the
spatial and time averaged Casimir-Polder force (Antezza et al., 2004).

In thermal equilibrium the temperature of the fused silica substrate $T_S$
was equal to the environment temperature $T_E$: $T_S=T_E=310\,$K.
The frequency shift (\ref{eq6p1}) was measured for a number of wall-atom
(center-of-mass of the condensate) separations from 7 to 11$\,\mu$m.
In Fig.\ \ref{fg6p1} the experimental data obtained by Obrecht et al. (2007)
in thermal equilibrium at separations below 10$\,\mu$m are shown
as crosses.
Harber et al. (2005) carefully estimated random, systematic and total
errors in the measured values of $\gamma_z$ in such experiments at
a 66\% confidence level. Obrecht et al. (2007) performed respective
analysis for each experimental point separately. The main sources of
systematic errors discussed by Harber et al. (2005) are connected with
possible presence of spatially inhomogeneous electric or magnetic
surface contaminations or uniform magnetic and electric fields.
Special investigations were performed to obtain upper bounds on all
systematic errors. In Fig.~30
the absolute total errors in the measurement of the separations and
$\gamma_z$ are presented in true scales at each individual data point.

The comparison between the experimental data and the theory is done as in
Fig.\ 14 (see Sec.\ IV.C) where the frequency shift due to the Casimir
pressure between two parallel plates was shown.
The measured quantity
(in this case the relative frequency shift) is plotted as crosses on
the same graph as the theoretical lines computed using
different approaches. The theoretical values of the relative frequency
shift as a function of separation are calculated using
Eqs.\ (\ref{eq6p1}), (\ref{eq22}) with the Casimir-Polder force
(\ref{eq21}). The mass of a Rb atom is $m=1.443\times 10^{-25}\,$kg.
The dymanic polarizability of a Rb atom in Eq.\ (\ref{eq21}) can
be considered as frequency-independent, and the static value
$\alpha({\rm i}\xi_l)\approx\alpha(0)=4.73\times 10^{-23}\,\mbox{cm}^{-3}$
was used in the computations. This allows one to obtain highly accurate
results for the separations under consideration (Babb et al., 2004).

Fused silica is a good insulator. However, like any insulator, it
possesses a nonzero dc conductivity at nonzero temperature. Electrical
conductivity in fused silica is ionic in nature and is  determined by the
concentration of the impurities (alkali ions) which are always present as
trace constituents. At $T_S=T_E=310\,$K this conductivity varies within
a wide region from $10^{-9}\,\mbox{s}^{-1}$ to $10^{2}\,\mbox{s}^{-1}$
(Bansal, Doremus, 1986; Shackelford, Alexander, 2001).
When neglecting the dc conductivity, the dielectric permittivity
of fused silica $\varepsilon({\rm i}\xi_l)$ as a function of $\xi_l$ can be
calculated (Caride et al., 2005) using the tabulated optical data
(Palik, 1985) and the dispersion relation (\ref{eq45}). In this case
the static dielectric permittivity has a finite value
$\varepsilon_0=3.81$. The respective computational results for
$\gamma_z$ with the dc conductivity of fused silica neglected are
shown in Fig.\ \ref{fg6p1} by the solid line (Obrecht et al., 2007).
Note that the results computed using
$\varepsilon({\rm i}\xi_l)=\varepsilon_0$ (Obrecht et al., 2007) and
frequency-dependent dielectric permittivity (Antezza et al., 2004;
Klimchitskaya, Mostepanenko, 2008b) are almost coincident at large
separation distances (only small deviations are observed at
$a< 8\,\mu$m).
The theoretical computations are in excellent agreement
with the data, as was stated by Obrecht et al. (2007).

The inclusion of the dc conductivity of fused silica in the model of the
dielectric response, as in Eq.\ (\ref{eq38}), dramatically affects the
calculational results. This changes the value of the reflection coefficient
$r_{\rm TM}(0,k_{\bot})$ and, consequently, the Casimir-Polder force
(\ref{eq21}) which leads to a change in the magnitude of $\gamma_z$
computed using Eqs.\ (\ref{eq22}), (\ref{eq6p1}).
The respective computational results for $\gamma_z$ are shown in
Fig.\ \ref{fg6p1} as the dashed line.
As is seen in the figure, the first two experimental points are in clear
disagreement with theory taking into account the conductivity of fused
silica.

\subsubsection{The force out of thermal equilibrium}
\label{sec:noneq}

Recently a nonequilibrium situation was considered where the substrate is at
a temperature $T_S$ but the environment (remote wall) is at another
temperature $T_E$ (Antezza et al., 2005). In this case the Casimir-Polder
force has two additional terms due to the thermal fluctuations
from the nearest wall at  a temperature $T_S$ and from remote walls
(environment) at a temperature $T_E$.

The force acting between an atom and a substrate out of thermal
equilibrium depends on both temperatures and is given by (Antezza et al., 2005)
\begin{equation}
F(a,T_S,T_E)=F(a,T_E)+F_n(a,T_S)-F_n(a,T_E).
\label{eq6p2}
\end{equation}
\noindent
The first term on the right-hand side of Eq.\ (\ref{eq6p2}) is the
Casimir-Polder force (\ref{eq21}). The nonequilibrium contribution
$F_n(a,T)$ obtained in the approximation of static atomic polarizability
is defined as (Antezza et al., 2005)
\begin{eqnarray}
&&
F_n(a,T)=-K\int_{0}^{\infty}\!\! d\omega\int_{0}^{\infty}\!\! dx
f(\omega,x){\rm e}^{-\frac{2\omega xa}{c}},
\label{eq6p3} \\
&&
f(\omega,x)=\frac{\omega^4x^2}{{\rm e}^{\frac{\hbar\omega}{k_BT}}-1}
\left[|p(\omega,x)|+{\rm Re}\,\varepsilon(\omega)-1-x^2\right]^{1/2}
\nonumber \\
&&~~~
\times\left[\frac{1}{\left|\sqrt{p(\omega,x)}+{\rm i}x\right|^2}+
\frac{(2x^2+1)(x^2+1+|p(\omega,x)|}{\left|\sqrt{p(\omega,x)}+
{\rm i}\varepsilon(\omega)x\right|^2}\right],
\nonumber \\
&&
K\equiv\frac{2\sqrt{2}\hbar\alpha(0)}{\pi c^4}, \qquad
p(\omega,x)\equiv\varepsilon(\omega)-1-x^2.
\nonumber
\end{eqnarray}
\noindent
The frequency shift of the condensate oscillations out of thermal
equilibrium can be calculated using Eqs.\ (\ref{eq6p1}) and (\ref{eq22})
where $F(a,T)$ is replaced with $F(a,T_S,T_E)$ given by
Eqs.\ (\ref{eq6p2}), (\ref{eq6p3}).

In the experiment by Obrecht et al. (2007) the fused-silica substrate
was heated by the absorption with laser light. The experimental data
obtained out of thermal equilibrium for $T_E=310\,$K and two different
values of substrate temperature, $T_S=479\,$K and $T_S=605\,$K,
are shown as crosses in Figs.\ \ref{fg6p2}(a) and \ref{fg6p2}(b),
respectively. Similar to Fig.\ \ref{fg6p1}, the absolute errors are
presented in true scales at each data point. The computational results
for $\gamma_z$ in a nonequilibrium situation, obtained by neglecting the dc
conductivity of fused silica (Obrecht et al., 2007) are presented in
Fig.\ \ref{fg6p2} as the solid lines. Note that the frequency dependence
of $\varepsilon(\omega)$ in Eq.\ (\ref{eq6p3}) does not affect the
contributions to the frequency shift from the nonequilibrium
terms in the total atom-wall force (\ref{eq6p2}).

Direct computations show that in the nonequilibrium situation the disagreement
between the experimental data and the theory with the dc conductivity of the
substrate material included widens further. The respective results are
presented in Fig.\ \ref{fg6p2} as the dashed lines (Klimchitskaya,
Mostepanenko, 2008b). As is seen in Fig.\ \ref{fg6p2}(a), the three
experimental points for $T_S=479\,$K exclude the dashed line and the
other two only touch it. The dashed line in Fig.\ \ref{fg6p2}(b)
demonstrates that all data for $T_S=605\,$K exclude the theoretical
prediction calculated with the inclusion of the dc conductivity of fused
silica. Thus, the confidence at which the theoretical approach based
on Eq.\ (\ref{eq38}) is excluded by data increases with the increase
of substrate temperature.

It is notable that the inclusion of the dc conductivity of fused silica
in the model of the dielectric response (\ref{eq38}) does not affect the
contributions to the frequency shift arising from the nonequilibrium
terms $F_n(a,T)$ in Eq.\ (\ref{eq6p2}) (Klimchitskaya,
Mostepanenko, 2008b). Thus, the conductivity influences the computational
results only through the equilibrium Casimir-Polder force (see Sec.\ VI.A.1).
The important role of the Obrecht et al. (2007) experiment
is that, not only was the model of the dielectric response
taking the dc conductivity into account excluded, but the thermal
effect, as predicted by the Lifshitz theory
with the dc conductivity omitted, was measured
for the first time.

\subsection{Future prospects to measure the Casimir-Polder force
in quantum reflection}
\label{sec:persp}

The magnitude and the distance dependence of the Casimir-Polder force
were confirmed by Sukenik et al. (1993) when studying the deflection of
ground-state Na atoms passing through a micron-sized parallel-plate
cavity. The intensity of an atomic beam transmitted through the cavity
was measured as a function of the plate separation. The comparison of
the experimental data with the theoretical position-dependent potential
for an atom between parallel ideal mirrors (Barton 1987a, 1987b)
allowed one to
investigate the atom-wall interaction at separation distances below $3\,\mu$m.
This experiment gave impetus to the investigation of the Casimir-Polder
forces in scattering experiments. Of special interest are situations when
the wave nature of atom becomes dominant with respect to its classical
behavior as a particle. Such a pure quantum effect is what is referred to
as {\it quantum reflection}, i.e., a process in which a particle moving
through a classically allowed region is reflected by a potential
without reaching a classical turning point. The possibility of
reflection of an ultracold atom under the influence of an {\it attractive}
atom-wall interaction was predicted long ago on quantum-mechanical grounds
[see, e.g., the paper by Lennard-Jones, Devonshire (1936)].
However, the experimental observation of this phenomenon has become possible
only recently due to the success in the production of ultracold atoms.
First it was investigated using liquid surfaces, as the reflection of He and H
atoms on liquid He (Nayak et al., 1983; Berkhout et al., 1989) and from
the sticking coefficient of H atoms on liquid He (Doyle et al., 1991;
Yu et al., 1993). Later a specular reflection of very slow metastable Ne atoms
on Si and BK7 glass surfaces was studied (Shimizu, 2001). The observed
velocity dependence was explained by the quantum reflection which is caused
by the attractive Casimir-Polder interaction.

Quantum reflection becomes efficient when the motion of the particle
can no longer be treated semiclassically (Friedrich, Trost, 2004).
The behavior of the particle is of a quantum character when
\begin{equation}
\frac{\partial\lambda_B(z)}{\partial z}\geq 1,
\label{eq6p4}
\end{equation}
\noindent
where $\lambda_B(z)=2\pi\hbar/\sqrt{2m[E-V(z)]}$ is the local de Broglie
wavelength for a particle of mass $m$ and
initial kinetic energy $E$ moving in the
potential $V(z)$. The same can be formulated as a condition that the
variation of the local wave vector $k=2\pi/\lambda_B(z)$,
perpendicular to the surface,
within the distance of the atomic de Broglie wavelength, is larger than
$k$ itself (Shimizu, 2001),
\begin{equation}
\Phi=\frac{1}{k^2}\,\frac{dk}{dz}> 1.
\label{eq6p5}
\end{equation}
\noindent

The reflection amplitude depends critically on the energy $V(z)$ of
the atom-wall
interaction. This has attracted considerable attention to the theoretical
investigation of the reflection amplitude depending on atomic energy
and the form of the interaction potential (Friedrich et al., 2002;
Jurisch, Friedrich, 2004; Madro\~{n}ero, Friedrich, 2007; Voronin,
Froelich, 2005; Voronin et al., 2005).
The reflection probability tends to unity,
as the incident velocity tends to zero. Thus, a high probability for the
quantum reflection calls for small incident velocities, i.e., for cold atoms.

Advances in cooling techniques in the past decade have made it
possible to perform experiments with cold atoms interacting with solid
surfaces. The quantum reflection of ${}^3$He atoms in the scattering from
an $\alpha$-quartz crystal was observed at energies far from
$E\to 0$ (Druzhinina, DeKieviet, 2003). The observation of the large
reflection amplitudes for the dilute Bose-Einstein condensate of ${}^{23}$Na
atoms on a Si surface (Pasquini et al., 2004, 2006)
allows the possibility of
using quantum reflection as a trapping mechanism.
The corresponding trap model
has been developed theoretically (Jurisch, Friedrich, 2006;
Madro\~{n}ero, Friedrich, 2007).
The quantum reflection of He${}^{\ast}$ atoms
on a flat Si surface and on microfabricated surface structures
were also investigated (Oberst et al., 2005a, 2005b). The latter experiment
has stimulated the development of adequate theoretical models for the
description of the scattering of atomic matter waves on ridged surfaces
(Kouznetsov, Oberst, 2005).

For theoretical calculations of the reflection amplitude, a simple attractive
atom-wall interaction potential is commonly used (Friedrich et al., 2002;
Druzhinina, DeKieviet, 2003; Oberst et al., 2005b) which is an interpolation
between the nonretarded van der Waals energy (\ref{eq18}) and the
Casimir-Polder energy (\ref{eq19}). The comparison of computational
results with the measurement data for the reflection amplitudes allows
one to estimate the parameters of the potential (Druzhinina, DeKieviet, 2003;
Oberst et al., 2005b). The increased precision of the measurements opens new
opportunities for the comparison with the more exact Eq.\ (\ref{eq15}) for
the free energy of atom-wall interactions. Such a comparison could yield
important new information on the role of the atomic and material properties
in dispersion forces.
As shown by Bezerra et al. (2008), for an atom of metastable
He${}^{\ast}$ and an Au wall at zero temperature, the largest deviations
between the phenomenological potential and the exact interaction energy
of about 10\% are achieved at separations of 300--500\,nm. However, at room
temperature the free energy deviates from the phenomenological potential
by 31\% at $a=5\,\mu$m.
There is also related theoretical research on Casimir physics in atom-wall
interaction devoted to the influence of nonzero temperature on atomic
polarizability (Buhmann, Scheel, 2008), the use of the lateral Casimir-Polder
force between an atom and a corrugated surface to study nontrivial
geometrical effects (Dalvit et al., 2008) and to the interaction of atoms
with conducting microstructures (Eberlein, Zietal, 2007).

\subsection{ Casimir-Polder interaction of atoms with
carbon nanotubes}
\label{sec:hydr}

The experiments with ultracold atoms have
generated interest on the scattering
and trapping of such atoms by different nanostructures (Arnecke et al., 2007;
Fermani et al., 2007). In this case, the probability of quantum reflection
is sensitive to the shape of the potential
(Friedrich et al., 2002; Arnecke et al., 2007). Thus,
it needs to be ascertained if the general theory of dispersion forces
can describe the interaction between microparticles and
nanostructures. An additional question to be answered is to what extent the
macroscopic concept of the dielectric permittivity can be used as an
adequate characteristic of nanostructure material properties.

As a first step in this direction, the Lifshitz theory of the van der Waals
force has been extended for the case of a microparticle interacting with
a plane surface of a uniaxial crystal or with a cylindrical shell
made of such crystal (Blagov et al., 2005). An approximate expression
for the free energy of a microparticle--cylinder interaction was obtained
using the PFA which is of high precision at microparticle--cylinder
separations smaller than the cylinder radius. The reflection coefficients
from the surface of a uniaxial crystal (with the optical axis $z$
perpendicular to the surface) can be expressed in terms of two dissimilar
dielectric permittivities $\varepsilon_x(\omega)$ and $\varepsilon_z(\omega)$.
For example, the multilayer graphite cylinder was considered as a
simple model of a multiwalled carbon nanotube. Using this model,
the Casimir-Polder interaction between hydrogen atoms (molecules) and
multiwalled carbon nanotubes was investigated (Blagov et al., 2005;
Klimchitskaya et al., 2006a). Later, Lifshitz-type formulas have been
obtained which describe the van der Waals and Casimir-Polder interaction
between a graphene sheet and a material plate or a microparticle
(Bordag et al., 2006) and a microparticle and a single-walled carbon
nanotube (Blagov et al., 2007). For this purpose graphene was considered
as an infinitesimally thin positive charged flat sheet, carrying a
continuous fluid with some mass and negative charge densities. This plasma
sheet is characterized by a typical wave number $K$ determined by the
parameters of the hexagonal structure of graphite. The interaction of the
electromagnetic oscillations with such a sheet was considered and the
normal modes and reflection coefficients were found (Barton, 2004, 2005).

A single-walled carbon nanotube can be approximately modeled as a
cylindrical sheet carrying a two-dimensional free-electron gas with the
same reflection coefficients as for a plane sheet (Blagov et al., 2007).
Note that the above approximate models do not take into account nanotube
chirality (Saito et al., 1998). This can be included by using the optical
data for the nanotube index of refraction. Regarding single-walled carbon
nanotubes, it  remains unclear if it is possible to describe nanotubes with
surfaces like metals or like semiconductors by varying only one parameter,
the typical wave number $K$, in the reflection coefficients.

By comparing the calculational results for the multiwalled nanotubes with
those for single-walled, it was demonstrated that the macroscopic
desciption using the concept of dielectric permittivity is
applicable for nanotubes with only two or three layers at separation
distances larger than 2\,nm (Blagov et al., 2007; Klimchitskaya et al., 2008a).
It is well to bear in mind that the description of multiwalled nanotubes
by the graphite dielectric permittivity and single-walled nanotubes as a
two-dimensional free-electron gas are only models and are not a
complete description of all the nanotube properties. Nevertheless, they are
good approximations for
the estimation of microparticle-nanotube interaction on
the basis of the Lifshitz theory.

\section{ LATERAL CASIMIR FORCE AND CASIMIR TORQUES}
\label{sec:lat}

\subsection{ Lateral Casimir force between corrugated surfaces}
\label{sec:corr}

The Casimir force considered in all the preceding sections is the {\it normal}
one, i.e., it leads to an attraction perpendicular to the surfaces of the
interacting bodies. However, as mentioned in Sec.\ I.B, there is a geometry
dependence for the Casimir force on the shape of the surface, especially
for corrugated surfaces. Thus, the nontrivial boundary dependence of the
normal Casimir force acting between a plate with periodic uniaxial sinusoidal
corrugations and a large sphere was demonstrated by Roy and Mohideen (1999).
It follows also from Eq.\ (\ref{eq53}) that the pressure between two
corrugated plates with parallel uniaxial sinusoidal corrugations of the same
period harmonically depends on the phase shift between the corrugations
(Bordag et al., 1995a). This should result in the lateral Casimir force
in addition to the normal one.

Using the path integral method, Golestanian and Kardar (1997, 1998) have
predicted the existence of the lateral Casimir force in the configuration
of two corrugated plates. Later the geometry dependence of the Casimir
energy for sinusoidally corrugated ideal metals plates was studied in
more detail by the path integral quantization of the electromagnetic
field (Emig et al., 2001, 2003). The general expressions for both the normal
and the lateral Casimir force were obtained in the lowest order of
the perturbation
theory with respect to the relative corrugation amplitudes $A_1$ and $A_2$.
Here, the interacting surfaces were described by Eq.\ (\ref{eq49}) with
$f_1(x,y)=\sin(2\pi x/\Lambda)$, $f_2(x,y)=\sin[2\pi (x+x_0)/\Lambda]$.
A harmonic dependence of the lateral force on the phase shift
$\varphi=2\pi x_0/\Lambda$ was obtained (Emig et al., 2001, 2003).
Later the exact solution was found for the lateral Casimir force
in the configuration of two parallel ideal metal plates with laterally
shifted uniaxial rectangular gratings (B\"{u}scher, Emig, 2005).

The influence of the lateral force acting on a sphere above a
corrugated plate can explain (Klimchitskaya et al., 2001) the nontrivial
character of the normal Casimir force in this configuration
experimentally demonstrated by Roy and Mohideen (1999).

All theoretical results on the lateral Casimir force which are mentioned
above were obtained for uniaxial corrugations of equal period. If the
periods of corrugations are different the lateral force becomes equal
to zero. The uniaxial character of the corrugations is also
important for the observation of the lateral Casimir force.
Under the assumption that there is a nonzero angle $\vartheta$ between the
axes of the corrugations, the phase shift along the $x$ axis becomes the
periodic function of $y$ with a period $\Lambda_y=\Lambda{\rm cot}\vartheta$.
Within one period $\varphi(y)$ depends on $y$ linearly, taking on values
from 0 to $2\pi$. The resulting lateral force should be averaged over the
period $\Lambda_y$. In the case of an infinite plate this leads to a zero
value at any $\vartheta\neq 0$. For real bodies of finite size the
lateral Casimir force will be measurable only for small deviations of the
corrugation axes from parallelity such that $\Lambda{\rm cot}\vartheta$
is much larger than the size of the smallest body. For example, for
$\Lambda\approx 1\,\mu$m and a body size of $10\,\mu$m the lateral Casimir
force is observable only if $\vartheta\ll 0.1\,$rad.

The existence of the lateral Casimir force opens new opportunities for
the application of the Casimir force in micromachines
(Ashourvan et al., 2007a, 2007b; Emig, 2007).
It is also important
that this force can be well controlled by specific shapes
of corrugations (Blagov et al., 2004).
Based on the path integral theory (Emig et al., 2003) it appears
that the lateral Casimir force is a promising venue for the
demonstration of shape dependences and diffraction-like effects.
For these reasons an experimental
investigation of the lateral Casimir force is of great interest.

\subsection{ Demonstration of the lateral Casimir force}
\label{sec:dem}

The lateral Casimir force between an Au plate and a sphere, both
sinusoidally corrugated, was measured for surface separations between 200 to
300\,nm using an AFM (Chen et al., 2002a, 2002b). The experiment was
performed at a pressure below 50\,mTorr and at room temperature.
A schematic diagram of the experiment is shown in Fig.\ \ref{fg7p1}.
To implement this experiment a diffraction grating with uniaxial
sinusoidal corrugations of period $\Lambda=1.2\,\mu$m and an amplitude
of 90\,nm was used as the template. In order to obtain perfect orientation
and phase between the corrugated surfaces, a special imprinting procedure
was developed (Chen et al., 2002a, 2002b). The
amplitude of the corrugations on the plate and that imprinted on the
sphere were measured using the AFM, to be $59\pm 7\,$nm and $8\pm 1\,$nm,
respectively. The calibration of the cantilever and the measurement of the
residual potential voltage between the sphere and the plate were done
by electrostatic means in a manner similar to experiments on measuring
the normal Casimir force (see Sec.\ IV.A).

The corrugated plate was mounted on two piezos that allowed independent
movement of the plate in the vertical and the horizontal directions with
the help of the $x$-piezo and $z$-piezo, respectively. Movement in the
$x$ direction is necessary to achieve a lateral phase shift between the
corrugations on the sphere and the plate. Independent movement in the $z$
direction is necessary for control of the surface separation. The lateral
Casimir force was measured at four different separations starting from
221\,nm with steps of 12\,nm. At each separation the measurement was repeated
60 times and the averaged force was determined. The averaged lateral force
measured at a separation 221\,nm is shown as solid squares in
Fig.\ \ref{fg7p2}. The sinusoidal oscillations as a function of the phase
difference between the two corrugations are clearly observed.
The periodicity of the lateral force oscillations is in agreement
with the corrugation period of the plate. A sine curve fit to the observed
data is shown in Fig.\ \ref{fg7p2} as the solid line and corresponds to an
amplitude 0.32\,pN. The detailed error analysis (Chen et al., 2002b) led
to the total absolute experimental error of 0.077\,pN at a 95\% confidence
level. The resulting precision of the amplitude measurements at the
closest separation is around 24\%.

The measurement data were compared with theoretical results. Note, that
the measurements were performed at separations from 200 to 300\,nm
where the corrections due to the skin depth are rather large (see Sec.\ IV.A).
In addition, the ratio of the corrugation amplitude on the plate to the
separation distance is not a small parameter ($A_1/a\approx 0.27$).
For this reason, it is not appropriate to compare experimental results
with theoretical predictions obtained for ideal metal test bodies in the
lowest order perturbation theory. However, separations are several times
smaller than the corrugation period $\Lambda$. Because of this,
diffraction-like effects do not contribute (see Sec.\ III.B).
In this case the theoretical expression for the lateral Casimir force
including corrections to the skin-depth can be obtained
in the form
\begin{equation}
F^{\rm lat}(a,\varphi)=
\frac{\pi^4 R\hbar c}{120a^4\Lambda}\,A_1A_2
G(a,\Lambda,A_1,A_2)\,\sin\varphi.
\label{eq7p1}
\end{equation}
\noindent
The explicit form of the function $G$ is different based on whether
the proximity force approximation or the pairwise summation
method is used.
In the framework of the PFA $G$ does not depend on $\Lambda$
 (Chen et al. 2002a, 2002b). The resulting theoretical values for the
amplitude of the lateral Casimir
force at $a=221\,$nm are 0.27\,pN and 0.31\,pN
when calculated up to the second and fourth perturbation orders in the small
parameters $A_i/a$, respectively. Thus, the fourth order term
is 15\% of the second order one and cannot be neglected.

The pairwise summation method is applicable to larger ratios
of $a/\Lambda$ and
the obtained results depend on $\Lambda$. Using this method,
the explicit expression for the function $G$ in Eq.~(\ref{eq7p1}) is
\begin{equation}
G=\kappa_E^{(sp)}(a)\left[S_{\rm pw}^{(1)}\left(2\pi\frac{a}{\Lambda}\right)+
\frac{5}{2}S_{\rm pw}^{(2)}\left(2\pi\frac{a}{\Lambda}\right)
\frac{A_1^2+A_2^2}{a^2}-5\cos\delta
S_{\rm pw}^{(2)}\left(4\pi\frac{a}{\Lambda}\right)
\frac{A_1A_2}{a^2}\right].
\label{eq7p2}
\end{equation}
\noindent
Here, the coefficients depending on $a/\Lambda$ are given by
\begin{eqnarray}
&&
S_{\rm pw}^{(1)}(x)={\rm e}^{-x}\left(1+x+\frac{x^2}{6}-
\frac{x^3}{6}\right)+\frac{x^4}{6}\Gamma(0,x),
\label{eq7p3} \\
&&
S_{\rm pw}^{(2)}(x)={\rm e}^{-x}\left(1+x+\frac{31}{80}x^2
+\frac{13}{240}x^3-
\frac{x^4}{480}+\frac{x^5}{480}\right)-\frac{x^6}{480}\Gamma(0,x)
\nonumber
\end{eqnarray}
\noindent
and $\Gamma(\alpha,x)$ is the incomplete gamma function.
The correction factor $\kappa_E^{(sp)}$ to the Casimir energy
due to the nonzero skin depth in the configuration of a sphere above a plate
is defined from
\begin{equation}
-\int_{a}^{\infty}E(a^{\prime})da^{\prime}=\frac{\pi^2\hbar c}{1440a^2}
\kappa_E^{(sp)}(a),
\label{eq7p4}
\end{equation}
\noindent
where $E(a)$ is the Casimir energy per unit area of two parallel
plates presented in Eq.~(\ref{eq8}).
Using Eqs.~(\ref{eq7p1}) and (\ref{eq7p2})
at $a=221\,$nm, one obtains the amplitude
of the lateral Casimir force equal to 0.24\,pN and 0.30\,pN in the
second and fourth perturbation orders in $A_i/a$,
respectively. Here, the fourth
order term is 25\% of the second order result. It is notable that
the amplitudes of the lateral Casimir force computed by using both
approximate methods up to the fourth perturbation order
in $A_i/a$ are almost
equal and in good agreement with the measurement result.
The force amplitudes
measured at different separations demonstrated the $a^{-4}$ dependence
in agreement with theoretical predictions.

Recently a theoretical approach was developed (Rodrigues et al., 2006a, 2007b)
which allows one to calculate the lateral Casimir force between
sinusoidally corrugated plates made of metals described by the plasma
model (\ref{eq27}) without using the PFA or pairwise summation,
but only in the lowest, second
perturbation order under the conditions
$A_1,\,A_2\ll\lambda_p,\,a,\,\Lambda$. These conditions are not satisfied
in the experiment discussed above (specifically, in that experiment
$A_1/a\approx 0.27$, $A_1/\lambda_p\approx 0.43$).
As we have seen, the contributions of the fourth perturbation orders
in $A_i/a$ are important.
Because of this, the
comparison of the experimental data with
the theoretical predictions
(Rodrigues et al., 2006a, 2007b) is not appropriate (Chen et al., 2007c).

\subsection{The Casimir torque}
\label{sec:torque}

Another interesting effect which was predicted long ago but has not yet
been observed is the Casimir torque. Such a torque arises due to the
zero-point electromagnitic oscillations in the configuration of anisotropic
or asymmetric bodies. In the case of plates made of a uniaxial crystal
described by the dielectric permittivities $\varepsilon_x(\omega)$ and
$\varepsilon_y(\omega)=\varepsilon_z(\omega)$, the torque arises if there is
a nonzero angle $\vartheta$ between the optical axes of the plates.
In the nonretarded limit the Casimir torque was investigated by Parsegian
and Weiss (1972). The retardation effects were taken into account by
Barash (1973). Recently these results were  again obtained using
another method (Philbin, Leonhardt, 2008).
The torque leads to the rotation of the plates until their
optical axes are aligned. In the nonretarded limit and for small
anisotropies
$|\varepsilon_x-\varepsilon_y|/\varepsilon_y\ll 1$ the torque
between the test bodies of area $S$ at a
separation $a$ is $M\sim -S\sin(2\vartheta)/a^2$ (Munday et al., 2005).
In the relativistic limit $M\sim -S\sin(2\vartheta)/a^3$ (Mostepanenko,
Trunov, 1997).

Recently an interesting experimental scheme was proposed to observe this
torque when a barium titanate plate is immersed in ethanol and an
anisotropic disk is placed above (Munday et al., 2005).
The dielectric permittivities are chosen in such a way that the Casimir
force between the anisotropic bodies is repulsive (see Sec.\ V.E.3).
The disk would float parallel to the plate at a distance where its weight is
counterbalanced by the Casimir repulsion, being free to rotate in response
to a small torque. Detailed numerical calculations are made
demonstrating the feasibility of this experiment (Munday et al., 2005).

Another possibility to observe the Casimir torque was theoretically
investigated in the configuration of two corrugated plates with a small
angle between the corrugation directions (Rodrigues et al., 2006b).
As discussed in Sec.\ VII.A, the lateral Casimir force and related
torque differ from zero for an
angle between corrugation axes $\vartheta\ll\Lambda/L$, where $L$ is the
size of the plate along the corrugation axis. The optimum values of the
corrugation period $\Lambda$ and plate separations were found
in order to get larger values of the Casimir torque in comparison with
those predicted for anisotropic test bodies of the same area
(Rodrigues et al., 2006b).

Both experiments, if successfully performed, will provide
important new information about dispersion forces between
real materials. {}From an application
standpoint, the Casimir torque may serve as one more mechanism of
control in micromachines.

\section{ CONCLUSIONS AND OUTLOOK}
\label{sec:conc}

In the foregoing, the intersection of experiment and theory in the
investigation of the Casimir force between real material bodies has
been reviewed. In the last ten years the field has undergone many
rapid advances. Many new results in the Casimir effect
of both a fundamental and applied character were obtained and have been
presented above.
They are related to condensed matter physics, statistical physics,
atomic physics and nanotechnology. This list should be supplemented with
elementary particle physics, quantum field theory, gravitation and cosmology,
and mathematical physics if one keeps in mind that many important results
on the Casimir effect are not
directly connected with experiments using real materials and, consequently,
are not covered in this
review.

Although experimental Casimir physics is still a young field, an
important period in its development has been completed and the related
conclusions can be formulated.
The most important of these conclusions is that the
Casimir force between real materials has been measured by many
experimental groups using different techniques and the obtained data
clearly demonstrate the role of real material properties and the geometrical
shape of the test bodies (see Secs.\ IV.A-C, V.B-D, VII.B). Although
there are still different opinions in the literature on the
measure of agreement between experiment and theory that has been achieved,
the statistical procedures for data processing and for the comparison
between the experimental and theoretical results in precise
force-distance measurements have been developed (see Sec.\ III.C) and
successfully used in several experiments (see Secs.\ IV.A,\,B, V.B,\,C).
Another important conclusion is that
 the application of the general Lifshitz theory of
dispersion forces to real materials at nonzero temperature
results in puzzles.
These puzzles are of both theoretical and experimental character.
On the theoretical side, an enormously large thermal effect was
predicted which arises when physically real but seemingly negligible
properties of real dielectrics are taken into account.
A related thermal
effect also arises for Drude metals, and both lead to a thermodynamic
inconsistency (Sec.\ II.D). Experimentally, the predicted large thermal
effects were excluded by the measurements of different groups performed
with metal, dielectric and semiconductor test bodies (see Secs.\ IV.B,
VI.A and V.B, respectively). If some specific properties of
the real bodies
(the dc conductivity of dielectrics which vanishes with vanishing
temperature or relaxation properties of conduction electrons in metals)
are disregarded, the Lifshitz theory is found to be in excellent
agreement with all performed experiments and with thermodynamics.
As was argued in Sec.~II.D.1, the dielectric permittivities used to describe
the drift and diffusion currents in dielectrics and metals are not
compatible with the Lifshitz theory because the existence of these
currents violate the state of thermal equilibrium.
This suggests that one should not include the drift and diffusion currents
in the model of the dielectric response
when applying the Lifshitz theory to real materials. There is no
consensus in the present literature with respect to this inference.

One can conclude that at the moment the Lifshitz theory, if applied in
accordance with its basic postulates, is consistent with the principles
of thermodynamics and all available experimental data.
Some claims may be raised because this theory does not include the
conductivity properties of real materials. However, the experimental
data testifies that these properties have nothing to do with
dispersion forces.
It is the authors' opinion that
a future theory of dispersion forces will not use reflection coefficients
expressed in terms of dielectric permittivity but
will deal with more general characteristics of scattering
processes allowing, in particular, the inclusion of spatial nonlocality.

Many other future applications of dispersion forces in real systems
are discussed in the literature, but
have not yet been investigated experimentally.
Because of this, they are not reflected in our review. One could mention
the van der Waals and Casimir interatomic
interaction in a magnetodielectric medium
(Spagnolo et al., 2007), the Casimir-Polder forces of excited atoms
(Buhmann, Welsch, 2008), the Casimir interaction between plates with
dielectric permittivity and magnetic permeability (Kenneth et al., 2002;
Pirozhenko, Lambrecht, 2008b).
Artificially microstructured metamaterials (Pendry et al., 1999)
are also promising for use in studies of the Casimir effect.
For example, the theoretical possibility that metamaterials might
lead to repulsive Casimir forces in some distance ranges has
been investigated (Henkel, Joulain, 2005;
Leonhardt, Philbin, 2007; Rosa et al., 2008).
It was pointed out that a strong modification of the Casimir force
is possible for surface separations around the resonance
wavelength of the magnetic response. Note that natural materials
are all physically limited to magnetic permeabilities
$\mu$ of order of unity in the visible spectrum.
However, artificially designed micrometer and submicrometer split
ring resonators and fish net arrays with $\mu>1$ have been
reported with resonances in the infrared and near optical
frequencies (Shalaev, 2007). Rosa et al. (2008) simulated
the Casimir force between a metamaterial slab and a half space
of Ag. A repulsive Casimir force was found for a short
band of separations only if there is no background metallic
response component. It was pointed out that the large
dissipation which is present in metamaterials (Chettiar et al.,
2007) would substantially reduce any Casimir repulsion.
It should also be noted that presently fabricated
metamaterilas in the visible frequency range have a very
narrow frequency region where $\mu>1$ (Chettiar et al.,
2007). Thus many engineering challenges remain for designing
appropriate materials. Nevertheless, they show great potential
for exploring new Casimir effect phenomena. Future prospects
to measure the thermal Casimir force, the Casimir force
between semiconductor surfaces and the Casimir-Polder force
in quantum reflection have been already discussed in
Secs.~IV.D, V.E and VI.B, respectively.
 This means that we can look forward to many
exciting applications of the Casimir force
in the near future.

\section*{Acknowledgments}

The authors are grateful to many colleagues for numerious
helpful discussions. We express
special gratitude to  J.\ F. Babb, V.\ B.\ Bezerra,
G.\ Bimonte, E.\ V.\ Blagov, M.\ Bordag, F.\ Chen, R.\ S.\ Decca,
E.\ Fischbach, B.\ Geyer,  D.\ E.\ Krause,
D.\ L\'{o}pez,  and C.\ Romero.
We are also grateful to M.\ Antezza, G.\ Carugno, G.\ Jourdan and R.\ Wild
for important additional information about their experiments.
This work was supported by the National Science Foundation (Grant No
PHY0653657), Department of Energy (Grant No DE-FG02-04ER46131) and
Deutsche Forschungsgemeinschaft (Grant No 436\,RUS\,113/789/0-4).

\bibliographystyle{apsrmp}
%%%%%%%%%%%%%%%%%%%%%%%%%%%%%%%%%%%%%%%%%%%%%%%
%%%%%%%%%%%%%%%%%%%%%%%%%%%%%%%%%%%%%%%%%%%%%%%
\thebibliography{999}
\bibliographystyle{apsrmp}
\item[]
Agranovich, V. M. and V.~L.~Ginzburg, 1984,
{\it Crystal Optics with Spatial Dispersion}
(Springer, Berlin).
\item[]
Antezza, M., L.~P.~Pitaevskii, and S.~Stringari, 2004,
Phys. Rev. A {\bf 70}, 053619.
\item[]
Antezza, M., L.~P.~Pitaevskii, and S.~Stringari, 2005,
Phys. Rev. Lett. {\bf 95}, 113202.
\item[]%{A51}
Antonini, P., G.~Bressi, G.~Carugno, G.~Galeazzi,
G.~Messineo, and G.~Ruoso, 2006,
New J. Phys. {\bf 8}, 239.
\item[]
Arnecke, F., H.~Friedrich, and J.~Madro\~{n}ero, 2007,
Phys. Rev. A {\bf 75}, 042903.
\item[]%{39}
Arnold, W., S.~Hunklinger, and K.~Dransfeld, 1979,
Phys. Rev. B {\bf 19}, 6049.
\item[]
Ashcroft, N. W. and N. D. Mermin, 1976,
{\it Solid State Physics}
(Saunders Col\-le\-ge, Philadelphia).
\item[]
Ashourvan, A., M.~Miri, and R.~Golestanian, 2007a,
Phys. Rev. Lett. {\bf 98}, 140801.
\item[]
Ashourvan, A., M.~Miri, and R.~Golestanian, 2007b,
Phys. Rev. E {\bf 75}, 040103.
\item[]%{9}
Babb, J.~F., G.~L.~Klimchitskaya, and V.~M.~Mostepanenko, 2004,
Phys. Rev. A {\bf 70}, 042901.
\item[]
Balian, R. and B.~Duplantier, 1977, Ann. Phys. (N.Y.) {\bf 104}, 300.
\item[]
Balian, R. and B.~Duplantier, 1978, Ann. Phys. (N.Y.) {\bf 112}, 165.
\item[]
Bansal, N. P. and R.~H.~Doremus, 1986,
{\it Handbook of Glass Properties} (Academic, New York).
\item[]
Barash, Yu.~S., 1973,
Izv. Vuzov., Radiofiz. {\bf 16}, 1227
(Sov. Radiophys. {\bf 16}, 945).
\item[]%{25}
Barash, Yu.~S. and V.~L.~Ginzburg, 1975,
Usp. Fiz. Nauk {\bf 116}, 5
(Sov. Phys. Uspekhi {\bf 18}, 305).
\item[]
Barton, G., 1987a,
Proc. Roy. Soc. Lond. A {\bf 410}, 141.
\item[]
Barton, G., 1987b,
Proc. Roy. Soc. Lond. A {\bf 410}, 175.
\item[]%{B1}
Barton, G., 2004,
J. Phys. A: Math. Gen. {\bf 37}, 1011.
\item[]%{B1-2}
Barton, G., 2005,
J. Phys. A: Math. Gen.
{\bf 38}, 2997.
\item[]
Berger, J., 1993,
{\it Statistical Decision Theory and Bayesian Analysis}
(Springer, New York).
\item[]
Bergstr\"{o}m, L., 1997,
Adv. Colloid Interface Sci. {\bf 70}, 125.
\item[]
Berkhout, J. J., O.~J.~Luiten, I.~D.~Setija, T.~W.~Hijmans, T.~Mizusaki,
and J.~T.~M.~Walraven, 1989,
Phys. Rev. Lett. {\bf 63}, 1689.
\item[]
Bezerra, V.~B., G.~L.~Klimchitskaya,
and C.~Romero, 1997,
Mod. Phys. Lett. A {\bf 12}, 2613.
\item[]
Bezerra, V.~B., G.~L.~Klimchitskaya,
and V.~M.~Mostepanenko, 2000a,
Phys. Rev. A
{\bf 62}, 014102.
\item[]
Bezerra, V.~B., G.~L.~Klimchitskaya, and
C.~Romero, 2000b,
{Phys. Rev. A} {\bf 61}, 022115.
\item[]
Bezerra, V.~B., G.~L.~Klimchitskaya,
and V.~M.~Mostepanenko, 2002a,
Phys. Rev. A
{\bf 65}, 052113.
\item[]% {z7}
Bezerra, V.~B., G.~L.~Klimchitskaya,
and V.~M.~Mostepanenko, 2002b,
Phys. Rev. A
{\bf 66}, 062112.
\item[]%{z9}
Bezerra, V.~B., G.~L.~Klimchitskaya, and
C.~Romero, 2002c,
{Phys. Rev. A} {\bf 65}, 012111.
\item[]% {z8}
Bezerra, V.~B.,
G.~L.~Klimchitskaya, V.~M.~Mostepanenko,
and C.~Romero, 2004,
Phys. Rev. A {\bf 69}, 022119.
\item[]%{50}
Bezerra, V.~B., R.~S.~Decca, E.\ Fischbach, B.\ Geyer,
G.\ L.\ Klimchitskaya, D.\ E.\ Krause, D.\ L\'opez,
V.\ M.\ Mostepanenko, and C.\ Romero, 2006,
Phys. Rev. E {\bf 73}, 028101.
\item[]% {z8b}
Bezerra, V.~B., G.~Bimonte,
G.~L.~Klimchitskaya, V.~M.~Mostepanenko,
and C.~Romero, 2007,
Eur. Phys. J. C {\bf 52}, 701.
\item[]% {z8}
Bezerra, V.~B.,
G.~L.~Klimchitskaya, V.~M.~Mostepanenko,
and C.~Romero, 2008,
Phys. Rev. A {\bf 78}, 042901.
\item[]%{z10}
Bimonte, G., 2006a,
Phys. Rev. Lett. {\bf 96}, 160401.
\item[]%{z11}
Bimonte, G., 2006b,
Phys. Rev. E {\bf 73}, 048101.
\item[]%{z12}
Bimonte, G., 2007,
New J. Phys. {\bf 9}, 981.
\item[]%{z11}
Bimonte, G., 2008,
Phys. Rev. A {\bf 78}, 062101.
\item[]
Bimonte, G., E.~Calloni, G.~Esposito, L.~Milano, and L.~Rosa,
2005a, Phys. Rev. Lett. {\bf 94}, 180402.
\item[]
Bimonte, G., E.~Calloni, G.~Esposito,  and L.~Rosa, 2005b,
Nucl. Phys. B {\bf 726}, 441.
\item[]
Bingqian, L., Z. Changchun, and L. Junhua, 1999,
J. Micromech. Microeng. {\bf 9}, 319.
\item[]
Blagov, E.~V., G.~L.~Klimchitskaya, U.~Mohideen, and V.~M.~Mostepanenko,
2004, Phys. Rev. A {\bf 69}, 044103.
\item[]%{B2}
Blagov, E.~V., G.~L.~Klimchitskaya, and V.~M.~Mostepanenko,
2005, Phys. Rev. B {\bf 71}, 235401.
\item[]%{z3}
Blagov, E.~V., G.~L.~Klimchitskaya, and V.~M.~Mostepanenko,
2007, Phys. Rev. B {\bf 75}, 235413.
\item[]%{28}
Blocki, J., J.~Randrup, W.~J.~Swiatecki, and C.~F.~Tsang,
1977,  Ann. Phys. (N.Y.) {\bf 105}, 427.
\item[]%{Boh}
Bohren, C. F. and D. R. Huffmann, 1998,
{\it Absorption and Scattering of Light by Small Particles}
(Wiley, New York).
\item[]%{31}
Bordag, M., 2006,
Phys. Rev. D {\bf 73}, 125018.
\item[]%{10}
Bordag, M., B.~Geyer, G.~L.~Klimchitskaya, and
V.\ M.\ Mos\-te\-pa\-nen\-ko, 1998,
{Phys. Rev. D}  {\bf 58}, 075003.
\item[]%{10-2}
Bordag, M., B.~Geyer, G.~L.~Klimchitskaya, and
V.\ M.\ Mos\-te\-pa\-nen\-ko, 1999,
{Phys. Rev. D}
{\bf 60}, 055004.
\item[]%{10-3}
Bordag, M., B.~Geyer, G.~L.~Klimchitskaya, and
V.\ M.\ Mos\-te\-pa\-nen\-ko, 2000a,
{Phys. Rev. D}
{\bf 62}, 011701(R).
\item[]%{z5}
Bordag, M., B.~Geyer, G.~L.~Klimchitskaya, and
V.\ M.\ Mos\-te\-pa\-nen\-ko, 2000b,
{Phys. Rev. Lett.} {\bf 85}, 503.
\item[]%{B3}
Bordag, M., B.~Geyer, G.~L.~Klimchitskaya, and V.~M.~Mostepanenko,
2006, Phys. Rev. B {\bf 74}, 205431.
\item[]
Bordag, M.,  G.~L.~Klimchitskaya, and
V.\ M.\ Mos\-te\-pa\-nen\-ko, 1994,
{Mod. Phys. Lett. A} {\bf 9}, 2515.
\item[]
Bordag, M.,  G.~L.~Klimchitskaya, and
V.\ M.\ Mos\-te\-pa\-nen\-ko, 1995a,
{Int. J. Mod. Phys. A} {\bf 10}, 2661.
\item[]
Bordag, M.,  G.~L.~Klimchitskaya, and
V.\ M.\ Mos\-te\-pa\-nen\-ko, 1995b,
{Phys. Lett. A} {\bf 200}, 95.
\item[]%{21}
Bordag, M., U.~Mohideen, and V.~M.~Mostepanenko, 2001,
{ Phys. Rep.} {\bf 353}, 1.
\item[]%{48}
Bostr\"{o}m, M. and B.~E.~Sernelius, 2000,
Phys. Rev. Lett. {\bf 84}, 4757.
\item[]%{z4}
Bostr\"{o}m, M. and B.~E.~Sernelius, 2004,
Physica A {\bf 339}, 53.
\item[]
Boyer, T. H., 1968,
Phys. Rev. {\bf 174}, 1764.

\item[]
Brandt, S., 1976,
{\it Statistical and Computational Methods in Data Analysis}
(North-Holland, Amsterdam).
\item[]%{19}
Bressi, G., G.\ Carugno, R.~Onofrio, and G.~Ruoso, 2002,
Phys. Rev. Lett. {\bf 88}, 041804.
\item[]%{49}
Brevik, I., J.~B.~Aarseth, J.~S.~H{\o}ye, and K.~A.~Milton, 2005,
Phys. Rev. E {\bf 71}, 056101.
\item[]
Brevik, I., S.~A.~Ellingsen, J.~S.~H{\o}ye, and K.~A.~Milton, 2008,
J. Phys. A: Math. Theor. {\bf 41}, 164017.
\item[]%{z14}
Brown, L.~S.  and G.~J.~Maclay, 1969,
{Phys. Rev.} {\bf 184}, 1272.
\item[]%{A50}
Brown-Hayes, M., D.~A.~R.~Dalvit, F.~D.~Mazzitelli,
W.~J.~Kim, and R.~Onofrio, 2005,
Phys. Rev. A {\bf 72}, 052102.
\item[]
Brownlee, K. A., 1965,
{\it Statistical Theory and Methodology in Science and
Engineering} (Wiley, New York).
\item[]
Bryksin, V. V. and P. M. Petrov, 2008,
 Fiz. Tverdogo Tela {\bf 50}, 222
(Phys. Solid State {\bf 50}, 229).
\item[]
Buhmann, S.~Y. and S.~Scheel, 2008,
Phys. Rev. Lett. {\bf 100}, 253201.
\item[]
Buhmann, S.~Y. and D.-G.~Welsch, 2007,
Progr. Quant. Electronics {\bf 31}, 51.
\item[]
Buhmann, S.~Y. and D.-G.~Welsch, 2008,
Phys. Rev. A {\bf 77}, 012110.
\item[]
Buhmann, S.~Y., D.-G.~Welsch, and T.\ Kampf, 2005,
Phys. Rev. A {\bf 72}, 032112.
\item[]%{2}
Buks, E. and M.~L.~Roukes, 2001,
Phys. Rev. B {\bf 63}, 033402.
\item[]%{33}
Buks, E. and M.~L.~Roukes, 2002,
Nature {\bf 419}, 119.
\item[]%{30}
Bulgac,  A., P.~Magierski, and A.~Wirzba, 2006,
Phys. Rev. D {\bf 73}, 025007.
\item[]
B\"{u}scher, R. and T.~Emig, 2004,
Phys. Rev. A {\bf 69}, 062101.
\item[]
B\"{u}scher, R. and T.~Emig, 2005,
Phys. Rev. Lett. {\bf 94}, 133901.
\item[]
Capasso, F., J.~N.~Munday, D.~Iannuzzi, and H.~B.~Chan, 2007,
IEEE J. Quant. Electr. {\bf 13}, 400.
\item[]%{46}
Caride, A.~O.,  G.~L.~Klimchitskaya, V.~M.~Mostepanenko,
and S.~I.~Zanette, 2005,
{Phys. Rev.} A {\bf 71} 042901.
\item[]
Carlin, B. P. and T. A. Louis, 2000,
{\it Bayes and Empirical Bayes Methods of Data Analysis}
(CRC Press, Boco Raton).
\item[]% {1}
Casimir, H.~B.~G., 1948,
{ Proc. K. Ned. Akad. Wet. B}
{\bf 51}, 793.
\item[]% {z13}
Casimir, H.~B.~G. and D.~Polder, 1948,
Phys. Rev. {\bf 73}, 360.
\item[]%{cg}
Castillo-Garza, R., C.-C.~Chang, D.~Jimenez, G.~L.~Klimchitskaya,
V.\ M.\ Mos\-te\-pa\-nen\-ko, and U.~Mohideen, 2007,
Phys. Rev. A {\bf 75}, 062114.
\item[]%{3}
Chan, H.~B., V.~A.~Aksyuk, R.~N.~Kleiman, D.~J.~Bishop, and F.~Capasso,
2001a, Science, {\bf 291}, 1941.
\item[]%{3-2}
Chan, H.~B., V.~A.~Aksyuk, R.~N.~Kleiman, D.~J.~Bishop, and F.~Capasso,
2001b, Phys. Rev. Lett. {\bf 87}, 211801.
\item[]%{
Chan, H.~B., Y. Bao, J. Zou, R. A. Cirelli, F. Clemens, W. M. Mansfield,
and C. S. Pai, 2008,
Phys. Rev. Lett. {\bf 101}, 030401.
\item[]
Chazalviel, J.-N., 1999.
{\it Coulomb Screening of Mobile Charges: Applications to
Material Science, Chemistry and Biology}
(Birkhauser, Boston).
\item[]
Chen, F. and U.~Mohideen, 2001,
Rev. Sci. Instrum. {\bf 72}, 3100.
\item[]%{20}
Chen, F., U.~Mohideen, G.~L.~Klimchitskaya, and
V.\ M.\ Mos\-te\-pa\-nen\-ko, 2002a,
Phys. Rev. Lett. {\bf 88}, 101801.
\item[]%{20-2}
Chen, F., U.~Mohideen, G.~L.~Klimchitskaya, and
V.\ M.\ Mos\-te\-pa\-nen\-ko, 2002b,
Phys. Rev. A {\bf 66}, 032113.
\item[]%{35}
Chen, F.,  G.~L.~Klimchitskaya, U.~Mohideen, and
V.\ M.\ Mos\-te\-pa\-nen\-ko, 2003,
Phys. Rev. Lett. {\bf 90}, 160404.
\item[]%{18-2}
Chen, F., G.~L.~Klimchitskaya, U.~Mohideen, and
V.\ M.\ Mos\-te\-pa\-nen\-ko, 2004a,
Phys. Rev. A {\bf 69}, 022117.
\item[]%{36}
Chen, F., U.~Mohideen, and
P.~W.~Milonni, 2004b,
In: Quantum Field Theory Under the Influence of External
Conditions, ed. K.~A.~Milton (Rinton Press, Princeton).
\item[]%{37}
Chen, F., U.~Mohideen, G.~L.~Klimchitskaya, and
V.\ M.\ Mos\-te\-pa\-nen\-ko, 2005,
Phys. Rev. A {\bf 72}, 020101(R).
\item[]%{380}
Chen, F.,  G.~L.~Klimchitskaya,
V.\ M.\ Mos\-te\-pa\-nen\-ko, and U.~Mohideen, 2006a,
Phys. Rev. Lett.  {\bf 97}, 170402.
\item[]%{38}
Chen, F., U.~Mohideen, G.~L.~Klimchitskaya, and
V.\ M.\ Mos\-te\-pa\-nen\-ko, 2006b,
Phys. Rev. A  {\bf 74}, 022103.
\item[]%{40}
Chen, F.,  G.~L.~Klimchitskaya,
V.\ M.\ Mos\-te\-pa\-nen\-ko, and U.~Mo\-hi\-deen, 2007a,
Optics Express {\bf 15}, 4823.
\item[]%{40a}
Chen, F.,  G.~L.~Klimchitskaya,
V.\ M.\ Mos\-te\-pa\-nen\-ko, and U.~Mo\-hi\-deen, 2007b,
Phys. Rev. B {\bf 76}, 035338.
\item[]%{z12}
Chen, F., U.~Mohideen, G.~L.~Klimchitskaya, and
V.\ M.\ Mos\-te\-pa\-nen\-ko, 2007c,
Phys. Rev. Lett. {\bf 98}, 068901.
\item[]
Chettiar, U. K., A. V. Kildishev, H.-K. Yuan,
W.-S. Cai, S.-M. Xiao, V. P. Drachev, and V. M.
Shalaev, 2007, Opt. Lett. {\bf 321}, 1671.
\item[]
Chiu, H.-C., C.-C.~Chang, R.~Castillo-Garza, F.~Chen,
and U.~Mohideen, 2008,
J. Phys. A: Math. Theor. {\bf 41}, 164022.
\item[]
Chumak, A. A., P. W. Milonni, and G. P. Berman, 2004,
Phys. Rev. B {\bf 70}, 085407.

\item[]
Cochran, W. G., 1954,
Biometrics {\bf 10}, 101.
\item[]
Contreras-Reyes, A. M. and W.~L.~Moch\'{a}n, 2005,
Phys. Rev. A {\bf 72}, 034102.
\item[]
C\^{o}t\'{e}, R., B.~Segev, and M.~G.~Raizen, 1998,
Phys. Rev. A {\bf 58}, 3999.
\item[]
Dalvit, D. A. R. and S. K. Lamoreaux, 2008,
Phys. Rev. Lett. {\bf 101}, 163203.
\item[]
Dalvit, D. A. R., P. A. Maia Neto, A. Lambrecht, and S. Reynaud, 2008,
Phys. Rev. Lett. {\bf 100}, 040405.
\item[]
Decca, R.~S., E.~Fischbach, G.~L.~Klimchitskaya, D.~E.~Krause,
D.~L\'opez, and V.~M.~Mostepanenko, 2003a,
Phys. Rev. D {\bf 68}, 116003.
\item[]
Decca, R.~S.,  D.~L\'opez, E.~Fischbach, and
 D.~E.~Krause, 2003b,
Phys. Rev. Lett. {\bf 91}, 050402.
\item
Decca, R. S., D.~L\'opez, E.~Fischbach, G.~L.~Klimchitskaya,
 D.~E.~Krause, and V.~M.~Mostepanenko, 2004,
J. Low Temp. Phys. {\bf 135}, 63.
\item[]
Decca, R.~S.,  D.~L\'opez, H.~B.~Chan, E.~Fischbach,
 D.~E.~Krause, and C.~R.~Jamell, 2005a,
Phys. Rev. Lett. {\bf 94}, 240401.
\item[]%{13}
Decca, R.~S., D.~L\'opez, E.~Fischbach, G.~L.~Klimchitskaya,
 D.~E.~Krause, and V.~M.~Mostepanenko, 2005b,
 Ann. Phys. (N.Y.) {\bf 318}, 37.
\item[]%{13a}
Decca, R.~S.,  D.~L\'opez, E.~Fischbach, G.~L.~Klimchitskaya,
 D.~E.~Krause, and V.~M.~Mostepanenko, 2007a,
Phys. Rev. D {\bf 75}, 077101.
\item[]%{13ab}
Decca, R.~S., D.~L\'opez, E.~Fischbach, G.~L.~Klimchitskaya,
 D.~E.~Krause, and V.~M.~Mostepanenko, 2007b,
Eur. Phys. J. C {\bf 51}, 963.
\item[]
Decca, R.~S., E.~Fischbach, G.~L.~Klimchitskaya, D.~E.~Krause,
D.~L\'opez, U. Mohideen, and V.~M.~Mostepanenko, 2008,
arXiv:0809.3576 (Phys. Rev. A, to appear).
\item[]
Derjaguin, B., 1934, Kolloid. Z. {\bf 69}, 155.
\item[]
Derjaguin, B., N.~V.~Chumakov, and Ya.~I.~Rabinovich,
1988, Adv. Colloid Intefrace Sci. {\bf 28}, 197.
\item[]
Doyle, J. M., J.~C.~Sandberg, I.~A.~Yu, C.~L.~Cesar, D.~Kleppner,
and T.~J.~Greytak, 1991,
Phys. Rev. Lett. {\bf 67}, 603.
\item[]
Druzhinina, V. and M.~DeKieviet, 2003,
Phys. Rev. Lett. {\bf 91}, 193202.
\item[]
Duraffourg, L. and P.~Andreucci, 2006,
Phys. Lett. A {\bf 359}, 406.
\item[]%{DLP}
Dzyaloshinskii, I.~E., E.~M.~Lifshitz,
and L.~P.~Pitaevskii, 1961,
{Usp. Fiz. Nauk} {\bf 73}, 381
(Adv. Phys. {\bf 38}, 165).
\item[]
Eberlein, C. and R. Zietal, 2007,
Phys. Rev. A {\bf 75}, 032516.
\item[]%{Ederth}
Ederth, T., 2000,
Phys. Rev. A {\bf 62}, 062104.
\item[]
Emig, T., 2007, Phys. Rev. Lett.
{\bf 98}, 160801.
\item[]
Emig, T., 2008,
J. Stat. Mech. P04007.
\item[]%{EmigB}
Emig, T. and R.~B\"{u}scher, 2004,
 Nucl. Phys. B {\bf 696}, 468.
\item[]
Emig, T., N.~Graham, R.~L.~Jaffe, and M.~Kardar, 2007,
Phys. Rev. Lett. {\bf 99}, 170403.
\item[]
Emig, T., N.~Graham, R.~L.~Jaffe, and M.~Kardar, 2008,
Phys. Rev. D {\bf 77}, 025005.
\item[]%{zz3}
Emig, T., A.~Hanke, R.~Golestanian, and M.~Kardar, 2001,
Phys. Rev. Lett. {\bf 87}, 260402.
\item[]%{zz4}
Emig, T., A.~Hanke, R.~Golestanian, and M.~Kardar, 2003,
Phys. Rev. A {\bf 67}, 022114.
\item[]%{29}
Emig, T., R.~L.~Jaffe, M.~Kardar, and A.~Scardicchio, 2006,
Phys. Rev. Lett. {\bf 96}, 080403.
\item[]
Esquivel, R. and V.~B.~Svetovoy, 2004,
Phys. Rev. A {\bf 69}, 062102.
\item[]%{Mexica-1}
Esquivel, R., C.~Villarreal, and W.~L.~Moch\'{a}n, 2003,
Phys. Rev. A {\bf 68}, 052103; {\bf 71}, 029904(E)

\item[]%{Rez}
Feinberg, J., A.~Mann, and M.~Revzen, 2001,
{Ann. Phys.} (N.Y.) {\bf 288}, 103.
\item[]
Fermani, R., S.~Scheel, and P.~L.~Knight, 2007,
Phys. Rev. A {\bf 75}, 062905.
\item[]
Fischbach, E., D.~E.~Krause, V.~M.~Mostepanenko, and
M.~Novello, 2001,
Phys. Rev. D {\bf 64}, 075010.
\item[]
Friedrich, H., G.~Jacoby, and C.~G.~Meister, 2002,
Phys. Rev. A {\bf 65}, 032902.
\item[]
Friedrich, H. and J.~Trost, 2004,
Phys. Rep. {\bf 397}, 359.

\item[]%{z24}
Genet, C., A.~Lambrecht, P.~Maia Neto, and S.~Reynaud, 2003a,
Europhys. Lett. {\bf 62}, 484.
\item[]%{z23}
Genet, C., A.~Lambrecht, and S.~Reynaud, 2000,
Phys. Rev. A {\bf 62}, 012110.
\item[]%{z30}
Genet, C., A.~Lambrecht, and S.~Reynaud, 2003b,
Phys. Rev. A {\bf 67}, 043811.
\item[]
Gerlach, E., 1971,
Phys. Rev. B {\bf 4}, 393.

\item[]
Geyer, B., G.~L.~Klimchitskaya, and V.~M.~Mostepanenko, 2001,
Int. J. Mod. Phys. A {\bf 16}, 3291.
\item[]%{45bb}
Geyer, B., G.~L.~Klimchitskaya, and V.~M.~Mostepanenko, 2003,
Phys. Rev. A {\bf 67}, 062102.
\item[]
Geyer, B., G.~L.~Klimchitskaya, and V.~M.~Mostepanenko, 2004,
Phys. Rev. A {\bf 70}, 016102.
\item[]%{45b}
Geyer, B., G.~L.~Klimchitskaya, and V.~M.~Mostepanenko, 2005a,
Phys. Rev. A {\bf 72}, 022111.
\item[]%{7}
Geyer, B., G.~L.~Klimchitskaya, and V.~M.~Mostepanenko, 2005b,
Phys. Rev. D {\bf 72}, 085009.
\item[]%{IJMPA}
Geyer, B., G.~L.~Klimchitskaya, and V.~M.~Mostepanenko, 2006,
Int. J. Mod. Phys. A {\bf 21}, 5007.
\item[]
Geyer, B., G.~L.~Klimchitskaya, and V.~M.~Mostepanenko, 2007,
J. Phys. A: Math. Theor. {\bf 40}, 13485.
\item[]%{48a}
Geyer, B., G.~L.~Klimchitskaya, and V.~M.~Mostepanenko, 2008a,
Ann. Phys. (N.Y.) {\bf 323}, 291.
\item[]
Geyer, B., G.~L.~Klimchitskaya,  U.~Mohideen,
and V.~M.~Mostepanenko, 2008b,
Phys. Rev. A {\bf 77}, 036102.
\item[]%{32}
Gies, H. and K.~Klingm\"{u}ller, 2006a,
Phys. Rev. Lett. {\bf 96}, 220401.
\item[]%{32-2}
Gies, H. and K.~Klingm\"{u}ller, 2006b,
Phys. Rev. D {\bf 74}, 045002.
\item[]%{32}
Gies, H. and K.~Klingm\"{u}ller, 2006c,
Phys. Rev. Lett. {\bf 97}, 220405.
\item[]
Ginzburg, V. L., 1985,
{\it Physics and Astrophysics} (Pergamon Press, Oxford).
\item[]
Golestanian, R. and M.~Kardar, 1997,
Phys. Rev. Lett. {\bf 78}, 3421.
\item[]
Golestanian, R. and M.~Kardar, 1998,
Phys. Rev. A {\bf 58}, 1713.
\item[]
Greenaway, D. L., G. Harbeke, F. Bassani, and E. Tosatti, 1969,
Phys. Rev. {\bf 178}, 1340.
\item[]
Harber, D.~M., J.~M.~McGuirk, J.~M.~Obrecht, and E.~A.~Cornell, 2003,
J. Low Temp. Phys. {\bf 133}, 229.
\item[]
Harber, D.~M., J.~M.~Obrecht, J.~M.~McGuirk, and E.~A.~Cornell, 2005,
Phys. Rev. A {\bf 72}, 033610.
\item[]
Hargreaves, C.~M., 1965, Proc. K. Ned. Akad. Wet. B
{\bf 68}, 231.
\item[]%{18}
Harris, B.~W., F.~Chen, and U.~Mohideen, 2000,
Phys. Rev. A {\bf 62}, 052109.
\item[]%{Henk1}
Henkel, C. and K.~Joulain, 2005,
Europhys. Lett. {\bf 72}, 929.
\item[]%{Henk2}
Henkel, C., K.~Joulain, J.-P.~Mulet, and J.-J.~Greffet, 2004,
Phys. Rev. A {\bf 69}, 023808.
\item[]
Hough, D. B. and L.~R.~White, 1980,
Adv. Colloid Interface Sci. {\bf 14}, 3.
\item[]% {z22}
H{\o}ye, J.~S., I.~Brevik, J.~B.~Aarseth,
and K.~A.~Milton, 2003,
{Phys. Rev.} E {\bf 67}, 056116.
\item[]%{z17}
H{\o}ye, J.~S., I.~Brevik, J.~B.~Aarseth, and K.~A.~Milton, 2006,
{J. Phys. A}: Math. Gen. {\bf 39}, 6031.
\item[]% {z22}
H{\o}ye, J.~S., I.~Brevik, S.~A.~Ellingsen, and J.~B.~Aarseth,
2007, {Phys. Rev.} E {\bf 75}, 051127.
\item[]%
H{\o}ye, J.~S., I.~Brevik, S.~A.~Ellingsen, and J.~B.~Aarseth,
2008, {Phys. Rev.} E {\bf 77}, 023102.
\item[]%{zzz1}
Iannuzzi, D., I.~Gelfand, M.~Lisanti, F.~Capasso, 2004a,
in: {\it Quantum Field Theory under the Influence of
External conditions}, ed. K.\ A.\ Milton
(Rinton Press, Princeton).
\item[]%{34}
Iannuzzi, D., M.~Lisanti, and F.~Capasso, 2004b,
Proc. Nat. Acad. Sci. USA {\bf 101}, 4019.
\item[]
Intravaia, F. and C.~Henkel, 2008,
J. Phys. A: Math. Theor. {\bf 41}, 164018.
\item[]%
Intravaia, F. and A.~Lambrecht, 2005,
Phys. Rev. Lett. {\bf 94}, 110404.
\item[]%{zzz2}
Intravaia, F., C.~Henkel, and A.~Lambrecht, 2007,
Phys. Rev. A {\bf 76}, 033820.

\item[]%{In-1}
Inui, N., 2003,
J. Phys. Soc. Jap. {\bf 72}, 2198.
\item[]%{In-2}
Inui, N., 2004,
J. Phys. Soc. Jap. {\bf 73}, 332.
\item[]%{In-3}
Inui, N., 2006,
J. Phys. Soc. Jap. {\bf 75}, 024004.
\item[]
Israelashvili, J., 1992,
{\it Intermolecular and Surface Forces} (Academic, New York).
\item[]
Jackson, J. D., 1999,
{\it Classical Electrodynamics} (John Wiley, New York).
\item[]
Jaffe, R. L., 2005,
Phys. Rev. D {\bf 72}, 021301.
\item[]
Jourdan, G., A.~Lambrecht, F.~Comin, and J.~Chevrier, 2009,
Europhys. Lett. {\bf 85}, 31001.
\item[]
Jurisch, A. and H.~Friedrich, 2004,
Phys. Rev. A {\bf 70}, 032711.
\item[]
Jurisch, A. and H.~Friedrich, 2006,
Phys. Lett. A {\bf 349}, 230.
\item[]%{4a}
Kardar, M. and R.~Golestanian, 1999,
Rev. Mod. Phys. {\bf 71}, 1233.
\item[]
Kats, E. I., 1977,
Zh. Eksp. Teor. Fiz. {\bf 73}, 212
(Sov. Phys. JETP {\bf 46}, 109).
\item[]%
Kenneth, O. and I.~Klich, 2006,
Phys. Rev. Lett. {\bf 97}, 160401.
\item[]%
Kenneth, O. and I.~Klich, 2008,
Phys. Rev. B {\bf 78}, 014103.
\item[]%{magnit}
Kenneth, O., I.~Klich, A.~Mann, and M.~Revzen, 2002,
Phys. Rev. Lett. {\bf 89}, 033001.
\item[]
Kim, W. J., M. Brown-Hayes, D. A. R. Dalvit, J. H. Brownell,
and R. Onofrio, 2008,
Phys. Rev. A {\bf 78}, 020101(R).
\item[]
Kittel, C., 1996, {\it Introduction to Solid State Physics}
(John Wiley, New York).
\item[]
Klimchitskaya, G.~L., 2008, arXiv:0811.4398
(J. Phys. Conf. Series, to appear).
\item[]
Klimchitskaya, G.~L. and B.~Geyer,
2008, J. Phys. A: Math. Theor. {\bf 41}, 164032.
\item[]
Klimchitskaya, G.~L. and V.~M.~Mostepanenko, 2001,
Phys. Rev. A {\bf 63}, 062118.
\item[]
Klimchitskaya, G.~L. and V.~M.~Mostepanenko, 2006,
Contemp. Phys. {\bf 47}, 131.
\item[]%{PRBcom}
Klimchitskaya, G.~L. and V.~M.~Mostepanenko, 2007,
Phys. Rev. B {\bf 75}, 036101.
\item[]
Klimchitskaya, G.~L. and V.~M.~Mostepanenko, 2008a,
Phys. Rev. E {\bf 77}, 023101.
\item[]
Klimchitskaya, G.~L. and V.~M.~Mostepanenko, 2008b,
J. Phys. A: Math. Theor. {\bf 41}, 312002(F).
\item[]
Klimchitskaya, G.~L. and Yu.~V.~Pavlov, 1996,
Int. J. Mod. Phys. A {\bf 11}, 3723.
\item[]%{15-2}
Klimchitskaya, G.~L., A.~Roy, U.~Mohideen, and
V.\ M.\ Mos\-tepanenko, 1999,
{ Phys. Rev. A} {\bf 60}, 3487.
\item[]%{27}
Klimchitskaya, G.~L., U.~Mohideen, and V.~M.~Mostepanenko,
2000, { Phys. Rev. A} {\bf 61}, 062107.
\item[]
Klimchitskaya, G.~L., S.~I.~Zanette, and A.~O.~Caride, 2001,
{ Phys. Rev. A} {\bf 63}, 014101.
\item[]%{z15}
Klimchitskaya, G.~L., R. S. Decca, E.
Fischbach,  D. E. Krause, D. L\'opez, and V.
M. Mostepanenko, 2005,
Int. J. Mod. Phys. A {\bf 20}, 2205.
\item[]%{z20}
Klimchitskaya, G.~L., E.~V.~Blagov,  and V.~M.~Mostepanenko,
2006a, J. Phys. A: Math. Gen. {\bf 39}, 6481.
\item[]%{45a}
Klimchitskaya, G.~L., F.~Chen, R.\ S.\ Decca,
E.\ Fischbach,  D.\ E.\ Krause, D.\ L\'opez,
U.~Mo\-hi\-deen, and V.~M.~Mostepanenko, 2006b,
J. Phys. A: Math. Gen. {\bf 39}, 6485.
\item[]%{JPA}
Klimchitskaya, G.~L., B.~Geyer,  and V.~M.~Mostepanenko,
2006c, J. Phys. A: Math. Gen. {\bf 39}, 6495.
\item[]%{z16}
Klimchitskaya, G.~L.,  U.~Mohideen, and V.~M.~Mostepanenko,
2007a, J. Phys. A: Math. Theor. {\bf 40}, 339(F).
\item[]
Klimchitskaya, G.~L.,  U.~Mohideen, and V.~M.~Mostepanenko,
2007b, J. Phys. A: Math. Theor. {\bf 40}, 841(F).
\item[]
Klimchitskaya, G.~L., E.~V.~Blagov,  and V.~M.~Mostepanenko,
2008a, J. Phys. A: Math. Theor. {\bf 41}, 164012.
\item[]
Klimchitskaya, G.~L.,  U.~Mohideen, and V.~M.~Mostepanenko,
2008b, J. Phys. A: Math. Theor. {\bf 41}, 432001(F).
\item[]
Kondepugi, D. and I. Prigogine, 1998,
{\it Modern Thermodynamics}
(John Wiley, New York).
\item[]
Kouznetsov, D. and H.~Oberst, 2005,
{ Phys. Rev. A} {\bf 72}, 013617.
\item[]%{34aaaa}
Krause,  D.~E., R.~S.~Decca, D.~L\'opez, and E.~Fischbach,
2007, Phys. Rev. Lett. {\bf 98}, 050403.
\item[]
Krech, M., 1994,
{\it The Casimir Effect in Critical Systems}
(World Scientific, Singapore).
\item[]%{0001}
Lambrecht, A. and  V. N. Marachevsky, 2008,
Phys. Rev. Lett. {\bf 100}, 160403.
\item[]%{0001}
Lambrecht, A. and  S.~Reynaud, 2000,
Eur. Phys. J. D {\bf 8}, 309.
\item[]
Lambrecht, A., V.~V.~Nesvizhevsky, R.~Onofrio, and
S.~Reynaud, 2005,
Class. Quant. Grav. {\bf 22}, 5397.
\item[]
Lambrecht, A., P. A. Maia Neto, and
S.~Reynaud, 2006,
New J. Phys. {\bf 8}, 243.
\item[]%{0002}
Lambrecht, A., I.~Pirozhenko, L.~Duraffourg, and P.~Andreucci,
2007, Eur. Phys. Lett. {\bf 77}, 44006; {\bf 81}, 19901(E).

\item[]%{14}
Lamoreaux, S.~K., 1997,{ Phys. Rev. Lett.}
{\bf 78}, 5.
\item[]
Lamoreaux, S.~K., 1998,{ Phys. Rev. Lett.}
 {\bf 81}, 5475(E).
\item[]
Lamoreaux, S.~K., 1999,{ Phys. Rev. A}
{\bf 59}, 3149(R).
\item[]%{22}
Lamoreaux, S.~K., 2005,
{Rep. Progr. Phys.}
{\bf 68}, 201.
\item[]%{A49}
Lamoreaux, S.~K. and W.~T.~Buttler, 2005,
Phys. Rev. E {\bf 71}, 036109.
\item[]
Landau, L.~D. and E.~M.~Lifshitz, 1980,
{\it Statistical Physics}
(Pergamon Press, Oxford), Pt.I.
\item[]
Landau, L.~D., E.~M.~Lifshitz, and L.~P.~Pitaevskii, 1984,
{\it Electrodynamics of Continuous Media}
(Pergamon Press, Oxford).
\item[]
Lehmann, E. L. and J. P. Romano, 2005,
{\it Testing Statistical Hypotheses}
(Springer, New York).
\item[]
Lennard-Jones, J. E. and A.~F.~Devonshire, 1936,
Proc. R. Soc. London A {\bf 156}, 6.
\item[]
Leonhardt, U. and T. G. Philbin, 2007,
New J. Phys. {\bf 9}, 254.
\item[]%{23}
Lifshitz, E.~M., 1956,
Zh. Eksp. Teor. Fiz. {\bf 29}, 94
(Sov. Phys. JETP  {\bf 2}, 73).
\item[]%{24}
Lifshitz, E.~M. and L.~P.~Pitaevskii, 1980,
{\it Statistical Physics} (Pergamon Press, Oxford), Pt.II.
\item[]
Lifshitz, E.~M. and L.~P.~Pitaevskii, 1981,
{\it Physical Kinetics}
(Pergamon Press, Oxford).
\item[]
Lifshitz, I.~M., M.~Ya.~Azbel', and M.~I.~Kaganov, 1973,
{\it Electron Theory of Metals} (Consultans Bureau, New York).
\item[]%{skin}
Lisanti, M., D.~Iannuzzi, and F.~Capasso, 2005,
Proc. Nat. Acad. Sci. USA {\bf 102}, 11989.
\item[]
London, F, 1930, Zs. Phys. {\bf 63}, 245.
\item[]%{11}
Long, J.~C., H.~W.~Chan, and J.~C.~Price, 1999,
Nucl. Phys. B {\bf 539}, 23.
\item[]
Lukosz, W., 1971, Physica {\bf 56}, 109.
\item[]
Madro\~{n}ero, J. and H.~Friedrich, 2007,
Phys. Rev. A {\bf 75}, 022902.
\item[]%{MahNinh}
Mahanty, J. and B.~W.~Ninham, 1976,
{\it Dispersion Forces}
(Academic, New York).
\item[]%{z31a}
Maia Neto, P. A., A.~Lambrecht,  and S.~Reynaud, 2005,
Phys. Rev. A {\bf 72}, 012115.
\item[]%
Maia Neto, P. A., A.~Lambrecht,  and S.~Reynaud, 2008,
Phys. Rev. A {\bf 78}, 012115.
\item[]
Maradudin, A.~A. and P.~Mazur, 1980,
Phys. Rev. B {\bf 22}, 1677.
\item[]
Mazur, P. and  A.~A. Maradudin, 1981,
Phys. Rev. B {\bf 23}, 695.
\item[]
McGuirk, J. M., D.~M.~Harber, J.~M.~Obrecht,  and E.~A.~Cornell, 2004,
Phys. Rev. A {\bf 69}, 062905.

\item[]
Mehra, J., 1967, Physica (Amsterdam) {\bf 37}, 145.
\item[]
Meurk, A., P.~F.~Luckham, and L.~Bergstr\"{o}m, 1997,
Langmuir {\bf 13}, 3896.
\item[]%{z27}
Milonni, P.~W., 1994,
{\it The Quantum Vacuum. An Introduction to Quantum Electrodynamics}
(Academic Press, San Diego).
\item[]
Milonni, P.~W., 2007,
Phys. Scr. {\bf 76}, C167.
\item[]%{z32}
Milton, K.~A., 2001,
{\it The Casimir Effect: Physical Manifestation of Zero-Point Energy}
(World Scientific, Singapore).
\item[]%{z33}
Milton, K.~A., 2004,
J. Phys. A: Math. Gen. {\bf 37}, R209.
\item[]%{15}
Mohideen, U. and A.~Roy, 1998,
{ Phys. Rev. Lett.}
{\bf 81}, 4549.
\item[]
Mostepanenko, V.~M. and B.~Geyer, 2008,
{J. Phys. A}: Math. Theor. {\bf 41}, 164014.
\item[]
Mostepanenko, V. M. and M.~Novello, 2001,
Phys. Rev. D {\bf 63}, 115003.
\item[]
Mostepanenko, V.~M. and N.~N.~Trunov, 1985,
Yadern. Fiz. {\bf 42}, 1297
(Sov. J. Nucl. Phys. {\bf 42}, 818).
\item[]
Mostepanenko, V. M. and N.~N.~Trunov, 1988,
{ Usp. Fiz. Nauk\/} {\bf 156}, 385
({Sov. Phys. Uspekhi} {\bf 31}, 965).
\item[]%{z18}
Mostepanenko, V.~M. and N.~N.~Trunov, 1997,
{\it The
Casimir Effect and its Applications}
(Clarendon Press, Oxford).
\item[]
Mostepanenko, V.~M., J.~F.~Babb, A.~O.~Caride, G.~L.~Klimchitskaya,
and S.~I.~Zanette, 2006a,
{J. Phys. A}: Math. Gen. {\bf 39}, 6583.
\item[]%{statusJPA06}
Mostepanenko, V.~M., V.~B.~Bezerra,
R.~S.~Decca, E.~Fischbach,
B.~Geyer, G.~L.~Klimchitskaya, D.~E.~Krause,
D.~L\'opez, and C.~Romero, 2006b,
{J. Phys. A}: Math. Gen. {\bf 39}, 6589.
\item[]%
Mostepanenko, V.~M., R.~S.~Decca, E.~Fischbach,
G.~L.~Klimchitskaya, D.~E.~Krause, and
D.~L\'opez,  2008,
{J. Phys. A}: Math. Theor. {\bf 41}, 164054.
\item[]%
Mostepanenko, V.~M., R.~S.~Decca, E.~Fischbach,
G.~Geyer, G.~L.~Klimchitskaya, D.~E.~Krause,
D.~L\'opez,  and U. Mohideen, 2009,
{Int. J. Mod. Phys. A}, to appear.
\item[]
Mott, N. F., 1990,
{\it Metal-Insulator Transitions}, 2nd ed. (Taylor and Francis, London).
\item[]
Munday, J.~N. and F.~Capasso, 2007,
Phys. Rev. A {\bf 75}, 060102(R).
\item[]
Munday, J.~N. and F.~Capasso, 2008,
Phys. Rev. A {\bf 77}, 036103.
\item[]
Munday, J.~N., D.~Iannuzzi, Yu.~Barash, and F.~Capasso, 2005,
Phys. Rev. A {\bf 71}, 042102.
\item[]
Munday, J.~N., F.~Capasso, V. A. Parsegian, and S. M. Bezrukov,
2008, Phys. Rev. A {\bf 78}, 032109.
\item[]
Munday, J.~N., F.~Capasso, and V. A. Parsegian,
2009, Nature {\bf 457}, 170.
\item[]
Nagai, M. and M.~Kuwata-Gonokami, 2002,
J. Phys. Soc. Jap. {\bf 71}, 2276.
\item[]
Nayak, V. U., D.~E.~Edwards, and N.~Masuhara, 1983,
Phys. Rev. Lett. {\bf 50}, 990.
\item[]
Ninham, B. W., V.~A.~Parsegian,  and J.~H.~ Weiss, 1970,
J. Stat. Phys. {\bf 2}, 323.
\item[]
Oberst, H., D.~Kouznetsov, K.~Shimizu, J.-I.~Fujita,
and F.~Shimizu, 2005a,
Phys. Rev. Lett. {\bf 94}, 013203.
\item[]
Oberst, H., Y.~Taashiro, K.~Shimizu, and F.~Shimizu, 2005b,
Phys. Rev. A {\bf 71}, 052901.
\item[]%{64a}
Obrecht, J.~M.,  R.~J.~Wild, M.~Antezza, L.~P.~Pitaevskii,
S.~Stringari, and E.~A.~Cornell, 2007,
Phys. Rev. Lett. {\bf 98}, 063201.

\item[]% {41}
Palik, E. D. (ed.), 1985,
{\it Handbook of Optical Constants of Solids}
(Academic, New York).
\item[]%{zz1}
Parsegian, V.~A., 2005,
{\it Van der Waals Forces: A Handbook for Biologists,
Chemists, Engineers, and Physicists}
(Cambridge University Press, Cambridge).
\item[]%{zz2}
Parsegian, V.~A. and G.~H.~Weiss, 1972,
J. Colloid Interface Sci. {\bf 40}, 35.
\item[]
Pasquini, T.~A., M.~Saba, G.-B.~Jo, Y.~Shin, W.~Ketterle, and
D.~E.~Pritchard, 2006,
Phys. Rev. Lett. {\bf 97}, 093201.
\item[]
Pasquini, T.~A., Y.~Shin, C.~Sanner, M.~Saba, A.~Schirotzek,
D.~E.~Pritchard, and W.~Ketterle, 2004,
Phys. Rev. Lett. {\bf 93}, 233201.
\item[]
Pendry, J. B., A. J. Holden, D. J. Robbins, and W. J. Stewart, 1999,
IEEE Trans. Microw. Theory Tech. {\bf 47}, 2075,
\item[]
Petrov, V., M.~Petrov, V.~Bryksin, J.~Petter,
and T.~Tschudi, 2006,
Opt. Lett. {\bf 31}, 3167.
\item[]
Philbin, T. G. and U. Leonhardt, 2008,
Phys. Rev. A {\bf 78}, 042107.
\item[]
Pirozhenko, I. and A.~Lambrecht, 2008a,
Phys. Rev. A {\bf 77}, 013811.
\item[]
Pirozhenko, I. and A.~Lambrecht, 2008b,
J. Phys. A: Math. Theor. {\bf 41}, 164015.
\item[]
Pirozhenko, I., A.~Lambrecht, and V.~B.~Svetovoy, 2006,
New J. Phys. {\bf 8}, 238.
\item[]
Pitaevskii, L. P., 2008,
Phys. Rev. Lett. {\bf 101}, 163202.
\item[]
Plunien, G., B.~M\"{u}ller, and W.~Greiner, 1986,
Phys. Rep. {\bf 134}, 87.
\item[]
Raabe, C. and D.-G.~Welsch, 2006,
Phys. Rev. A {\bf 73}, 063822.
\item[]
Rabinovich, S.~G., 2000,
{\it Measurement Errors and Uncertainties.
Theory and Practice}
(Springer, New York).
\item[]
Rabinovich, Ya.~I. and N.~V.~Churaev, 1989,
Kolloid Zh. {\bf 51}, 83
(Colloid J. USSR {\bf 51}, 65).
\item[]
Rodrigues, A., M. Ibanescu, D. Iannuzzi, J. D. Iannopoulos,
and S. G. Johnson, 2007a,
Phys. Rev. A {\bf 76}, 032106.
\item[]
Rodrigues, R.~B., P.~A.~Maia Neto, A.~Lambrecht, and S.~Reynaud,
2006a, Phys. Rev.  Lett. {\bf 96}, 100402.
\item[]
Rodrigues, R.~B., P.~A.~Maia Neto, A.~Lambrecht, and S.~Reynaud,
2006b, Europhys. Lett. {\bf 76}, 822.
\item[]
Rodrigues, R.~B., P.~A.~Maia Neto, A.~Lambrecht, and S.~Reynaud,
2007b, Phys. Rev. A {\bf 75}, 062108.
\item[]
Rosa, F. S. S., D. A. R. Dalvit, and P. W. Milonni, 2008,
Phys. Rev. Lett. {\bf 100}, 183602.
\item[]%{17}
Roy, A., C.-Y.~Lin, and U.~Mohideen, 1999,
{ Phys. Rev. D}
{\bf 60}, 111101(R).
\item[]%{16}
Roy, A. and U.~Mohideen, 1999,
{ Phys. Rev. Lett.}
{\bf 82}, 4380.
\item[]
Rumer, Yu.~B. and M.~S.~Ryvkin, 1980, {\it Thermodynamics,
Statistical Physics, and Kinetics} (Mir, Moscow).
\item[]
Sabbah, A. J. and D.~M.~Riffe, 2002,
Phys. Rev. B {\bf 66}, 165217.
\item[]
Saharian, A. A., 2006,
Phys. Rev. D {\bf 73}, 064019.
\item[]
Saito, R., G.~Dresselhaus, and M.~S.~Dresselhaus, 1998,
{\it Physical Properties of Carbon Nanotubes}
(Imperial College Press, London).
\item[]%{Jaffe}
Scardicchio, A. and R.~L.~Jaffe, 2006,
Nucl. Phys. B {\bf 743}, 249.
\item[]
Schram, K., 1973, Phys. Lett. A {\bf 43}, 283.
\item[]
Schwinger, J., L.~L.~DeRaad, and K.~A.~Milton, 1978,
Ann. Phys. (N.Y.) {\bf 115}, 1.
\item[]%{z34}
Sernelius, B.~E., 2005,
Phys. Rev. B {\bf 71}, 235114.
\item[]
Shackelford, J. F. and W.~Alexander (eds.), 2001,
{\it Material Science and Engineering Handbook}
(CRC Press, Boca Raton).
\item[]
Shalaev, V. M., 2007,
Nature Photonics {\bf 1}, 41.
\item[]
Shih, A. and V.~A.~Parsegian, 1975,
Phys. Rev. A {\bf 12}, 835.
\item[]
Shimizu, F., 2001,
Phys. Rev. Lett. {\bf 86}, 987.
\item[]
Shklovskii, B. I. and A.~L.~Efros, 1984,
{\it Electronic Properties of Doped Semiconductors.
Solid State Series,} v.45 (Springer, Berlin),
\item[]% {42}
Smythe, W.~R., 1950,
{\it Electrostatics and Electrodynamics}
(McGraw-Hill, New York).
\item[]
Soltani, M., M.~Chaker, E.~Haddad, R.~V.~Kruzelecky, and
D.~Nikanpour, 2004,
J. Vac. Sci. Technol. A {\bf 22}, 859.
\item[]
Sotelo, J., J.~Ederth, and G.~Niklasson, 2003,
Phys. Rev. B {\bf 67}, 195106.
\item[]
Spagnolo, S., D.~A.~R.~Dalvit, and P.~W.~Milonni, 2007,
Phys. Rev. A {\bf 75}, 052117.
\item[]
Sparnaay, M. J., 1958,
Physica {\bf 24}, 751.
\item[]
Sparnaay, M. J., 1989,
In: Physics in the Making, eds. A.~Sarlemijn, M.~J.~Sparnaay
(North-Holland, Amsterdam).
\item[]
Speake, C. C. and C. Trenkel, 2003,
Phys. Rev. Lett. {\bf 90}, 160403.
\item[]
Suh, J. Y., R.~Lopez, L.~C.~Feldman, and R.~F.~Haglund, 2004,
J. Appl. Phys. {\bf 96}, 1209.
\item[]
Sukenik, C. I., M.~G.~Boshier, D.~Cho, V.~Sandoghbar,
and E.~A.~Hinds, 1993,
Phys. Rev. Lett. {\bf 70}, 560.
\item[]
Svetovoy, V.~B., 2008,
Phys. Rev. Lett. {\bf 101}, 163603.
\item[]
Svetovoy, V.~B. and R.~Esquivel, 2005,
Phys. Rev. E {\bf 72}, 036113.
\item[]
Svetovoy, V.~B. and M.~V.~Lokhanin, 2003,
Phys. Rev. A {\bf 67}, 022113.
\item[]
Svetovoy, V.~B., P. J. van Zwol, G. Palasantzas, and
J. Th. M. De Hosson, 2008,
Phys. Rev. B {\bf 77}, 035439.
\item[]%{Tom1}
Toma\v{s}, M.~S., 2002,
Phys. Rev. A {\bf 66}, 052103.
\item[]%{Tom2}
Toma\v{s}, M.~S., 2005,
Phys. Lett. A {\bf 342}, 381.
\item[]%{Tor-Lam}
Torgerson, J.~R. and S.~K.~Lamoreaux, 2004,
Phys. Rev. E {\bf70}, 047102.

\item[]
van Blockland, P. H. G. M. and J.~T.~G.~Overbeek,
1978,
{J. Chem. Soc. Faraday Trans.\/} {\bf 74}, 2637
\item[]
Van Kampen, N. G., B.~R.~A.~Nijboer, and K.~Schram, 1968,
Phys. Lett. A {\bf 26}, 307.
\item[]
van Zwol, P. J., G.~Palasantzas, and J. Th. M. De Hosson, 2007,
Appl. Phys. Lett. {\bf 91}, 144108.
\item[]
van Zwol, P. J., G.~Palasantzas, M. van de Schootbrugge,
and J. Th. M. De Hosson, 2008,
Appl. Phys. Lett. {\bf 92}, 054101.
\item[]
Verleur, H. W., A.~S.~Barker, and C.~N.~Berglund, 1968,
Phys. Rev. {\bf 172}, 788.
\item[]
Visser, J., 1981,
Adv. Coll. Interface Sci. {\bf 15}, 157.
\item[]%{38b}
Vogel, T., G.~Dodel, E.~Holzhauer, H.~Salzmann, and A.~Theurer,
1992, Appl. Opt. {\bf 31}, 329.
\item[]%{VP}
Volokitin, A.~I. and B.~N.~J.~Persson, 2007,
Rev. Mod. Phys.  {\bf 79}, 1291.
\item[]
Voronin, A. Yu. and P.~Froelich, 2005,
J. Phys. B: At. Mol. Opt. Phys. {\bf 38}, L301.
\item[]
Voronin, A. Yu., P.~Froelich, and B.~Zygelman, 2005,
Phys. Rev. A {\bf 72}, 062903.
\item[]
Yu, I. A., J.~M.~Doyle, J.~C.~Sandberg, C.~L.~Cesar, D.~Kleppner,
and T.~J.~Greytak, 1993,
Phys. Rev. Lett. {\bf 71}, 1589.
\item[]%{zs95}
Zhou, F. and L.~Spruch, 1995,
Phys. Rev. A {\bf 52}, 297.
\item[]%{zs04}
Zurita-S\'{a}nchez,  J.~R., J.-J.~Greffet, and L.~Novotny,
2004, Phys. Rev. A {\bf 69}, 022902.
\item[]
Zylbersztejn, A. and N.~F.~Mott, 1975,
Phys. Rev. B {\bf 11}, 4383.

%%%%%%%%%%%%%%%%%%%%%%%%%%%%%%%%%%%%%%%%%%%%%

\endthebibliography
%%%%%%%%%%%%%%%%%%%%%%%
%\end{document}%%%_FIGURES
%%%%%%%%%%%%%%%%%%%%%%%
\newpage
\begin{figure}
\vspace*{-8cm}
\centerline{
\includegraphics{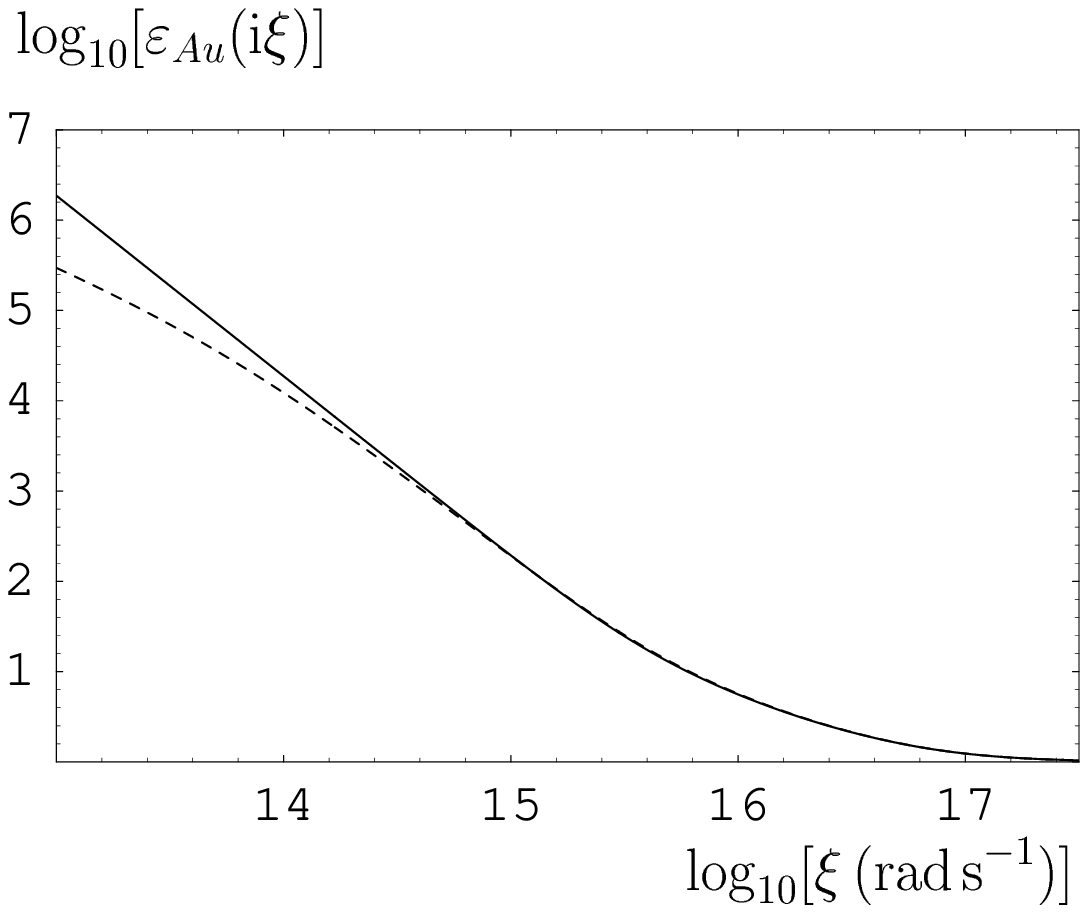}
}
\vspace*{-9cm}
\caption{The dielectric permittivity of Au along the imaginary
frequency axis.
The solid line is obtained using the generalized plasma-like model
taking into account the optical data related to the core electrons.
The dashed line is obtained using the optical data extrapolated to low
frequencies by the Drude model.}
\end{figure}
%%%
\begin{figure}
\vspace*{-3cm}
\centerline{
\includegraphics{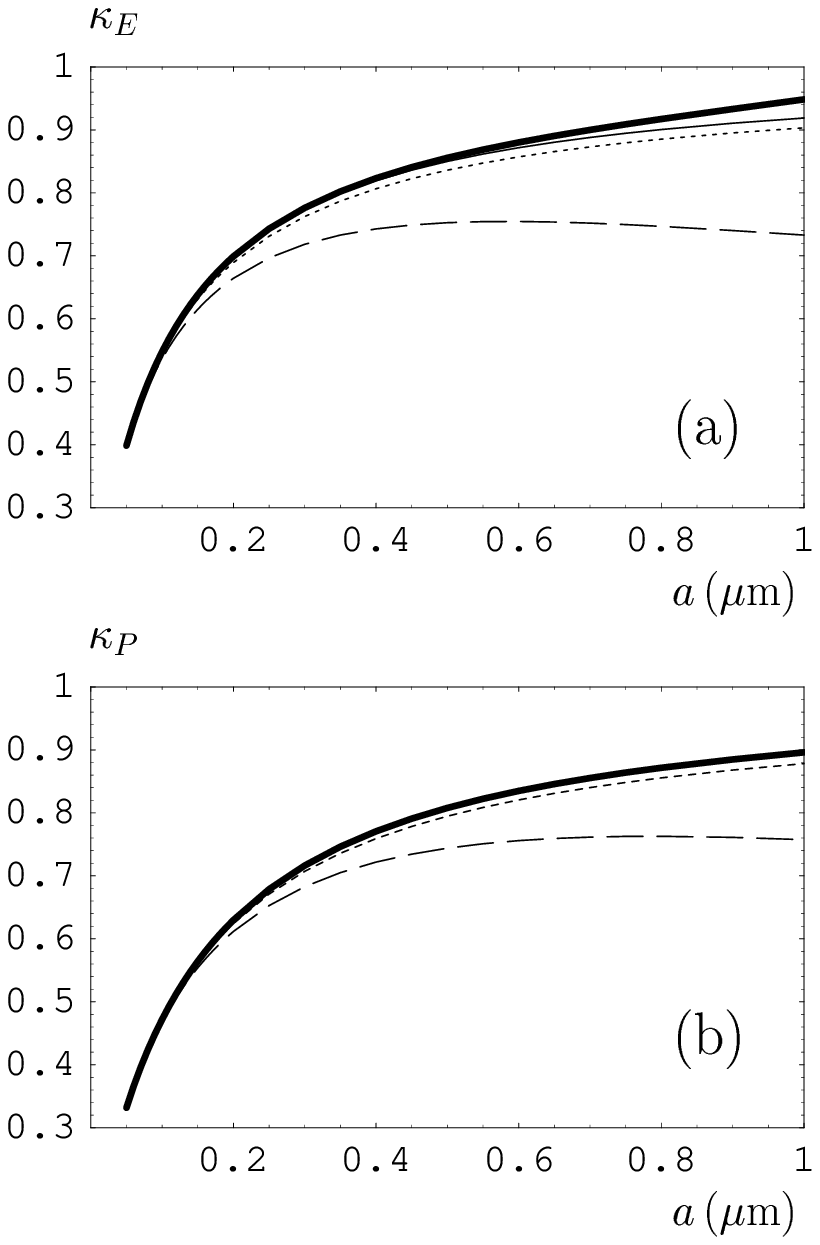}
}
\vspace*{-13cm}
\caption{Correction factors (a) to the Casimir energy per unit area
and (b) pressure versus separation between Au plates. Bold and fine
solid lines are computed using the generalized plasma-like model
at $T=300\,$K and $T=0$, respectively. Long- and short-dashed
lines are calculated using
the extrapolation of optical data by the Drude model
at $T=300\,$K and $T=0$, respectively.}
\end{figure}
%%%
\begin{figure}
\vspace*{-5cm}
\centerline{
\includegraphics{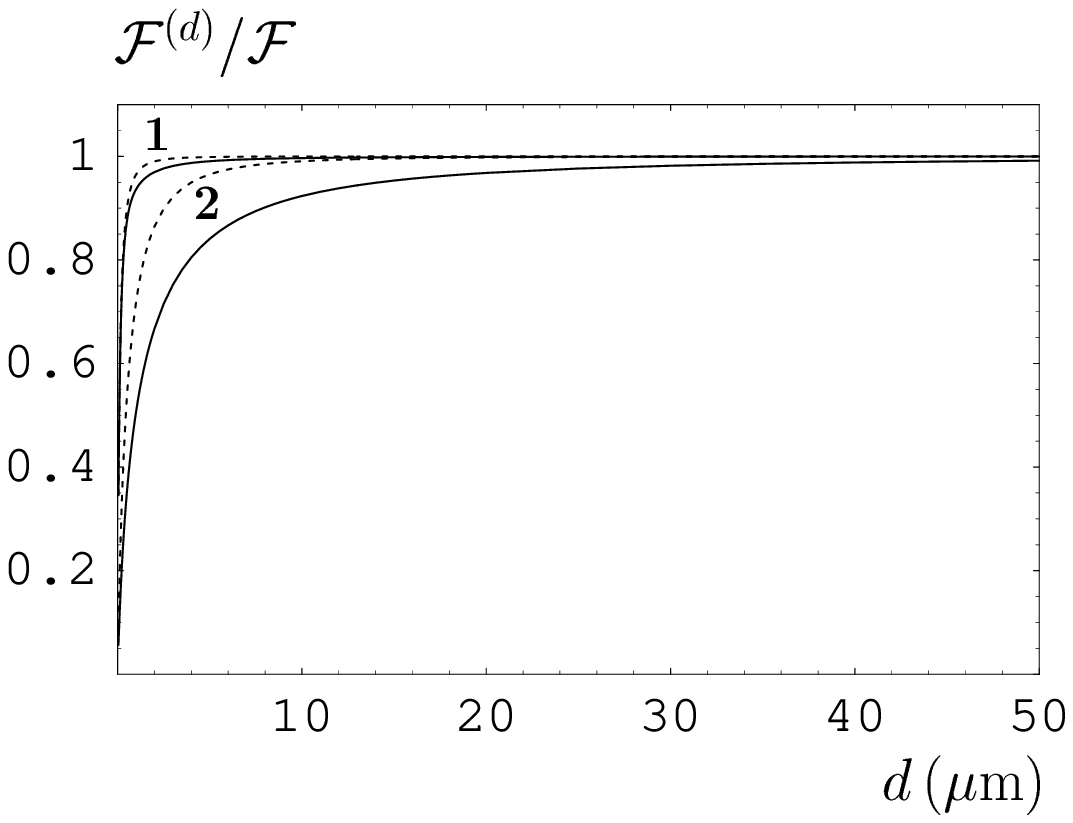}
}
\vspace*{-10cm}
\caption{The ratio of the
Casimir free energy for a Si plate of thickness $d$ at
a separation $a=1\,\mu$m (lines labeled 1) and
$a=5\,\mu$m (lines labeled 2) from Au semispace to the
free energy in the configuration of Si and Au semispaces at the respective
separations. Solid lines are computed at $T=300\,$K. Dashed
lines are computed at $T=0$.}
\end{figure}
%%%
\begin{figure}
\vspace*{2cm}
\centerline{
\includegraphics{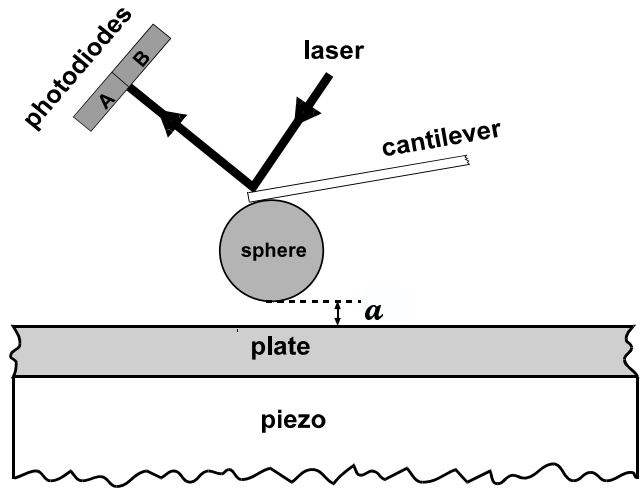}
}
\vspace*{-7cm}
\caption{Schematic diagram of the measurement of the Casimir force
using an atomic force microscope.}
\end{figure}
%%%
\begin{figure}
\vspace*{1cm}
\centerline{
\includegraphics{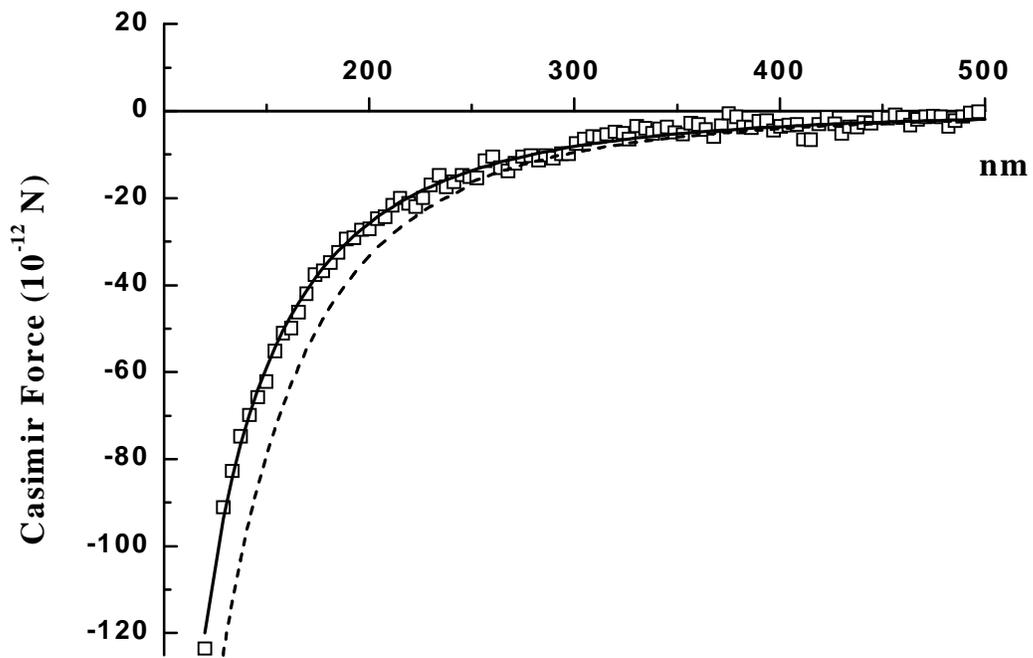}
}
\vspace*{-13cm}
\caption{Measured mean Casimir force
between Al surfaces versus separation is shown as open
squares. The theoretical Casimir force with corrections due to skin-depth
and surface roughness and for ideal metal surfaces is shown by the
solid and dashed lines, respectively.}
\end{figure}
%%%
\begin{figure}
\vspace*{-6cm}
\centerline{
\includegraphics{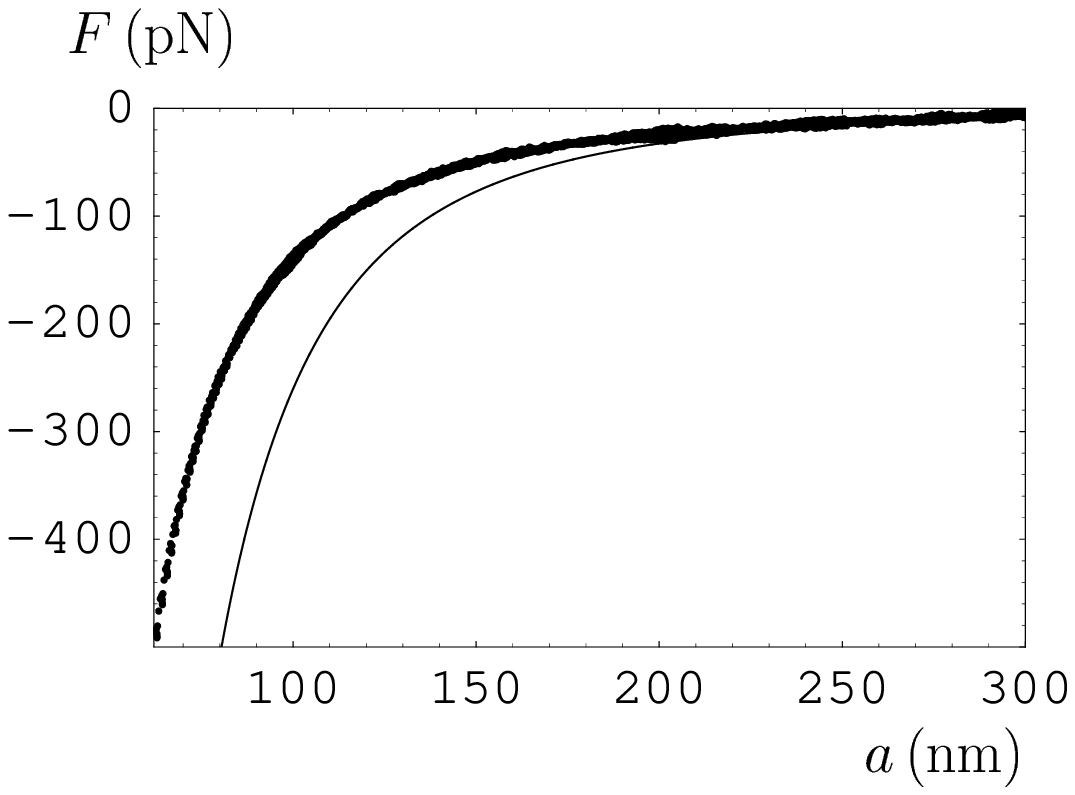}
}
\vspace*{-10cm}
\caption{The measured mean Casimir force between Au surfaces
versus separation is shown as dots. Theoretical Casimir force for
ideal metal surfaces is shown by the solid line.}
\end{figure}
%%%
\begin{figure}
\vspace*{-6cm}
\centerline{
\includegraphics{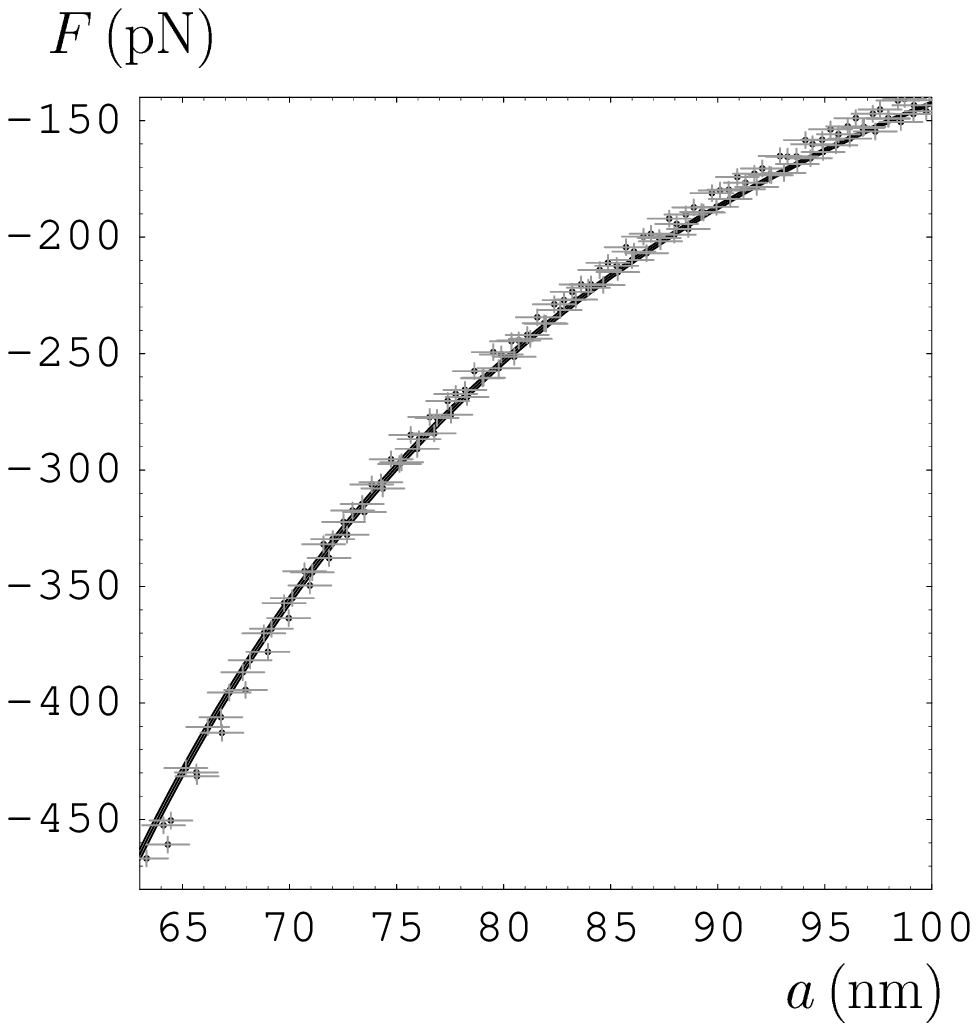}
}
\vspace*{-9cm}
\caption{The measured mean Casimir force between Au surfaces
versus separation is shown as crosses. The arms of the crosses represent
in true scale absolute errors determined at 95\% confidence. Theoretical
Casimir force calculated using the generalized plasma-like model is
shown by the solid band
.}
\end{figure}
%%%
\begin{figure}
\vspace*{-6cm}
\centerline{
\includegraphics{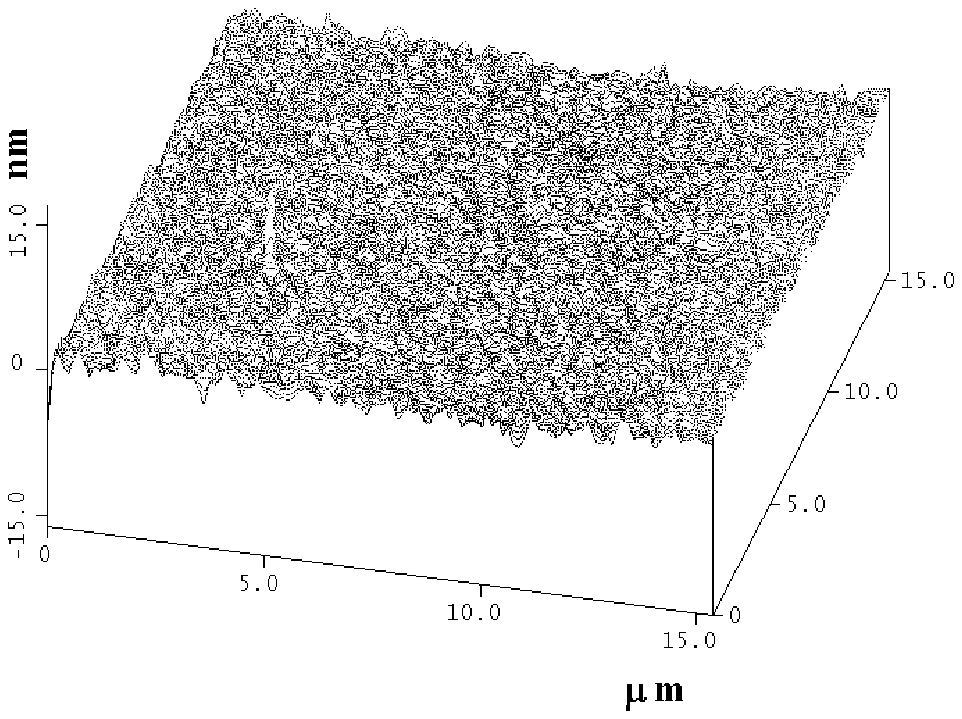}
}
\vspace*{-12cm}
\caption{Typical atomic force microscope image of the Au
coating on the plate with roughness height $h$ and lateral
scale $l$.}
\end{figure}
%%%
\begin{figure}
\vspace*{-2cm}
\centerline{
\includegraphics{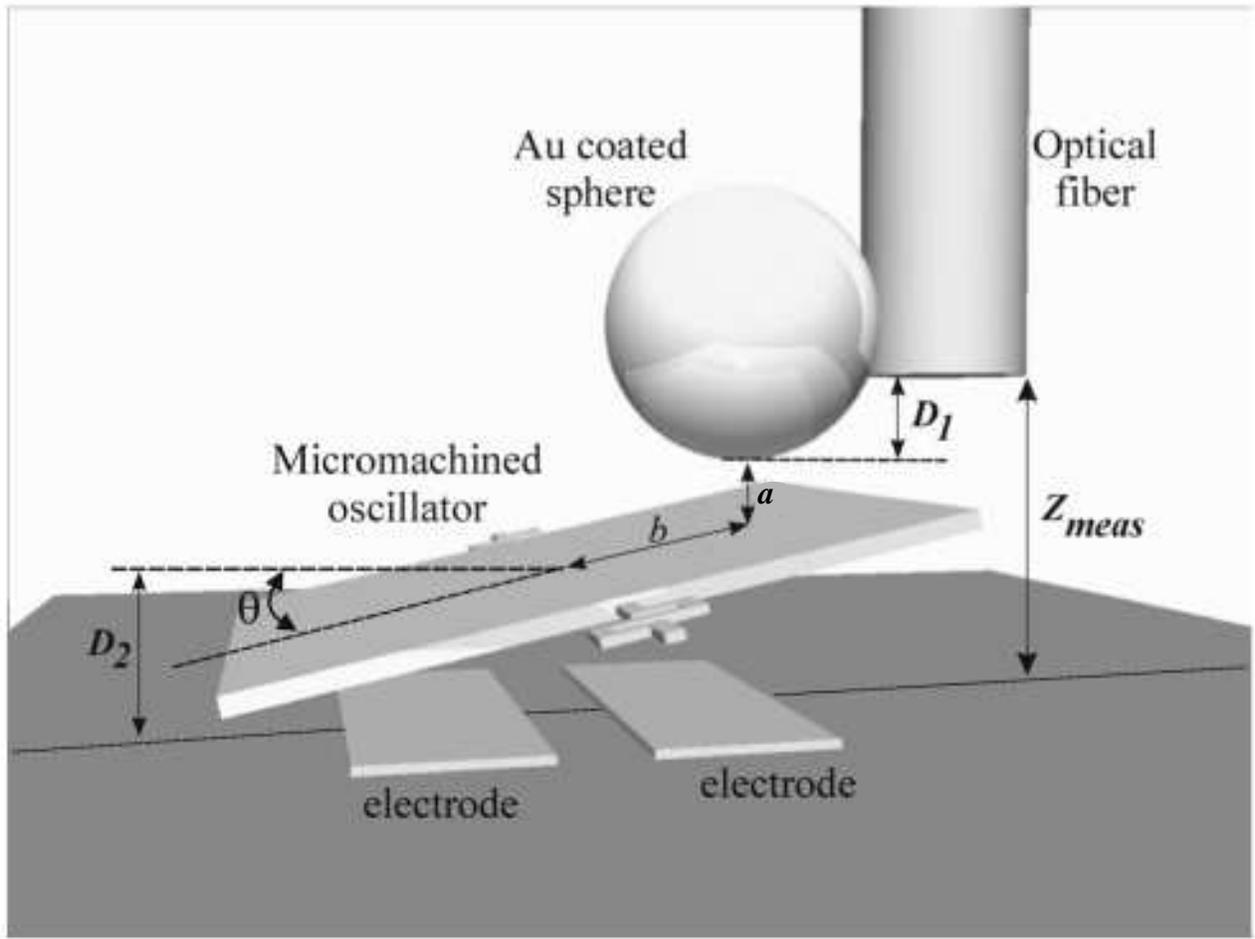}
}
\vspace*{-12cm}
\caption{Schematic diagram of the measurement of the Casimir
force using a micromachined oscillator.}
\end{figure}
%%%
\begin{figure}
\vspace*{-11cm}
\centerline{
\includegraphics{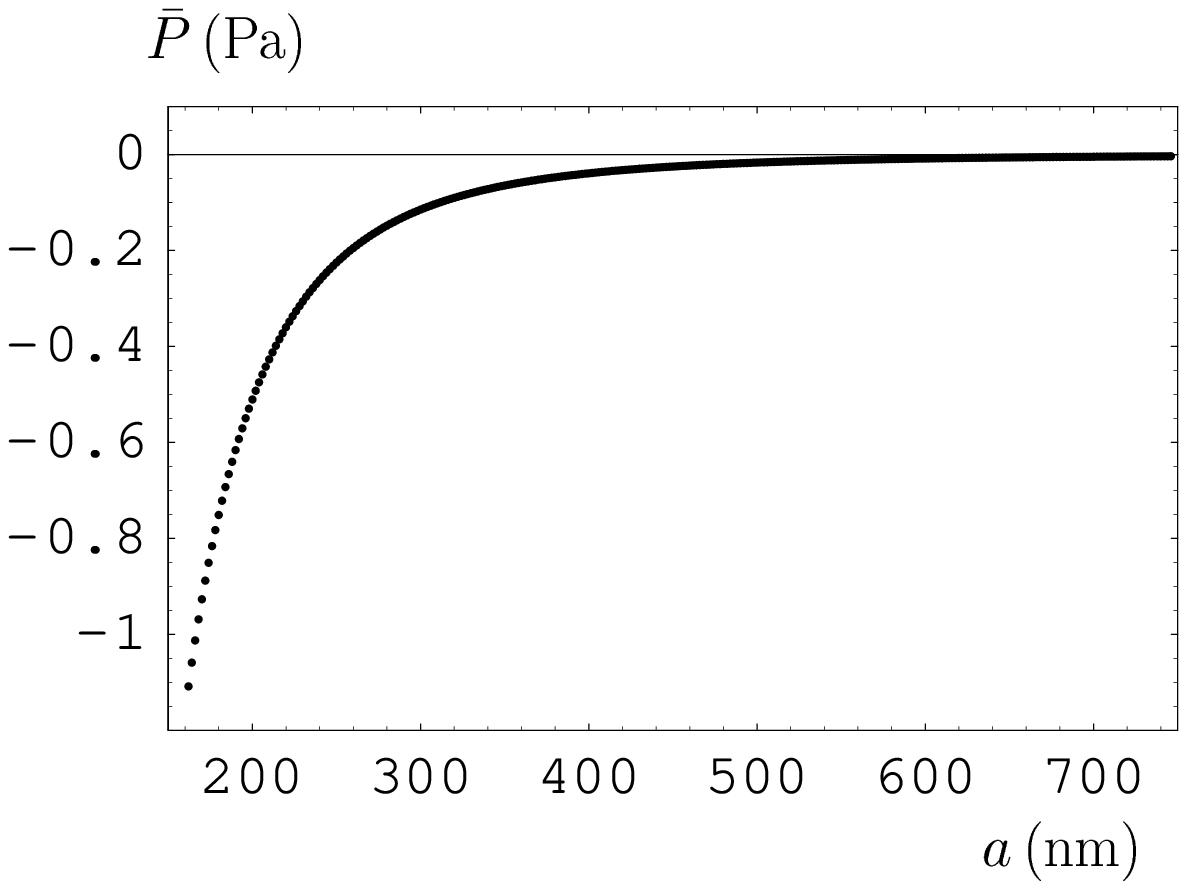}
}
\vspace*{-6cm}
\caption{The measured mean Casimir pressure between two parallel plates
versus separation
is shown as dots.}
\end{figure}
%%%
\begin{figure}
\vspace*{-4cm}
\centerline{
\includegraphics{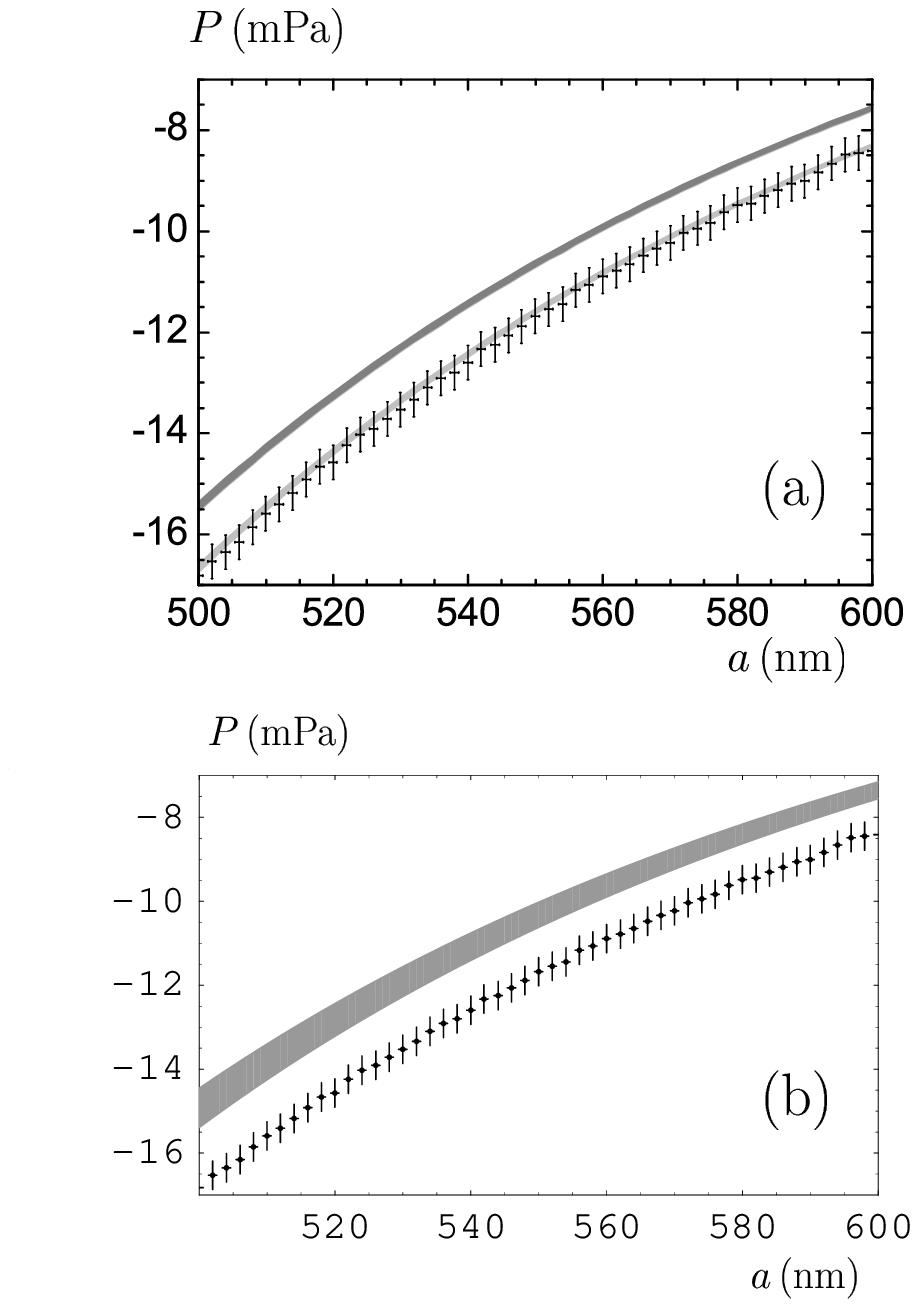}
}
\vspace*{-11cm}
\caption{The measured mean Casimir pressure together with the absolute errors
in the separation and pressure versus separation
is shown as crosses. (a) The theoretical Casimir pressure computed using
the generalized plasma-like model and the optical data extrapolated by
the Drude model is shown by the light-gray and dark-gray bands,
respectively. (b) The theoretical Casimir pressure
computed using different sets
of optical data available in the literature versus separation is shown
as the dark-gray band.}
\end{figure}
%%%
\begin{figure}
\vspace*{-1cm}
\centerline{
\includegraphics{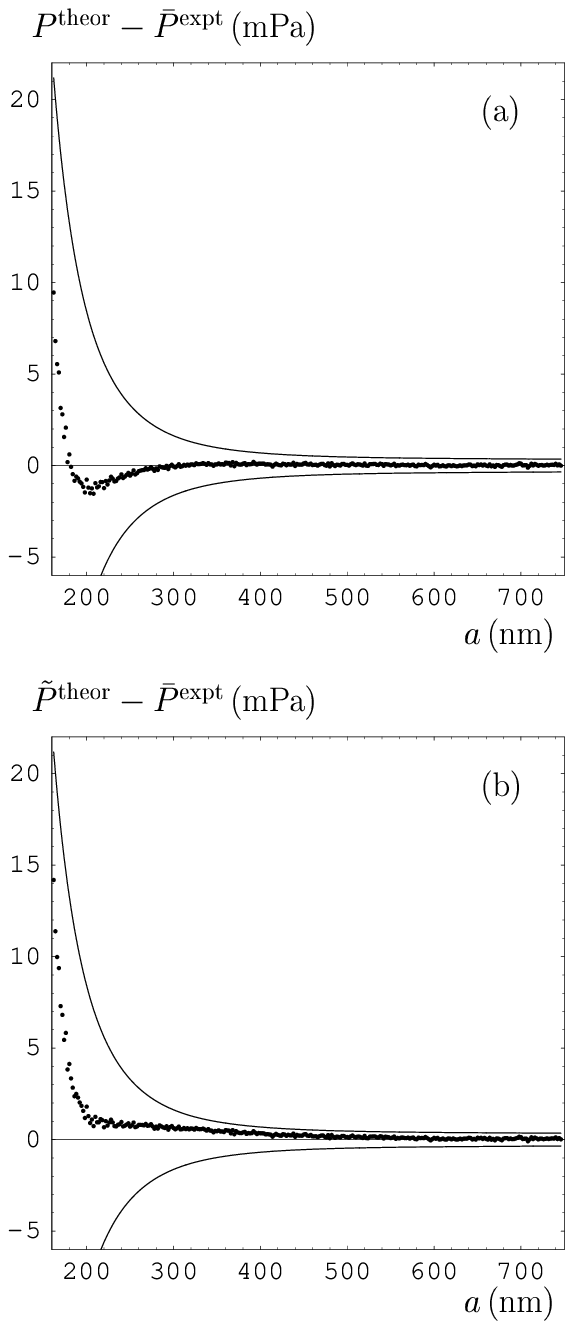}
}
\vspace*{-15cm}
\caption{The difference between theoretical and mean experimental Casimir
pressures versus separation is shown as dots, and the 95\% confidence
intervals are shown as the solid lines. The theoretical Casimir
pressures are computed (a) using the generalized plasma-like model and
(b) using the Leontovich surface impedance approach.}
\end{figure}
%%%
\begin{figure}
\vspace*{-11cm}
\centerline{
\includegraphics{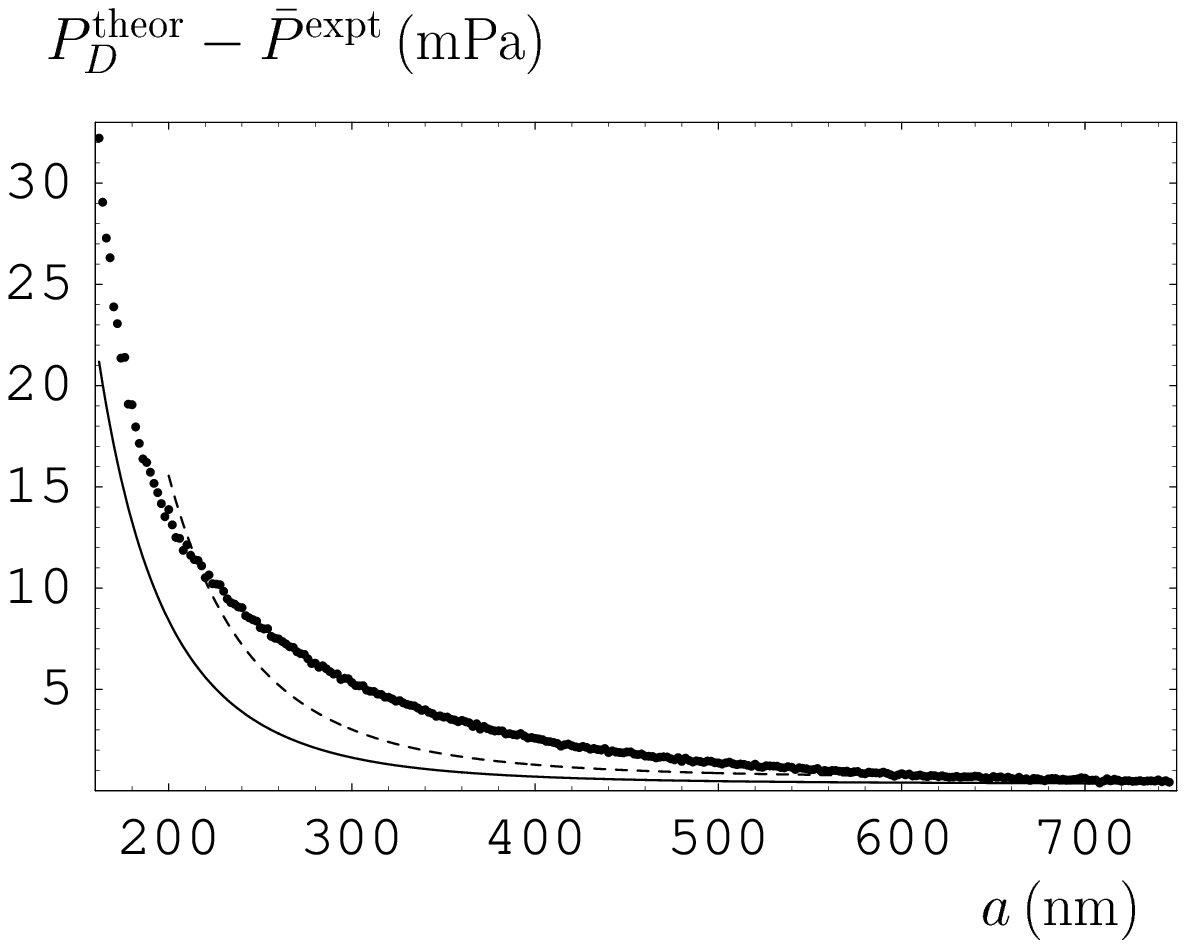}
}
\vspace*{-6cm}
\caption{The difference between theoretical Casimir pressure calculated
with the optical data extrapolated using the Drude model and mean
exprimental Casimir pressure versus separation. The solid and dashed lines
indicate the borders of the 95\% and 99.9\% confidence intervals,
respectively.}
\end{figure}
%%%
\begin{figure}
\vspace*{-13cm}
\centerline{
\includegraphics{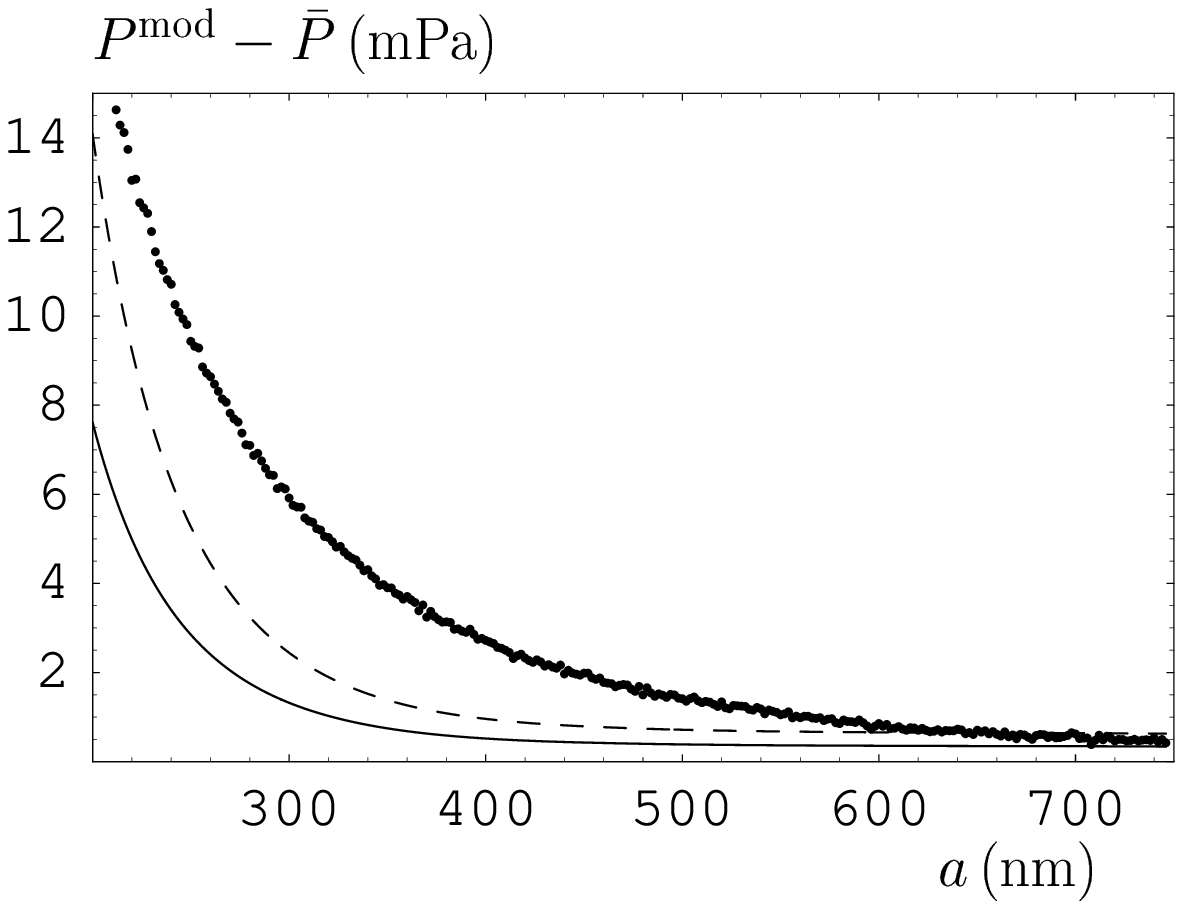}
}
\vspace*{-5cm}
\caption{\label{fgLam1}The difference between theoretical
Casimir pressures using the effect
of charge screening for the transverse magnetic mode and mean
experimental data versus separation is shown as dots.
The solid and dashed lines indicate the borders of the 95\%
and 99.9\% confidence intervals, respectively.}
\end{figure}
%%%
\begin{figure}
\vspace*{-8cm}
\centerline{
\includegraphics{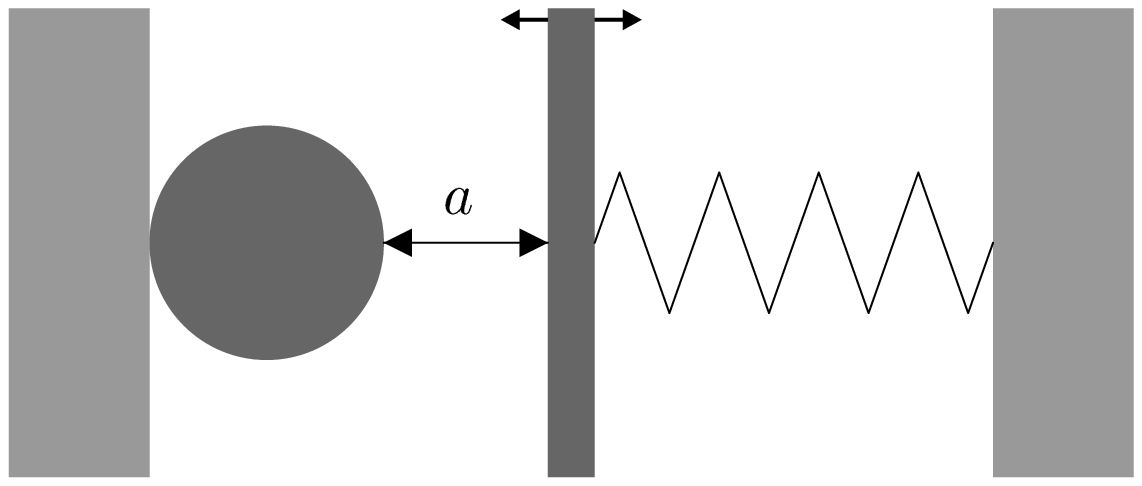}
}
\vspace*{-11cm}
\caption{Schematic diagram for the model of oscillator actuated
by the Casimir force.}
\end{figure}
%%%
\begin{figure}
\vspace*{-4cm}
\centerline{
\includegraphics{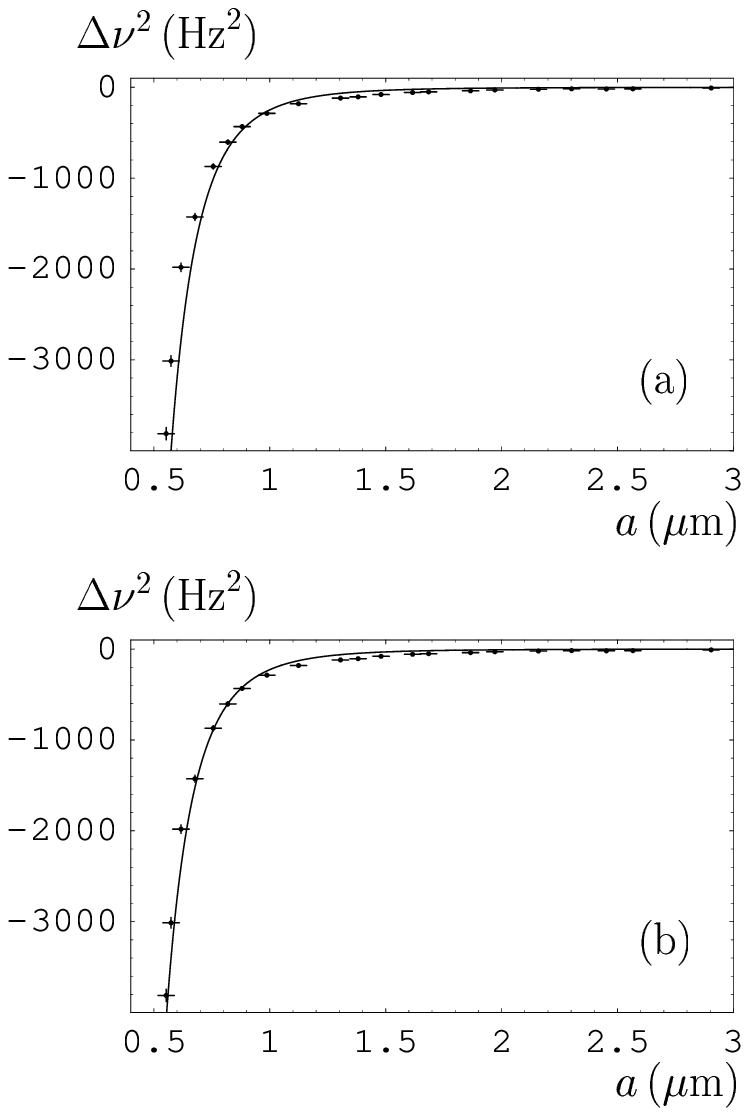}
}
\vspace*{-13cm}
\caption{The measured frequency shift versus separation together with the
absolute errors in separation and $\Delta\nu^2$ is shown as crosses.
Solid lines indicate the theoretical frequency shift calculated for
(a) ideal metal plates and (b) real metals with inclusion of the
skin-depth correction.}
\end{figure}
%%%
\begin{figure}
\vspace*{-1cm}
\centerline{
\includegraphics{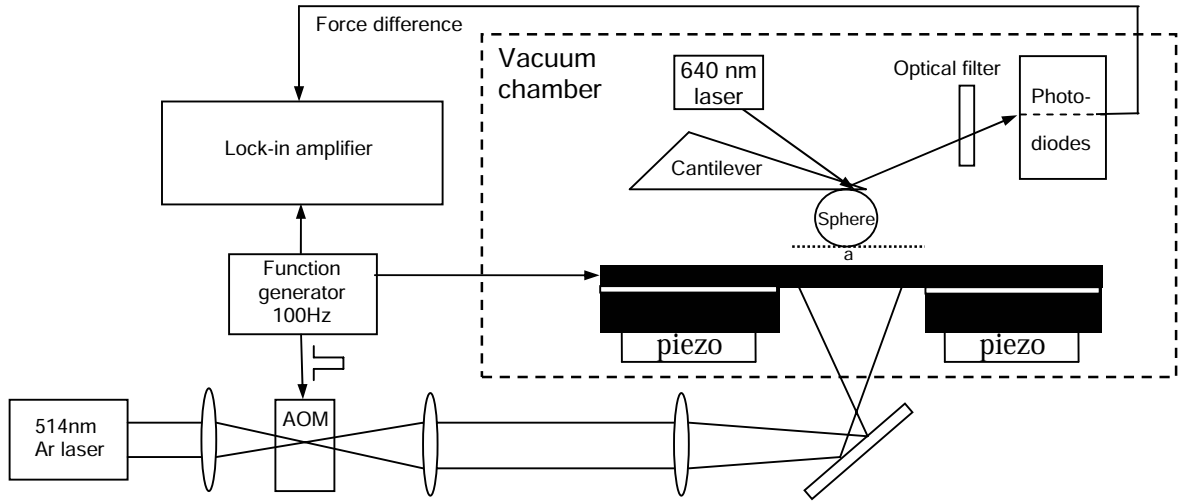}
}
\vspace*{-15cm}
\caption{\label{fgVp1} Schematic of the experimental setup on measuring
the difference Casimir force between Au sphere and Si plate illuminated
with laser pulses.}
\end{figure}
%%%
\begin{figure}
\vspace*{-5cm}
\centerline{
\includegraphics{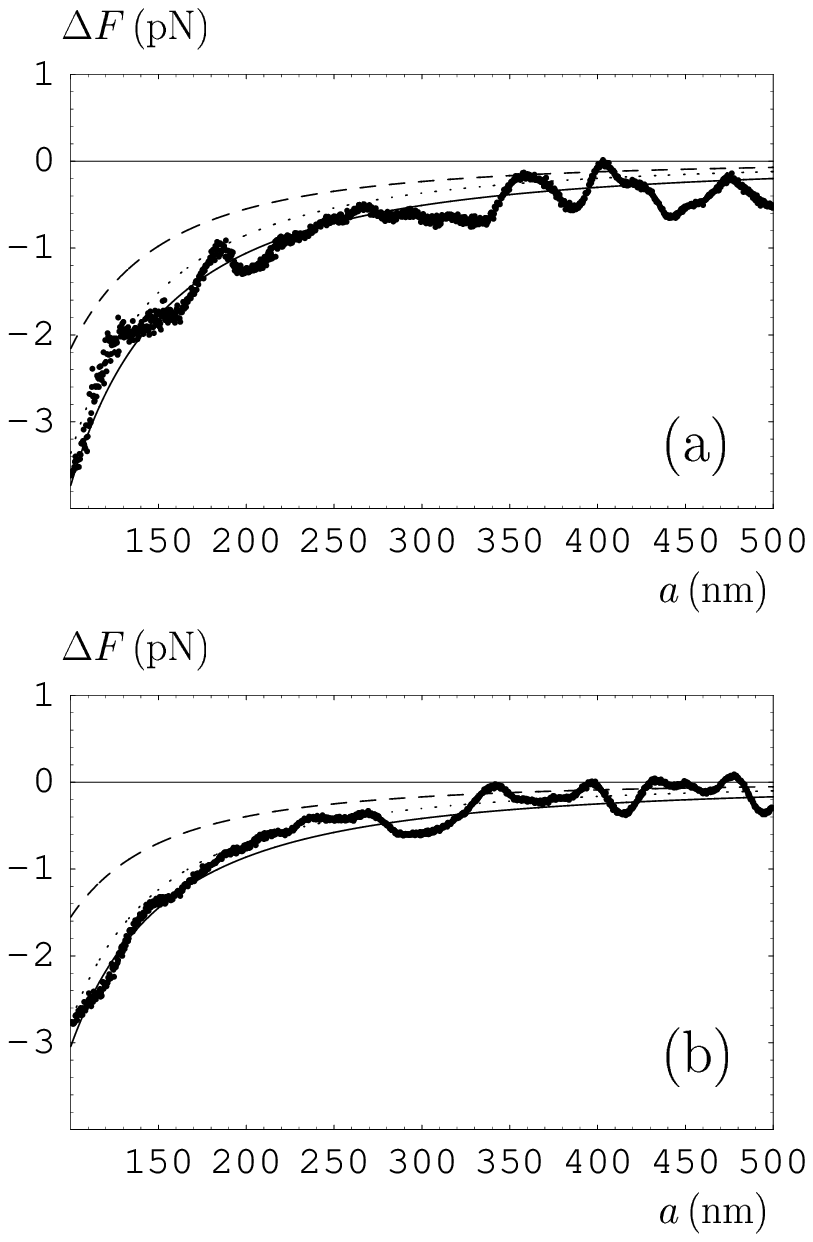}
}
\vspace*{-12cm}
\caption{\label{fgVp3} The difference of the Casimir forces in the
presence and in the absence of light versus separation for different
absorbed powers: (a) 9.3\,mW and (b) 4.7\,mW. The measured differences
$\langle F^{\rm expt}\rangle$ are shown as dots. See text for the
description of theoretical solid, dotted and dashed lines computed
using the different approaches.}
\end{figure}
%%%
\begin{figure}
\vspace*{-8cm}
\centerline{
\includegraphics{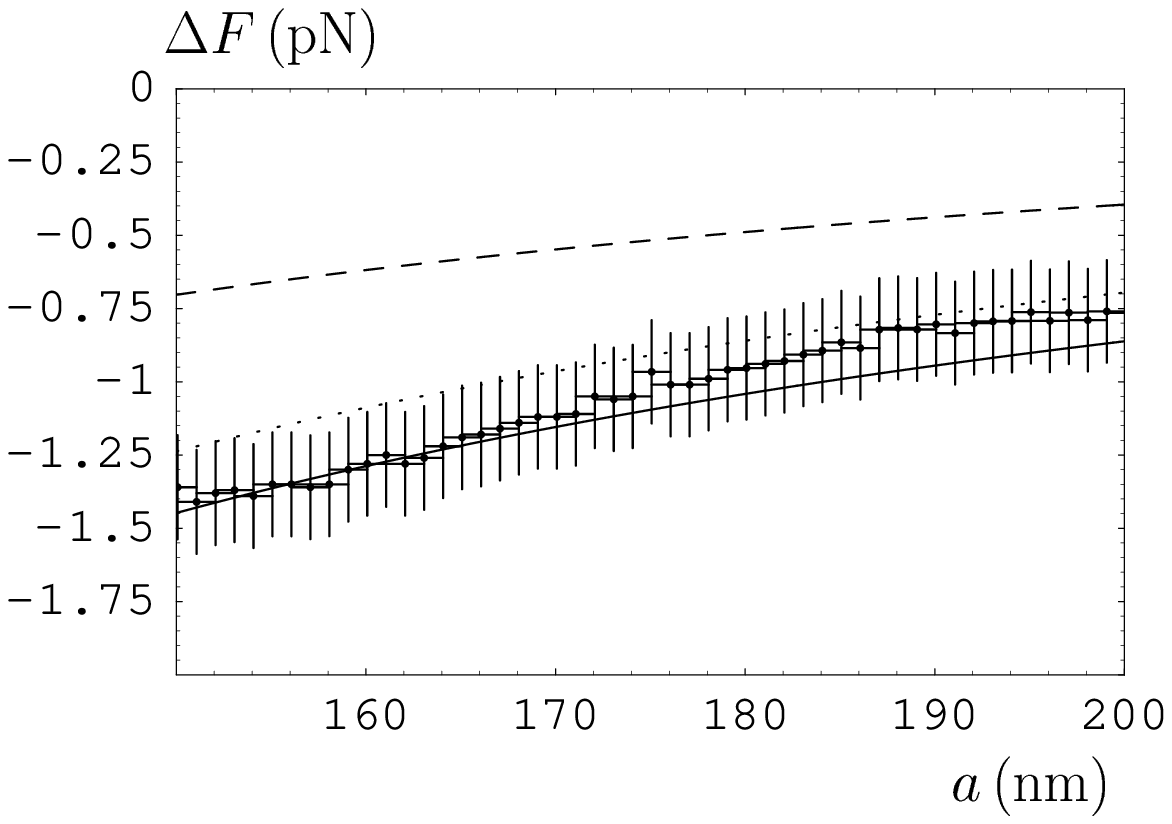}
}
\vspace*{-10cm}
\caption{\label{fgVp4} The experimental differences in the Casimir
force with their experimental errors are shown as crosses (the absorbed
power is equal to 4.7\,mW). The solid and dotted lines
 represent the theoretical differences
computed at $T=300\,$K using the model with a finite static
permittivity of high-resistivity Si, but different models for Si in
the presence of light (see text for further discussion).
The dashed line represents the theoretical differences
computed at the same temperature including the dc
conductivity.}
\end{figure}
\begin{figure}
\vspace*{-8cm}
\centerline{
\includegraphics{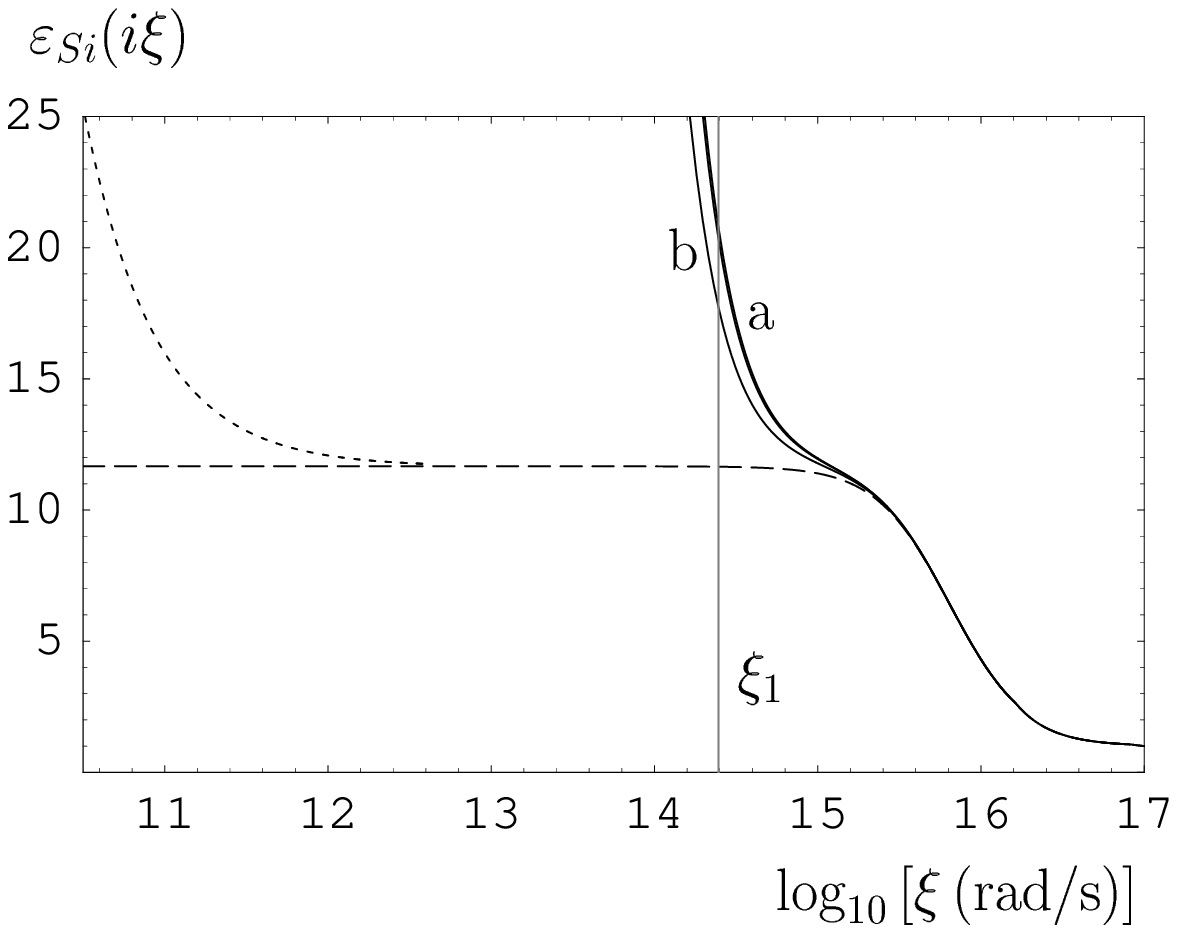}
}
\vspace*{-9cm}
\caption{\label{fgVp5} The dielectric permittivity of the Si membrane
along the imaginary frequency axis for Si with different concentration
of charge carriers (see text for further discussion).}
\end{figure}
%%%
\begin{figure}
\vspace*{-4cm}
\centerline{
\includegraphics{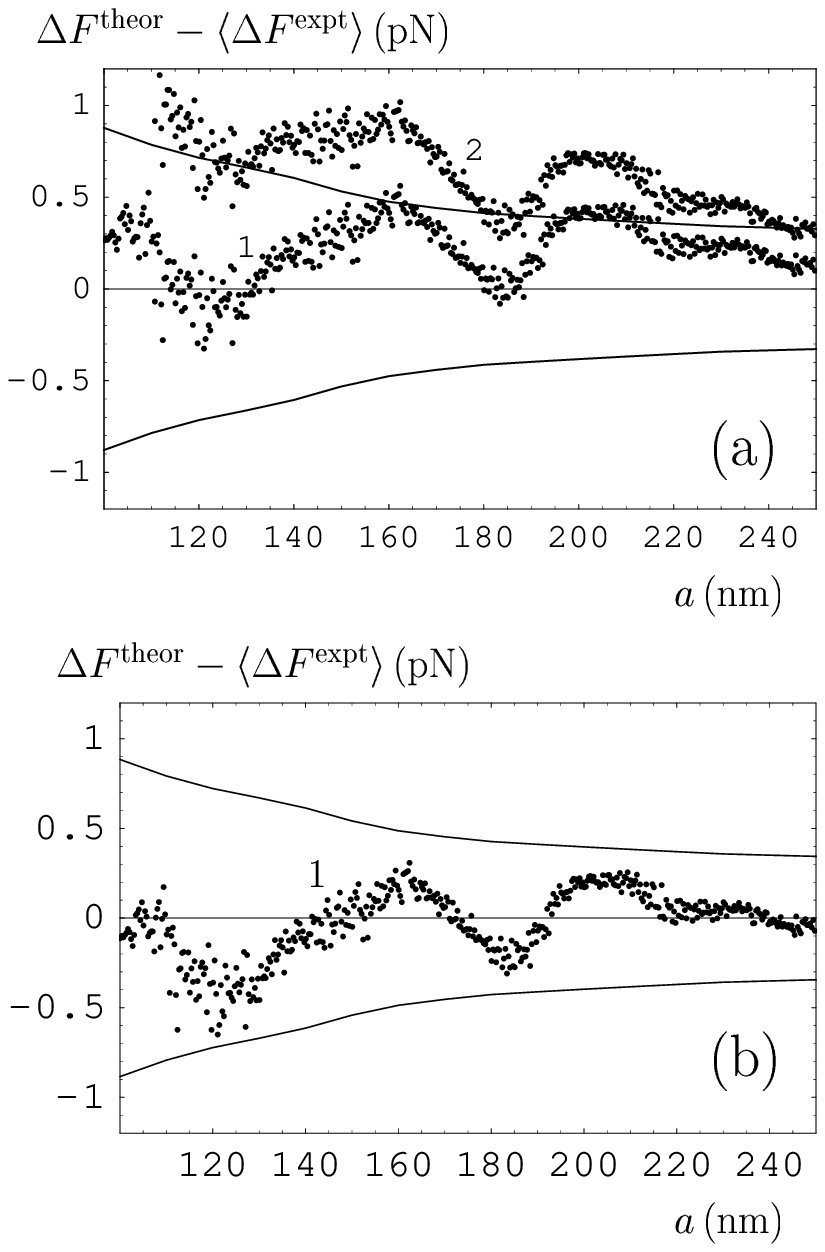}
}
\vspace*{-11cm}
\caption{\label{fgVp6}Theoretical minus experimental differences of the
Casimir force versus separation are shown as dots. In the absence of
the laser pulse the theoretical results for dots labeled 1 are computed with
a finite static dielectric permittivity of high resistivity Si and
for dots labeled 2 taking the dc conductivity of high resistivity
Si into account. When the laser pulse is on, the charge carriers are
described by (a) the Drude and (b) the plasma model. Solid lines show
the 95\% confidence interval. }
\end{figure}
%%%
\begin{figure}
\vspace*{-1cm}
\centerline{
\includegraphics{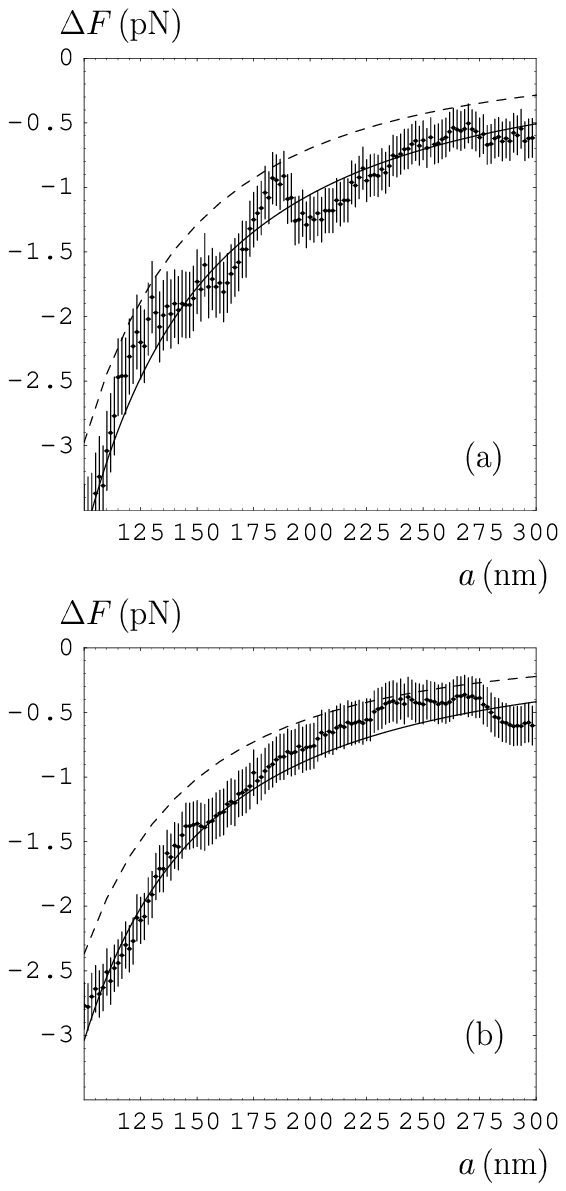}
}
\vspace*{-14cm}
\caption{\label{fgVpLP} Difference of the Casimir forces between an
Au-coated sphere and Si plate in the presence and in the absence of laser
light on the plate versus separation for the absorbed powers of
(a) 9.3\,mW and (b) 4.7\,mW. The experimental data are shown as
crosses. The solid and dashed lines are computed using the standard Lifshitz
theory with the dc conductivity of Si in the dark phase neglected and
taking into account the effect of charge screening for the transverse
magnetic mode, respectively.}
\end{figure}
%%%
\begin{figure}
\vspace*{-6cm}
\centerline{
\includegraphics{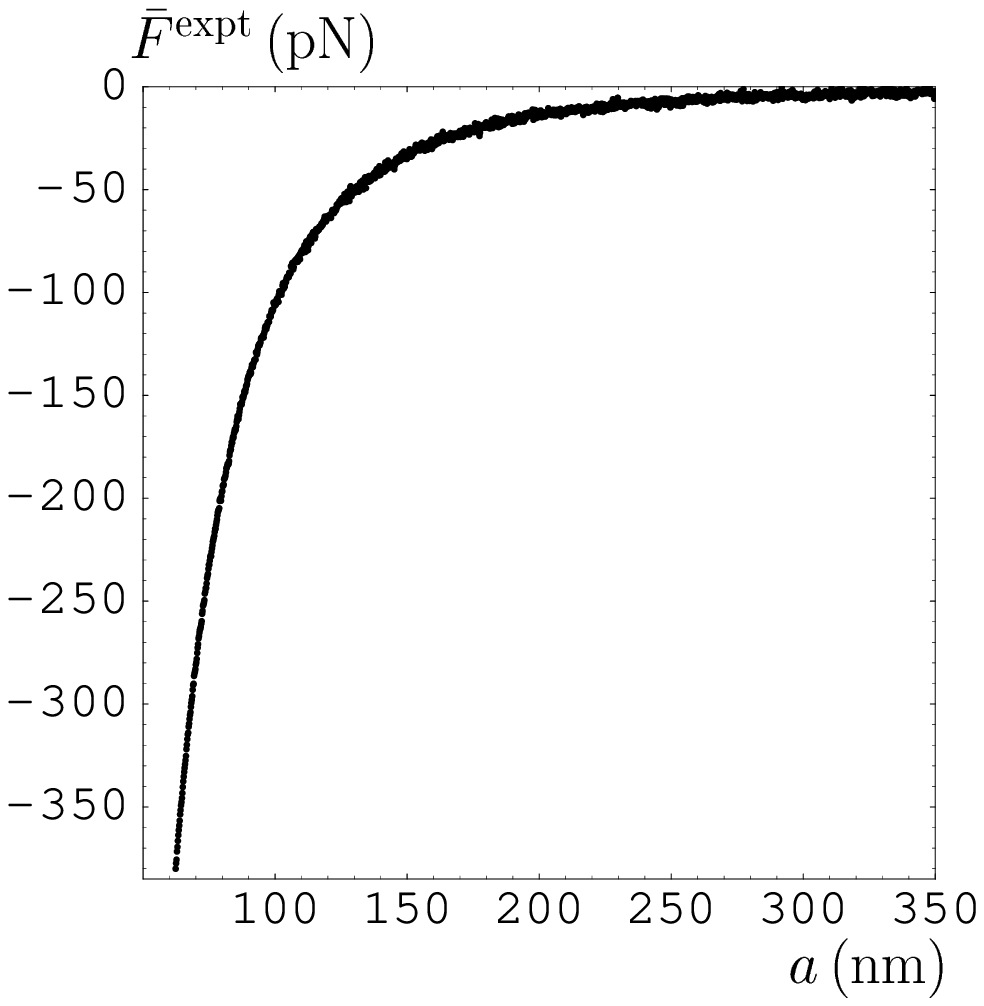}
}
\vspace*{-9cm}
\caption{\label{fgVp7} The mean measured Casimir force versus separation
between an Au sphere and B-doped Si plate.}
\end{figure}
%%%
\begin{figure}
\vspace*{-6cm}
\centerline{
\includegraphics{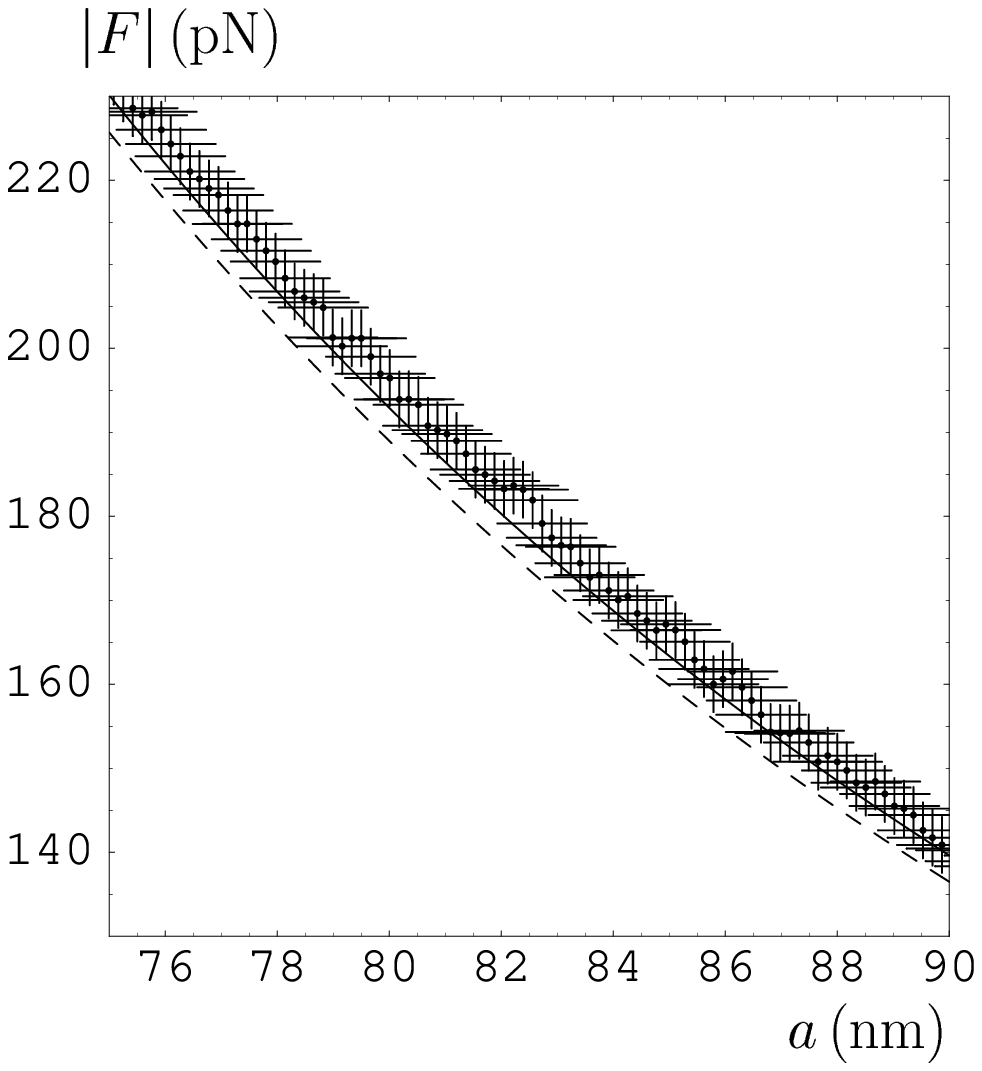}
}
\vspace*{-10cm}
\caption{\label{fgVp8} The magnitude of the experimental Casimir force with
the errors shown as crosses versus separation. Solid line shows the
theoretical dependence calculated for the sample used in the experiment
and dashed line for a dielectric Si.}
\end{figure}
%%%
\begin{figure}
\vspace*{-6cm}
\centerline{
\includegraphics{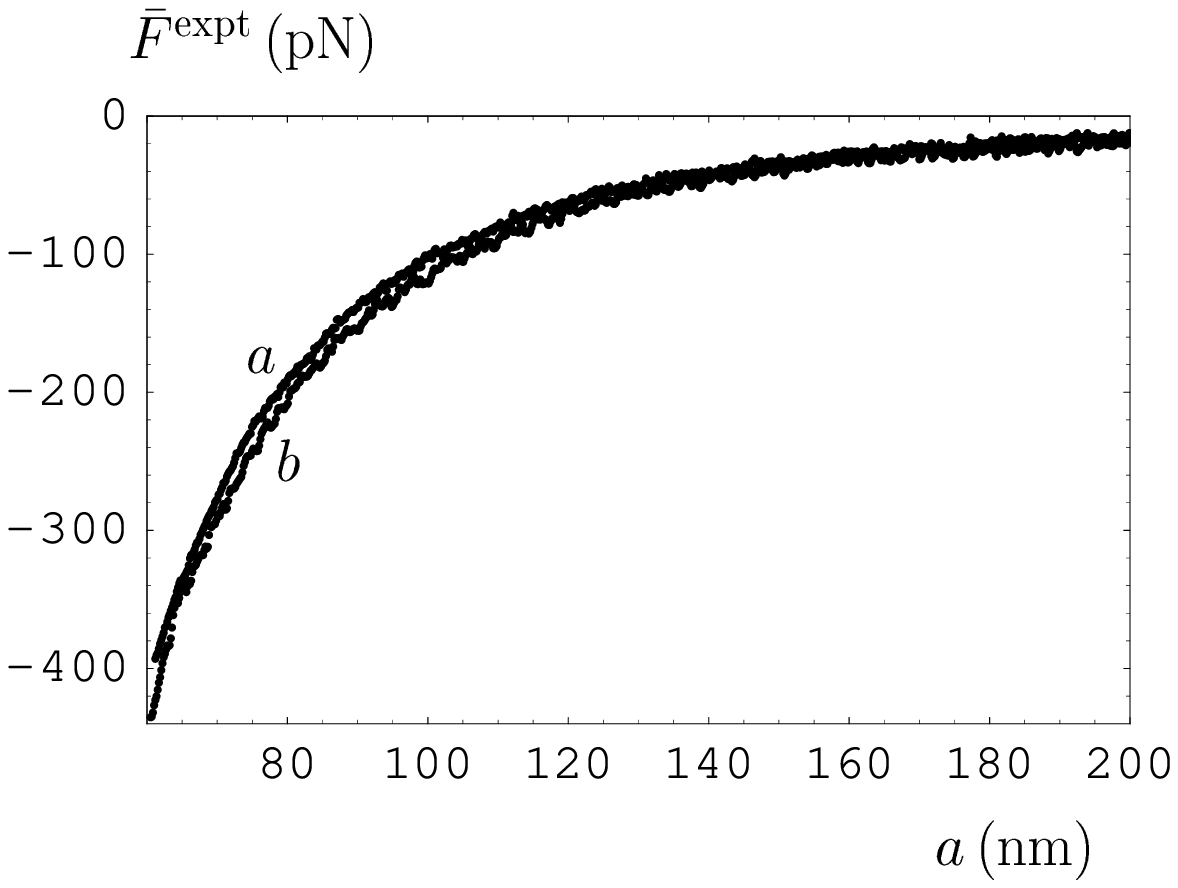}
}
\vspace*{-10cm}
\caption{\label{fgVp9} The mean measured Casimir force versus separation
between an Au sphere and two N-doped Si plates of (a) higher
and (b) lower resistivities.}
\end{figure}
%%%
\begin{figure}
\vspace*{-6cm}
\centerline{
\includegraphics{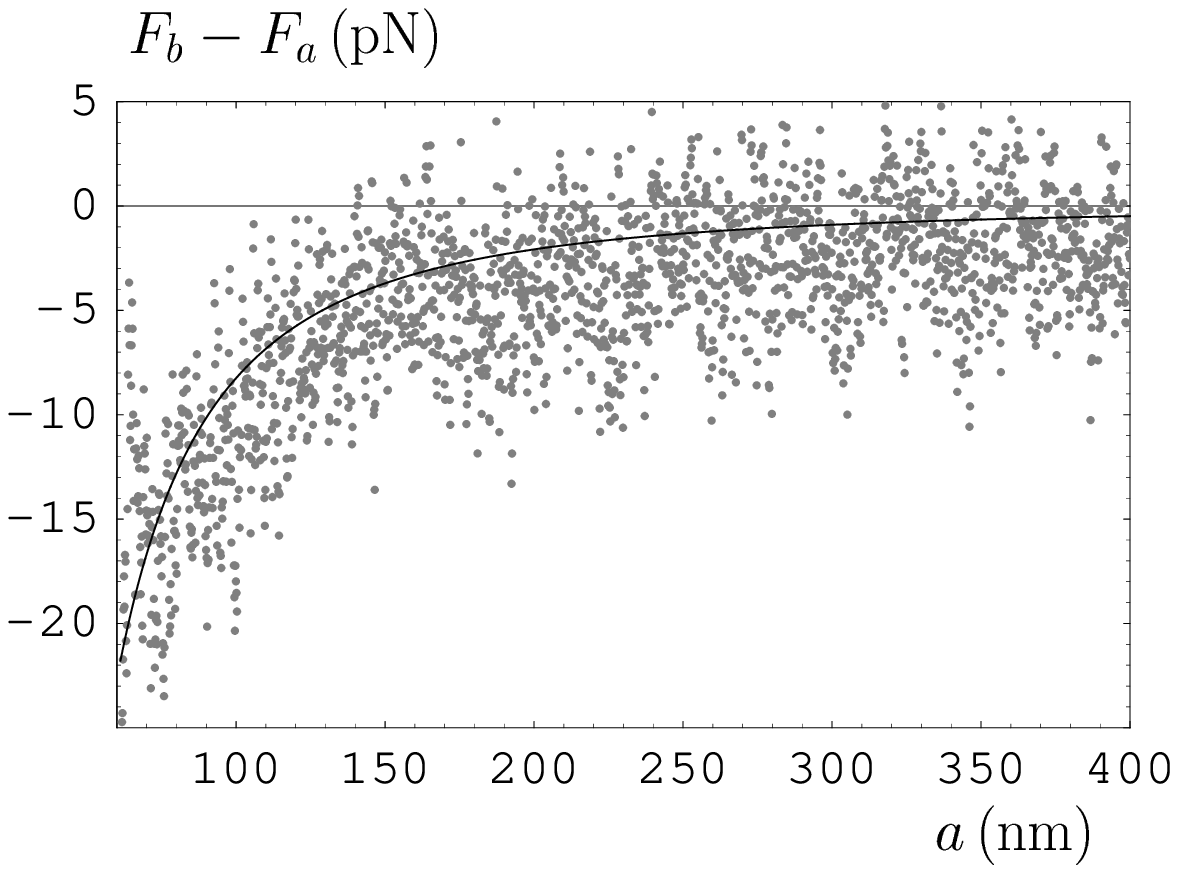}
}
\vspace*{-10cm}
\caption{\label{fgVp11} The differences of the mean measured Casimir
forces acting between an Au sphere and lower and higher resistivity Si
samples versus separation is shown as dots. The respective theoretical
difference is shown by the solid line. }
\end{figure}
%%%
\begin{figure}
\vspace*{-12cm}
\centerline{
\includegraphics{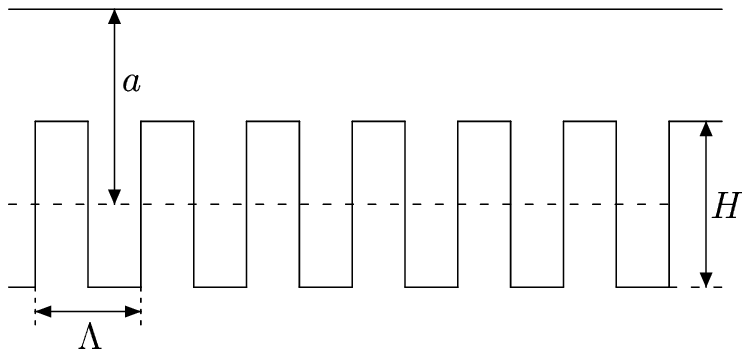}
}
\vspace*{-6cm}
\caption{\label{trenches} The periodical uniaxial rectangular
corrugations on one of the plates.}
\end{figure}
%%%
\begin{figure}
\vspace*{7cmcm}
\centerline{
\includegraphics{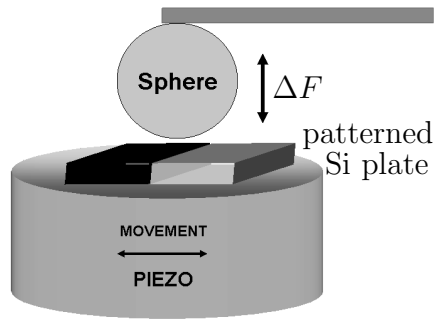}
}
\vspace*{-18cm}
\caption{\label{fgVp12} A schematic diagram of the
experimental setup for the measurement of the difference Casimir force
between an Au coated sphere and patterned Si plate.}
\end{figure}
%%%
\begin{figure}
\vspace*{-10cm}
\centerline{
\includegraphics{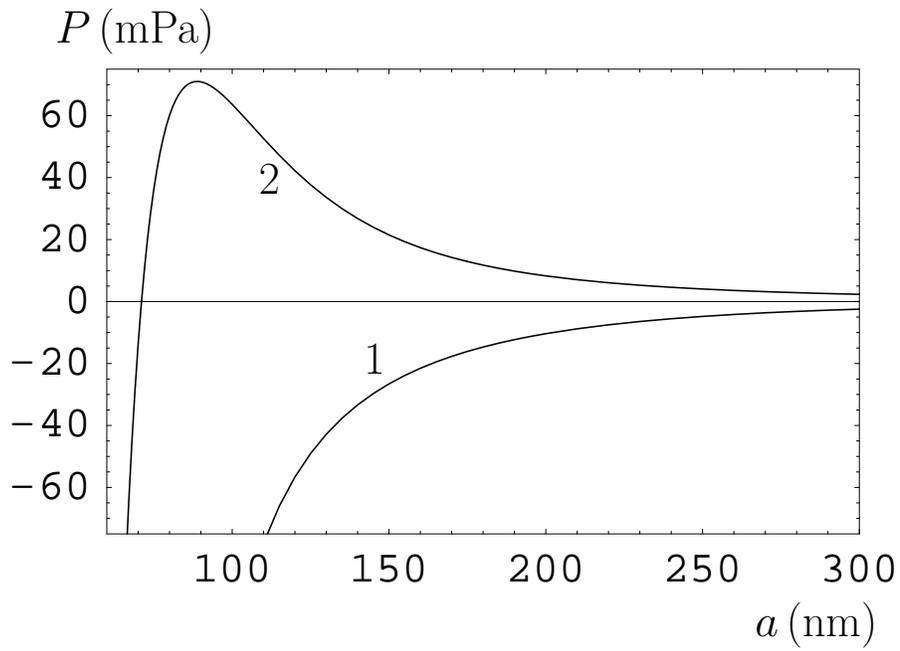}
}
\vspace*{-6cm}
\caption{\label{fgVp13} The Casimir pressure versus separation in a
three-layer system $\alpha$-Al${}_2$O${}_3$-ethanol-Si with no
light on the Si plate (line 1) and with the illuminated Si
plate (line 2).}
\end{figure}
%%%
\begin{figure}
\vspace*{-8cm}
\centerline{
\includegraphics{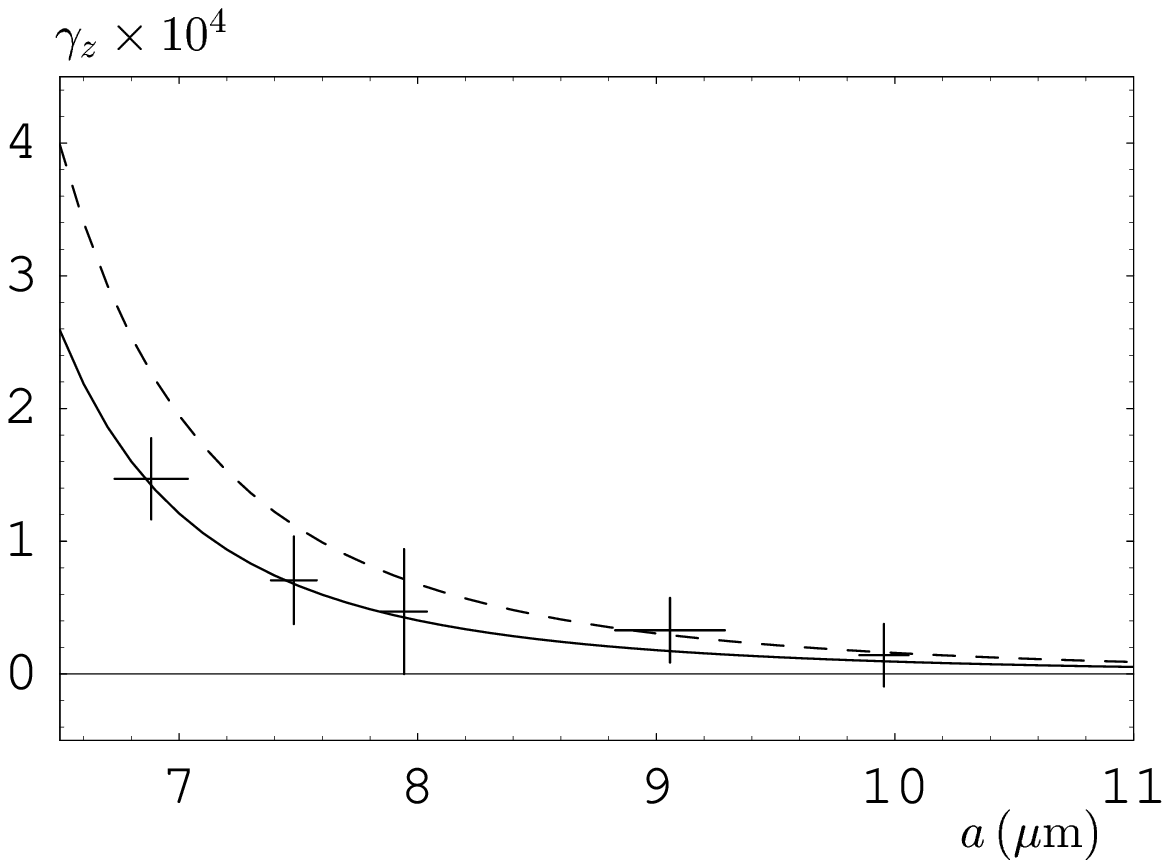}
}
\vspace*{-6cm}
\caption{\label{fg6p1} The fractional change in the trap frequency
versus separation in thermal
equilibrium with $T_S=T_E=310\,$K computed by neglecting (solid line)
and including  (dashed line) the conductivity of the dielectric
substrate. The experimental data are shown as crosses.}
\end{figure}
%%%
\begin{figure}
\vspace*{-6cm}
\centerline{
\includegraphics{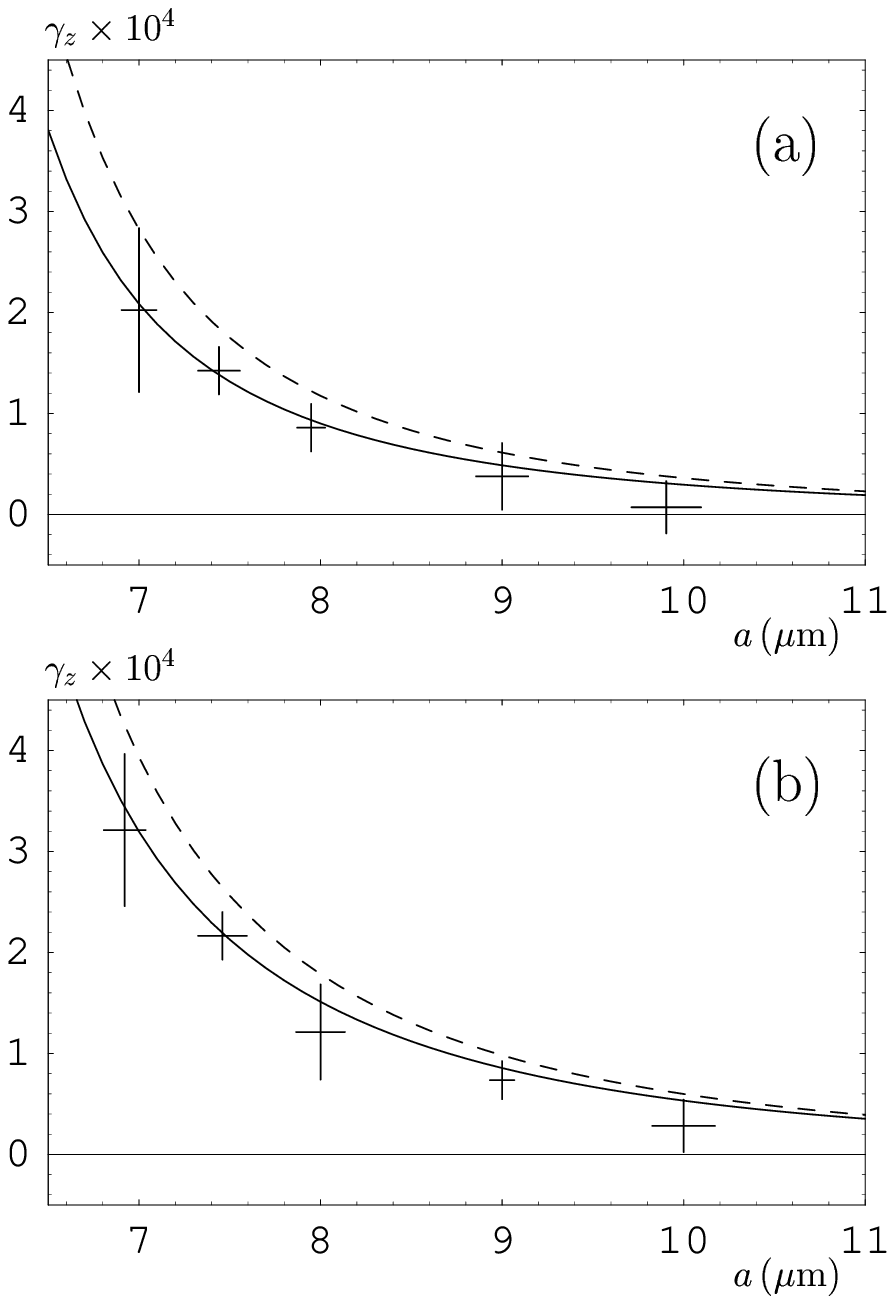}
}
\vspace*{-8cm}
\caption{\label{fg6p2} The fractional change in the trap frequency versus
separation
out of  thermal equilibrium
(a) with $T_S=479\,$K and $T_E=310\,$K
and (b) $T_S=605\,$K, $T_E=310\,$K.
Computations are done by neglecting (solid line)
and including (dashed line) the conductivity of the dielectric
substrate. The experimental data are shown as crosses.}
\end{figure}
%%%
\begin{figure}
\vspace*{-5cm}
\centerline{
\includegraphics{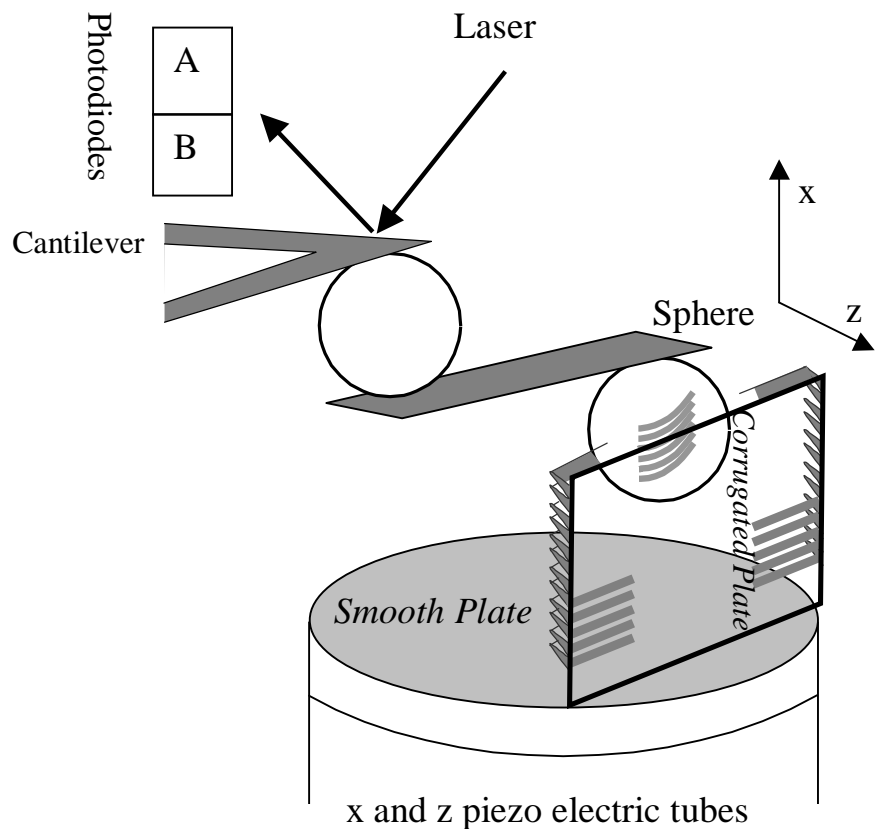}
}
\vspace*{-7cm}
\caption{\label{fg7p1} A schematic of the experimental setup for
measuring the lateral Casimir force. The $x$-piezo and $z$-piezo
are independent.}
\end{figure}
%%%
\begin{figure}
\vspace*{-5cm}
\centerline{
\includegraphics{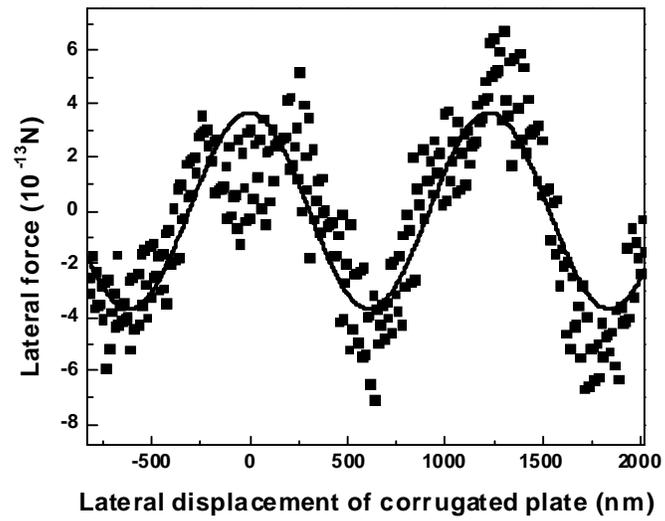}
}
\vspace*{-9cm}
\caption{\label{fg7p2} The average measured lateral Casimir force as
a function of the lateral displacement is shown as solid squares.
The solid line is the best fit sine curve to the data leading to a
lateral force amplitude of 0.32\,pN.}
\end{figure}
%%%
%%%
\end{document}